\documentclass[12pt]{article}
\usepackage[dvips,letterpaper,margin=1in,bottom=1in]{geometry}
\usepackage[toc,page]{appendix}
\usepackage{amsmath} 
\usepackage{amsthm} 
\usepackage{amssymb}	
\usepackage{amsfonts}
\usepackage{amsbsy}
\usepackage{authblk}
\usepackage{booktabs}
\usepackage[font=small]{caption}
\usepackage{color}
\usepackage{enumerate}
\usepackage{epstopdf}
\usepackage[T1]{fontenc}
\usepackage{graphicx}
\usepackage{hyperref}
\usepackage[utf8]{inputenc}
\usepackage{listings}
\usepackage{mathtools, cuted}
\usepackage{multicol}
\usepackage{paralist}
\usepackage{physics}
\usepackage{setspace}
\usepackage{subcaption}
\usepackage[disable]{todonotes}
\usepackage{wrapfig}
\newcommand{\exb}{$\vb*{E} \times \vb*{B}$ }

\newcommand{\ensav}[2]{\left<{#1}\right>_{#2}}
\newcommand{\vareps}{\varepsilon}

\newcommand{\figref}[1]{figure~\ref{fig:#1}}
\newcommand{\figsref}[2]{figures~\ref{fig:#1} and~\ref{fig:#2}}
\newcommand{\figsdash}[2]{figures~\ref{fig:#1}--\ref{fig:#2}}
\newcommand{\Figref}[1]{Figure~\ref{fig:#1}}
\newcommand{\Figsref}[2]{Figures~\ref{fig:#1} and~\ref{fig:#2}}

\graphicspath{{figures/plots/}, {figures/ext_plots/}}   

\title{Ion-scale turbulence in MAST: anomalous transport, subcritical
transitions, and comparison to BES measurements}
\author[1,2,3]{F. van Wyk\thanks{ferdinand.vanwyk@physics.ox.ac.uk}}
\author[1,4]{E. G. Highcock\thanks{highcock@chalmers.se}}
\author[2]{A. R. Field}
\author[2]{C. M. Roach}
\author[1,5]{A. A. Schekochihin\thanks{alex.schekochihin@physics.ox.ac.uk}}
\author[1]{F. I. Parra}
\author[1,6]{W. Dorland}
\affil[1]{Rudolf Peierls Centre for Theoretical Physics, University of Oxford, Oxford OX1 3NP, UK}
\affil[2]{CCFE, Culham Science Centre, Abingdon OX14 3DB, UK}
\affil[3]{STFC Daresbury Laboratory, Daresbury WA4 4AD, UK}
\affil[4]{Chalmers University of Technology, Department of Physics, SE-412 96, Got{\"e}borg, Sweden}
\affil[5]{Merton College, Oxford OX1 4JD, UK}
\affil[6]{Department of Physics, University of Maryland, College Park, MD 20742-4111, USA}
\date{}

\begin{document}

\maketitle

\begin{abstract}
  We investigate the effect of varying the ion temperature gradient (ITG) and
  toroidal equilibrium scale sheared flow on ion-scale turbulence in the outer
  core of MAST by means of local gyrokinetic simulations. We show that
  nonlinear simulations reproduce the experimental ion heat flux and that the
  experimentally measured values of the ITG and the flow shear lie close to the
  turbulence threshold. We demonstrate that the system is subcritical in the
  presence of flow shear, i.e., the system is formally stable to small
  perturbations, but transitions to a turbulent state given a large enough
  initial perturbation. We propose that the transition to subcritical
  turbulence occurs via an intermediate state dominated by low number of
  coherent long-lived structures, close to threshold, which increase in number
  as the system is taken away from the threshold into the more strongly
  turbulent regime, until they fill the domain and a more conventional
  turbulence emerges.  We show that the properties of turbulence are
  effectively functions of the distance to threshold, as quantified by the ion
  heat flux. We make quantitative comparisons of correlation lengths, times,
  and amplitudes between our simulations and experimental measurements using
  the MAST BES diagnostic. We find reasonable agreement of the correlation
  properties, most notably of the correlation time, for which significant
  discrepancies were found in previous numerical studies of MAST turbulence.

\end{abstract}

\listoftodos
\newpage
\section{Introduction}
\label{sec:introduction}

  Understanding and controlling turbulence is crucial to the realisation of
  fusion as an energy source~\cite{Krushelnick2005}. Turbulence at
  perpendicular length scales of the order of the ion Larmor radius can
  be driven by: the ion temperature gradient (ITG) $\kappa_T \equiv - \dv*{\ln
  T_i}{r}$ ($T_i$ is the ion temperature and $r$ is a dimensionless
  radial coordinate defined later), which drives the well-known ITG
  instability~\cite{Coppi1967,Cowley1991}; the electron temperature gradient
  (ETG) $\kappa_{Te} \equiv - \dv*{\ln T_e}{r}$ (where $T_e$ is the
  electron temperature), which at sufficient $\beta$ (ratio of plasma pressure
  to magnetic pressure) can drive microtearing modes
  (MTMs)~\cite{Applegate2007}; and a combination of electron temperature and
  density gradients, which drive trapped-electron modes
  (TEMs)~\cite{Dannert2005}.  The electron temperature gradient also drives ETG
  modes that create plasma turbulence at finer electron
  scales~\cite{Dorland2000,Jenko2000}. Recent experimental~\cite{Mantica2009,
  Mantica2011} and numerical~\cite{Citrin2014} studies of JET plasmas have
  demonstrated that ion-scale turbulence is ``stiff'' with respect to changes
  in $\kappa_T$, i.e., small changes in $\kappa_T$ can lead to large changes in
  the turbulent transport. Similarly, there is
  experimental evidence that small-scale electron turbulence exhibits similar
  behaviour due to changes in $\kappa_{Te}$~\cite{Hillesheim2013}. Power-balance
  calculations for the Mega Ampere Spherical Tokamak (MAST) indicate that heat
  transport is usually carried predominantly through the electron
  channel~\cite{Akers2003}, and gyrokinetic simulations have shown that it can
  be due to a combination of microtearing modes, which have been shown to be
  unstable at $\beta \gtrsim 0.1$, and fine-scale ETG-driven
  turbulence~\cite{Roach2005, Joiner2006, Applegate2004, Applegate2007,
  Dickinson2012, Hillesheim2016}. Similar findings have also been reported for
  NSTX~\cite{Levinton2007, Wong2007, Wong2008, Guttenfelder2011,
  Guttenfelder2012, Guttenfelder2013}. The main reason for the dominance of the
  electron channel and relative weakness of the ion transport is believed to be
  the suppression of ion-scale turbulence by significant differential rotation
  present in spherical tokamaks. Understanding, based on measuring and modelling
  the structure of this weak ion-scale turbulence, the physics of the
  associated transport and of its suppression is a key challenge of fusion
  plasma theory, both for MAST and for tokamaks generally.

  Indeed, studies of many experiments have shown that turbulence can be
  affected by the profile of the toroidal rotation, which is driven by the
  neutral beam injection (NBI) heating system~\cite{Burrell1997,Mantica2009,
  Mantica2011, Field2011,Ghim2014}. The plasma flow associated with toroidal
  rotation, which is sheared, has components both parallel and perpendicular to
  the direction of the magnetic field.  Perpendicular flow shear, quantified by
  $\gamma_E = (r/q)\dv*{\omega}{r}$~$(a/v_{\mathrm{th}i})$ ($q$ is the safety
  factor, $\omega$ is the frequency of toroidal rotation, $a$ is the minor
  radius of the device, $v_{\mathrm{th}i} = \sqrt{2T_i/m_i}$ is the thermal
  velocity, and $m_i$ is the ion mass), has been shown to reduce, or even
  eliminate, turbulence in tokamaks~\cite{Burrell1997,Hahm1995}. Numerical
  studies of core turbulence in MAST~\cite{Akers2003, Roach2009, Field2011} and
  NSTX~\cite{Kaye2007b,Kaye2013} have confirmed that ion-scale turbulence is
  often suppressed by the perpendicular flow shear. Parallel flow shear has
  also been shown to drive a linear instability~\cite{Catto1973}, which can
  increase the level of turbulence, although, at the levels of flow shear
  considered in this work, we do not expect the destabilising effect of the
  parallel flow shear to be significant. Thus, at ion scales, there is a
  competition in fusion plasmas between the destabilising effects of the
  ITG/TEM instabilities and the parallel flow shear, and the stabilising effect
  of the perpendicular flow shear.

  It has been shown that perpendicular flow shear can render the
  plasma completely linearly stable~\cite{Barnes2011a}. However, this may still
  entail substantial transient growth of perturbations and, given a large
  enough initial perturbation, can lead to a saturated nonlinear state -- a
  phenomenon known as ``subcritical'' turbulence~\cite{Roach2009, Newton2010,
  Highcock2011, Schekochihin2012, Landreman2015}. We have previously studied
  this transition to subcritical turbulence in MAST in Ref.~\cite{VanWyk2016},
  and proposed the following transition scenario: close to the turbulence
  threshold, the nonlinear state is dominated by coherent, long-lived
  structures; as the system is taken away from the threshold, the number of
  these structures increases until they fill the simulation domain and a
  conventional turbulent state is recovered. In this paper, we will focus
  on the nature of ion-scale turbulence in MAST (driven by a combination of the
  ITG and trapped electron modes) and present a more comprehensive view
  of the changes in turbulence that occur as the system is taken away from the
  threshold. We do this via nonlinear gyrokinetic simulations by varying
  $\kappa_T$ and $\gamma_E$ and perform an analysis of the turbulence structure
  and make detailed comparisons with experimental measurements.

  At the temperatures and densities found in fusion experiments such as MAST,
  it can be shown that the conditions for a fluid description are rarely
  satisfied and that a kinetic description must be used.
  Gyrokinetics~\cite{Frieman1982, Sugama1998, Abel2013} has emerged as the most
  appropriate first-principles description in the context of plasma turbulence
  in the core of tokamaks. In this paper, we use the local gyrokinetic code
  GS2\footnote{\url{http://gyrokinetics.sourceforge.net}}~\cite{Kotschenreuther1995,
  Dorland2000}
  to solve the gyrokinetic equation.  GS2 includes a large number of physical
  effects relevant to experimental plasmas, such as realistic magnetic-surface
  geometries, arbitrary numbers of kinetic species, realistic Fokker-Planck
  collision operators, and so on. This has allowed simulations of sufficient
  realism to be compared quantitatively to experimental measurements.
  Local gyrokinetic codes, such as GS2, take as input the values and first
  derivatives of equilibrium quantities at a particular radial location and
  predict a host of quantities that could theoretically be measured by an
  experimental diagnostic, for example, the flux of particles, momentum, and
  heat, or, indeed, the full density and temperature fluctuation fields.

  In conjunction with increasingly realistic modelling, more sophisticated
  diagnostic techniques have been developed, which aid in our understanding of
  the conditions inside the reactor and allow us to make comparisons with
  modelling results. The beam-emission-spectroscopy (BES) diagnostic in MAST is
  one such diagnostic that measures ion-scale density
  fluctuations~\cite{Field2009, Field2012}. More specifically, the BES
  diagnostic infers ion-scale density fluctuations from D$_\alpha$
  emission (the emission of light resulting from the dominant (n=3-2) visible
  transition of ionised deuterium), which is generated as a result of the
  injection of neutral particles by the NBI system. The BES diagnostic takes
  measurements in a two-dimensional radial-poloidal plane. From the BES
  measurements, it is possible to estimate a number of useful correlation
  properties of the turbulence~\cite{Durst1992, McKee2003, Shafer2012,
  Ghim2012, Ghim2013, Fox2016}: the correlation time $\tau_c$, via the
  cross-correlation time delay (CCTD) method; the radial and poloidal
  correlation lengths $l_R$ and $l_Z$; and the relative density-fluctuation
  field $\delta n_i/n_i$.  Measurements of fluctuating quantities allow more
  extensive quantitative comparisons between experiment and simulations via the
  use of ``synthetic diagnostics'', which take account of the measurement
  characteristics of the particular diagnostic and modify the simulation output
  accordingly~\cite{Holland2009, Ghim2012, Field2014, Fox2016}.

  Previous studies of BES data and comparisons with ion-scale simulation
  data have been performed on DIII-D~\cite{McKee1999,
  McKee2003, Holland2009, Holland2011, Rhodes2011, Shafer2012, Gorler2014}
  and MAST~\cite{Field2014}. In the L-mode studies on DIII-D, good agreement
  was found between experimental measurements and synthetic results from
  local simulations in the mid-core region ($0.4 < r/a < 0.75$,
  where $r/a$ is the normalized radius), both in terms of transport and
  fluctuation characteristics. In the outer-core region ($r/a > 0.75$), GYRO
  simulations again showed good agreement for the fluctuation characteristics, but
  underpredicted the heat fluxes and fluctuation amplitudes by almost an
  order of magnitude~\cite{Holland2009, Holland2011, Rhodes2011, Shafer2012}.
  However, subsequent local simulations using the GENE
  code~\cite{Gorler2014} more closely matched the experimental measurements.
  The motivation for this work is the study performed in
  Ref.~\cite{Field2014} of MAST turbulence that used the BES
  diagnostic to measure turbulent density fluctuations in the outer core of an
  L-mode plasma and compared their correlation properties with those inferred
  from global gyrokinetic simulations. The discharge studied was specifically
  designed to have high flow shear at mid-radius to produce an internal
  transport barrier (ITB), and, as a consequence, had low flow shear in the
  outer core, where ion-scale turbulence would not, therefore be completely
  suppressed. There was some agreement at mid-radii, however, significant
  discrepancies were found in the ion heat flux and turbulence correlation time
  at outer radii. In this work we simulate ion-scale turbulence in the
  outer core of the same L-mode discharge as in~\cite{Field2014} using
  high-resolution local gyrokinetic simulations. While previous gyrokinetic
  modelling of similar MAST L-mode plasmas showed that electron-scale
  turbulence can play a significant role~\cite{Roach2009, Field2011} (as is
  the case for this discharge, where $Q_e \sim 8 Q_i$), it has been shown that
  the suppression of ion heat transport is due to the effect of flow shear and it
  is this phenomenon that we study further in this paper\footnotemark.
  Therefore, it is of interest to study purely ion-scale turbulence, as we do
  in this work, in order to make comparisons with BES data, which only covers
  turbulent fluctuations at ion scales. We shall see that local gyrokinetic
  modelling does produce turbulent fluctuations whose correlation properties
  are consistent with experimental measurements, in particular the turbulence
  correlation time. However, we also find that GS2 underpredicts the turbulence
  amplitude, similar to previous studies~\cite{Holland2009, Holland2011,
  Shafer2012, Gorler2014}.
  \footnotetext{The observed electron-scale turbulence may be driven by ETG
    and/or microtearing modes.  While ETG modes are not expected to contribute
    significantly to turbulence at ion scales~\cite{Guttenfelder2013},
    microtearing modes may play a role, however, we have not included these (or
    other electromagnetic effects) in our simulations due to the small value of
    $\beta$ compared to previous studies of these
    effects~\cite{Roach2005,Guttenfelder2013} and due to computational
    constraints. Simulations investigating electromagnetic effects may be
    attempted in future.}

  In simulating experimentally relevant plasmas using gyrokinetic
  codes, we aim to achieve the following. First, we want to understand better
  the physical mechanisms that most strongly influence turbulence and the
  associated enhanced transport. Specifically, we wish to know how turbulence
  characteristics (such as transport, spatial scales, time scales, etc.) change
  in the outer core of MAST with the equilibrium parameters $\kappa_T$ and
  $\gamma_E$. Secondly, in light of newly available experimental data from the
  MAST BES diagnostic~\cite{Field2014}, we want to establish whether the
  turbulence characteristics found in local gyrokinetic GS2 simulations agree
  with experimental BES measurements within the experimental uncertainties in
  measurements of $\kappa_T$ and $\gamma_E$. Such quantitative comparisons with
  experimental results are essential in developing confidence in our
  theoretical models and numerical implementations. In understanding the
  properties of turbulence, we ultimately aim to guide the optimisation and
  design of future experiments and fusion reactors by acquiring the ability to
  predict and control the turbulence.

  The rest of this paper is organised as follows. In section~\ref{sec:setup},
  we give an overview of the MAST discharge that we will be considering, as
  well as of gyrokinetics and of the numerical tools that we will use for our
  study. Our main results are split into two sections.

  In section~\ref{sec:nl},
  we study numerically the effect on turbulence in MAST of changing $\kappa_T$
  and $\gamma_E$ by performing a two-dimensional parameter scan in these two
  equilibrium parameters. We map out the turbulence threshold and show that the
  experimentally measured ion heat flux is close to the numerical values found
  near this threshold, thus suggesting that the turbulence in MAST is
  near-marginal (section~\ref{sec:heat_flux}). We demonstrate that the
  turbulence is subcritical (section~\ref{sec:subcritical}), with large
  initial perturbations required to ignite it (a phenomenon not
  previously observed for an experimentally relevant plasma) and estimate the
  conditions necessary for the onset of turbulence. We then show that the
  near-threshold state is one dominated by long-lived, coherent structures,
  which exist against a background of much smaller fluctuations
  (section~\ref{sec:struc_analysis}). These structures are shown to be regions
  of increased density, radial flow, and temperature fluctuations. Sufficiently
  far from the turbulence threshold in parameter space, we recover a more
  conventional turbulent state consisting of many strongly interacting eddies
  while being sheared apart by the perpendicular flow shear. We
  demonstrate that many of the properties of the system (e.g. the number of
  structures, their amplitude, shear due to zonal flows, etc.) are effectively
  functions only of the distance from the turbulence threshold, as quantified
  by the ion heat flux.

  In section~\ref{sec:struc_of_turb}, we make comparisons with experimental
  measurements from the BES diagnostic. We present two types of correlation
  analysis of our simulations: of the numerical data processed through a
  synthetic diagnostic (section~\ref{sec:corr_synth}) and of raw GS2 data,
  with no modelling of the diagnostic (section~\ref{sec:corr_gs2}). We show
  that there is reasonable agreement with experimental measurements in the case
  of the analysis with the synthetic diagnostic. However, radial correlation
  lengths predicted by GS2 are shown to be below the resolution threshold of
  the BES diagnostic in MAST (an issue discussed in detail in~\cite{Fox2016}). This
  conclusion stems from studying correlation parameters of the raw GS2 density
  fluctuations and suggests that care must be taken when interpreting BES
  measurements. Comparison between results of analysis with and without the
  synthetic diagnostic shows that the synthetic diagnostic has a measurable
  effect on several turbulence characteristics, including the poloidal
  correlation length and the fluctuation amplitude, consistent with the
  conclusions of Ref.~\cite{Fox2016}.  Finally, we present the correlation
  lengths and times as functions of the ion heat flux and again show that the
  structure of the turbulence in our simulations is effectively only a function
  of this parameter, which measures the distance to the turbulence threshold.

\section{Experimental and numerical details}
\label{sec:setup}

\subsection{MAST discharge \#27274}
  MAST is a medium-sized, low-aspect-ratio ($\approx 1.5$) tokamak with
  a major radius $R_0 \approx 0.9$~m and a minor radius $a \approx 0.6$~m.  In
  this work, we will focus on the MAST discharge \#27274, one of a
  set of three nominally identical experiments (i.e. having identical profiles
  and equilibria) previously reported in~\cite{Field2014} and differing only in
  the radial viewing location of the BES system. These three discharges were
  \#27272, \#27268, and \#27274, wherein the centre of the BES was located at
  $1.05$~m, $1.2$~m, and $1.35$~m, respectively.  The MAST BES
  diagnostic~\cite{Field2009, Field2012} observes an area of approximately
  $16\times8$~cm$^2$ in the radial and poloidal directions, respectively,
  corresponding approximately to one third of the minor radius of the plasma.
  Thus, the combination of these three discharges provided a complete radial
  profile of BES measurements on the outboard side of the plasma.

  Each discharge produced an L-mode plasma with strong toroidal rotation and,
  therefore, with mean flow shear perpendicular and parallel to the magnetic
  field~\cite{Field2014}. Previous investigations of MAST turbulence for
  similar configurations~\cite{Roach2009, Field2011} found that ion-scale
  turbulence is suppressed in the core region by strong flow shear. However,
  the flow shear is weaker in the outer-core region, where ITG modes are not
  completely suppressed, making it possible to study ion-scale turbulence. In
  this work, we will restrict our attention to the time window
  $t=0.250\pm0.002$~s and the radial location\footnotemark~$r = D/2a =
  0.8~(\equiv r_0)$ of discharge \#27274, where $D$ is the diameter of the flux
  surface and $a$ is the half diameter of the last closed flux surface (LCFS),
  both measured at the height of the magnetic axis.  Importantly, there is no
  large-scale and disruptive magnetohydrodynamic (MHD) activity at this time
  and radial location~\cite{Field2014}; such activity would interfere with the
  quality of BES measurements. The normalized radial location $r=0.8$
  corresponds to a major radius of approximately $1.32$~m and, therefore, falls
  within the viewing area covered by discharge \#27274 [see
  \figref{flux_surfaces}].
  \footnotetext{We use $r=D/2a$ as the definition of the radial location because
    it corresponds to the radial coordinate used by the Miller specification of
    the flux-surface geometry~\cite{Miller1998}. In terms of other commonly used
    radial coordinates, $r = 0.8$ corresponds to $\rho_\mathrm{tor} =
    \sqrt{\psi_\mathrm{tor}/\psi_\mathrm{tor,LCFS}} = 0.7$ and
    $\rho_\mathrm{pol} = \sqrt{\psi_\mathrm{pol}/\psi_\mathrm{pol,LCFS}} =
    0.87$, where $\psi_\mathrm{tor} = {(1/2\pi)}^2 \int_0^V dV \vb*{B} \cdot
    \nabla \phi$ is the toroidal magnetic flux, $V$ is the volume enclosed by
    the flux surface, $\vb*{B}$ is the magnetic field, $\phi$ is the toroidal
    angle, and $\psi_\mathrm{tor,LCFS}$ is the toroidal flux enclosed by the
    last closed flux surface [see \figref{flux_surfaces}], $\psi_\mathrm{pol} =
    {(1/2\pi)}^2 \int_0^V dV \vb*{B} \cdot \nabla \theta$ is the poloidal
    magnetic flux, $\theta$ is the poloidal angle, and $\psi_\mathrm{pol,LCFS}$
    is the poloidal flux enclosed by the LCFS.}

\subsection{Equilibrium profiles}
\label{sec:exp_profiles}

  \begin{figure}[t]
    \centering
    \begin{subfigure}[t]{0.55\linewidth}
      \includegraphics[width=\linewidth]{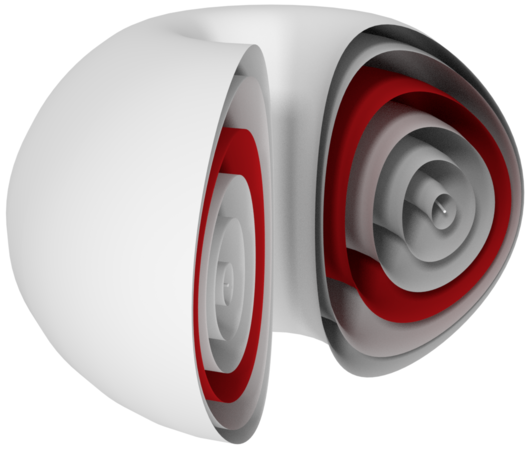}
      \caption{}
      \label{fig:nested_flux_surfaces}
    \end{subfigure}
    \begin{subfigure}[t]{0.37\linewidth}
      \includegraphics[width=\linewidth]{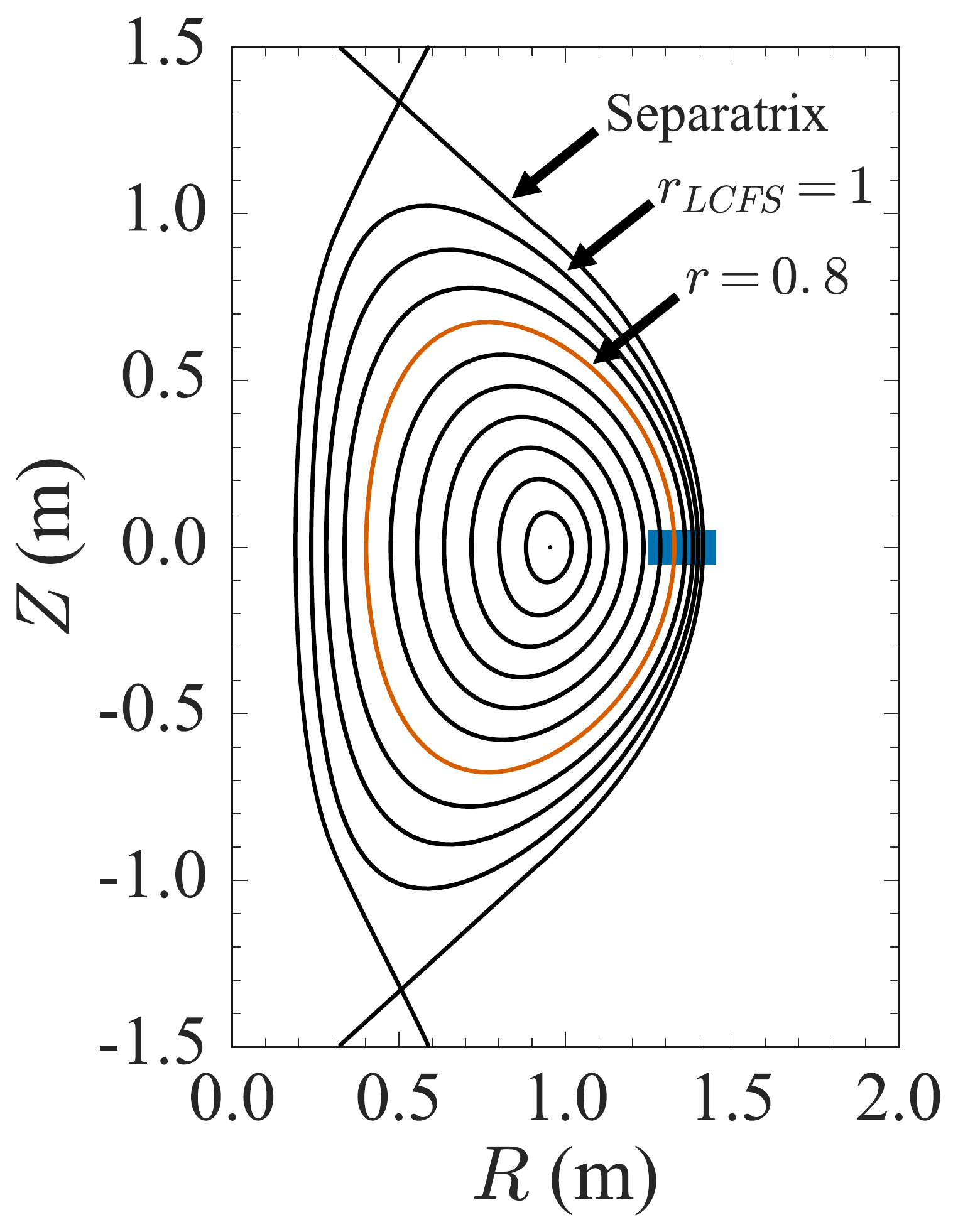}
      \caption{}
      \label{fig:flux_surfaces}
    \end{subfigure}
    \caption[MAST flux surfaces]{\subref*{fig:nested_flux_surfaces} A
      three-dimensional view of the nested flux surfaces.
      \subref*{fig:flux_surfaces} The poloidal cross-section of the magnetic
      geometry along with the LCFS and the separatrix, which separates closed
      field lines from open ones. The flux surface of interest is at $r = 0.8$,
      shown in red. It was chosen so that this surface intersects the BES
      measurement plane for discharge \#27274 (blue shaded region).
    }
  \end{figure}
  MAST has a range of diagnostics that allow us to extract the equilibrium
  parameters required to conduct a numerical transport study. The ion
  temperature, $T_i$, and toroidal flow velocity, $u_\phi = R \omega$, where
  $\omega$ is the toroidal angular rotation frequency, were obtained from
  charge-exchange-recombination spectroscopy (CXRS) measurements of C$^{+6}$
  impurity ions with a spatial resolution of $\sim 1$~cm~\cite{Conway2006}. The
  electron density, $n_e$, and temperature, $T_e$, were obtained from a
  Thomson-scattering diagnostic~\cite{Scannell2010} with resolution comparable
  to the CXRS system. These measured profiles were mapped onto flux-surface
  coordinates by the pre-processing code $MC^\mathit{3}$ using a
  motional-Stark-effect-constrained EFIT equilibrium~\cite{Lao1985}.  These
  equilibrium profiles served as input to the transport analysis code
  TRANSP\footnote{\url{http://w3.pppl.gov/transp/}}~\cite{Hawryluk1981}, which
  calculates the transport coefficients of particles, momentum, and heat.
  \Figref{nested_flux_surfaces} shows a three-dimensional view of the
  axisymmetric nested flux surfaces and \figref{flux_surfaces} shows the
  poloidal cross-section of the flux surfaces extracted from an EFIT
  equilibrium. The $r=0.8$ surface is highlighted in both plots. The
  measurement window of the BES diagnostic for discharge \#27274 is also shown
  in \figref{flux_surfaces}.  The chosen flux surface at $r = 0.8$ intersects
  the measurement window at the outboard midplane, allowing comparisons of
  turbulence characteristics between our numerical predictions of turbulence
  and experimental measurements.

  The important experimental quantities needed to conduct a numerical study are
  the radial profiles of $T_i$, $T_e$, $n_i$ (the ion density), $n_e$, and
  $\omega$. There are no direct measurements of $n_i$ in MAST, but we assume
  that it is equal to $n_e$, as measured by the Thomson-scattering diagnostic,
  due to quasineutrality (in MAST, we typically have an effective ion charge
  $Z_{\mathrm{eff}} \lesssim 1.5$). To conduct a numerical study of turbulence
  at $r=0.8$ (using the local formulation of gyrokinetics; see
  section~\ref{sec:gk_theory}), we need the equilibrium quantities listed above
  and their first derivatives (gradient length scales). The (normalised)
  gradient length scales of $T_i$, $T_e$, and $n_e$, and flow shear (gradient
  of $\omega$) are, by definition,
  \begin{align}
      \label{ti_prime}
      \frac{1}{L_{Ti}} &= - \dv{\ln T_i}{r} \equiv \kappa_T, \\
      \label{te_prime}
      \frac{1}{L_{Te}} &= - \dv{\ln T_e}{r}, \\
      \label{ne_prime}
      \frac{1}{L_{ne}} &= - \dv{\ln n_e}{r}, \\
      \label{flow_shear}
      \hspace{-2em}\mathrm{and}\hspace{2em}
      \gamma_E &= \frac{r_0}{q_0} \dv{\omega}{r} \frac{a}{v_{\mathrm{th}i}},
  \end{align}
  where $q(\psi) = \pdv*{\psi_\mathrm{tor}}{\psi_{\mathrm{pol}}}$ is the safety
  factor and $q_0$ is its value at $r_0$.  The flow-shear parameter $\gamma_E$
  can be interpreted as the (non-dimensionalised) shear of the component of the
  toroidal rotation that is perpendicular to the local magnetic field.

\begin{figure}
    \centering
    \begin{subfigure}[t]{0.35\textwidth}
      \includegraphics[width=\textwidth]{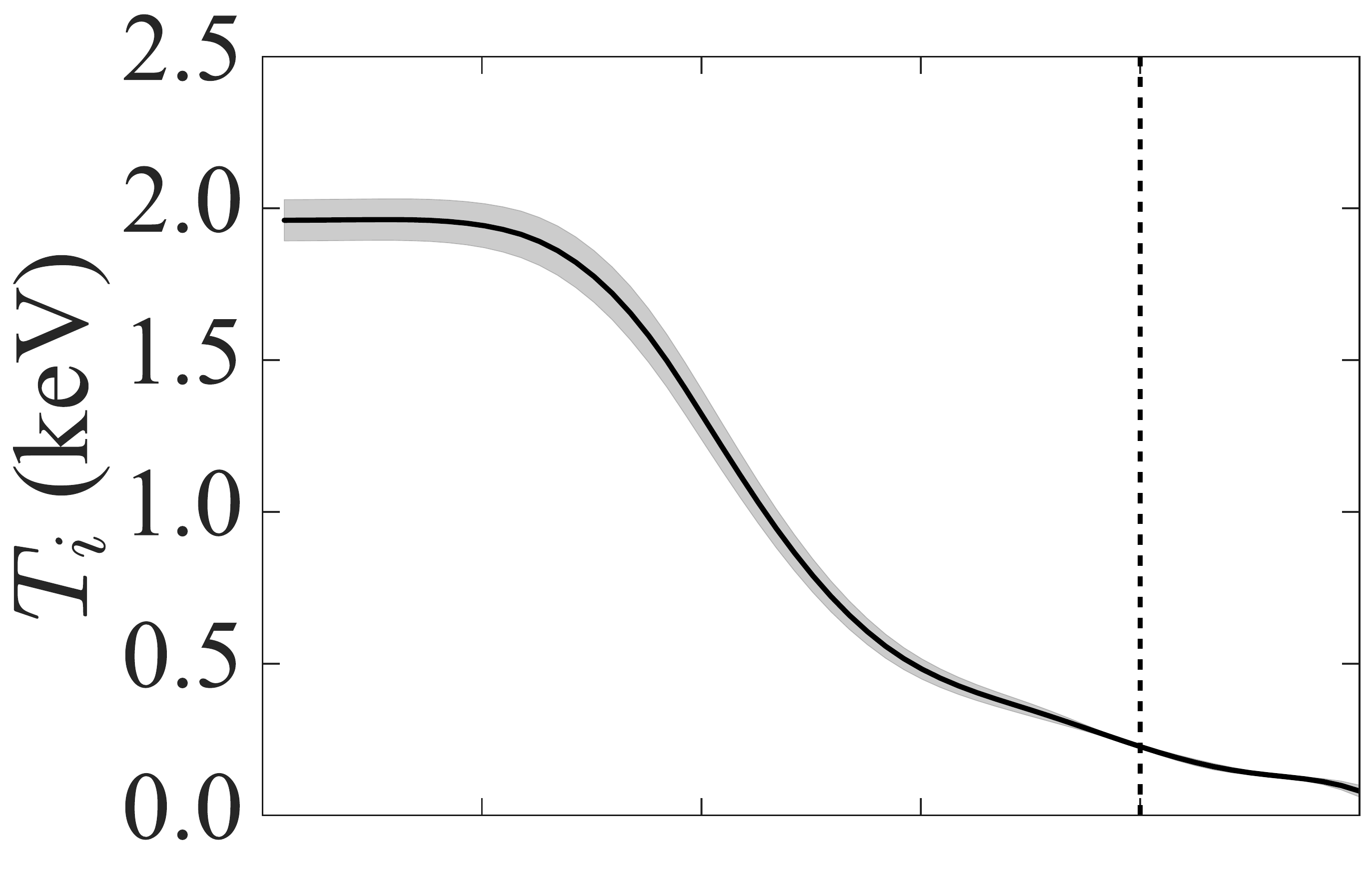}
      \caption{}
      \label{fig:ti}
    \end{subfigure}
    \begin{subfigure}[t]{0.35\textwidth}
      \includegraphics[width=\textwidth]{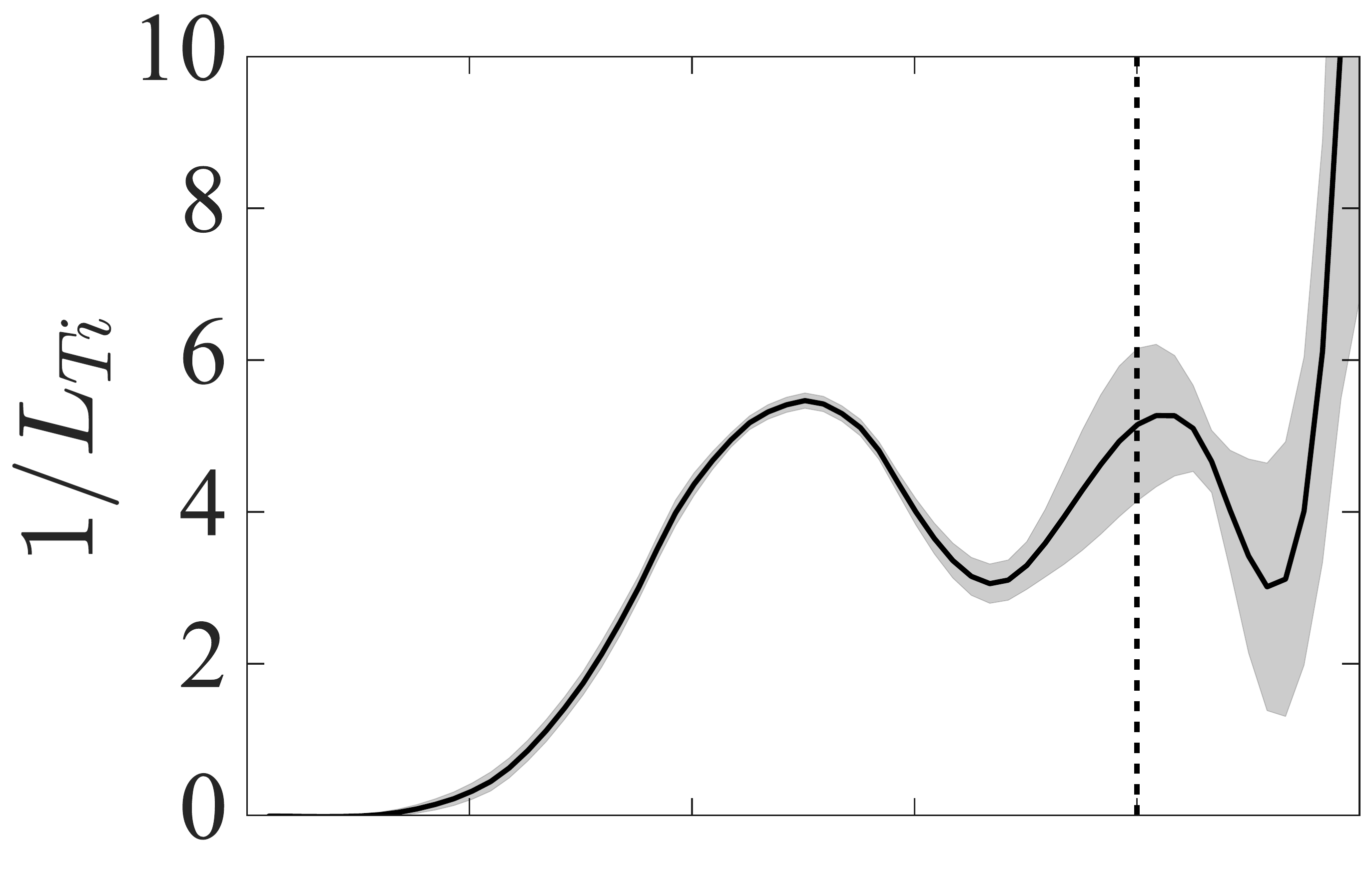}
      \caption{}
      \label{fig:ti_prime}
    \end{subfigure}
    \begin{subfigure}[t]{0.35\textwidth}
      \includegraphics[width=\textwidth]{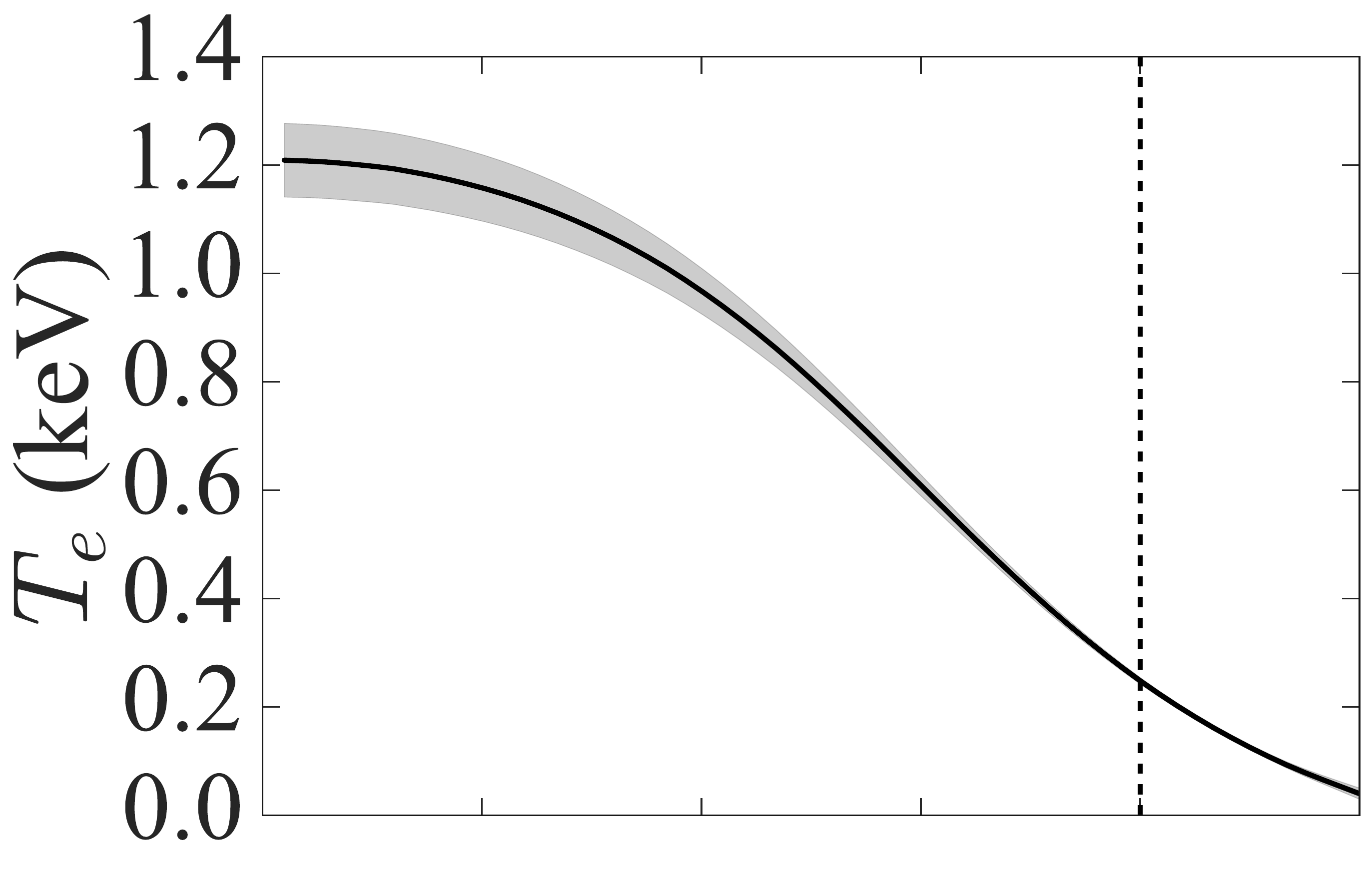}
      \caption{}
      \label{fig:te}
    \end{subfigure}
    \begin{subfigure}[t]{0.35\textwidth}
      \includegraphics[width=\textwidth]{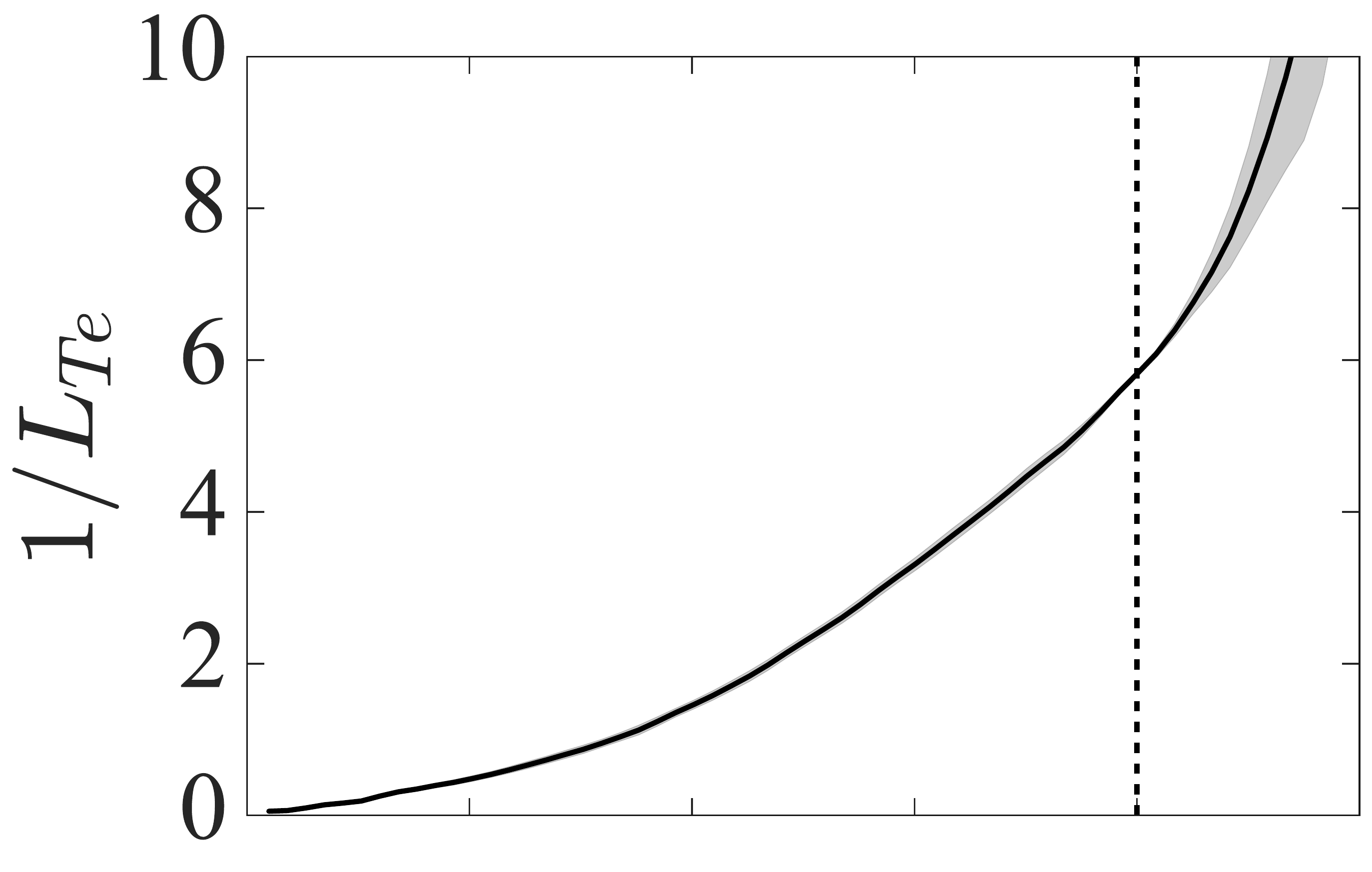}
      \caption{}
      \label{fig:te_prime}
    \end{subfigure}
    \begin{subfigure}[t]{0.35\textwidth}
      \includegraphics[width=\textwidth]{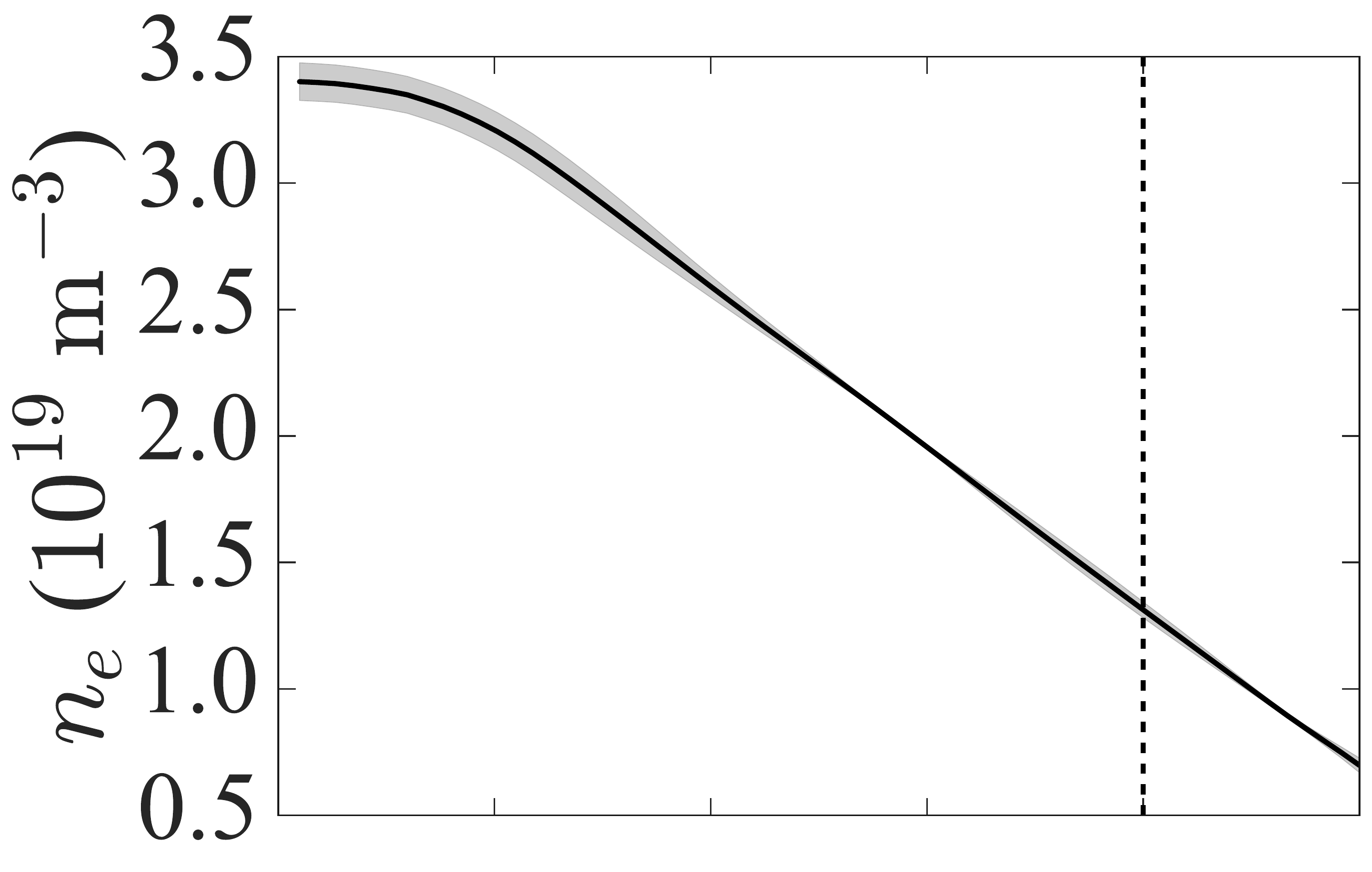}
      \caption{}
      \label{fig:ne}
    \end{subfigure}
    \begin{subfigure}[t]{0.35\textwidth}
      \includegraphics[width=\textwidth]{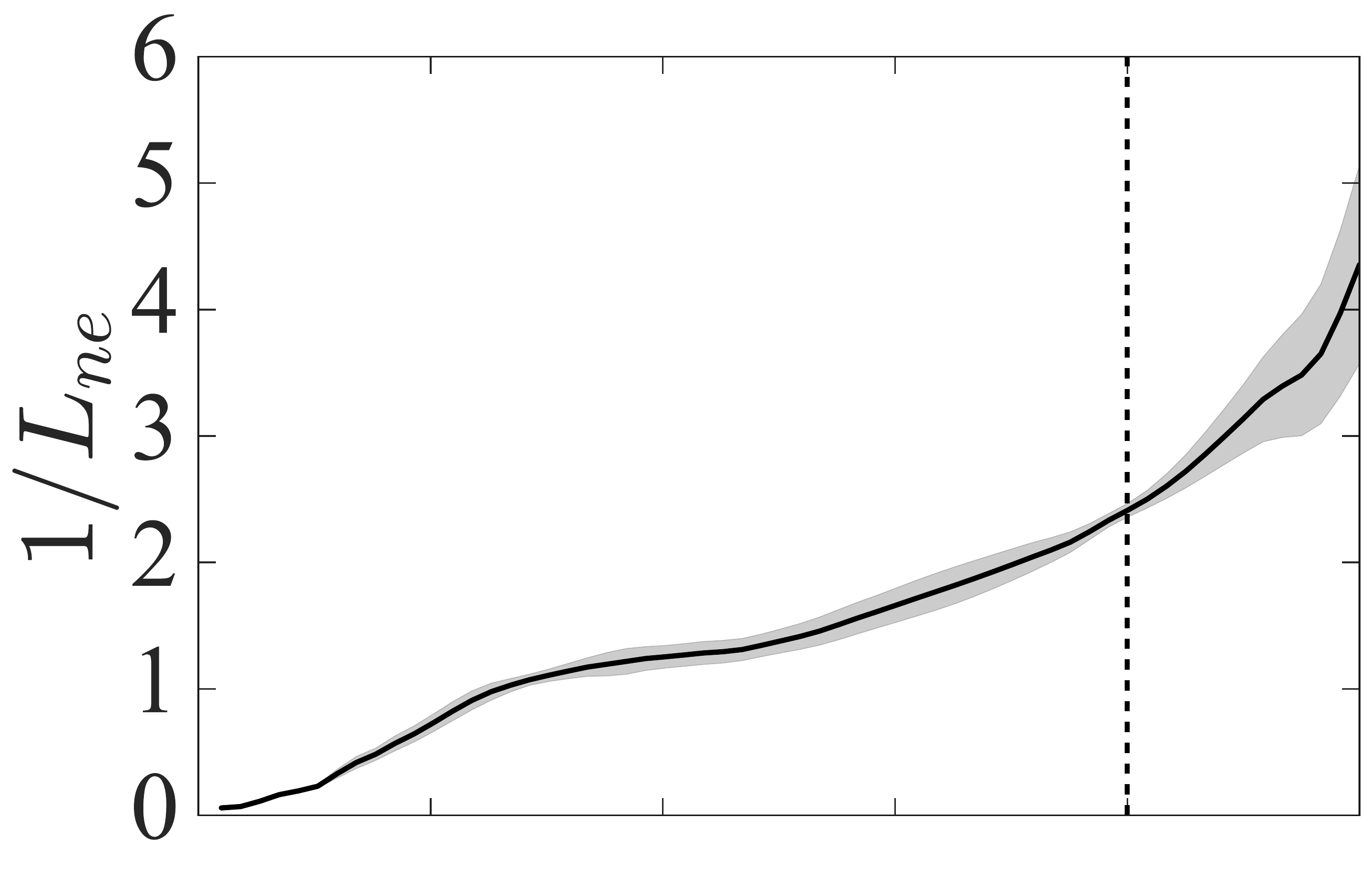}
      \caption{}
      \label{fig:ne_prime}
    \end{subfigure}
    \begin{subfigure}[t]{0.37\textwidth}
      \includegraphics[width=\textwidth]{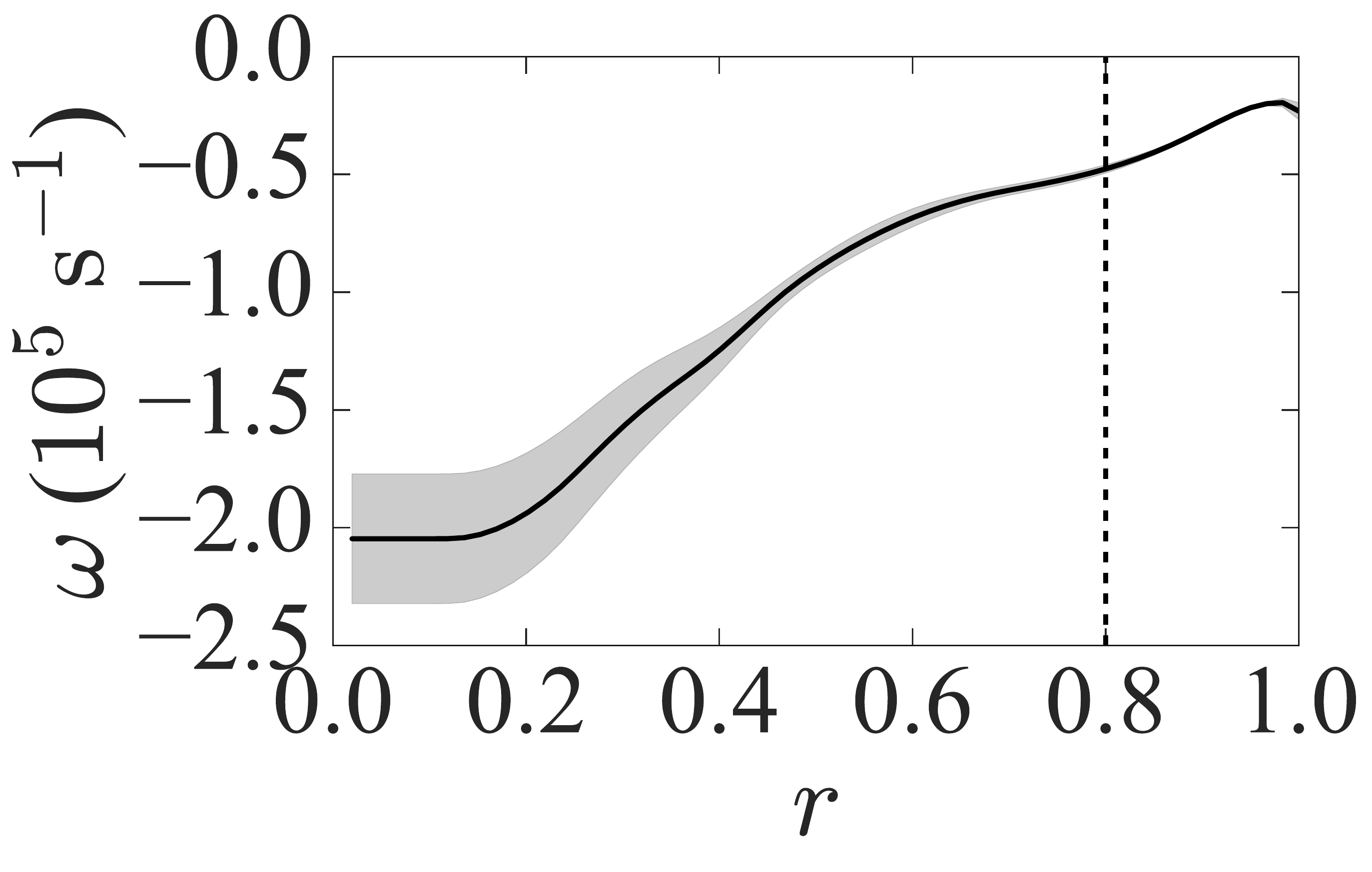}
      \caption{}
      \label{fig:omega}
    \end{subfigure}
    \begin{subfigure}[t]{0.37\textwidth}
      \includegraphics[width=\textwidth]{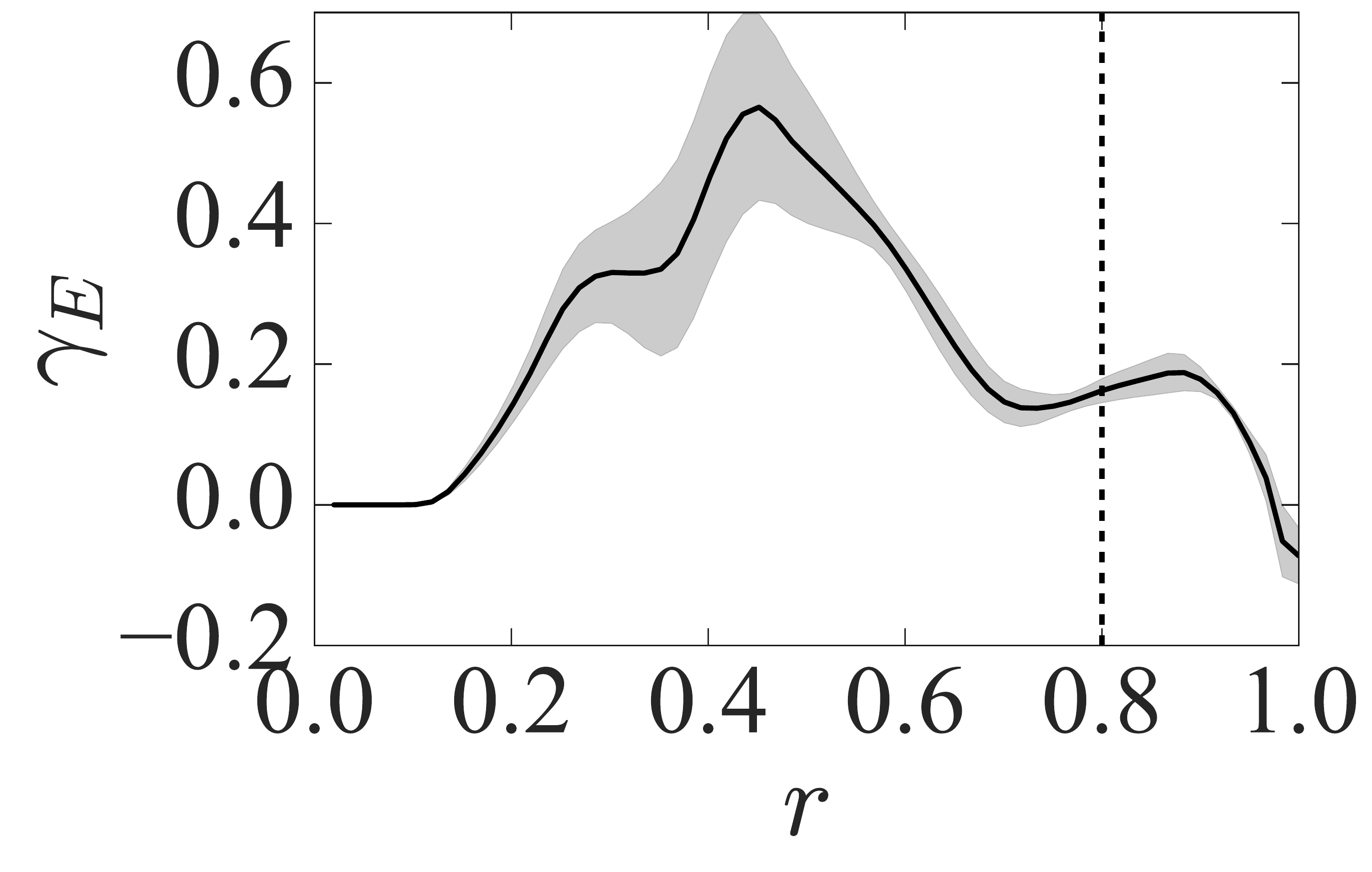}
      \caption{}
      \label{fig:g_exb}
    \end{subfigure}
    \caption[Experimental profiles]{Radial profile measurements from MAST
      discharge \#27274 of
      \subref*{fig:ti} the ion temperature, $T_i$,
      \subref*{fig:ti_prime} the ion temperature gradient, $1/L_{Ti}$,
      calculated using~\eqref{ti_prime},
      \subref*{fig:te} the electron temperature, $T_e$,
      \subref*{fig:te_prime} the electron temperature gradient, $1/L_{Te}$,
      calculated using~\eqref{te_prime},
      \subref*{fig:ne} the electron density, $n_e$,
      \subref*{fig:ne_prime} the electron density gradient, $1/L_{ne}$,
      calculated using~\eqref{ne_prime},
      \subref*{fig:omega} the toroidal angular frequency, $\omega$, and
      \subref*{fig:g_exb} the flow shear, $\gamma_E$, calculated
      using~\eqref{flow_shear}. The dashed line in each plot indicates $r=0.8$
      and the shaded regions indicate the standard deviation of the profiles
      over a $20$-ms time window around $t = 0.25$~s.
    }
    \label{fig:profiles}
  \end{figure}

  The left-hand column of \figref{profiles} shows the radial profiles of $T_i$,
  $T_e$, $n_e$, and $\omega$ as functions of $r$. The gradient scale
  lengths~\eqref{ti_prime}--\eqref{ne_prime} and flow shear~\eqref{flow_shear}
  are plotted as functions of $r$ in the right-hand column
  in~\figref{profiles}. The dashed lines indicate $r=0.8$. The
  profiles in \figref{profiles} (and in \figref{q_exp} below) represent a
  $20$-ms time average around $t = 0.25$~s and the shaded areas indicate the
  standard deviations. The nominal experimental values of the quantities that
  we will vary in this study are $\kappa_T = 5.1 \pm 1$ and $\gamma_E = 0.16
  \pm 0.02$.

  The profiles of the ion and electron heat fluxes ($Q_i^{\exp}$ and
  $Q_e^{\exp}$, respectively) were obtained from a transport
  analysis using TRANSP based on the equilibrium profiles shown in
  \figref{profiles}. The profiles are shown in~\figref{q_exp}. In this work, we
  normalise all heat fluxes to the ion gyro-Bohm value
  \begin{equation}
    Q_{\mathrm{gB}} = n_i T_i v_{\mathrm{th}i} \frac{\rho_i^2}{a^2},
    \label{q_gb}
  \end{equation}
  where $\rho_i$ is the ion gyroradius.  We see from~\figref{q_exp} that the
  main loss of heat in the system is via transport due to the electrons: the
  experimental level of ion heat flux at $r = 0.8$ is
  $Q^{\exp}_i/Q_{\mathrm{gB}} = 2 \pm 1$, while the electron heat flux is
  $Q^{\exp}_e/Q_{\mathrm{gB}} = 15.2 \pm 0.9$. This is partially due to the
  suppression of ion turbulence by flow shear (as we will show in this paper)
  and possibly also due to significant heat transport driven in the electron
  channel via the ETG instability and MTMs as has been observed in other
  studies~\cite{Roach2005, Joiner2006, Roach2009, Guttenfelder2011,
  Guttenfelder2012, Dickinson2012}. In this work, we will focus exclusively on
  ion-scale turbulence in order to make contact with ion-scale turbulence
  measurements from the BES. We will briefly comment on the transport predicted
  by our simulations in the electron channel, but leave the full investigation
  to future work.

  \begin{figure}[t]
    \centering
    \includegraphics[width=0.6\linewidth]{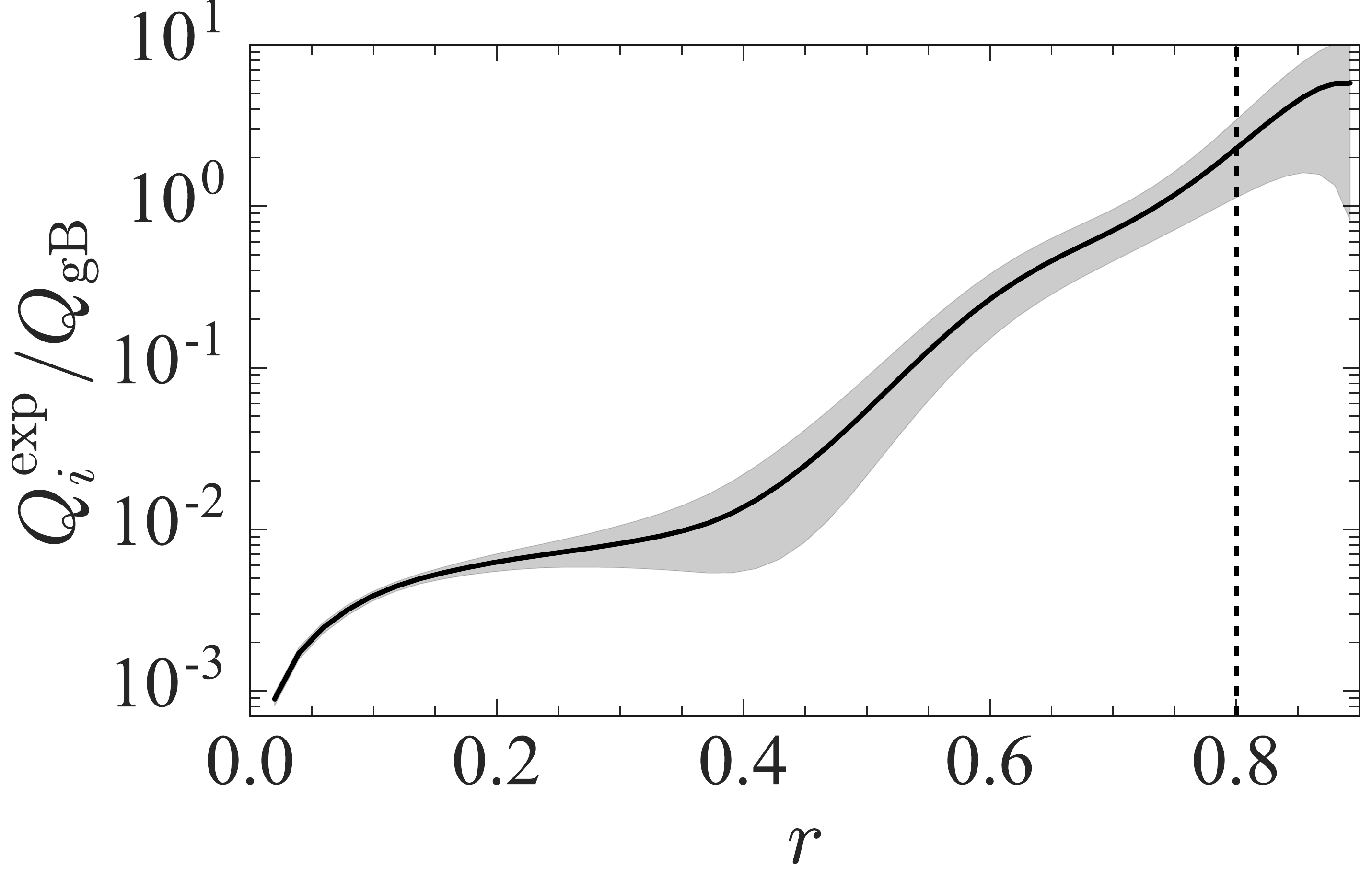}
    \caption{
      Experimental ion and electron heat fluxes as functions of $r$ determined
      from power balance by TRANSP analysis. Both fluxes are normalized to the
      local ion gyro-Bohm value~\eqref{q_gb}. The dashed line indicates
      $r=0.8$, where BES measurements were made, and the shaded regions
      indicate the uncertainty estimated by TRANSP.
    }
    \label{fig:q_exp}
  \end{figure}

\subsection{Local gyrokinetic description}
\label{sec:gk_theory}
  We model the turbulence in MAST using gyrokinetic theory~\cite{Frieman1982,
  Sugama1998, Abel2013}.  For a detailed review, the reader is referred
  to~\cite{Abel2013} and references therein, while only a brief overview is
  given here. The gyrokinetic equation describes the evolution of the
  non-Boltzmann part of the perturbed (from a background Maxwellian $F_s$)
  particle distribution function, $h_s(t, \vb*{R}_s, \vareps_s, \mu_s,
  \sigma)$, of a species $s$, where $\vb*{R}_s$ is the guiding-centre
  coordinate, $\vareps_s$ is the particle energy, $\mu_s$ is the magnetic
  moment of species $s$, and $\sigma$ is the sign of the (peculiar) parallel
  velocity $v_\parallel$. The gyrokinetic equation is\footnotemark
  \begin{equation}
    \begin{split}
      &\left(\pdv{}{t} + \vb*{u} \cdot \nabla \right) \left(h_s
      - \frac{Z_s e \ensav{\varphi}{\vb*{R}_s}}{T_s} F_s\right) +
      \left(v_\parallel \vb*{b} + \vb*{V\!}_{{\rm D}s} +
      \ensav{\vb*{V\!}_E}{\vb*{R}_s}\right)
      \cdot \nabla{h_s} - \ensav{C[h_s]}{\vb*{R}_s} \\
      &\quad=
      -\ensav{\vb*{V\!}_E}{\vb*{R}_s} \cdot \nabla r
      \left[\dv{\ln n_s}{r} + \left(\frac{\vareps_s}{T_s} -
        \frac{3}{2}\right)\dv{\ln T_s}{r}
      + \frac{m_s v_\parallel}{T_s}\frac{R B_\phi}{B}\dv{\omega}{r}\right]F_s,
      \label{gk}
    \end{split}
  \end{equation}
  where $\vb*{u} = \omega(r) R^2 \nabla \phi$ is the toroidal rotation
  velocity, $\omega(r)$ is the toroidal angular frequency, $\phi$ is the
  toroidal angle, $\varphi$ is the electrostatic potential perturbation,
  $\ensav{\ldots}{\vb*{R}_s}$ is an average over the particle orbit at constant
  $\vb*{R}_s$, $F_s = n_s(r) {\left[ m_s / 2 \pi T_s(r)\right]}^{3/2}
  e^{- \vareps_s/T_s(r)}$ is the background Maxwellian, $\vb*{V\!}_{{\rm D}s}
  = (c/Z_s e B)\vb*{b}\times \left[ m_s v_\parallel^2 \vb*{b} \cdot
  \nabla \vb*{b} + \mu \nabla B \right]$ is the magnetic drift velocity,
  \begin{equation}
    \vb*{V\!}_E = \frac{c}{B}\vb*{b}\times\nabla\varphi
    \label{v_exb}
  \end{equation}
  is the \exb drift velocity, $C[h_s]$ is the linearised collision
  operator~\cite{Abel2008a,Barnes2008}, and $B_\phi$ is the toroidal component
  of the magnetic field.
  \footnotetext{
    The equilibrium quantities $n_s$, $T_s$, and $\omega$ are functions only of
    the poloidal magnetic flux $\psi$. For the purposes of this work, we have
    converted this dependence from $\psi$ to the Miller coordinate $r = D/2a$
    introduced previously. Since $r$ is also a flux-surface label, we can use
    the following equation to relate gradients in $\psi$ and $r$:
    $\nabla r = \dv*{r}{\psi} \nabla \psi$.
  }

  To close our system of equations we use the quasineutrality condition
  \begin{equation}
    \sum_s Z_s\delta n_s = 0
    \quad\Rightarrow\quad
    \sum_s \frac{Z_s^2 e \varphi}{T_s} n_s = \sum_s Z_s \int d^3 \vb*{v}
    \ensav{h_s}{\vb*{r}},
    \label{quasineutrality}
  \end{equation}
  where $\ensav{\ldots}{\vb*{r}}$ indicates a gyroaverage at constant particle
  position $\vb*{r}$, to calculate $\varphi$ using $h_s$.

  In order for the local approximation to be valid, we require that
  $\rho_i/a \ll 1$, where we assume that other important length scales in the
  system, such as $L_{Ti}$, are of the same order as $a$.  The turbulence
  predicted by our simulations is, therefore, only representative of the
  turbulence at a single flux surface, even though our box sizes can be
  the size of MAST (several such formally overlapping simulations can be then
  used to model transport across the entire radial extent of the machine).
  For the MAST discharge and radial location studied in this work, one finds
  $\rho_i/a \sim 1/100$, where $\rho_i \approx 6 \times 10^{-3}$~m and
  $a\approx 0.6$~m. While this is a reasonably small number, previous studies
  of simpler geometries have suggested that non-local effects can reduce the
  level of turbulent transport by 50\% at values similar to
  $1/100$~\cite{McMillan2010, Saarelma2012}. A scan of different values of
  $\rho^*$ using a global gyrokinetic code would be required to test whether
  non-local effects change the level of turbulence for the MAST turbulence
  studied in this paper. In addition, the coherent structures described in
  section~\ref{sec:coherent_strucs} are similar in size to the gradient
  length scales and so global effects might affect their characteristics.
  However, the cost of using global simulations would be too large for the
  parameter scans performed in this paper. There is ongoing work to extend
  the GS2 code to include finite system-size effects~\cite{Parra2015}, such
  as profile variation, which may be used in future to test their effect on
  MAST turbulence.

  In adopting equations~\eqref{gk} and \eqref{quasineutrality} we have formally
  assumed that the Mach number $M$ of the plasma rotation is small, but that
  the flow shear is large enough to affect the plasma dynamics:
  \begin{equation}
    \frac{R\omega}{v_{{\mathrm{th}}i}} = M \ll 1,\quad
    |a\nabla\ln\omega| \sim \frac{1}{M}.
    \label{flow_scaling}
  \end{equation}
  This allows us to formulate local gyrokinetics in a rotating surface,
  neglecting effects such as the Coriolis and centrifugal force, but retaining
  the effect of flow shear~\cite{Abel2013}. We have also assumed that the
  fluctuations are purely electrostatic, i.e., that there are no fluctuating
  magnetic fields. Previous studies of
  MAST~\cite{Applegate2004,Applegate2007,Roach2005,Dickinson2012} and of
  NSTX~\cite{Levinton2007,Wong2007,Wong2008,Guttenfelder2011,Guttenfelder2012,
  Guttenfelder2013} have shown that including electromagnetic fluctuations
  affects core turbulence in regions of the plasma where $\beta \gtrsim 0.1$
  and $\beta \gtrsim 0.05$, for MAST and NSTX, respectively.
  On the peripheral MAST L-mode surface studied in this paper, $\beta$ is
  smaller than that level by an order of magnitude, viz. $\beta \approx 0.005$,
  and so electromagnetic effects are expected to be negligible.

\subsection{Numerical set-up}
\begin{figure}[t]
    \centering
    \includegraphics[width=0.6\linewidth]{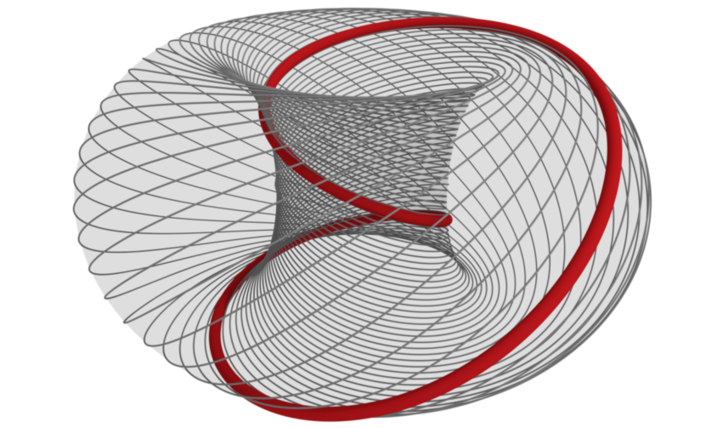}
    \caption[MAST magnetic field lines]{
      Magnetic field lines that lie on the flux
      surface at $r = 0.8$ (setting $q=2$ for visualisation purposes, so that
      field lines are closed). The field line marked in red is the centre line
      of the GS2 flux tube that we use as the simulation domain. The flux tube
      follows the field line once around the flux surface in the poloidal
      direction.
    }
    \label{fig:flux_tube}
  \end{figure}

  In this work, we have used the local gyrokinetic code
  GS2\footnote{\url{http://gyrokinetics.sourceforge.net}}~\cite{Kotschenreuther1995,
  Dorland2000,HighcockThesis} to solve the system of equations~\eqref{gk}
  and~\eqref{quasineutrality} to give us the time evolution of $h_s$ and
  $\varphi$. GS2 solves the gyrokinetic equation in a region known as a
  ``flux tube'', shown in~\figref{flux_tube}. The GS2 flux tube follows a
  central magnetic field line once around in the poloidal direction
  (represented in~\figref{flux_tube} by the field line highlighted in red).

  The MAST local equilibrium parameters used in our simulations, extracted
  from the MAST diagnostics and EFIT equilibrium, are given in
  table~\ref{tab:sim_params}. We have included electrons in our simulations as
  a kinetic species. Our GS2 simulations had resolution of $85 \times 32 \times
  20$ grid points in the radial $\times$ binormal $\times$ parallel directions,
  and $27 \times 16$ pitch-angle $\times$ energy-grid points, respectively. The
  corresponding box sizes were $L_x \approx 200 \rho_i$ in the radial direction
  (with maximum wavenumber $k_{x,\max} \rho_i \simeq 1.3 $) and $L_y \approx 62
  \rho_i$ in the binormal direction (with maximum wavenumber $k_{y,\max} \rho_i
  \simeq 3$). We note that the radial box size is larger than the minor radius
  of MAST, however, this is required to achieve sufficient $k_x$ resolution to
  resolve the effect of the flow shear (see appendix~\ref{App:flow_shear}).
  Artificial numerical dissipation was used to damp electron modes at small
  scales.
  \begin{table}[t]
    \footnotesize
    \centering
    \caption{GS2 equilibrium parameters calculated from diagnostic measurements
      and from the EFIT equilibrium of the MAST discharge \#27274 and
      appropriately normalised. The nominal experimental values for $\kappa_T$
      and $\gamma_E$ are $\kappa_T = 5.1 \pm 1$ and $\gamma_E = 0.16 \pm 0.02$.
      The reference magnetic field is the toroidal magnetic field strength at
      the magnetic axis, i.e., $B_{\mathrm{ref}} = B_{\phi}(r=0)$.
    }
    \begin{tabular}{r c c}
      \toprule

      Quantity & GS2 variable & Value \\
      \midrule

      $\beta = {8\pi n_i T_i}/{B_\mathrm{ref}^2}$ & \texttt{beta} &
      0.0047 \\

      $\beta' = \pdv*{\beta}{r}$ & \texttt{beta\_prime\_input}
      & -0.12 \\

      Eff. ion charge for collisions $Z_{\mathrm{eff}}
      = {\sum_i n_i Z_i^2/|\sum_i n_i Z_i|}$ & \texttt{zeff} & 1.59 \\

      Elec.-ion collisionality $\nu_{ei}$ & \texttt{vnewk\_2} & 0.59 \\

      Elec. density $n_{eN} = n_e/n_i$ &
      \texttt{dens\_2} & 1.00 \\

      Elec. density grad. $1/L_{ne} = - \dv*{\ln n_e}{r}$ &
      \texttt{fprim\_2} & 2.64 \\

      Elec. mass $m_{eN} = m_e/m_i$ &
      \texttt{mass\_2} & $1 / (2 \times 1836)$ \\

      Elec. temp. $T_{eN} = T_e/T_i$ &
      \texttt{temp\_2} & 1.09 \\

      Elec. temp. grad. $\kappa_{Te} \equiv 1/L_{Te} = - \dv*{\ln T_e}{r}$ &
      \texttt{tprim\_2} &  5.77\\

      Elongation $\kappa$ & \texttt{akappa} & 1.46 \\

      Elongation derivative $\kappa' = \dv*{\kappa}{r}$ &
      \texttt{akappri} & 0.45 \\

      Flow shear $\gamma_E = (r_0/q_0) \dv*{\omega}{r} (a/v_{{\mathrm{th}}i})$ &
      \texttt{g\_exb} & [0, 0.19] \\

      Ion collisionality $\nu_i$ & \texttt{vnewk\_1} & 0.02 \\

      Ion density $n_{iN} = n_i/n_i$ & \texttt{dens\_1} & 1.00 \\

      Ion density grad. $1/L_{ni} = - \dv*{\ln n_i}{r}$ &
      \texttt{fprim\_1} & 2.64 \\

      Ion mass $m_{iN} = m_i/m_i$ & \texttt{mass\_1} & 1.00 \\

      Ion temp. $T_{iN} = T_i/T_i$ & \texttt{temp\_1} & 1.00 \\

      Ion temp. grad. $\kappa_T \equiv 1/L_{Ti} = - \dv*{\ln T_i}{r}$ &
      \texttt{tprim\_1} & [4.3, 8.0] \\

      Magnetic field reference point $R_\mathrm{geo}$ & \texttt{r\_geo} & 1.64 \\

      Magnetic shear $\hat{s} = r/q\dv*{q}{r}$ & \texttt{s\_hat\_input} & 4.00\\

      Major radius $R_{N} = R/a$ & \texttt{rmaj} & 1.49 \\

      Miller radial coordinate $r = {D/2a}$ & \texttt{rhoc} & 0.80 \\

      Safety factor $q = \pdv*{\psi_\mathrm{tor}}{\psi_{\mathrm{pol}}}$ &
      \texttt{qinp} & 2.31\\

      Shafranov Shift $1/a \dv*{R}{r}$ & \texttt{shift} & -0.31 \\

      Triangularity $\delta$ & \texttt{tri} & 0.21 \\

      Triangularity derivative $\delta' = \dv*{\delta}{r}$ &
      \texttt{tripri} & 0.46 \\

      \bottomrule
    \end{tabular}
    \label{tab:sim_params}
  \end{table}

  GS2 solves \eqref{gk} for $h_s$, from which one can calculate a
  range of physical characteristics of the turbulence, e.g., the density-,
  flow-, temperature-fluctuation fields, as well as particle, momentum, and
  heat fluxes, and so on. Of particular importance in this work are the ion
  density fluctuation field,
  \begin{equation}
    \frac{\delta n_i}{n_i} =
    \frac{1}{n_i} \int \dd^3 \vb*{v} \ensav{h_i}{\vb*{r}},
    \label{delta_n}
  \end{equation}
  where $\delta n_i/n_i$ is an order-unity quantity, and the radially
  outward, time-averaged turbulent heat flux carried by the ions,
  \begin{equation}
    Q_i = \left\langle\frac{1}{V}\int \dd^3 \vb*{r} \int \dd^3 \vb*{v}
    \frac{m_i v^2}{2} h_i \vb*{V\!}_E \cdot \nabla r \right\rangle_{\mathrm{fs}},
    \label{q_def}
  \end{equation}
  where $V$ is the volume of the flux tube and $\langle \ldots
  \rangle_{\mathrm{fs}}$ denotes a flux-surface average. The heat flux $Q_i$
  can be normalised to the gyro-Bohm heat flux given by~\eqref{q_gb}.

\section{Numerical results}
\label{sec:nl}

In this section, we present the results of a two-dimensional scan in the two
local equilibrium parameters, $\kappa_T$ and $\gamma_E$, that have been
identified to have a strong effect on the properties of the turbulence. We
demonstrate that GS2 simulations are able to match the experimental ion heat flux
at equilibrium-parameter values within the experimental uncertainty and that
the experiment lies close to the turbulence threshold
(section~\ref{sec:heat_flux}). We find that the turbulence
is subcritical, meaning that it can be sustained in the absence of linearly
growing eigenmodes: it is driven instead by transiently growing modes, provided
the transient growth is sufficient and the initial amplitudes are large enough
(section~\ref{sec:subcritical}). We study the linear dynamics and estimate the
conditions necessary to ignite turbulence, namely the transient amplification
factor and time. Studying the real-space structure of turbulence
(section~\ref{sec:struc_analysis}), we detect coherent, long-lived structures
close to marginality, and summarise a novel structure-counting analysis of
these previously presented in~\cite{VanWyk2016}. Moving away from the
turbulence threshold into more strongly-driven regimes, the number of turbulent
structures increases rapidly. Far from the turbulence threshold, the turbulence
is similar to what is encountered in the absence of flow shear, characterised
by many interacting eddies. We estimate the \exb shear due to the zonal flows
(section~\ref{sec:zf_shear}) and show that it is small compared to the
background flow shear close to the turbulence threshold, but becomes comparable
to, and eventually dominates over, the flow shear far from the threshold,
resembling a system in the absence of flow shear.  This suggests that the
observed nonlinear state dominated by coherent structures is an intermediate
state between completely suppressed turbulence and the zonal-flow regulated
scenarios observed in conventional ITG-unstable plasmas~\cite{Dimits2000}.

\subsection{Heat flux}
\label{sec:heat_flux}

  A scan was performed in the parameters $\kappa_T$ and $\gamma_E$ to
  investigate the dependence of turbulent transport on them. The experimental
  values and associated measurement uncertainties were $\kappa_T = 5.1 \pm 1$
  and $\gamma_E = 0.16 \pm 0.02$. Because of the presence of these
  uncertainties and of the sensitive dependence of the heat flux on $\kappa_T$
  and $\gamma_E$, it was necessary to cover a range of their values even just
  to have a meaningful comparison with the experiment. We also performed
  simulations outside the experimental uncertainty ranges to aid our
  understanding of how the nature of the turbulence changes with $\kappa_T$ and
  $\gamma_E$ and, in particular, how it is different near to, versus far from,
  the (nonlinear) stability threshold.  Our entire study covered $\kappa_T \in
  [3.0, 8.0]$ and $\gamma_E \in [0, 0.19]$ and consisted of 76 simulations (see
  Appendix~\ref{App:parameter_scan} for a table of the parameter values). All
  simulations were run until they reached a statistical steady state, i.e.,
  until the running time average became independent of time. Averages were
  taken typically over a time period of approximately
  $200$--$400~(a/v_{\mathrm{th}i})$, which corresponds to $\sim
  800$--$1600~\mu$s, but in many cases longer. The error in these time
  averaged quantities represents the standard deviation from the average during
  the above time periods.

\begin{figure}[t]
    \centering
    \includegraphics[width=0.6\linewidth]{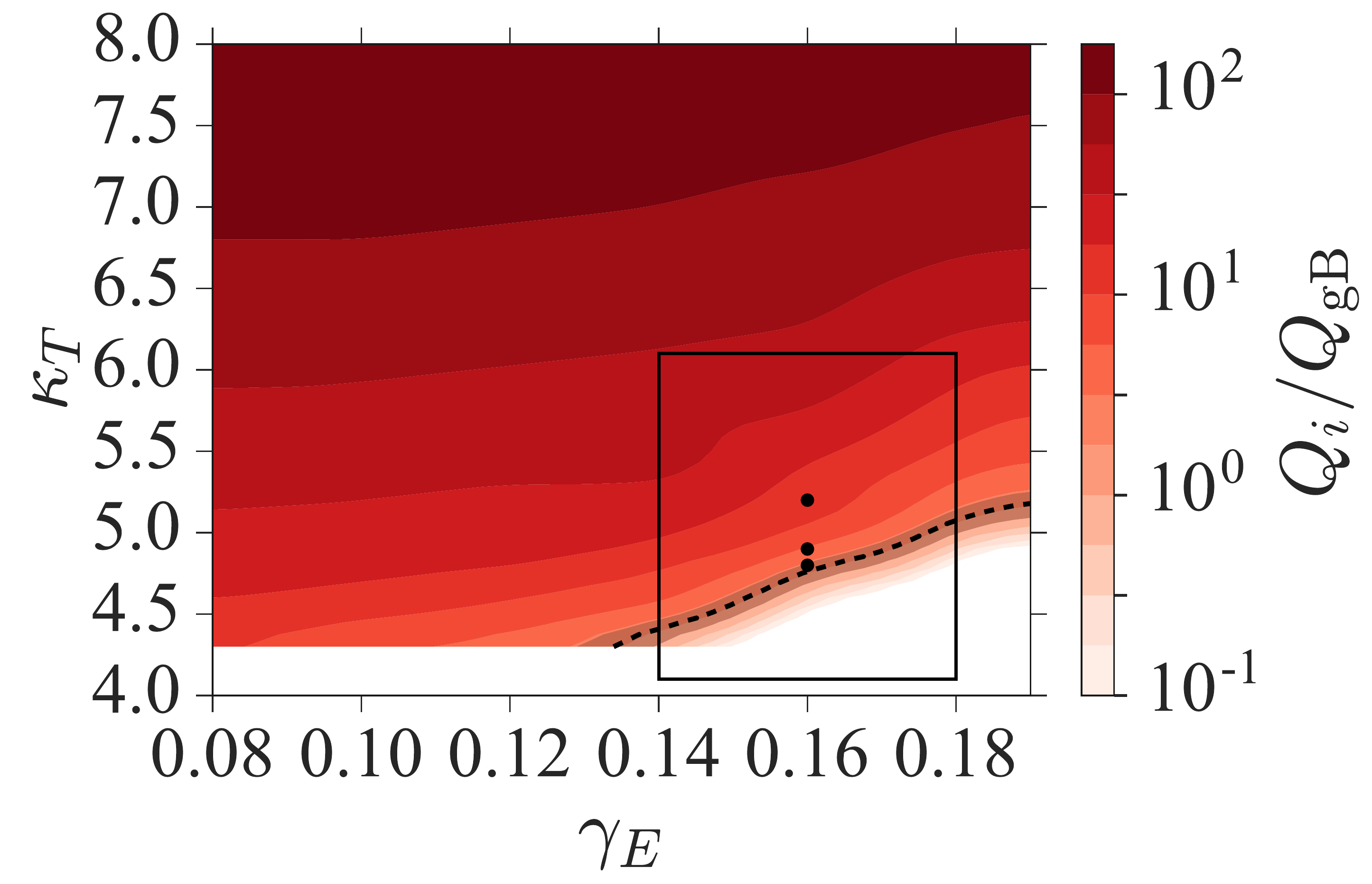}
    \caption{
      Ion heat flux $Q_{i}/Q_{\mathrm{gB}}$ as a
      function of $\kappa_T$ and $\gamma_E$ for all simulations with
      $\gamma_E>0$. The rectangular region indicates the range in $\kappa_T$
      and $\gamma_E$ consistent with the experiment and measurement
      uncertainties. The dashed line indicates the value of
      $Q_{i}^{\exp}/Q_{\mathrm{gB}}$ and the shaded area the experimental
      uncertainty. The experiment is clearly near the turbulence threshold
      defined by $(\kappa_T, \gamma_E)$. The points indicate the parameter
      values for which the density-fluctuation fields are shown
      in~\figref{density_fluctuations}.
    }
    \label{fig:contour_heatmap}
  \end{figure}

  \Figref{contour_heatmap} shows the turbulent ion heat flux versus $\kappa_T$
  and $\gamma_E$ found in our simulations for the full parameter scan with the
  rectangular region indicating the extent of the experimental errors in the
  equilibrium parameters. The dashed line indicates the value of experimental
  heat flux, $Q_i^{\exp}/Q_{\mathrm{gB}}$, and the shaded region the
  experimental uncertainty in its determination.  This figure demonstrates two
  key conclusions of this work:
  \begin{inparaenum}[(i)]
    \item GS2 is able to match the experimental heat flux within the
      experimental uncertainties of $\kappa_T$ and $\gamma_E$, and
    \item the experimental regime is located close to the turbulence threshold
      (defined as the separating line between the regions of parameter space
      with $Q_i = 0$ and $Q_i > 0$).
  \end{inparaenum}

  \begin{figure}[t]
    \centering
    \begin{subfigure}{0.46\linewidth}
      \includegraphics[width=\linewidth]{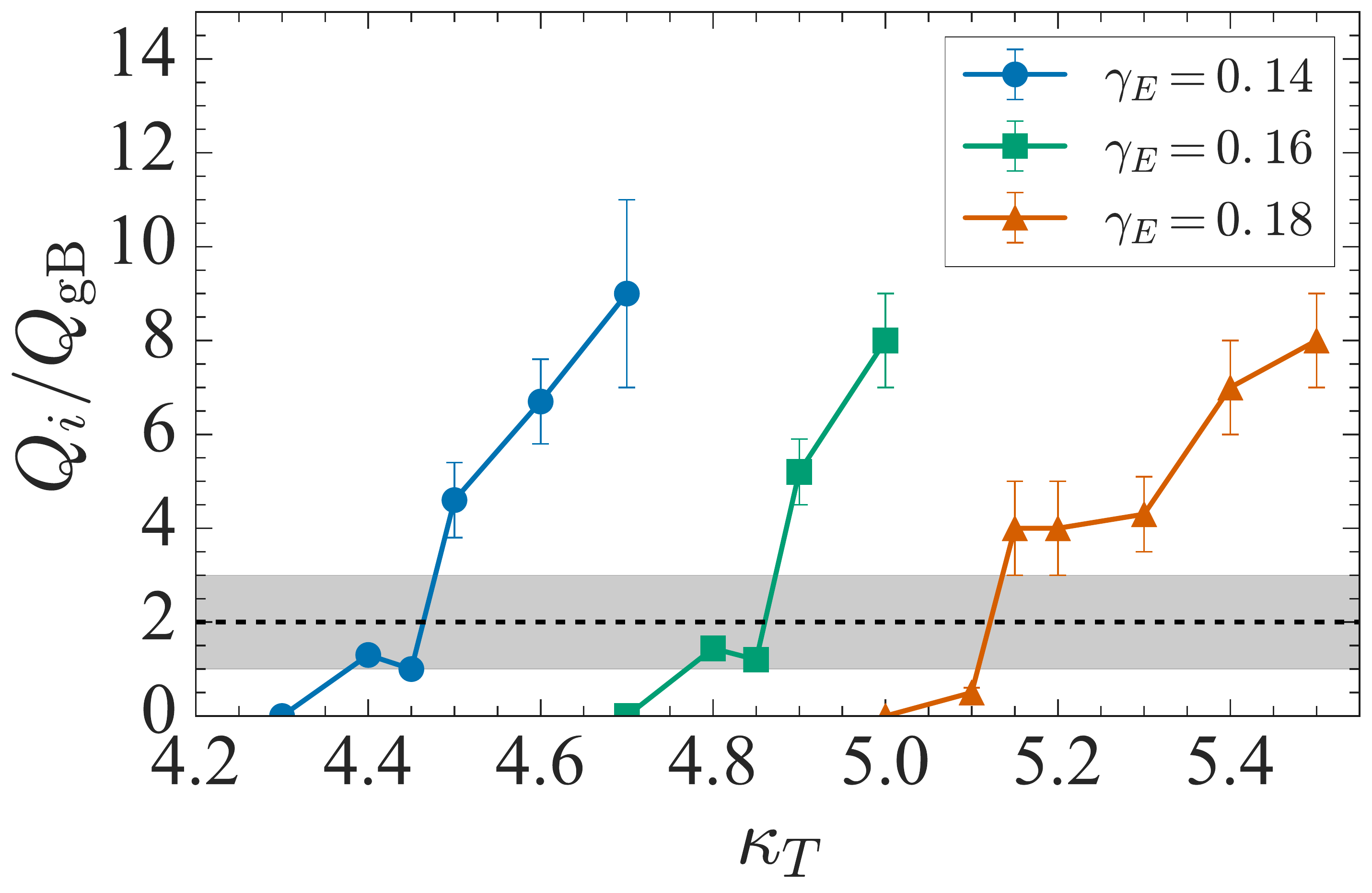}
      \caption{}
      \label{fig:q_vs_tprim_marginal}
    \end{subfigure}
    \hfill
    \begin{subfigure}{0.46\linewidth}
      \includegraphics[width=\linewidth]{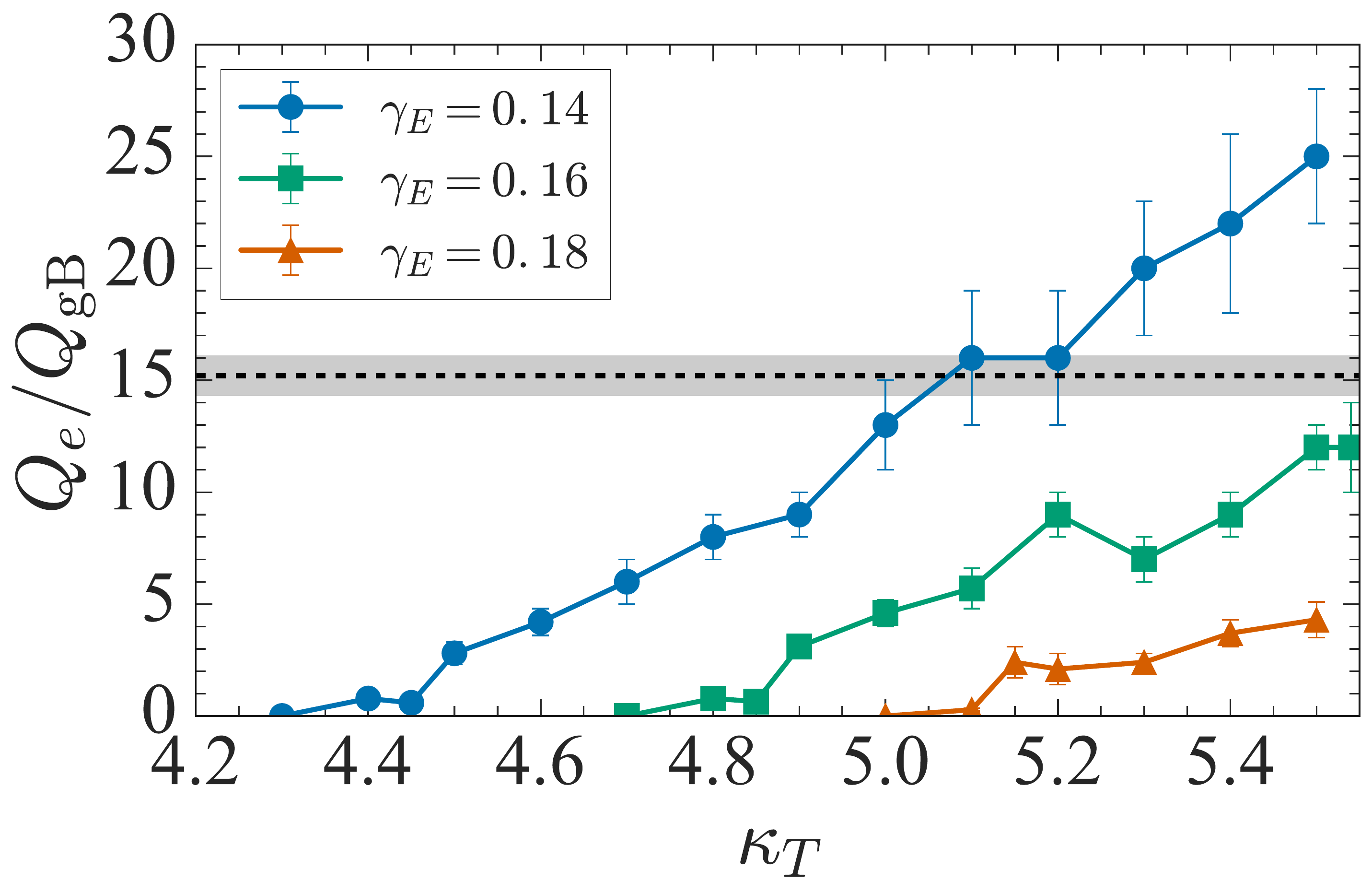}
      \caption{}
      \label{fig:qe_vs_tprim_marginal}
    \end{subfigure}
    \caption{
      \subref*{fig:q_vs_tprim_marginal} Ion heat flux
      $Q_i/Q_{\mathrm{gB}}$ and \subref*{fig:qe_vs_tprim_marginal} electron
      heat flux $Q_e/Q_{\mathrm{gB}}$ as a functions of $\kappa_T$ strictly
      within experimental uncertainty of $\kappa_T$ and $\gamma_E$, and close
      to the turbulence threshold.  The shaded region in each plot indicates the
      experimental heat fluxes $Q_i^{\exp}/Q_{\mathrm{gB}} = 2 \pm 1$ and
      $Q_e^{\exp}/Q_{\mathrm{gB}} = 15.2 \pm 0.9$, determined
      from~\figref{q_exp}.
    }
  \end{figure}

  \Figref{q_vs_tprim_marginal} shows $Q_i/Q_{\mathrm{gB}}$ as a function of
  $\kappa_T$ strictly within the region of measurement uncertainty of
  $\kappa_T$ and $\gamma_E$, close to the turbulence threshold. The dashed line
  and shaded region indicate $Q_i^{\exp}/Q_{\mathrm{gB}}$ and its associated
  uncertainty.  We see that there is a range of $\kappa_T$ and $\gamma_E$
  values where we might expect $Q_i/Q_{\mathrm{gB}}$ to match
  $Q_i^{\exp}/Q_{\mathrm{gB}}$.  From this figure, we can also
  identify several simulations that represent the marginally unstable cases in
  our parameter scan: $(\kappa_T, \gamma_E) = (4.4, 0.14), (4.8, 0.16), (5.1,
  0.18)$. We will consider these parameter values
  section~\ref{sec:subcritical}, when studying the conditions necessary to
  reach a saturated turbulent state.  Furthermore, we have a number of
  individual simulations that match the value of $Q_i^{\exp}/Q_{\mathrm{gB}}$.
  A list of these is given in table~\ref{tab:exp_match_sims}. We will
  investigate these simulations further when we make more detailed comparisons
  with the experiment, in section~\ref{sec:struc_of_turb}.

  \Figref{qe_vs_tprim_marginal} shows that the electron heat flux,
  $Q_e^{\exp}/Q_{\mathrm{gB}}$, is not fully captured by our nonlinear
  ion-scale simulations: namely, in our simulations, we
  observe $Q_e/Q_i \sim 0.6$, whereas from the experiment we expect
  $Q_e^{\exp}/Q_i^{\exp} \sim 7.6$ (see \figref{q_exp}). It is likely that
  electron-scale turbulence~\cite{Joiner2006, Roach2009, Guttenfelder2011,
  Guttenfelder2012, Guttenfelder2013, Colyer2017} is present in the real
  machine, while it cannot be resolved in our simulations, and that it
  dominates electron heat transport. Thus, given the likely
  existence of turbulence on both electron and ion scales, a programme of
  gyrokinetic simulations capturing electron and ion scales simultaneously
  would ideally be necessary. While individual such multiscale simulations have been
  performed~\cite{Howard2016a,Howard2016}, we cannot afford the number of
  such simulations that would be necessary to carry out a parameter scan as
  extensive as we present in this paper. Instead, we will focus on local
  simulations of ion-scale turbulence, and compare the results from these
  simulations with ion-scale BES measurements from MAST.

\begin{figure}[t]
    \centering
    \begin{subfigure}{0.49\linewidth}
      \includegraphics[width=\linewidth]{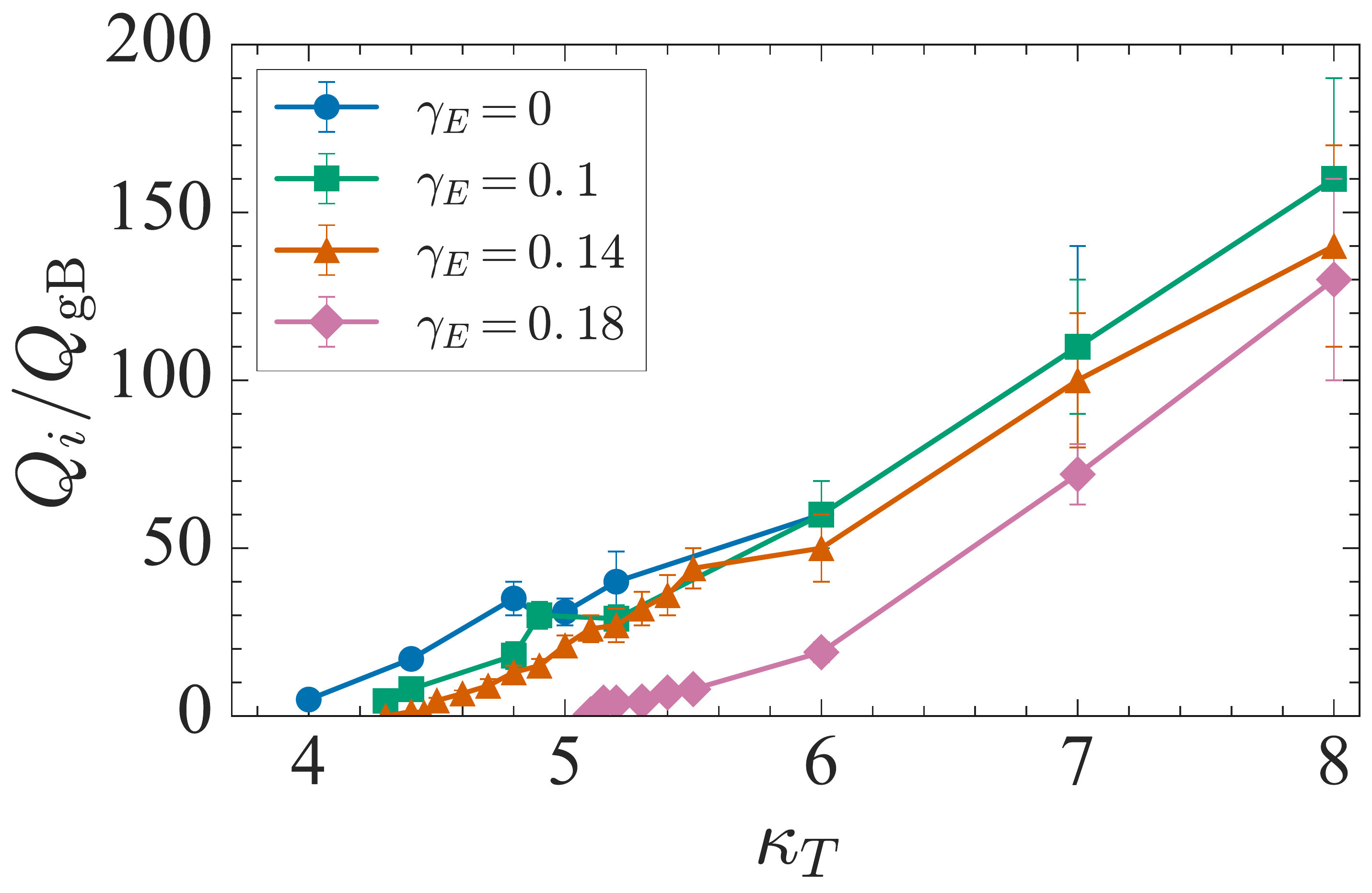}
      \caption{}
      \label{fig:q_vs_tprim}
    \end{subfigure}
    \hfill
    \begin{subfigure}{0.49\linewidth}
      \includegraphics[width=\linewidth]{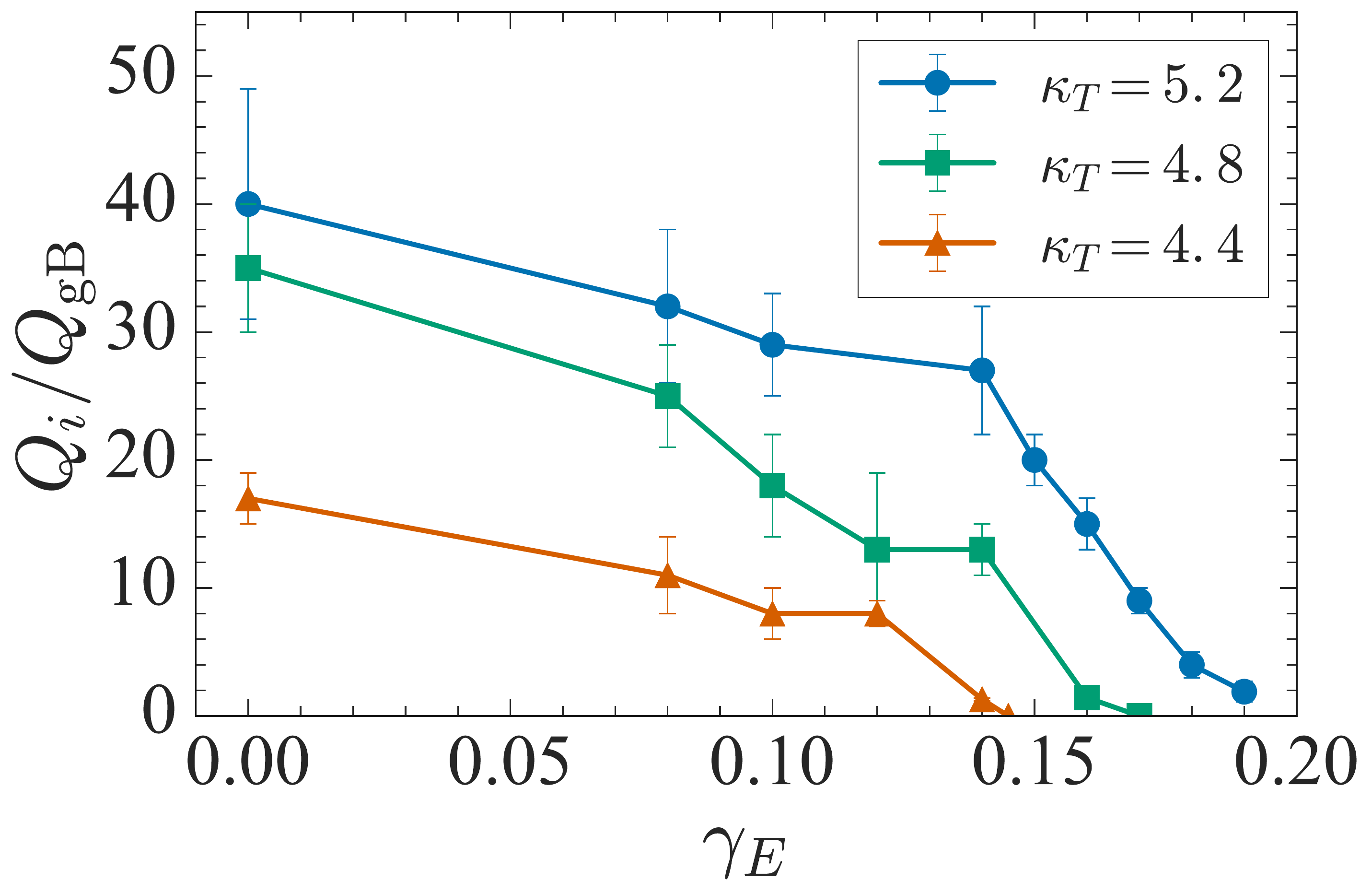}
      \caption{}
      \label{fig:q_vs_gexb}
    \end{subfigure}
    \caption{
      \subref*{fig:q_vs_tprim} Ion heat flux $Q_i/Q_{\mathrm{gB}}$ as a
      function of $\kappa_T$ for several values of $\gamma_E$.
      \subref*{fig:q_vs_gexb} $Q_i/Q_{\mathrm{gB}}$ as a function of $\gamma_E$
      for several values of $\kappa_T$.
    }
    \label{fig:q_line_plots}
  \end{figure}

  \Figref{q_vs_tprim} shows the values of $Q_i/Q_{\mathrm{gB}}$ from
  \figref{contour_heatmap} for several values of $\gamma_E$ as a function of
  $\kappa_T$, whereas \figref{q_vs_gexb} shows $Q_i/Q_{\mathrm{gB}}$ as a
  function of $\gamma_E$ for several values of $\kappa_T$.  We see that an
  $O(1)$ change in $\kappa_T$ gives rise to an $O(10)$ change in
  $Q_i/Q_{\mathrm{gB}}$, and even more dramatically for changes in $\gamma_E$,
  which requires only an $O(0.1)$ change to cause $O(10)$ changes in the ion
  heat flux. An important conclusion from this figure is that the
  presence of flow shear does not significantly affect the stiffness of the
  transport, i.e., the rate of increase of $Q_i/Q_{\mathrm{gB}}$ with respect
  to $\kappa_T$, but only changes the threshold value of $\kappa_T$ above which
  turbulence is present. This increase in critical $\kappa_T$ without a change in the
  stiffness of $Q_i/Q_{\mathrm{gB}}$ with respect to $\kappa_T$ has been
  observed in numerical simulations of simplified ITG-unstable plasmas in the
  presence of flow shear~\cite{Highcock2010, Barnes2011a}. It is also in
  agreement with experimental~\cite{Mantica2009,Mantica2011} and
  numerical~\cite{Citrin2014} findings in the outer core of the JET experiment,
  which also showed that ion heat transport's stiffness is not affected by an
  increase in $\gamma_E$, whereas the critical $\kappa_T$ threshold does
  increase with $\gamma_E$.

  \begin{table}[t]
    \centering
    \caption{Parameter values for simulations that match the experimental
      heat flux, $Q_i^{\exp}/Q_{\mathrm{gB}} = 2 \pm 1$.
    }
    \begin{tabular}{c c c}
      \toprule
      $\kappa_T$ & $\gamma_E$ & $Q_i/Q_{\mathrm{gB}}$ \\
      \midrule
      4.4 & 0.14 & $1.3 \pm 0.1$ \\
      4.45 & 0.14 & $1.0 \pm 0.1$ \\
      4.8 & 0.16 & $1.44 \pm 0.05$ \\
      4.85 & 0.16 & $1.2 \pm 0.1$ \\
      5.15 & 0.18 & $4 \pm 1$ \\
      5.2 & 0.18 & $4 \pm 1$ \\
      \bottomrule
    \end{tabular}
    \label{tab:exp_match_sims}
  \end{table}

\subsection{Subcritical turbulence}
\label{sec:subcritical}
  We have found that in all our simulations with $\gamma_E>0$, small amplitude
  initial perturbations decayed (i.e. the system was linearly stable) and a
  finite initial perturbation was always required in order to ignite turbulence
  and reach a saturated turbulent state. Turbulence in MAST in the equilibrium
  configuration that we study here belongs to the class of subcritical
  systems~\cite{Trefethen1993,Schekochihin2012,
  Highcock2012,Landreman2015}, where linear modes are formally stable, but may
  be transiently amplified by a given factor over a given time. If the
  transient amplification is sufficient for nonlinear interactions to become
  significant before the modes decay, then a turbulent state may emerge. This
  turbulent state persists provided the fluctuation amplitudes do not fall
  below some critical value (for example, by way of the chaotic evolution,
  with occasional large deviations from an average fluctuation level that
  characterises the turbulent state) below which they cannot be transiently
  amplified once again back to nonlinearly sustained saturated level.

  In this work, we have assumed that other activity in the experiment (e.g.
  large-scale MHD modes or more virulent turbulence on neighbouring flux
  surfaces) can generate arbitrarily large perturbations as an initial
  condition to our system. For this reason, we have used the largest initial
  perturbation allowed by the numerical algorithm used in GS2, i.e., as large
  as possible without forcing the system to evolve the distribution function
  with time steps so small that the simulations would require prohibitively
  long simulation times. All nonlinear simulations presented in
  section~\ref{sec:heat_flux} were run with such large initial conditions. For
  the regions where we have reported $Q_i = 0$, we could not ignite turbulence
  using even the largest initial condition tolerated by the GS2 algorithm. In
  this section we will demonstrate the subcritical nature of the turbulence by
  investigating the effect of changing the amplitude of the initial
  perturbation in both linear and nonlinear simulations.

  \subsubsection{Minimum initial perturbation amplitude}
  We start by considering the nonlinear time evolution of a simulation at the
  nominal equilibrium parameters $(\kappa_T, \gamma_E) = (5.1, 0.16)$.
  \Figref{phiinit} shows $Q_i/Q_{\mathrm{gB}}$ as a function of time for
  nonlinear simulations with increasing initial amplitude.  These parameter
  values represent a simulation somewhat away from the turbulence threshold
  [see~\figref{contour_heatmap}] and yet, for a range of initial amplitudes, we
  see that the system decays rapidly. This is a clear indication that the
  turbulence is subcritical. We see that there is a certain minimum initial
  perturbation amplitude starting from which it is possible for the system to
  reach a saturated state, rather than decay. Importantly, for simulations that
  do reach a saturated state, the level of saturation does not depend on the
  amplitude of the initial perturbation.

  \subsubsection{Finite life time of turbulence}
  \begin{figure}[t]
    \centering
    \begin{subfigure}[t]{0.49\textwidth}
      \includegraphics[width=\textwidth]{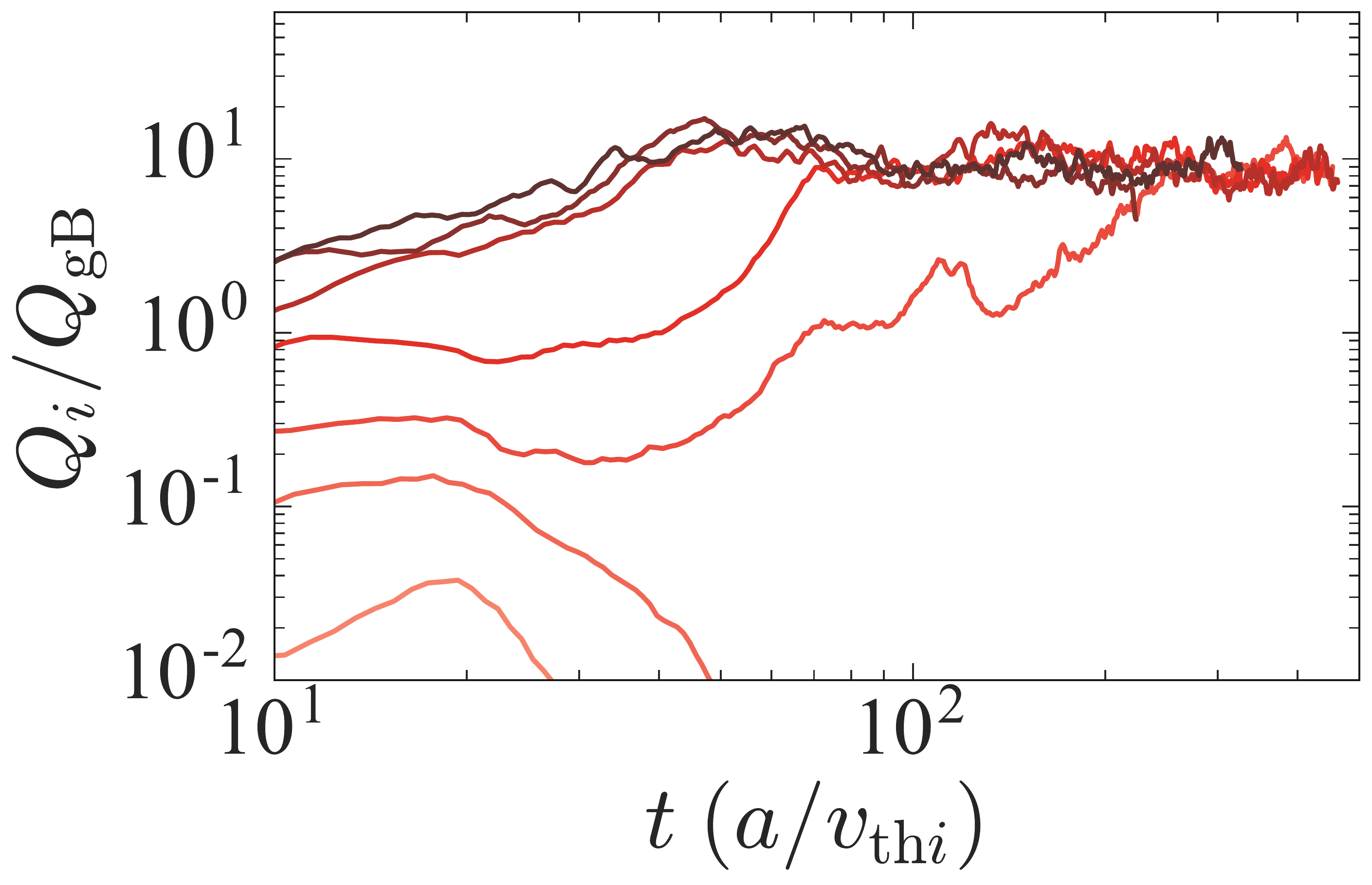}
      \caption{}
      \label{fig:phiinit}
    \end{subfigure}
    \begin{subfigure}[t]{0.49\textwidth}
      \includegraphics[width=\linewidth]{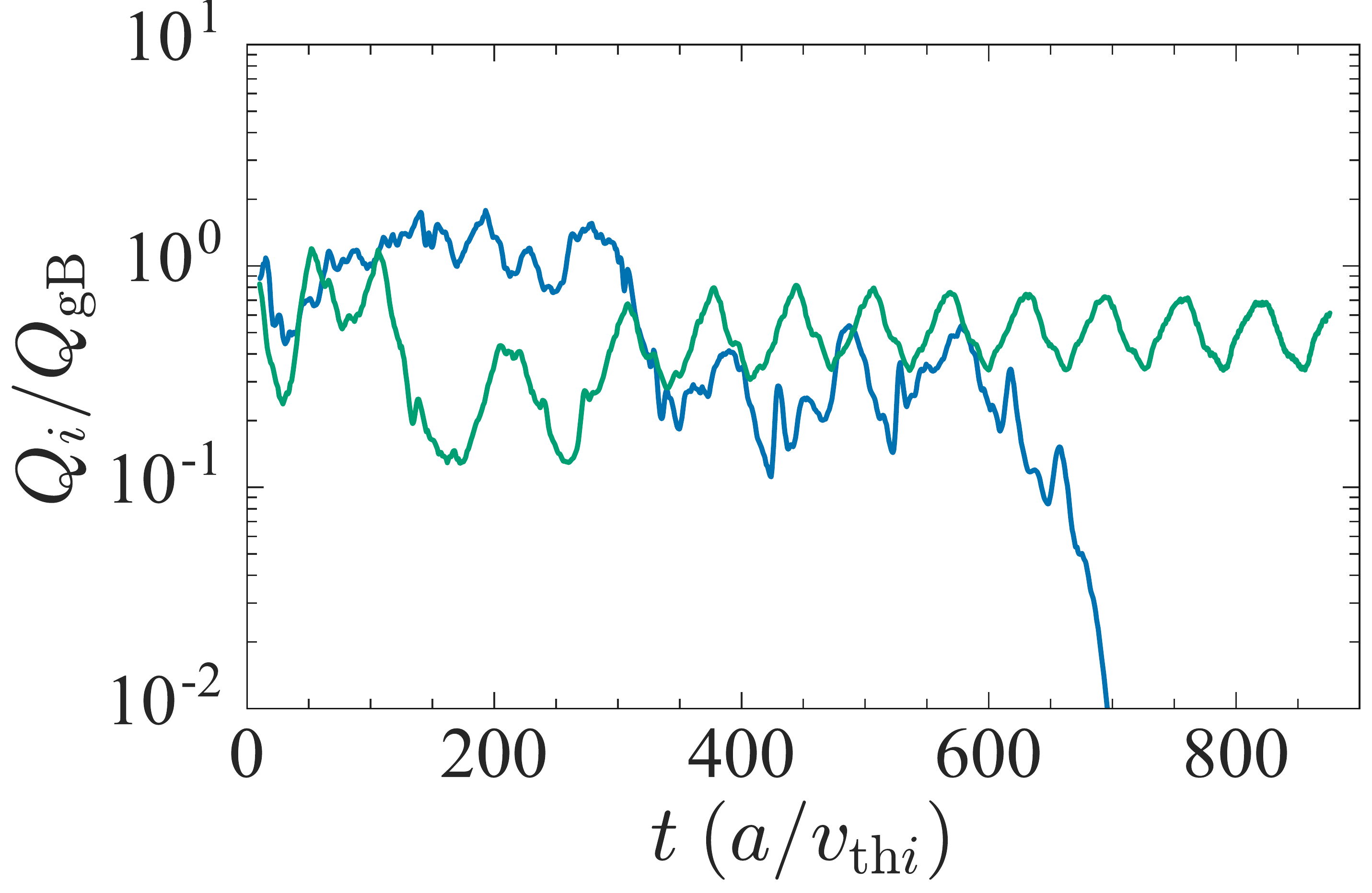}
      \caption{}
      \label{fig:subcrit_demo}
    \end{subfigure}
    \caption{
      \subref*{fig:phiinit} Ion heat flux $Q_i/Q_{\mathrm{gB}}$ as a
      function of time for different initial-perturbation amplitudes for
      $(\kappa_T, \gamma_E) = (5.1, 0.16)$, keeping all other parameters the
      same.
      \subref*{fig:subcrit_demo} $Q_i/Q_{\mathrm{gB}}$ as a function of
      time for two identical simulations at $(\kappa_T, \gamma_E ) = (5.1, 0.18)$.
      The difference between the time series shown as the blue and green lines
      is the realisation of the random noise with which GS2 initialised these
      simulation. Beyond $t=300$~$(a/v_{\mathrm{th}i})$, the simulations seem
      to converge to a similar average value before one is abruptly quenched
      due to the amplitudes falling below the critical values required to
      sustain a saturated state.
    }
  \end{figure}
  A large initial perturbation is not sufficient to guarantee that a
  subcritical system continues in a statistically steady state indefinitely. In
  simulations with equilibrium parameters close to the turbulence threshold, we
  found that turbulence could be quenched at a seemingly unpredictable time.
  For example, \figref{subcrit_demo} shows the time trace of
  $Q_i/Q_{\mathrm{gB}}$ for two identical simulations at the parameter values
  $(\kappa_T, \gamma_E) = (5.1, 0.18)$, close to the turbulence threshold.
  These simulations were initialised with random noise of a given amplitude in
  each Fourier mode and the only difference between the two simulations is the
  realisation of this random noise. We see the simulations saturate at a
  similar level beyond $t=300$~$(a/v_{\mathrm{th}i})$, but then one of them
  abruptly decays. This is another indication that the system is subcritical:
  the decaying simulation has fallen below the critical amplitude needed to
  sustain turbulence. Practically, in this study, we decided that a simulation
  reached a saturated state if the heat flux evolved at a roughly constant
  value for at least $200$~$(a/v_{\mathrm{th}i})$.

  The finite life time of turbulence in subcritical systems is well established
  in some hydrodynamic systems, such as fluid flow in a pipe~\cite{Faisst2004}.
  By running a large number of identical pipe-flow
  experiments~\cite{Peixinho2006, Hof2006, Avila2011} and numerical
  simulations~\cite{Faisst2004, Hof2006, Avila2010, Avila2011}, it was shown
  that the ``life time'' of subcritical turbulence (the characteristic time
  that elapses before turbulence decays to laminar flow) is a function of the
  Reynolds number.  The Reynolds number in pipe flows quantifies the ``distance
  from the turbulence threshold''.  In particular, it was shown that the larger
  its value (i.e., the further the system is from the turbulence threshold),
  the longer the turbulence is likely to persist. More recently, the same
  phenomenon of finite turbulence lifetime was observed in MHD simulations of
  astrophysical Keplerian shear flow systems~\cite{Rempel2010}, where the
  distance from threshold was characterised by the magnetic Reynolds number and
  the turbulence persists longer for large values of this parameter.

  Given the above considerations, we would also expect the subcritical
  turbulence considered here to persist for longer times at larger values of
  $Q_i/Q_{\mathrm{gB}}$. The pipe-flow and astrophysical studies referred to
  above relied on running many experiments and simulations in order to build up
  sufficient statistics to determine the dependence of the turbulence lifetimes
  on the system parameters. With the high resolutions demanded by nonlinear
  gyrokinetic simulations of plasmas in the core of tokamaks we are neither
  able to run a sufficient number of simulations nor to run them for a
  sufficient amount of time to determine the turbulence lifetimes for our
  system.  However, this may be possible in future, given advances in computing
  and numerics or through the use of reduced models.

  \subsubsection{Transient growth of perturbations}
  A system can reach a saturated turbulent state despite being stable to
  infinitesimal perturbations due to transient growth of (large enough) finite
  perturbations.  This transient growth can sustain turbulence
  provided perturbations reach an amplitude sufficient for nonlinear
  interaction. Having established the subcritical nature of the system, the
  question we would now like to address is how much transient growth is
  sufficient for the system to reach a turbulent state. We have already seen
  which values of $\kappa_T$ and $\gamma_E$ lead to a turbulent state [see
  \figref{contour_heatmap}] and we now investigate transient growth of
  perturbations via linear GS2 simulations at these values of $\kappa_T$ and
  $\gamma_E$.

\begin{figure}[t]
    \centering
    \begin{subfigure}[t]{0.49\textwidth}
      \includegraphics[width=\textwidth]{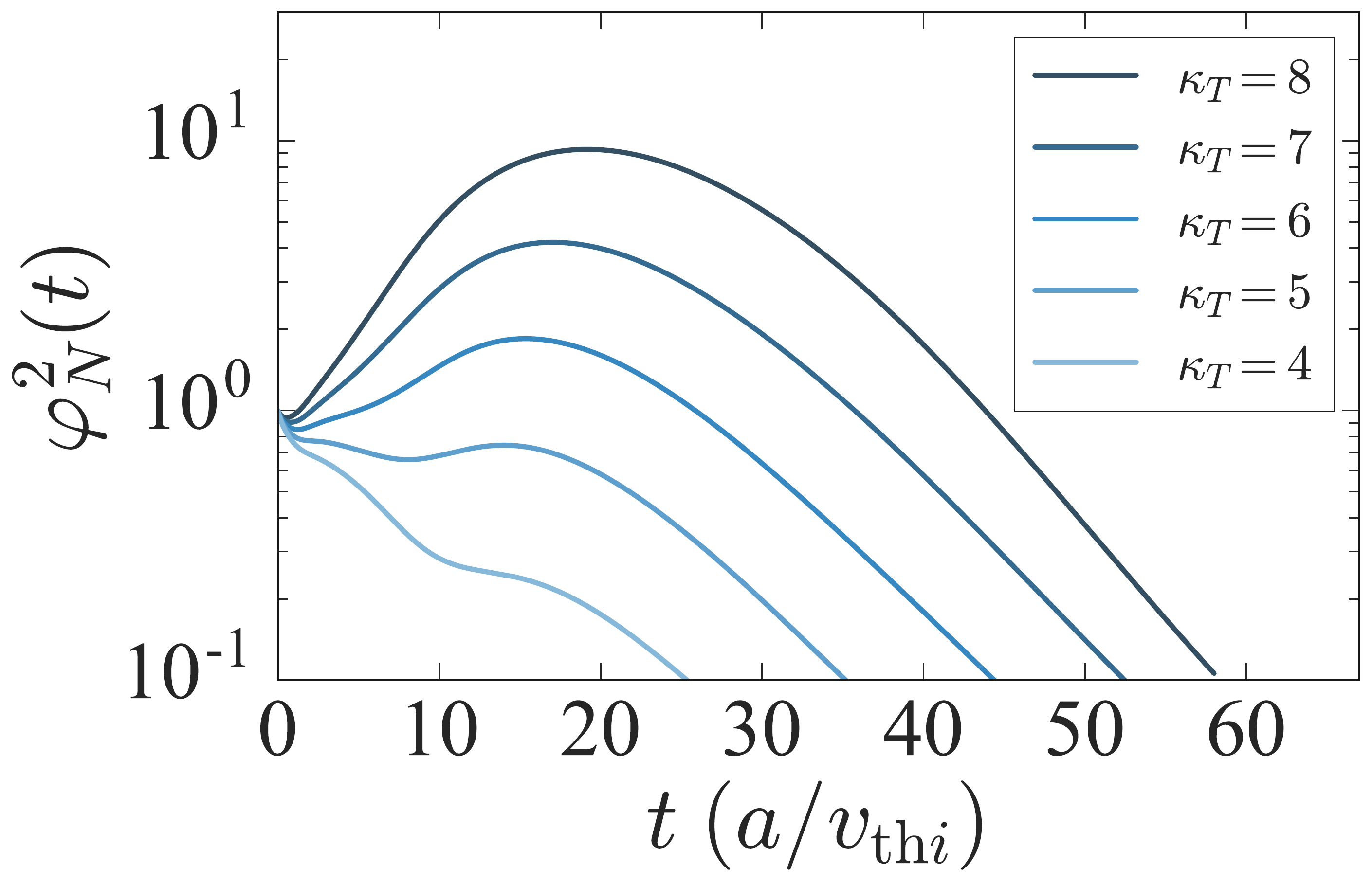}
      \caption{}
      \label{fig:transient}
    \end{subfigure}
    \hfill
    \begin{subfigure}[t]{0.49\textwidth}
      \includegraphics[width=\linewidth]{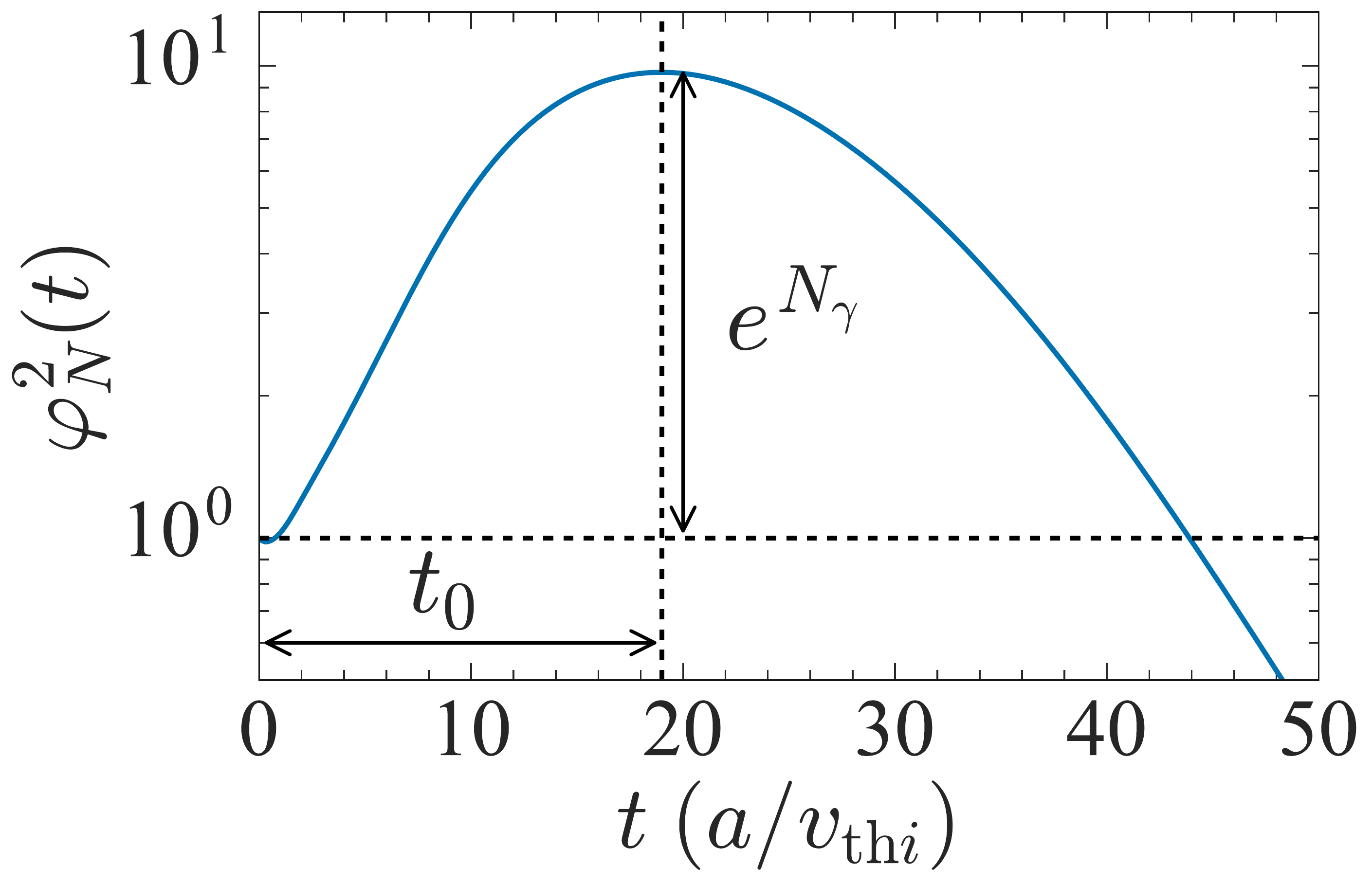}
      \caption{}
      \label{fig:phi_transient}
    \end{subfigure}
    \caption{
      \subref*{fig:transient}
      Transient growth of initial perturbations of the electrostatic potential
      $\varphi^2_N(t)$ (normalised to the time at which flow shear is switched
      on) at $\gamma_E = 0.16$, for a range of $\kappa_T$ values.
      These time evolutions were obtained from purely linear simulations for
      $k_y \rho_i = 0.2$, approximately the wavenumber that gives the largest
      transient growth [see~\figref{N_16}], and summed over $k_x$. As
      $\kappa_T$ is increased, the strength of the transient growth also
      increases.
      \subref*{fig:phi_transient} $\varphi^2_N(t)$ as a function of time for a
      strongly growing mode at $(\kappa_T, \gamma_E, k_y \rho_i) = (8,
      0.16, 0.2)$ further illustrating transient amplification.  The total
      amplification factor is $e^{N_\gamma}$ and the time taken to reach
      maximal amplification is $t_0$.
    }
  \end{figure}

  We performed an extensive series of linear simulations and calculated the
  time evolution of the electrostatic potential $\varphi$ as a function of $k_y
  \rho_i$, $\kappa_T$, and $\gamma_E$. \Figref{transient} shows an example of
  the time evolution of $\varphi$ (at $k_y \rho_i = 0.2$ and $\gamma_E=0.16$) for a
  range of $\kappa_T$, normalised to the value of $\varphi$ at the time (called
  $t=0$) when the flow shear is switched on, that is, $\varphi_N^2(t) =
  \varphi^2(t)/\varphi^2(0)$. We have averaged $\varphi$ over $k_x$.
  \Figref{transient} illustrates the phenomenon of transient growth in a
  subcritical system and we see that, as $\kappa_T$ is increased, the system
  exhibits stronger transient growth.  At $\gamma_E = 0.16$, we saw in
  \figref{contour_heatmap} that turbulence could be sustained at $\kappa_T
  \gtrsim 4.8$. Indeed, \figref{transient} shows that there is only a marginal
  amount of transient growth at $\kappa_T \approx 4.8$.

  \subsubsection{Characterising transient growth}
  For linear simulations exhibiting transient growth, one cannot define a
  ``linear growth rate'', as one does for linear simulations with $\gamma_E =
  0$ where $\varphi(t)$ grows exponentially.  However, methods for determining
  an ``effective'' linear growth rate have been outlined in
  Ref.~\cite{Roach2009} and~\cite{Schekochihin2012}. Here, we follow
  Ref.~\cite{Schekochihin2012} and use the ``transient-amplification factor''
  as a measure of the vigour of the transient growth.  For a total
  amplification factor $e^{N_\gamma}$, the amplification exponent $N_\gamma$ is
  defined by
  \begin{equation}
    N_\gamma = \frac{1}{2} \ln \frac{\varphi^2(t_0)}{\varphi^2(0)} =
    \int_0^{t_0} \dd t \gamma(t),
    \label{amp_exponent}
  \end{equation}
  where $t_0$ is the time taken to reach the maximum amplification, and
  $\gamma(t)$ is the time-dependent growth rate. These quantities are
  illustrated in \Figref{phi_transient}, which shows a typical linear
  simulation with strong amplification, with $e^{N_\gamma}$ and $t_0$
  indicated.

  It was argued in Ref.~\cite{Schekochihin2012} that the parameters $N_\gamma$
  and $t_0$ determine whether turbulence can be sustained, in the following way.
  Perturbations grow only transiently because flow shear leads to
  $k_x(t) = k_x(0) - \gamma_E k_y t$ being swept from the region where
  perturbations are unstable to larger values, where they are stabilised by
  dissipation. If nonlinear interactions scatter energy back into the unstable
  modes before perturbations decay they can be transiently amplified once
  again, and so on. In this way, a nonlinear saturated state can be sustained.
  The typical timescale for nonlinear interactions is the nonlinear
  decorrelation time $\tau_\mathrm{NL} \sim 1/k_\perp V_E$, where $k_\perp$ is
  the typical perpendicular wave number, and $V_E$ is given by \eqref{v_exb}.
  To sustain turbulence, transient growth should last at least as long as one
  nonlinear decorrelation time:
  \begin{align}
    \begin{split}
      t_0 &\gtrsim \tau_{\mathrm{NL}}.
      \label{schek_t0}
    \end{split}
  \end{align}
  At the same time, the rate of amplification should be at least comparable to
  the nonlinear decorrelation rate:
  \begin{equation}
    \frac{N_\gamma}{t_0} \gtrsim \frac{1}{\tau_{\mathrm{NL}}}.
    \label{schek_gamma_eff}
  \end{equation}
  Combining \eqref{schek_t0} and \eqref{schek_gamma_eff}, we see that a
  sustained turbulent state requires
  \begin{equation}
    N_\gamma \gtrsim 1.
    \label{schek_N}
  \end{equation}

  \subsubsection{Conditions for the onset of subcritical turbulence}
\begin{figure}[t]
    \centering
    \begin{subfigure}[t]{0.49\textwidth}
      \includegraphics[width=\textwidth]{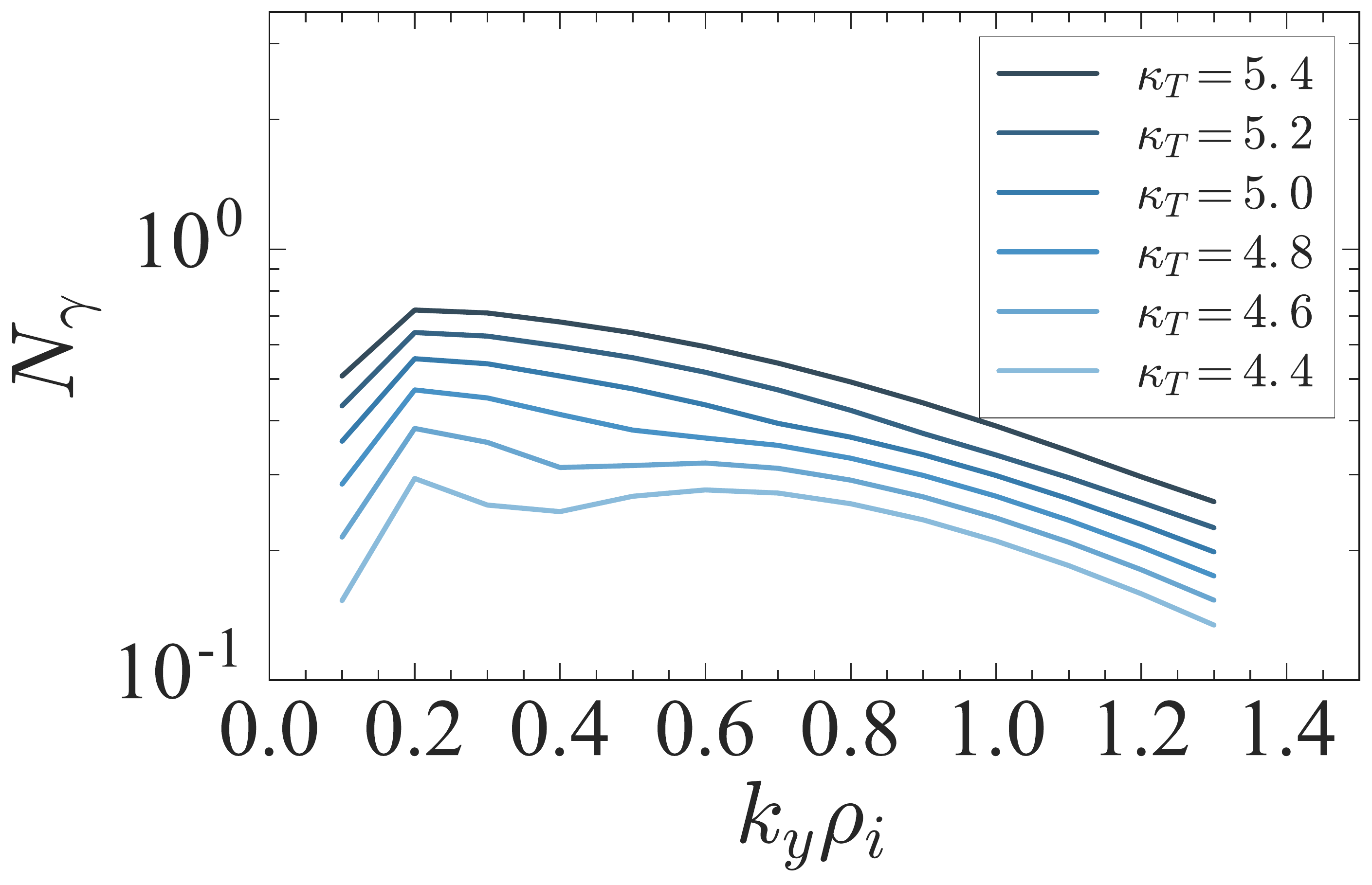}
      \caption{}
      \label{fig:N_16}
    \end{subfigure}
    \hfill
    \begin{subfigure}[t]{0.49\textwidth}
      \includegraphics[width=\textwidth]{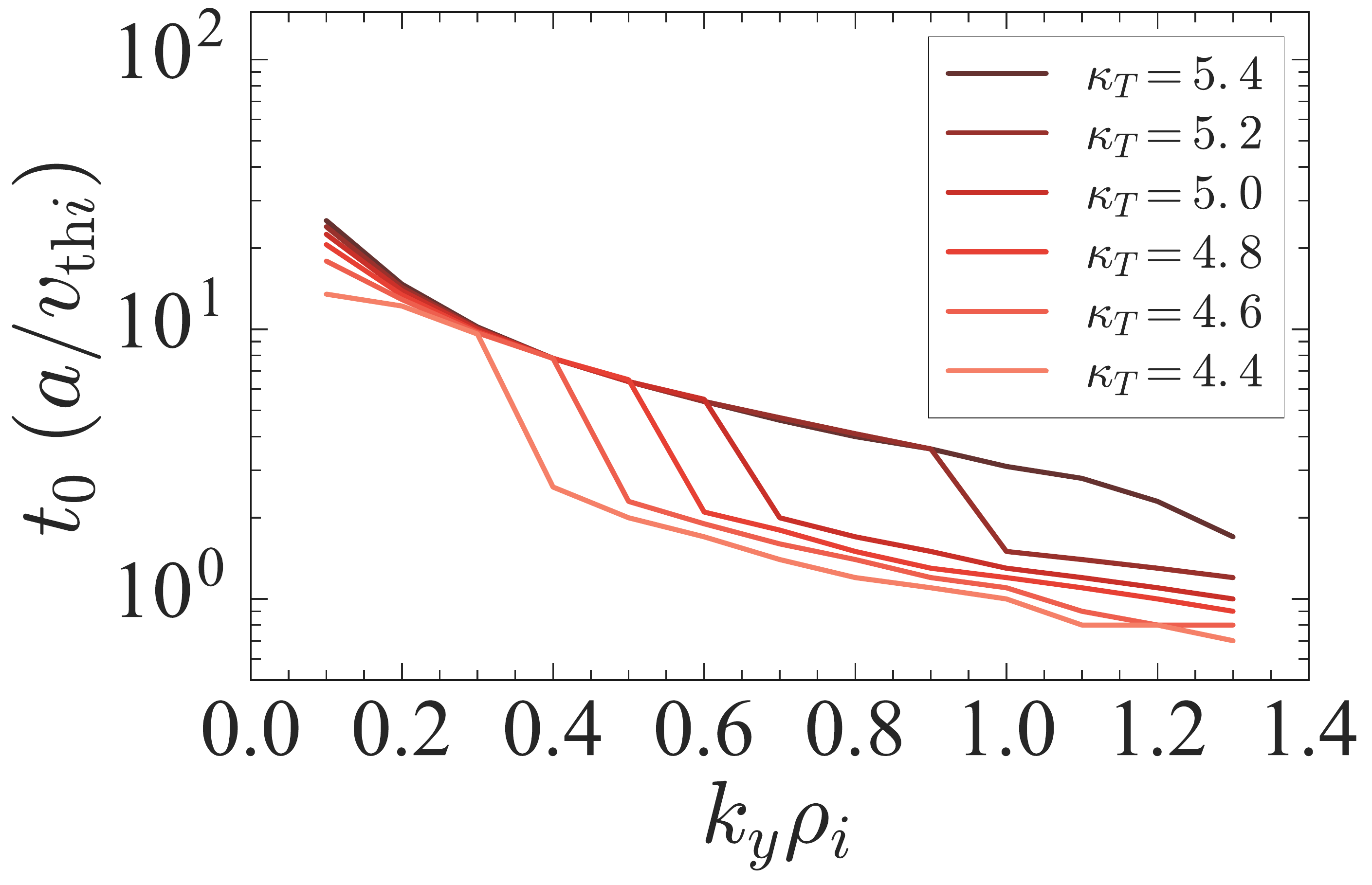}
      \caption{}
      \label{fig:t0_16}
    \end{subfigure}
    \caption{
      \subref*{fig:N_16} The transient-amplification factor $N_\gamma$, defined
      in \eqref{amp_exponent}, for a range of values of $\kappa_T$ at $\gamma_E =
      0.16$. $N_\gamma$ increases smoothly with increasing $\kappa_T$
      as the nonlinear threshold is passed.
      \subref*{fig:t0_16} Time $t_0$ taken to reach maximum amplification
      for a range of values of $\kappa_T$, also at $\gamma_E = 0.16$.
      Increasing $\kappa_T$ leads to transient amplification lasting for a
      longer time.
    }
    \label{fig:N_and_t0_16}
  \end{figure}

  We now want to
  estimate the critical values of $N_\gamma$ and $t_0$ above which turbulence
  is triggered and a saturated state can be established in our system.
  \Figref{N_and_t0_16} shows $N_\gamma$ and $t_0$ as functions of $k_y \rho_i$
  for a range of different $\kappa_T$ values at $\gamma_E = 0.16$ (only wave
  numbers up to $k_y \rho_i = 1.3$ are shown, because numerical dissipation
  effectively suppresses transient growth beyond this value). As a point of
  reference, for $\gamma_E = 0.16$, the transition to turbulence occurs at
  $\kappa_T \approx 4.8$ [see \figref{q_vs_tprim_marginal}]. For the linear
  simulations in \figref{N_and_t0_16}, we see a relatively smooth increase in
  $N_\gamma$ and $t_0$ as $\kappa_T$ is increased across this nonlinear
  threshold, with larger transient amplification and modes with smaller $k_y
  \rho_i$ experiencing amplification over a longer time period.

\begin{figure}[t]
    \centering
    \begin{subfigure}[t]{0.49\textwidth}
      \includegraphics[width=\textwidth]{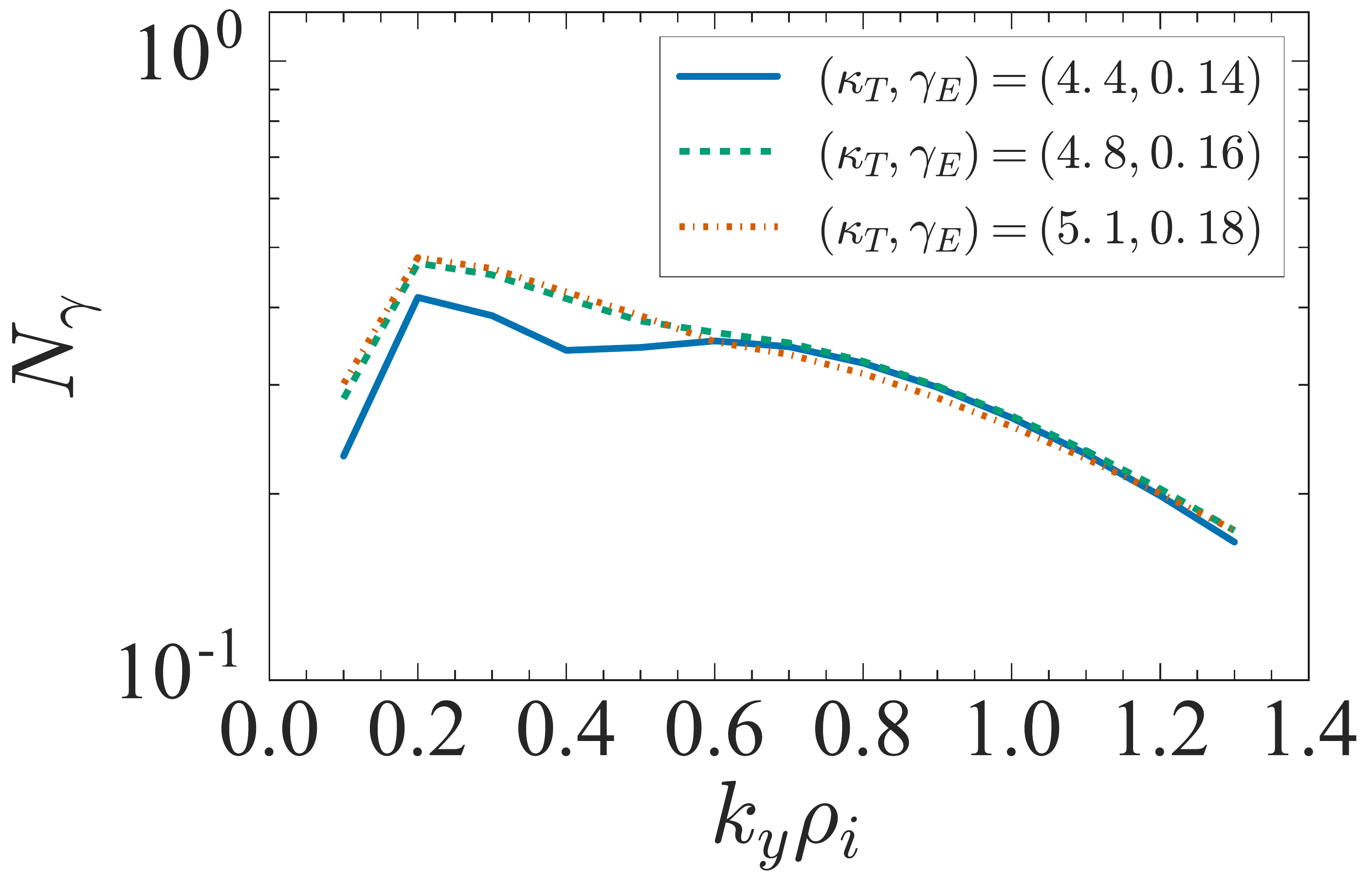}
      \caption{}
      \label{fig:N_marginal}
    \end{subfigure}
    \hfill
    \begin{subfigure}[t]{0.49\textwidth}
      \includegraphics[width=\textwidth]{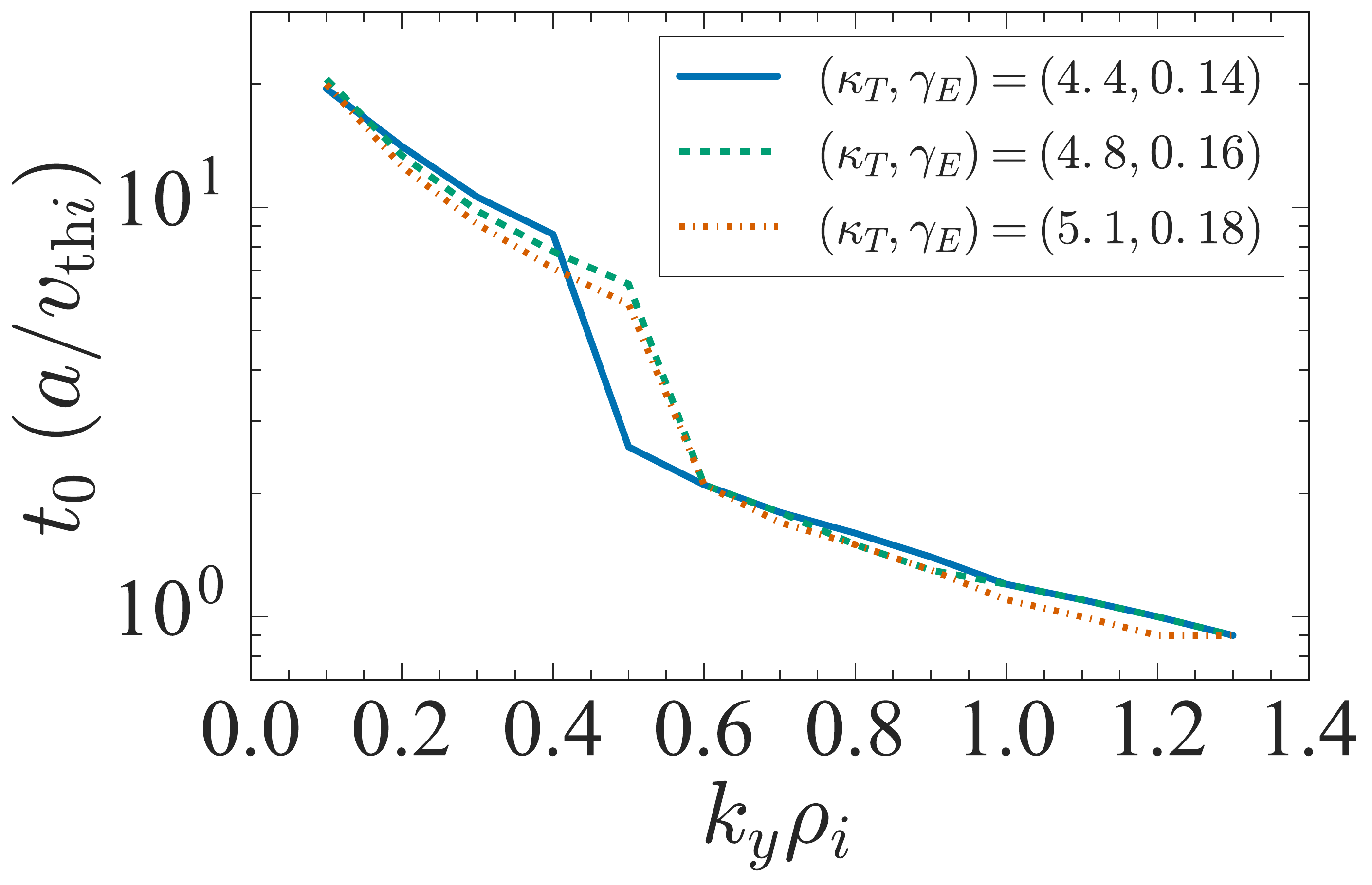}
      \caption{}
      \label{fig:t0_marginal}
    \end{subfigure}
    \caption[$N_\gamma$ and $t_0$ versus $k_y \rho_i$ for marginally unstable
             simulations]{
      \subref*{fig:N_marginal} Transient-amplification factor $N_\gamma$ [see
      equation~\eqref{amp_exponent}] and
      \subref*{fig:t0_marginal} transient-amplification time $t_0$
      for the three marginal simulations identified in
      section~\ref{sec:heat_flux}. The values of $N_\gamma$ and $t_0$ that
      correspond to the marginally unstable equilibria are approximately the
      same, suggesting that these represent the critical values required for
      the system to reach a saturated turbulent state.
    }
    \label{fig:N_and_t0_marginal}
  \end{figure}

  To investigate the conditions for the onset of turbulence, we consider
  $N_\gamma$ and $t_0$ for the marginally unstable simulations identified in
  section~\ref{sec:heat_flux}. Figures~\ref{fig:N_marginal} and
  \subref{fig:t0_marginal} show $N_\gamma$ and $t_0$ as functions of $k_y
  \rho_i$ for $(\kappa_T, \gamma_E) = (4.4, 0.14), (4.8, 0.16), (5.1, 0.18)$.
  We see that both $N_\gamma$ and $t_0$ are roughly the same for our marginally
  unstable simulations, suggesting that the values shown in
  Figures~\ref{fig:N_marginal} and \subref{fig:t0_marginal} are indeed the
  critical values necessary for the onset of turbulence.

\begin{figure}[t]
    \centering
    \includegraphics[width=0.6\linewidth]{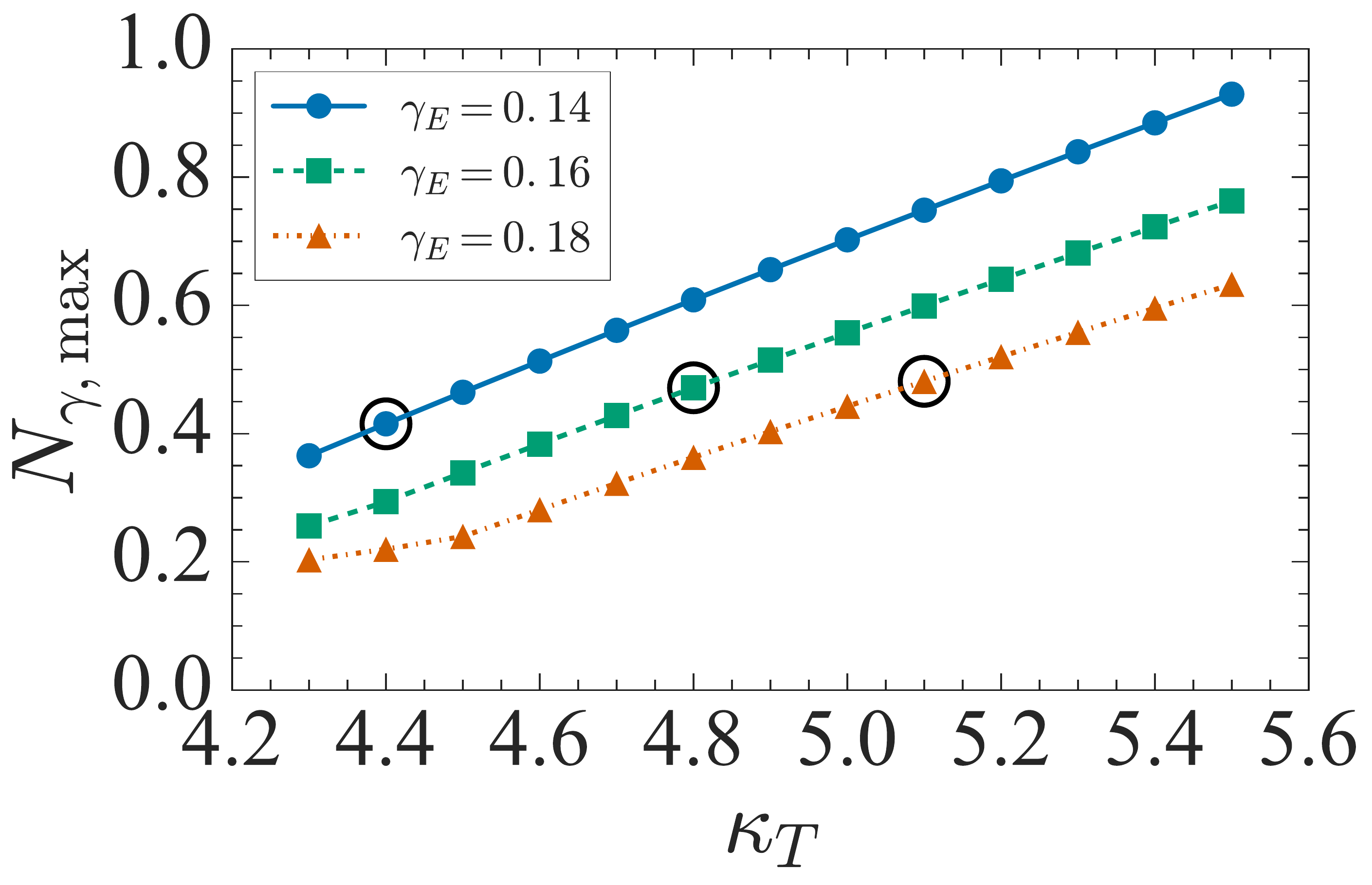}
    \caption[Maximum transient-amplification factor]{
      Maximum transient-amplification factor $N_{\gamma,\max}$ versus
      $\kappa_T$ for three values of $\gamma_E$ within the range of
      experimental uncertainty. The simulations circled in black represent the
      critical values of $\kappa_T$ above which turbulence can be sustained,
      suggesting the onset of turbulence occurs at $N_{\gamma,\max} \simeq 0.4$.
    }
    \label{fig:max_trans_amp}
  \end{figure}

  We see from \figsref{N_16}{N_marginal} that the maximum $N_\gamma$ is at $k_y
  \rho_i \approx 0.2$ and we consider its value here to to determine the
  critical condition. \Figref{max_trans_amp} shows the maximum value
  $N_{\gamma,\max}$ of the transient-amplification factor as a function of
  $\kappa_T$. The marked simulations are for the critical values of $\kappa_T$
  above which turbulence can be sustained, given a sufficiently large initial
  perturbation amplitude. \Figref{max_trans_amp} shows that $N_{\gamma,\max}$
  scales linearly with $\kappa_T$ for each $\gamma_E$, with higher values of
  $\gamma_E$ resulting in lower values of $N_{\gamma,\max}$.  The other
  important feature is that the values of $N_{\gamma,\max}$ at the critical
  values of $\kappa_T$ are similar, giving an approximate critical condition:
  $N_{\gamma,\max} \sim 0.4$. This  value of $N_{\gamma,\max}$ is comparable to
  that found in previous work~\cite{Schekochihin2012,Highcock2012}.

  Returning to \figref{t0_marginal}, and assuming that low-$k_y$ modes are the
  important ones for sustaining turbulence, it is reasonable to estimate that
  the onset of turbulence requires $t_0 \gtrsim 10$~$(a/v_{\mathrm{th}i})$.
  We will return to the comparison of $t_0$ with $\tau_{\mathrm{NL}}$ after
  estimating $\tau_{\mathrm{NL}}$ in section~\ref{sec:time_scales}, where we
  confirm that $t_0 \gtrsim \tau_{\mathrm{NL}}$ and, therefore, that a
  sustained turbulent state requires an amplification time comparable to (or
  greater than) the nonlinear decorrelation time.

  We have shown that the changes in $N_\gamma$ and $t_0$ are relatively smooth
  as the turbulence threshold is surpassed (determined from our simulations in
  section~\ref{sec:heat_flux}), suggesting nonlinear simulations are essential
  in predicting the \emph{exact} transition to turbulence.  In the next
  section, we will investigate the nature of this transition by considering the
  real-space structure of the turbulence in our nonlinear simulations.

\subsection{Structure of turbulence close to and far from the threshold}
\label{sec:struc_analysis}

  Having established the subcritical nature of the system, we now investigate
  the consequences for the structure of turbulence. We will argue that our
  subcritical system supports the formation of long-lived coherent structures
  close to the turbulence threshold. In this context, we take ``coherent''
  to mean turbulent structures that remain distinct in space as they move
  through the simulation domain and exist for (most of) the duration of
  the simulation (see section~\ref{sec:coherent_strucs}).
  We will also show that the heat flux is proportional to the
  product of number of these structures and their maximum amplitude, and that
  the properties of the turbulence are characterised by the ``distance from
  threshold'' (as opposed to the specific values of the stability parameters
  $\kappa_T$ and $\gamma_E$), as measured, for example, by the turbulent ion
  heat flux. We previously reported some of these results in
  Ref.~\cite{VanWyk2016}, based on the simulations in this study, and provide a
  more comprehensive description here.

  \subsubsection{Coherent structures in the near-marginal state}
  \label{sec:coherent_strucs}
  \begin{figure}[t]
    \centering
    \begin{subfigure}{0.49\linewidth}
      \includegraphics[width=\linewidth]{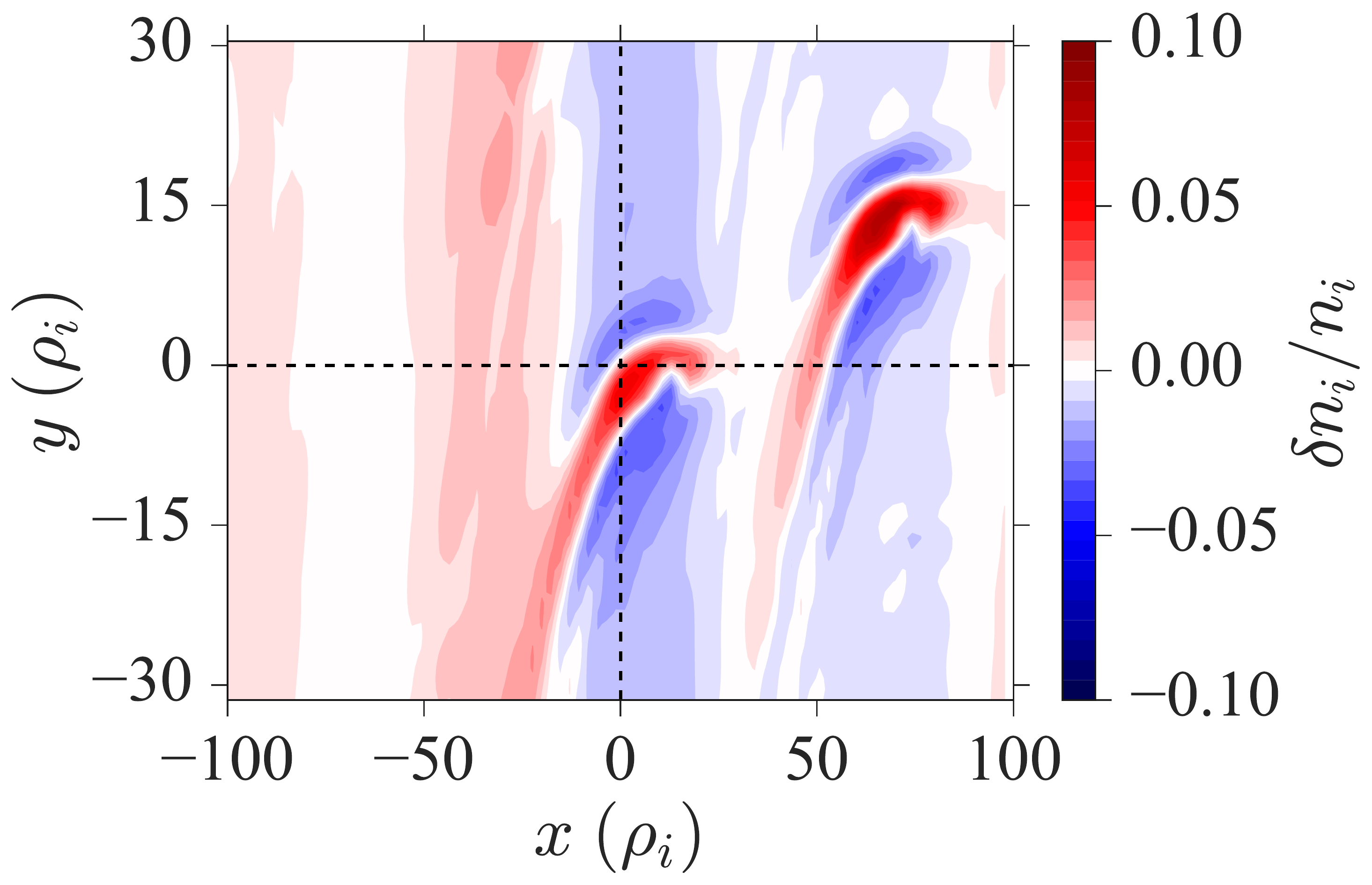}
      \caption{}
      \label{fig:marginal}
    \end{subfigure}
    \begin{subfigure}{0.49\linewidth}
      \includegraphics[width=\linewidth]{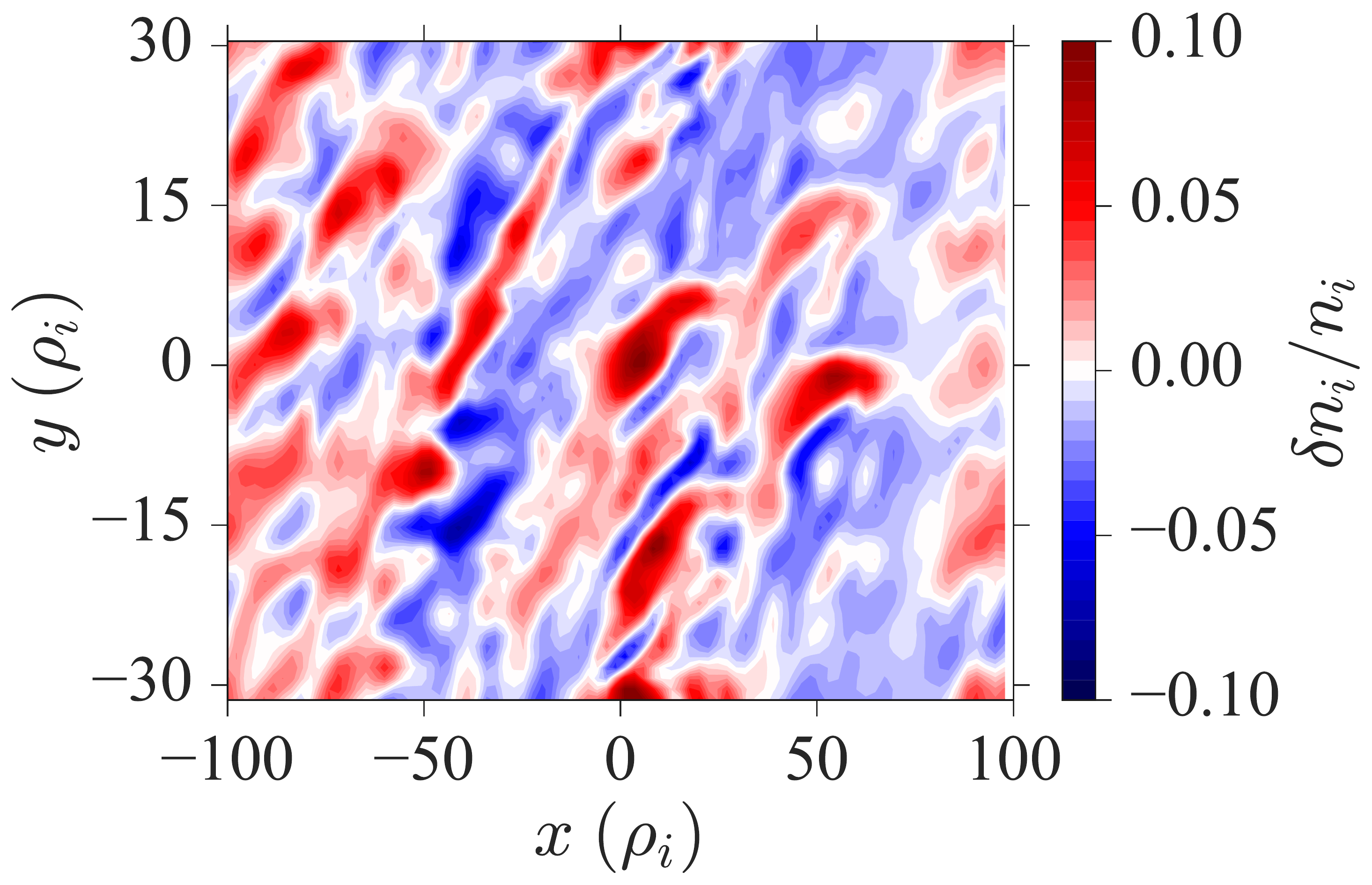}
      \caption{}
      \label{fig:intermediate}
    \end{subfigure}
    \begin{subfigure}{0.49\linewidth}
      \includegraphics[width=\linewidth]{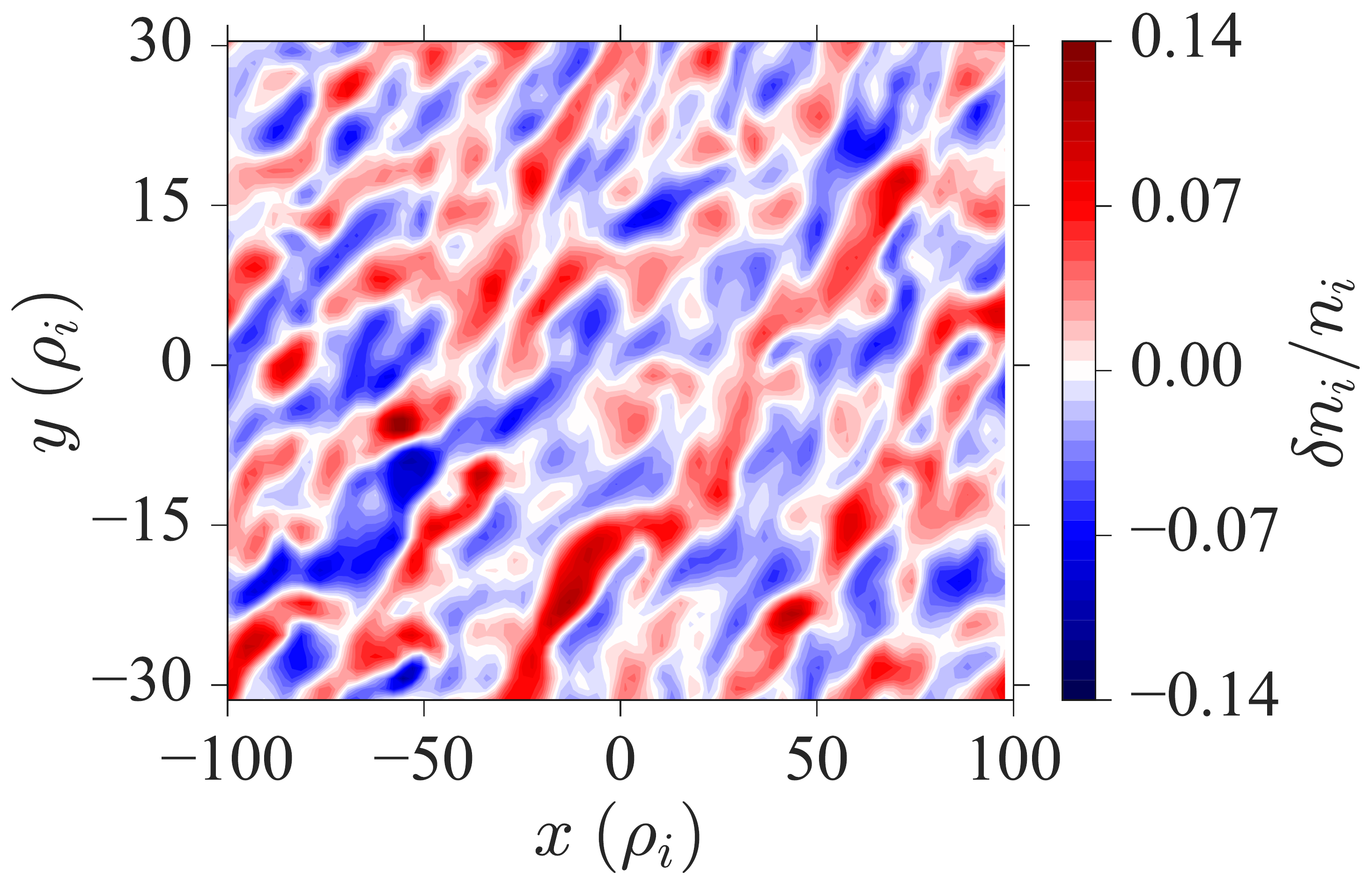}
      \caption{}
      \label{fig:strongly_driven}
    \end{subfigure}
    \begin{subfigure}{0.49\linewidth}
      \includegraphics[width=\linewidth]{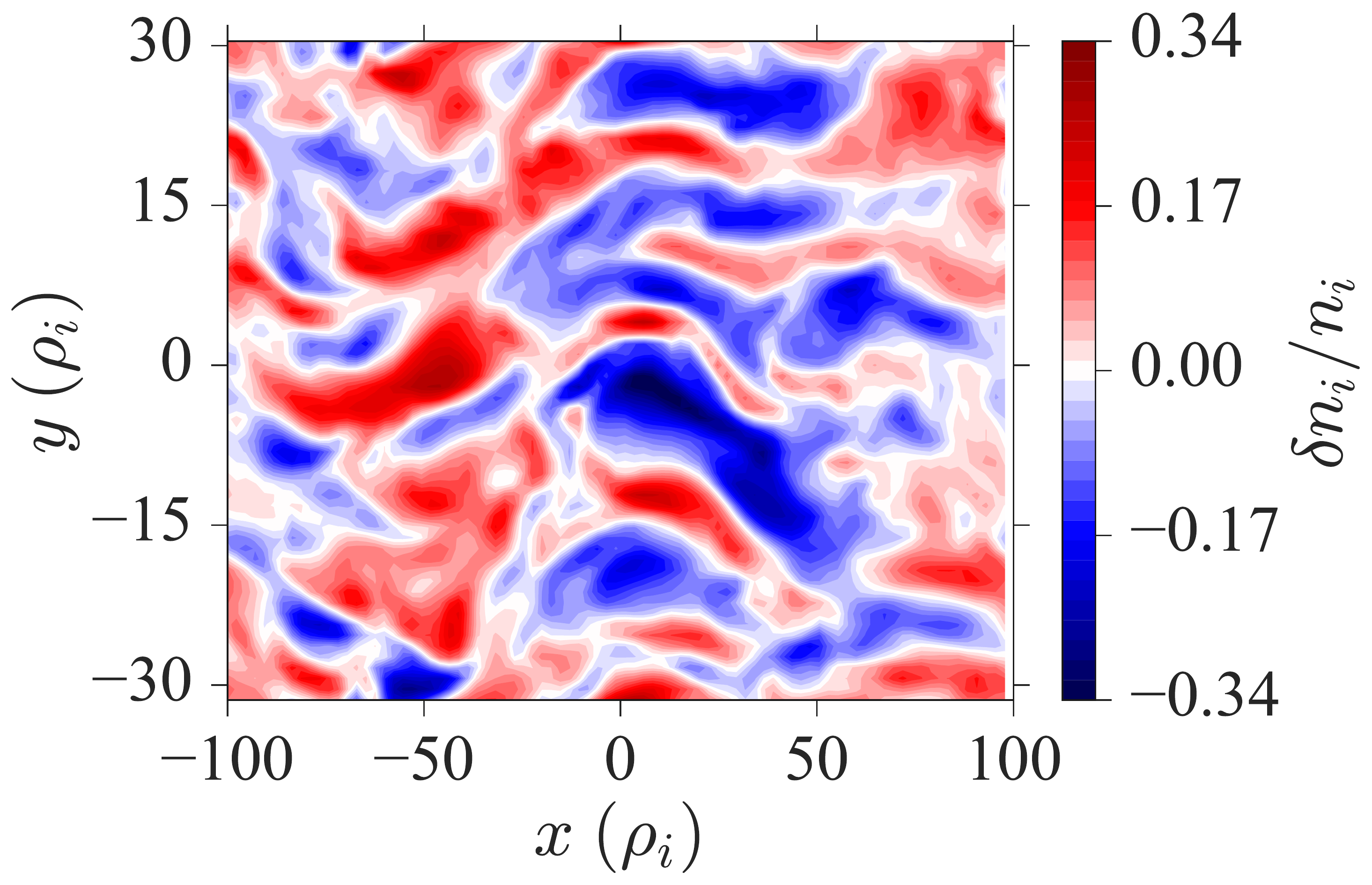}
      \caption{}
      \label{fig:no_shear}
    \end{subfigure}
    \caption[Real-space density-fluctuation fields on $(x,y)$ plane]{
      Density-fluctuation field $\delta n_i/n_i$ at the outboard midplane of
      MAST as a function of the local GS2 coordinates $x$ and $y$, for four
      combinations of stability parameters.
      \subref*{fig:marginal} Near-threshold turbulence, $(\kappa_T, \gamma_E) =
      (4.8, 0.16)$. The dashed lines indicate the planes of constant $x$ and
      $y$ used to demonstrate the parallel structure in
      \figref{parallel_density}.
      \subref*{fig:intermediate} Turbulence intermediate between the
      near-threshold and strongly driven cases, $(\kappa_T, \gamma_E) =
      (4.9,0.16)$.
      \subref*{fig:strongly_driven} Strongly driven turbulence, $(\kappa_T,
      \gamma_E) = (5.2,0.16)$.
      \subref*{fig:no_shear} Turbulence without flow shear, $(\kappa_T,
      \gamma_E) = (5.2,0)$, showing strong zonal flows.
	}
    \label{fig:density_fluctuations}
  \end{figure}
\begin{figure}[t]
    \centering
    \begin{subfigure}{0.49\linewidth}
      \includegraphics[width=\linewidth]{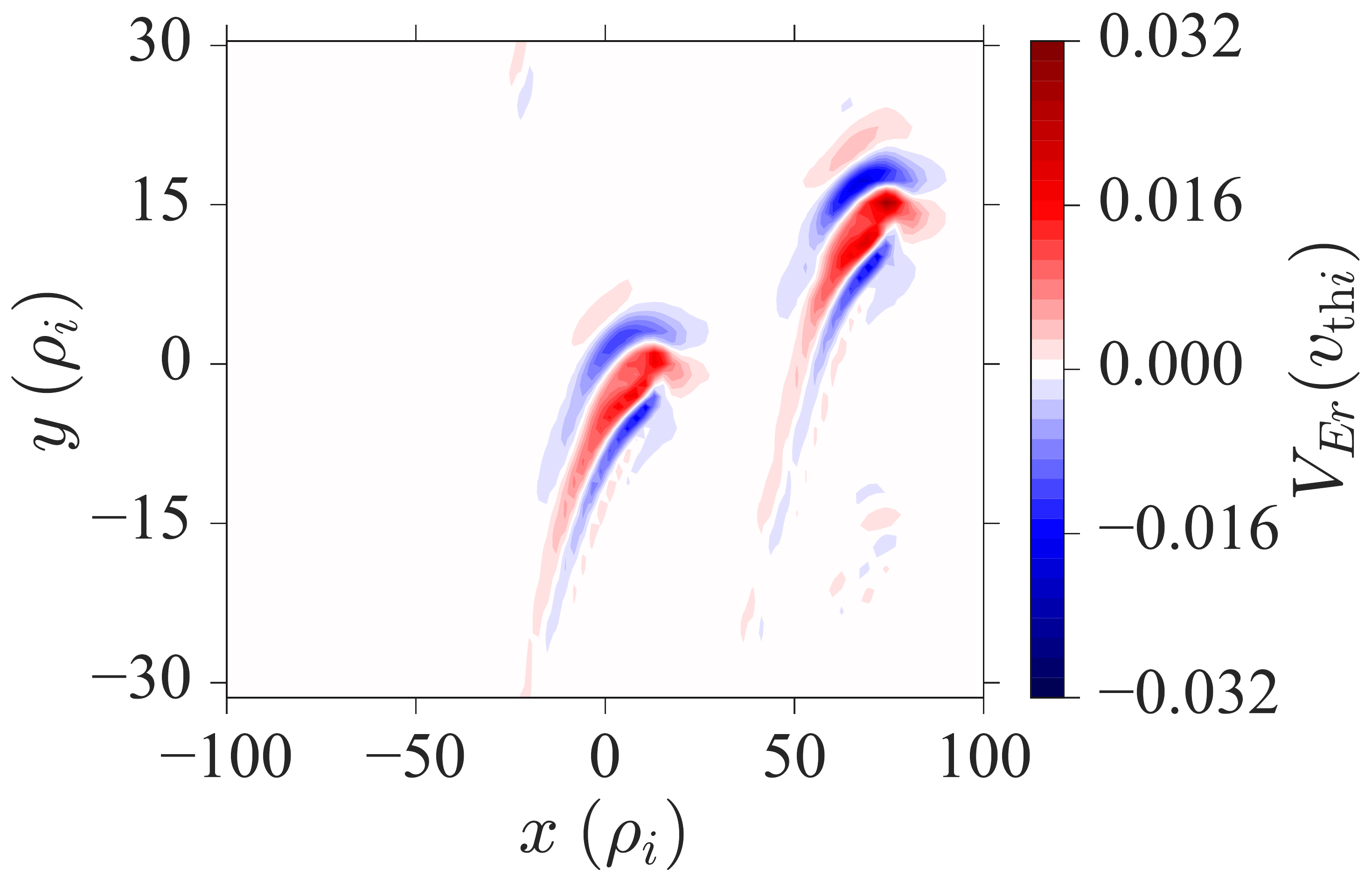}
      \caption{}
      \label{fig:vel_marginal}
    \end{subfigure}
    \begin{subfigure}{0.49\linewidth}
      \includegraphics[width=\linewidth]{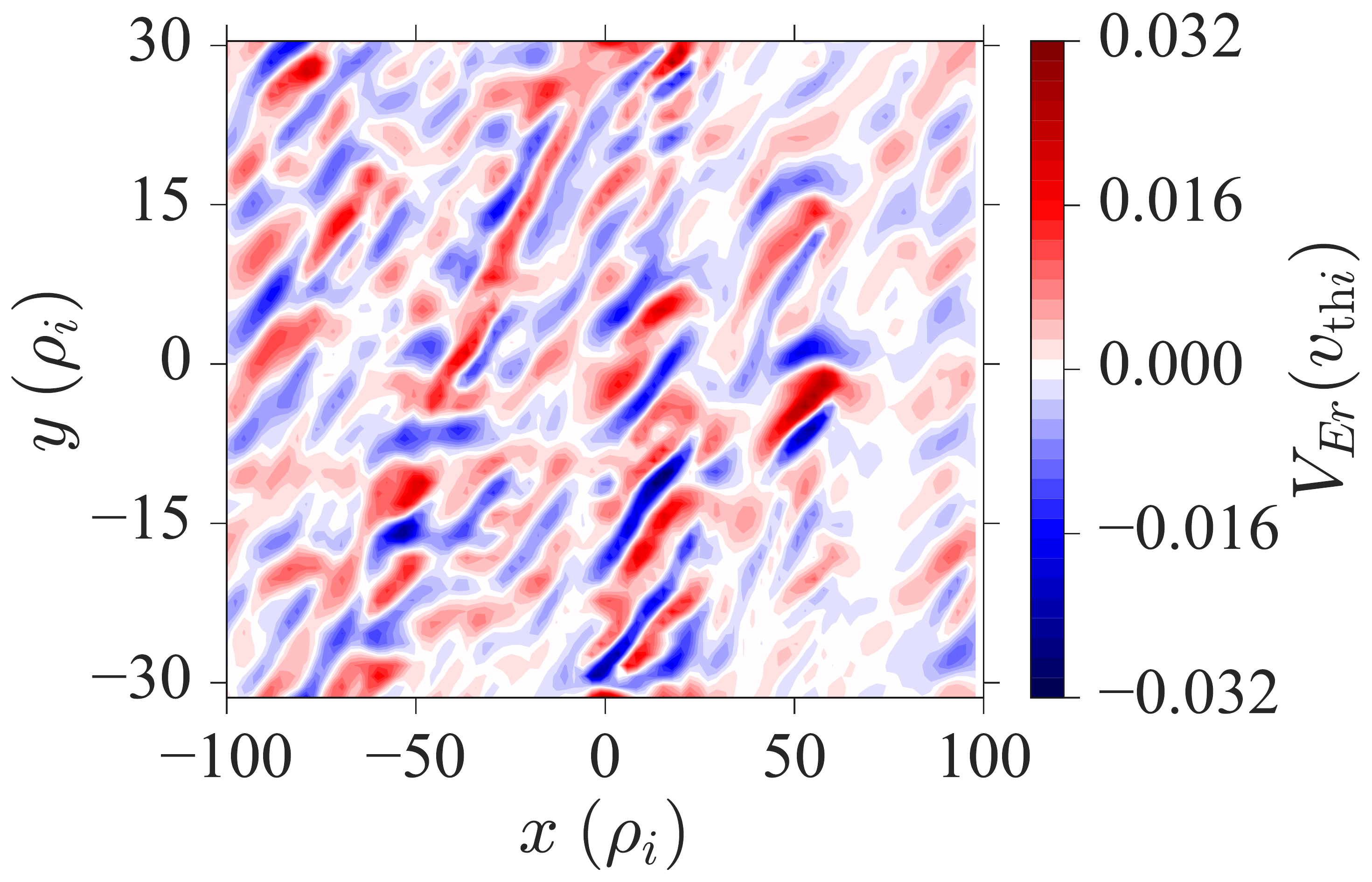}
      \caption{}
      \label{fig:vel_intermediate}
    \end{subfigure}
    \begin{subfigure}{0.49\linewidth}
      \includegraphics[width=\linewidth]{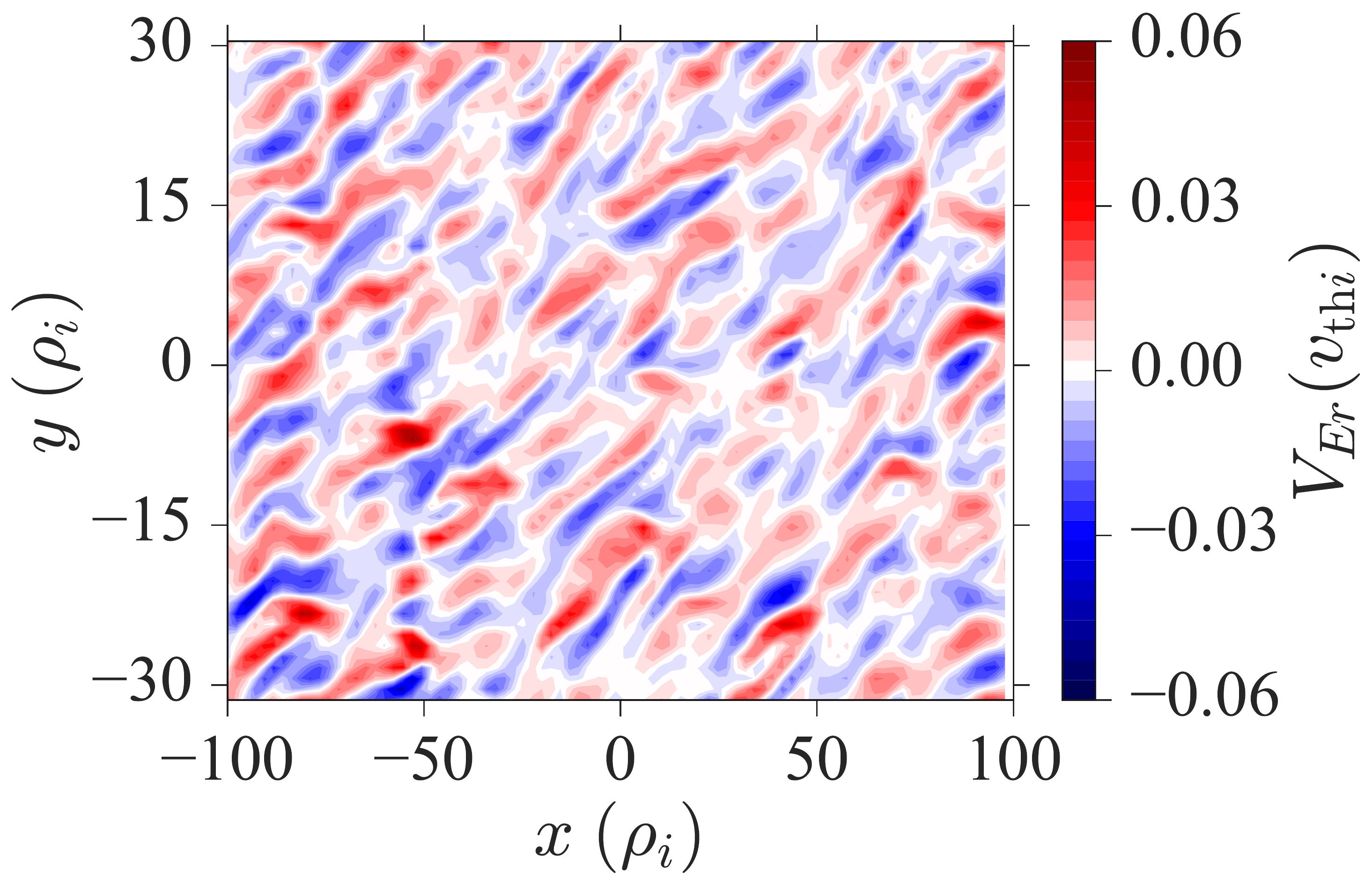}
      \caption{}
      \label{fig:vel_strongly_driven}
    \end{subfigure}
    \begin{subfigure}{0.49\linewidth}
      \includegraphics[width=\linewidth]{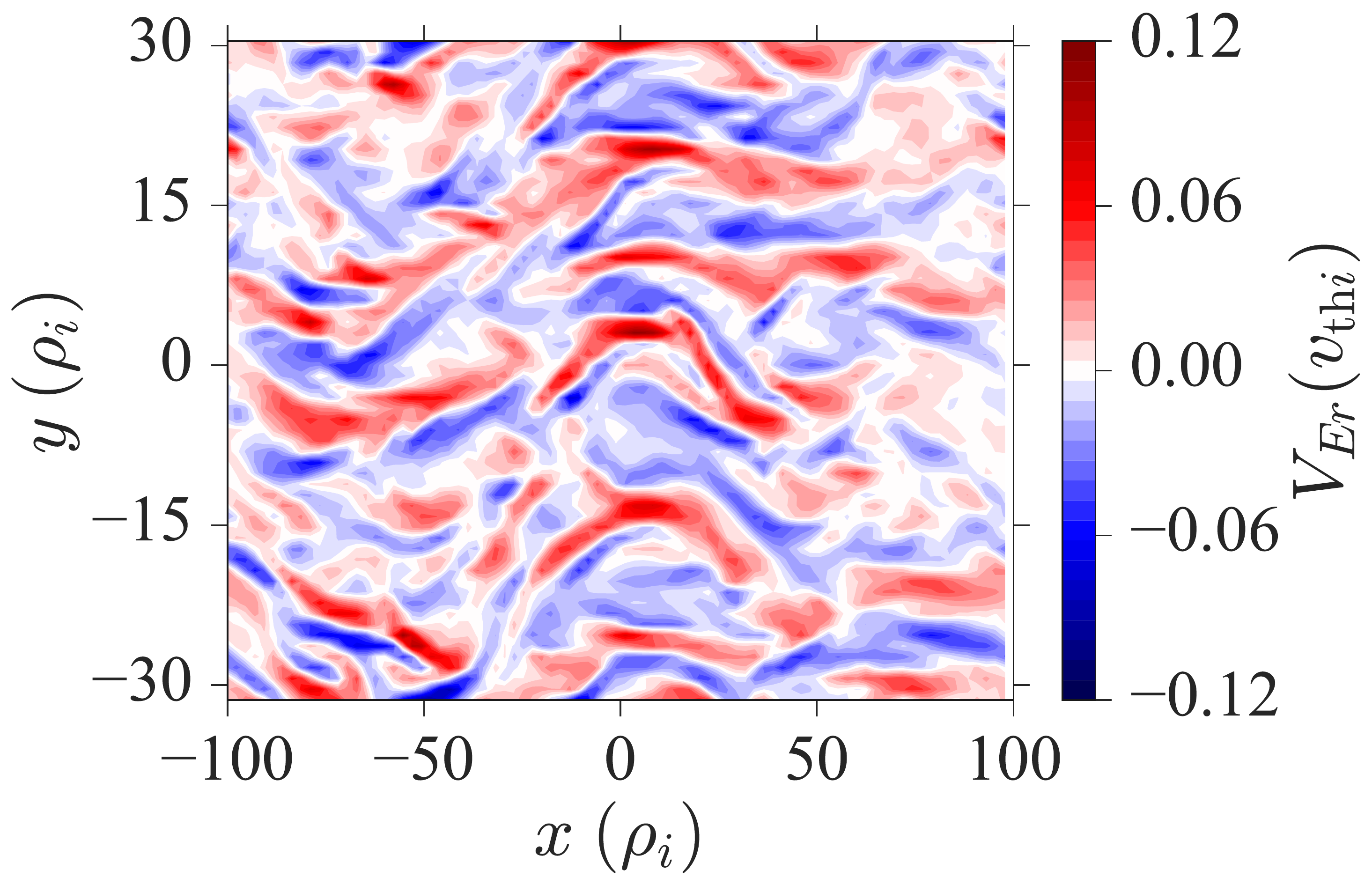}
      \caption{}
      \label{fig:vel_no_shear}
    \end{subfigure}
    \caption[Real-space radial \exb velocity on $(x,y)$ plane]{
      Radial \exb velocity $V_{Er}$ at the outboard midplane of MAST as a
      function of the local GS2 coordinates $x$ and $y$ for the same
      equilibrium parameters as in \figref{density_fluctuations}.
	}
    \label{fig:vel_fluctuations}
  \end{figure}

  \Figref{density_fluctuations} shows the density-fluctuation field $\delta n_i
  / n_i$ at the outboard midplane of MAST as a function of the local GS2
  coordinates $x$ and $y$. The simulations shown
  in \figsdash{marginal}{strongly_driven} are marked by
  points in~\figref{contour_heatmap} and, importantly, all three are well within
  the region of experimental uncertainty. We have chosen four combinations of the
  stability parameters $(\kappa_T, \gamma_E)$ as the system is taken away from
  the turbulence threshold: $(4.8, 0.16)$, which is close to the turbulence
  threshold [\figref{marginal}], $(4.9, 0.16)$, an intermediate case between
  the marginal and strongly driven turbulence [\figref{intermediate}], $(5.2,
  0.16)$, a strongly driven case further from the threshold
  [\figref{strongly_driven}], and $(5.2, 0)$, a case without flow shear
  [\figref{no_shear}], representative of the normal, supercritical ITG
  turbulence that has been thoroughly studied in the
  past~\cite{Waltz1988,Dimits1996,Rogers2000}.  For the same four cases,
  \figsref{vel_fluctuations}{tperp_fluctuations} show the perturbed radial \exb
  velocity $V_{Er}$ and the perpendicular temperature-fluctuation $\delta
  T_{\perp i}/T_{\perp i} \equiv a/\rho_i \delta T_{\perp i}/T_{\perp i}$
  fields.  We have calculated $V_{Er}$ velocity by taking the radial component
  of~\eqref{v_exb}, given by (see equation (3.42) in Ref~\cite{HighcockThesis})
  \begin{equation}
    V_{Er} = \frac{c}{a B_{\mathrm{ref}}} \frac{1}{|\nabla \psi|}
    \qty|\pdv{\psi}{r}|_{r_0} \pdv{\varphi}{y},
    \label{v_er}
  \end{equation}
  recalling that $a$ is the half diameter of the LCFS, $B_{\mathrm{ref}}$ is
  the toroidal magnetic field at the magnetic axis, $\psi$ is the poloidal
  magnetic flux, and $r=D/2a$.

\begin{figure}[t]
    \centering
    \begin{subfigure}{0.9\linewidth}
      \includegraphics[width=\linewidth]{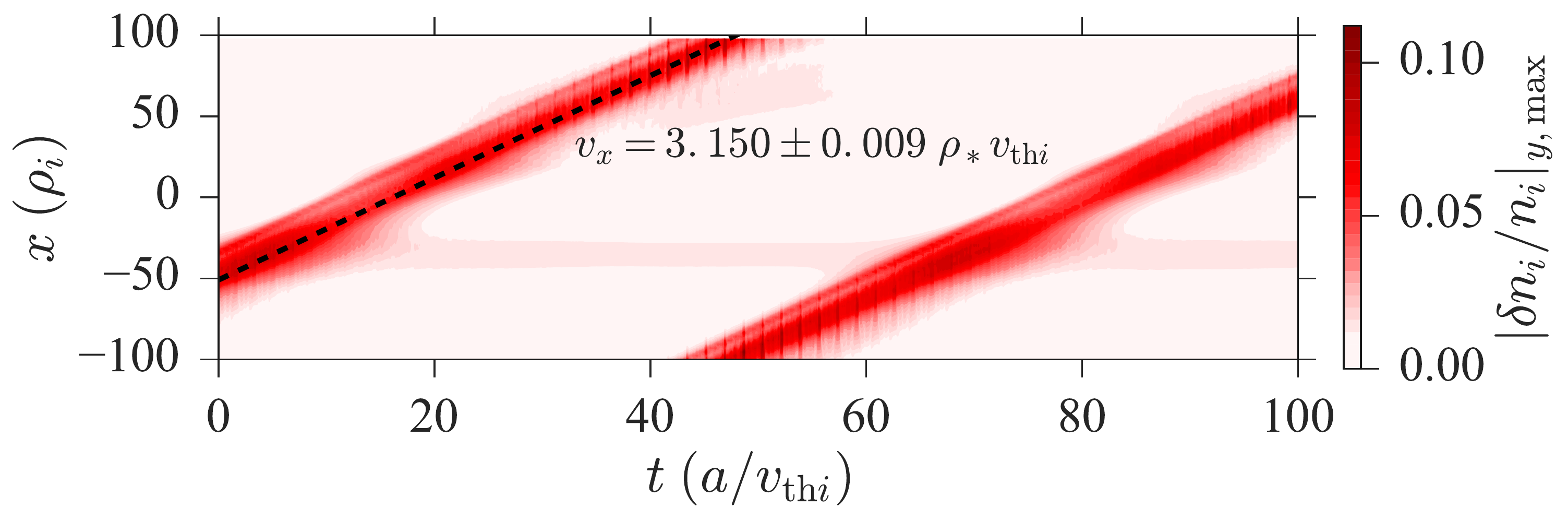}
      \caption{}
      \label{fig:marginal_xt}
    \end{subfigure}
    \begin{subfigure}{0.9\linewidth}
      \includegraphics[width=\linewidth]{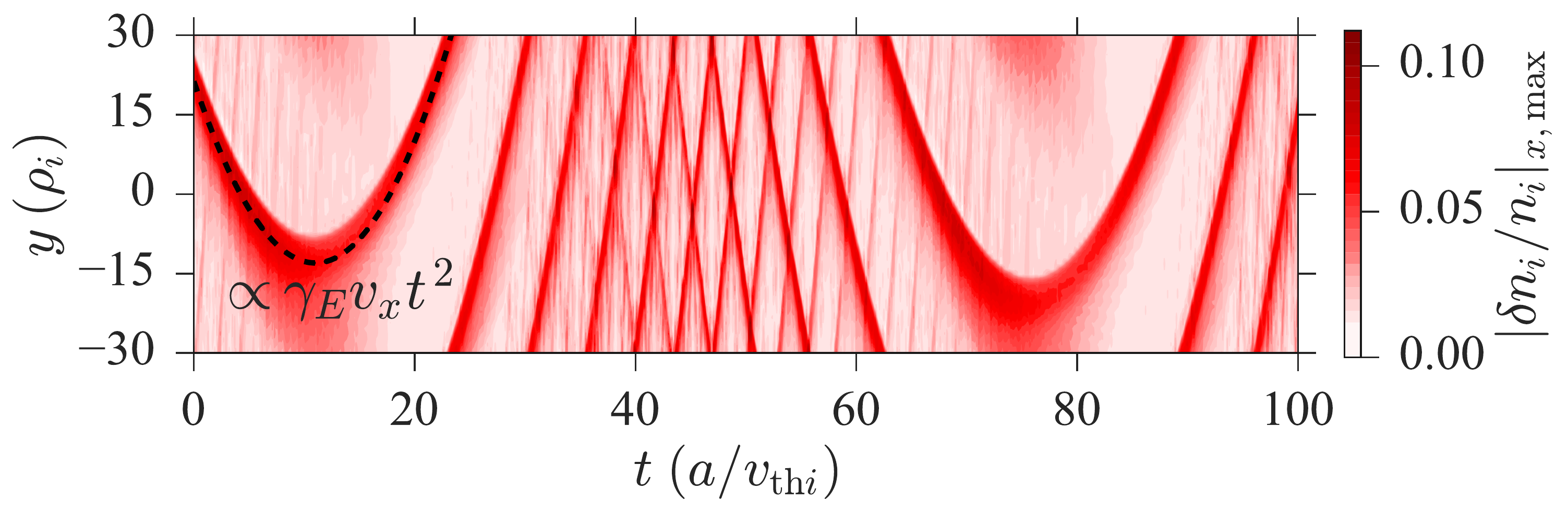}
      \caption{}
      \label{fig:marginal_yt}
    \end{subfigure}
    \caption{
      Density-fluctuation field $\delta n_i / n_i$ as a function of
      \subref*{fig:marginal_xt} $x$ and $t$ (taking the maximum in
      the $y$ direction) and \subref*{fig:marginal_yt} $y$ and $t$ (taking the
      maximum in the $x$ direction) for a marginally unstable case with
      $(\kappa_T, \gamma_E) = (5.1, 0.18)$, which contains only one coherent
      structure. The structure is advected both radially and poloidally.
      The GS2 domain is periodic in $x$ and $y$ and so this is the
      same structure throughout the entire time period shown. The dashed line
      in \subref*{fig:marginal_xt} indicates $x=v_x t$, and in
      \subref*{fig:marginal_yt} it indicates $y \propto \gamma_E v_x t$, showing
      that the poloidal advection is due to the flow associated with the shear
      $\gamma_E$.
    }
    \label{fig:marginal_vs_t}
  \end{figure}

  As the system is taken away from the threshold, the nature of the fluctuation
  field changes as follows. The near-threshold state [\figref{marginal}] is
  dominated by coherent, long-lived (see \figref{marginal_vs_t}) structures
  that are at high intensity compared to the background fluctuations. As
  $\kappa_T$ is slightly increased (in this case by only 0.1), these structures
  become more numerous [\figref{intermediate}], but have roughly the same
  maximum amplitude: ${(\delta n_i/n_i)}_{\max} \sim 0.08$. In contrast, the
  strongly driven state [far from threshold; \figref{strongly_driven}] exhibits
  a more conventional turbulence, characterised by many interacting eddies with
  larger amplitudes.

  These simulations are typical of the cases close to
  and far from the turbulence threshold, i.e.,\ in simulations near the
  threshold, we always find sparse but well-defined coherent structures that
  survive against a backdrop of weaker fluctuations [with the important
  exception of the case of $\gamma_E=0$ shown in \figref{no_shear}]. Likewise, for
  all cases where the system is taken away from the threshold by increasing
  $\kappa_T$, or decreasing $\gamma_E$, the transition from coherent structures
  to strongly driven interacting eddies occurs the same way: the structures
  become more numerous, while maintaining roughly the same amplitude, until
  they fill the entire domain, interact with each other, and break up.  For
  parameter values far from the threshold, we observe no discernible coherent
  structures, but rather strongly time-dependent fluctuations with amplitudes
  that increase with $\kappa_T$.

  \begin{figure}[t]
    \centering
    \begin{subfigure}{0.49\linewidth}
      \includegraphics[width=\linewidth]{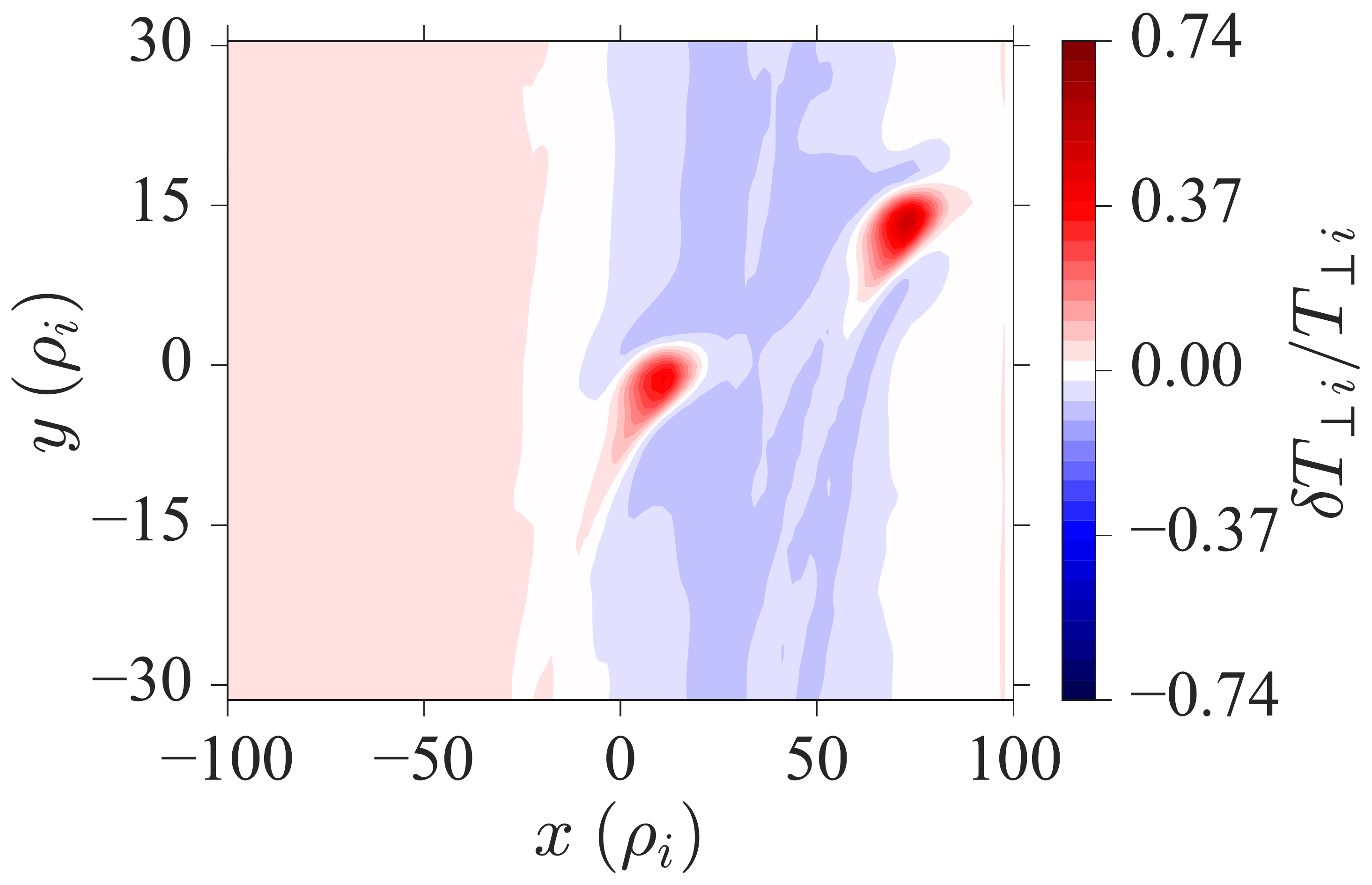}
      \caption{}
      \label{fig:tperp_marginal}
    \end{subfigure}
    \begin{subfigure}{0.49\linewidth}
      \includegraphics[width=\linewidth]{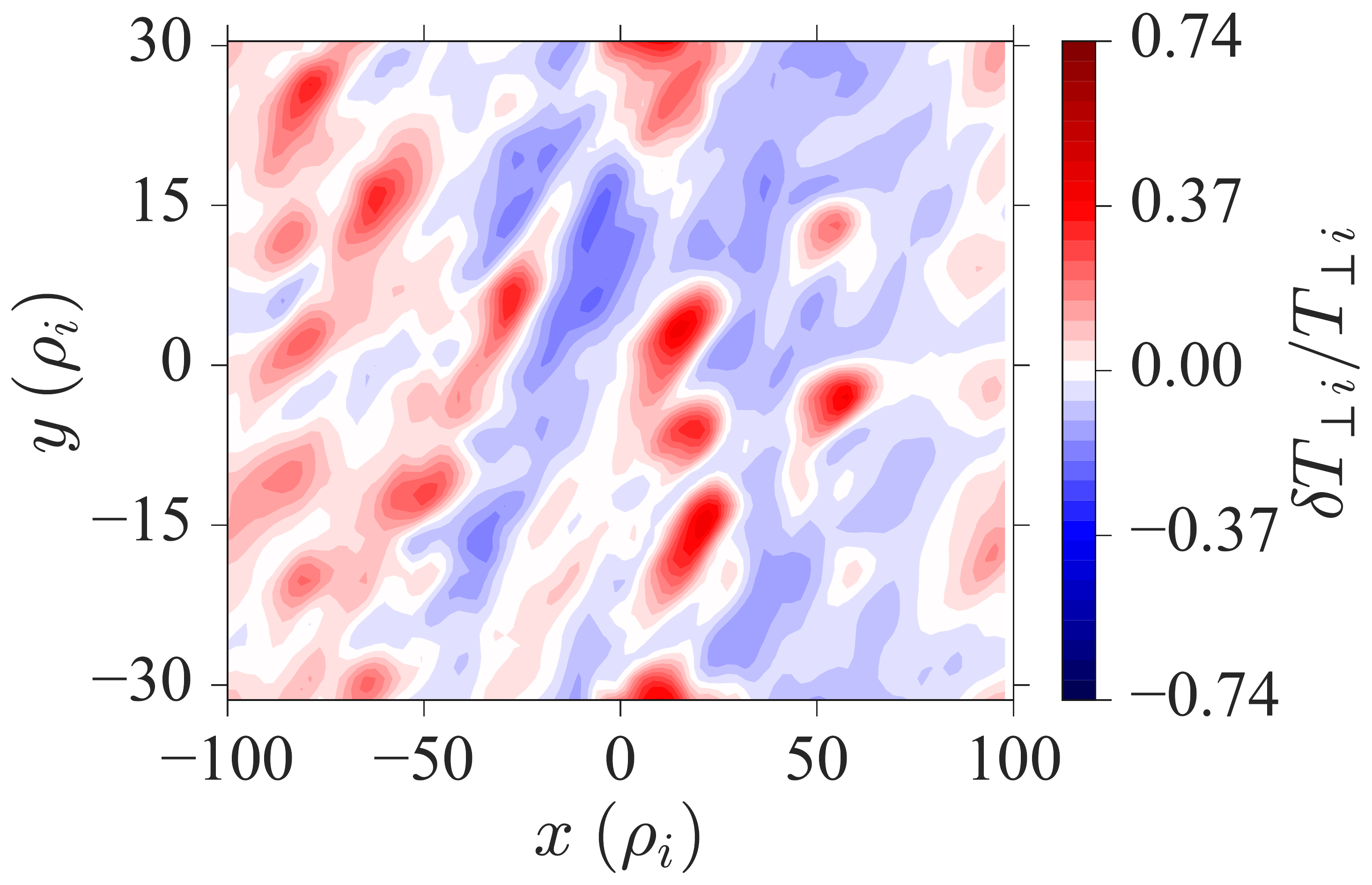}
      \caption{}
      \label{fig:tperp_intermediate}
    \end{subfigure}
    \begin{subfigure}{0.49\linewidth}
      \includegraphics[width=\linewidth]{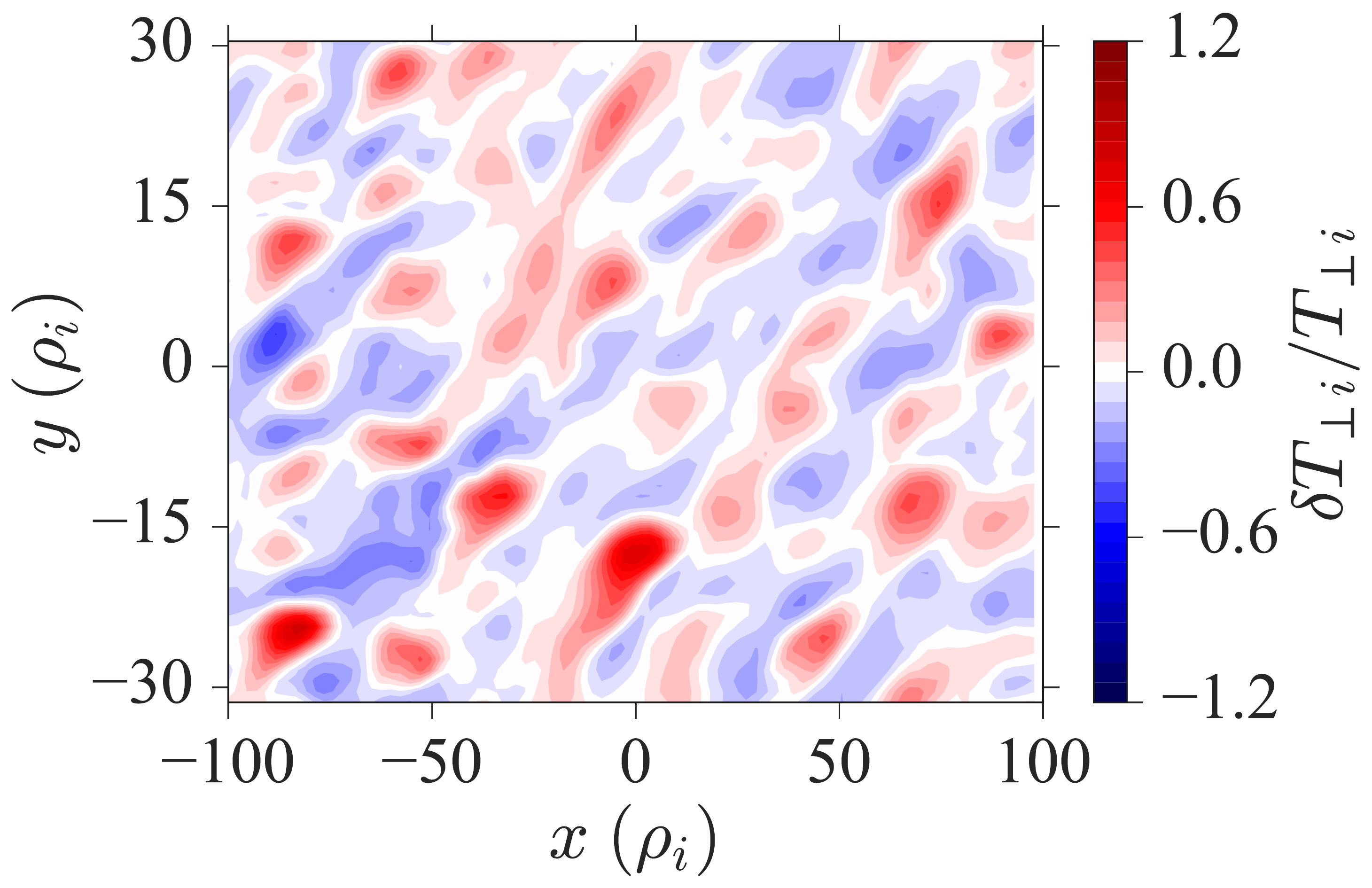}
      \caption{}
      \label{fig:tperp_strongly_driven}
    \end{subfigure}
    \begin{subfigure}{0.49\linewidth}
      \includegraphics[width=\linewidth]{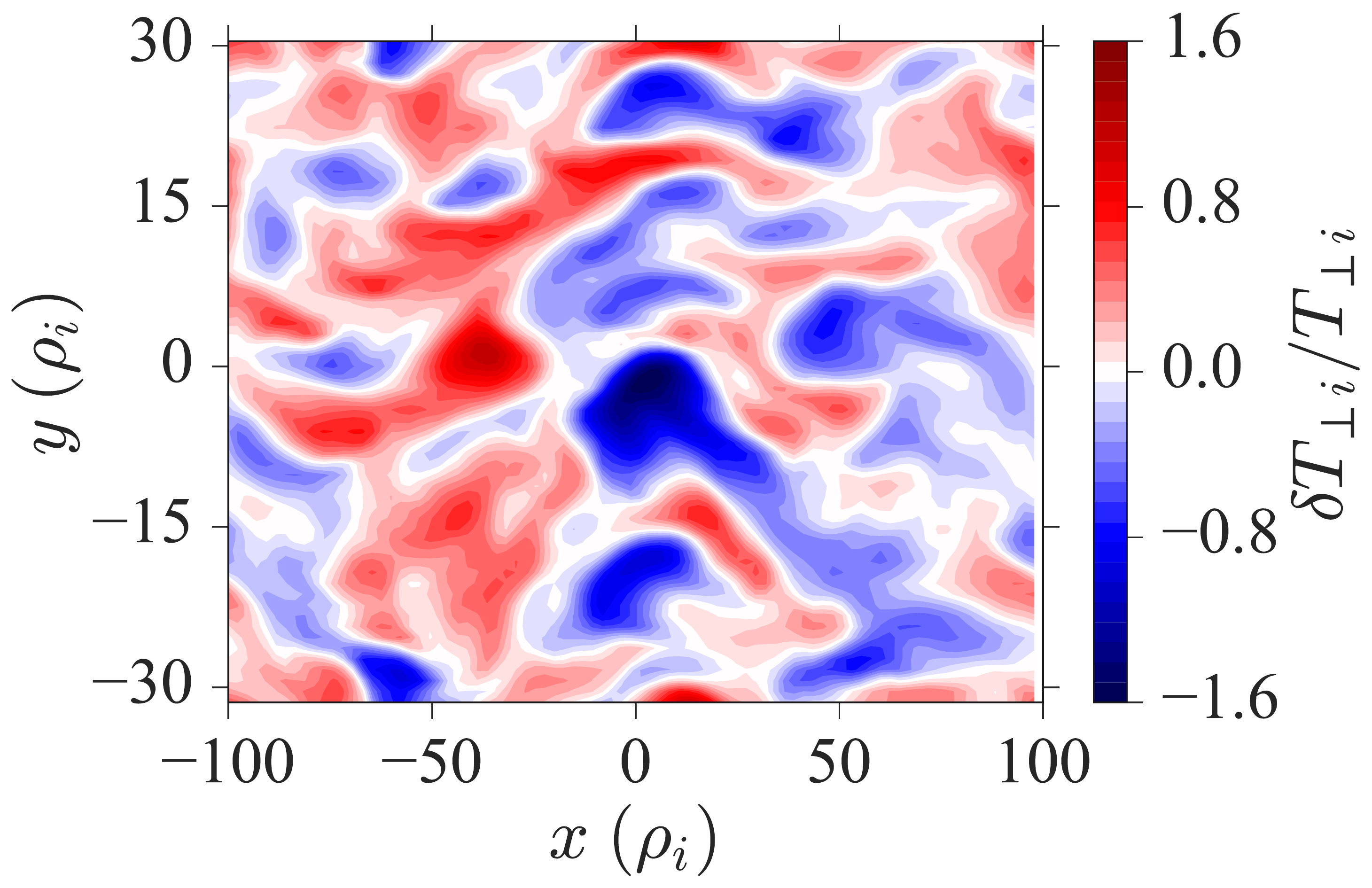}
      \caption{}
      \label{fig:tperp_no_shear}
    \end{subfigure}
    \caption{
      Perpendicular-temperature fluctuation field
      $\delta T_{\perp i}/T_{\perp i}$ at the outboard midplane of MAST as
      a function of the local GS2 coordinates $x$ and $y$ for the same
      equilibrium parameters as in \figref{density_fluctuations}.
	}
    \label{fig:tperp_fluctuations}
  \end{figure}

\begin{figure}[t]
    \centering
    \begin{subfigure}{0.49\linewidth}
      \includegraphics[width=\linewidth]{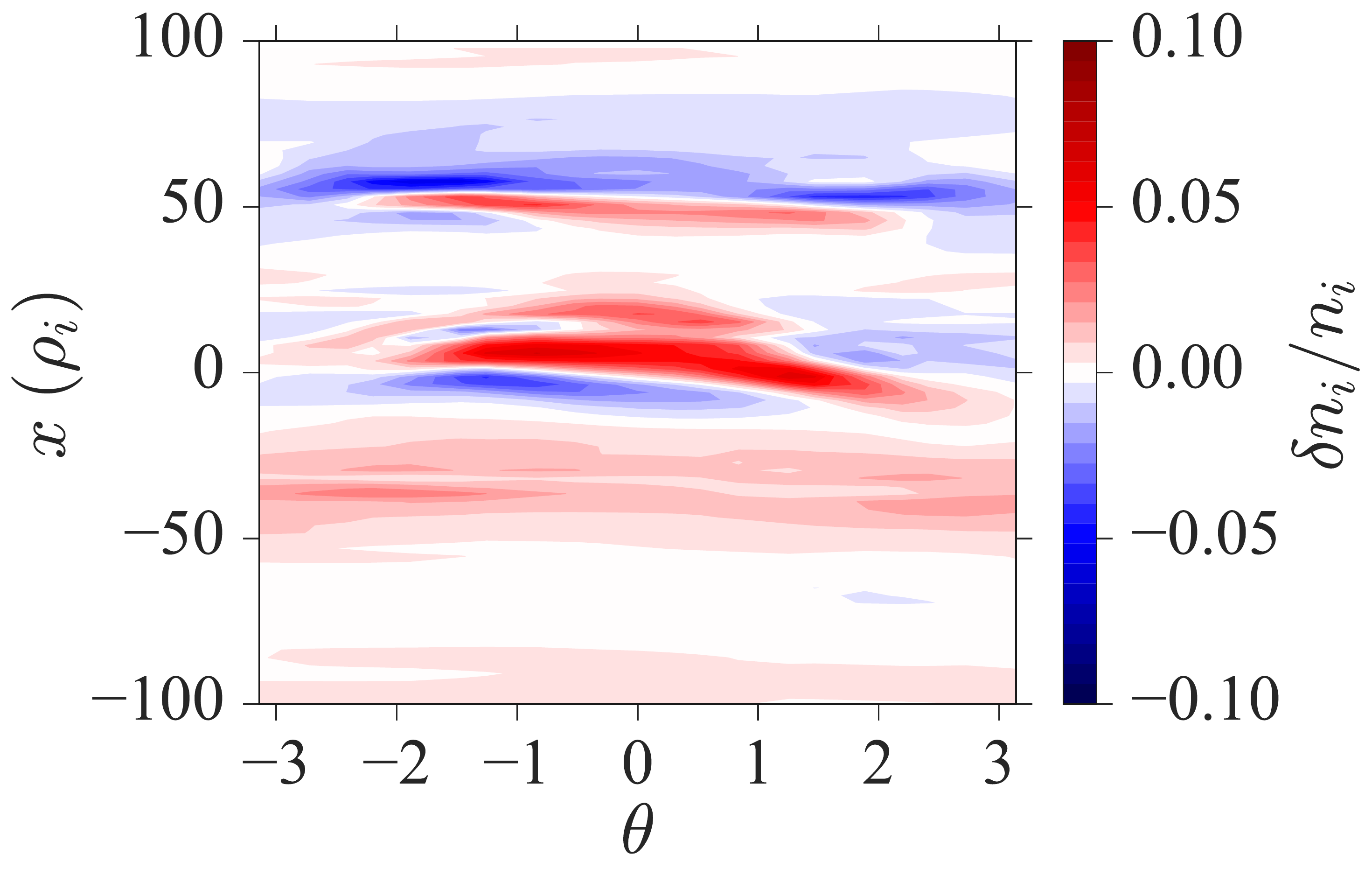}
      \caption{}
      \label{fig:marginal_xz}
    \end{subfigure}
    \hfill
    \begin{subfigure}{0.49\linewidth}
      \includegraphics[width=\linewidth]{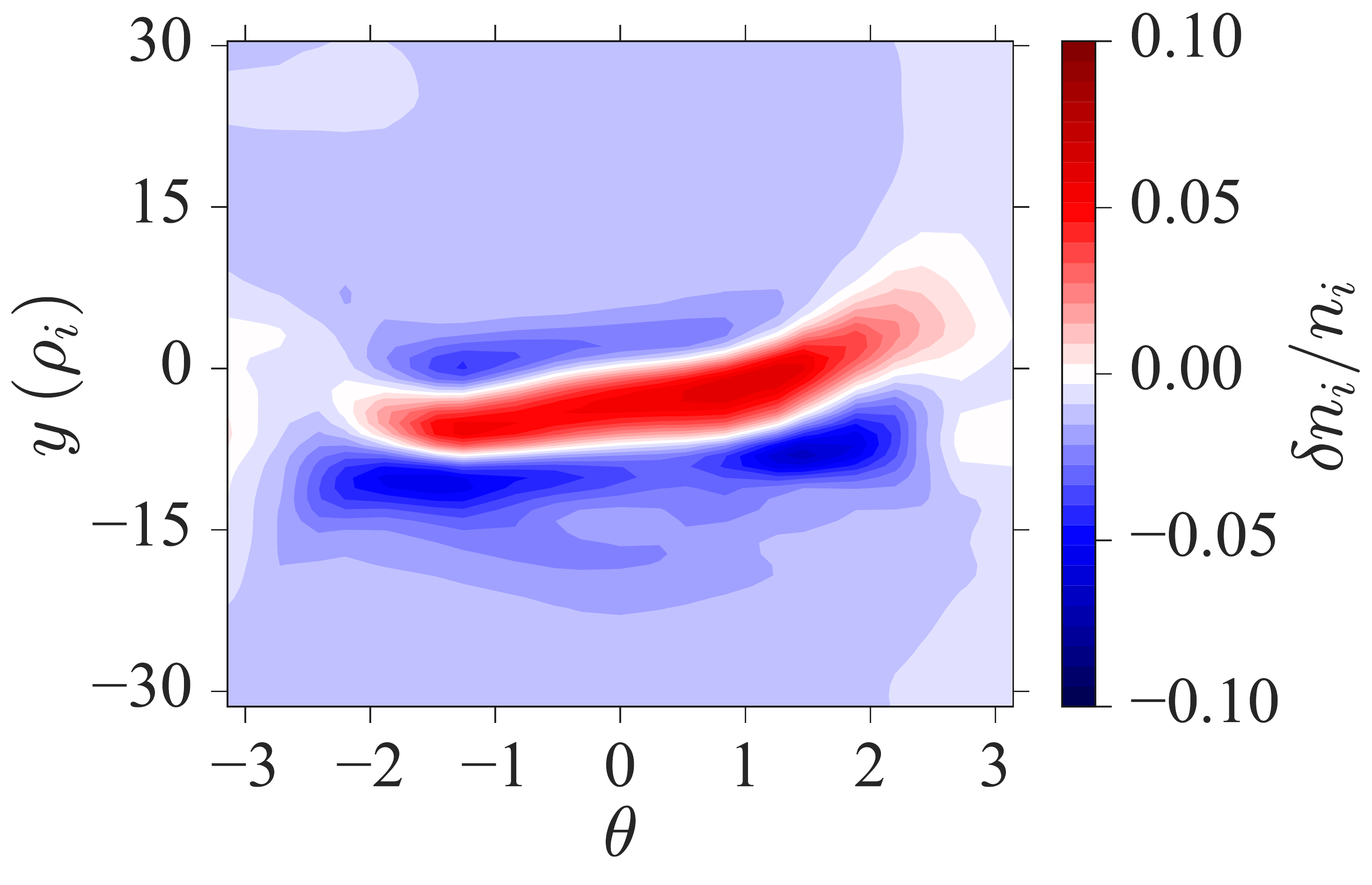}
      \caption{}
      \label{fig:marginal_yz}
    \end{subfigure}
    \caption[Real-space density-fluctuation fields in parallel direction]{
      \subref*{fig:marginal_xz} Density-fluctuation field $\delta n_i / n_i$ in
      the $x$-$z$ plane at $y=0$.
      \subref*{fig:marginal_yz} Density-fluctuation field $\delta n_i / n_i$ in
      the $y$-$z$ plane at $x=0$. Both plots are for the same simulation
      and at the same time as in \figref{marginal}, where the corresponding
      planes are indicated by the dashed lines. The
      parallel direction in GS2 is quantified by the poloidal angle $\theta$.
    }
    \label{fig:parallel_density}
  \end{figure}

  We complete our description of these coherent structures by examining their
  parallel extent and their motion.  \Figref{parallel_density} shows two views
  of the coherent structures from \figref{marginal} in the parallel direction
  (which in GS2 is quantified by the poloidal angle $\theta$): at constant $y$
  [\figref{marginal_xz}] and at constant $x$ [\figref{marginal_yz}]. It is
  clear that the coherent structures are elongated in the parallel direction
  and have an amplitude much larger than the background fluctuations.

  The motion of the coherent structures results from a combination of the
  background plasma flow, and a radial drift of the structures themselves.
  Importantly, they are long-lived as we will now show by looking at their
  motion in time. \Figsref{marginal_xt}{marginal_yt} show $\delta n_i /
  n_i$ for a marginal nonlinear simulation at $(\kappa_T, \gamma_E) = (5.1,
  0.18)$, which has only one coherent structure, as a function of $(t,x)$ and
  $(t,y)$ (taking the maximum value of $\delta n_i/n_i$ in the other
  direction), respectively.  \Figref{marginal_xt} shows the radial motion of
  the structure across the domain, which the structure crosses in a time of
  roughly $50~(a/v_{\mathrm{th}i})$, and illustrates the long-lived nature of
  coherent structures close to the turbulence threshold (recalling that the GS2
  domain is periodic in $x$ and $y$). We see that the structure exists for $t >
  100~(a/v_{\mathrm{th}i})$. The radial motion of the structure in
  \figref{marginal_xt} has a constant velocity: fitting its trajectory with a
  straight line (the dashed line) gives $v_x = 3.150 \pm
  0.009$~$\rho_* v_{\mathrm{th}i}$ (where $\rho_* = \rho_i/a = 1/100$, given
  that $\rho_i = 6.08 \times 10^{-3}$~m and $a = 0.58$~m). \Figref{marginal_yt}
  shows the poloidal advection of the structure with a much shorter poloidal
  crossing time of roughly $5~(a/v_{\mathrm{th}i})$. The poloidal motion of the
  structure is due to the combination of poloidal advection by the mean flow
  (remembering that, since we have moved to the rotating frame, the flow is
  zero at $x=0$) and the radial drift of the structures. As we saw in
  \figref{marginal_xt}, $v_x$ is constant and the radial position is given by
  $x(t) = v_x t$. The poloidal advection due to the flow shear is given by
  $v_y(t) = \gamma_E x(t)$ and so the direction of the flow shear reverses at
  $x=0$. Combining the expressions for $x(t)$ and $v_y(t)$ and integrating, we
  find that $y(t) \propto \gamma_E v_x t^2$, and, as shown by the dashed line
  in \figref{marginal_yt}, this describes the poloidal motion of the
  structures, which indeed reverses direction at $x=0$.

  The coherent structures in the marginal case, such as the one described
  above, are unlike the strongly interacting eddies in the cases far from the
  turbulence threshold and are more likely to constitute a nonlinear travelling
  wave (soliton-like) solution to the gyrokinetic equation. However, more work
  is needed to develop an analytic description of these structures.

  \subsubsection{$Q_i/Q_{\mathrm{gB}}$ as an order parameter}
  The results in section~\ref{sec:coherent_strucs} suggest that the nature of
  the turbulence is determined by how far the system is from the turbulence
  threshold. This means perhaps that the important
  metric that should be used to quantify the state of the system is the
  ``distance from threshold'' and not the specific values of
  $\kappa_T$ and $\gamma_E$ (although both can be used to control the distance
  from threshold). The ion heat flux $Q_i/Q_{\mathrm{gB}}$ is a strong function
  of $\kappa_T$ and $\gamma_E$, increasing monotonically as the system is taken
  away from the threshold [see \figref{contour_heatmap}],
  so we can use $Q_i/Q_{\mathrm{gB}}$ as a control parameter to measure the
  distance from the threshold. In sections~\ref{sec:max_amp}
  and~\ref{sec:struc_count}, we will quantify the change in the nature of the
  turbulence, namely, the change in the amplitude and number of coherent
  structures, for our parameter scan and show that the distance from threshold
  is indeed the relevant order parameter.

  \subsubsection{Maximum amplitude}
  \label{sec:max_amp}
  \begin{figure}[t]
    \centering
    \includegraphics[width=0.6\linewidth]{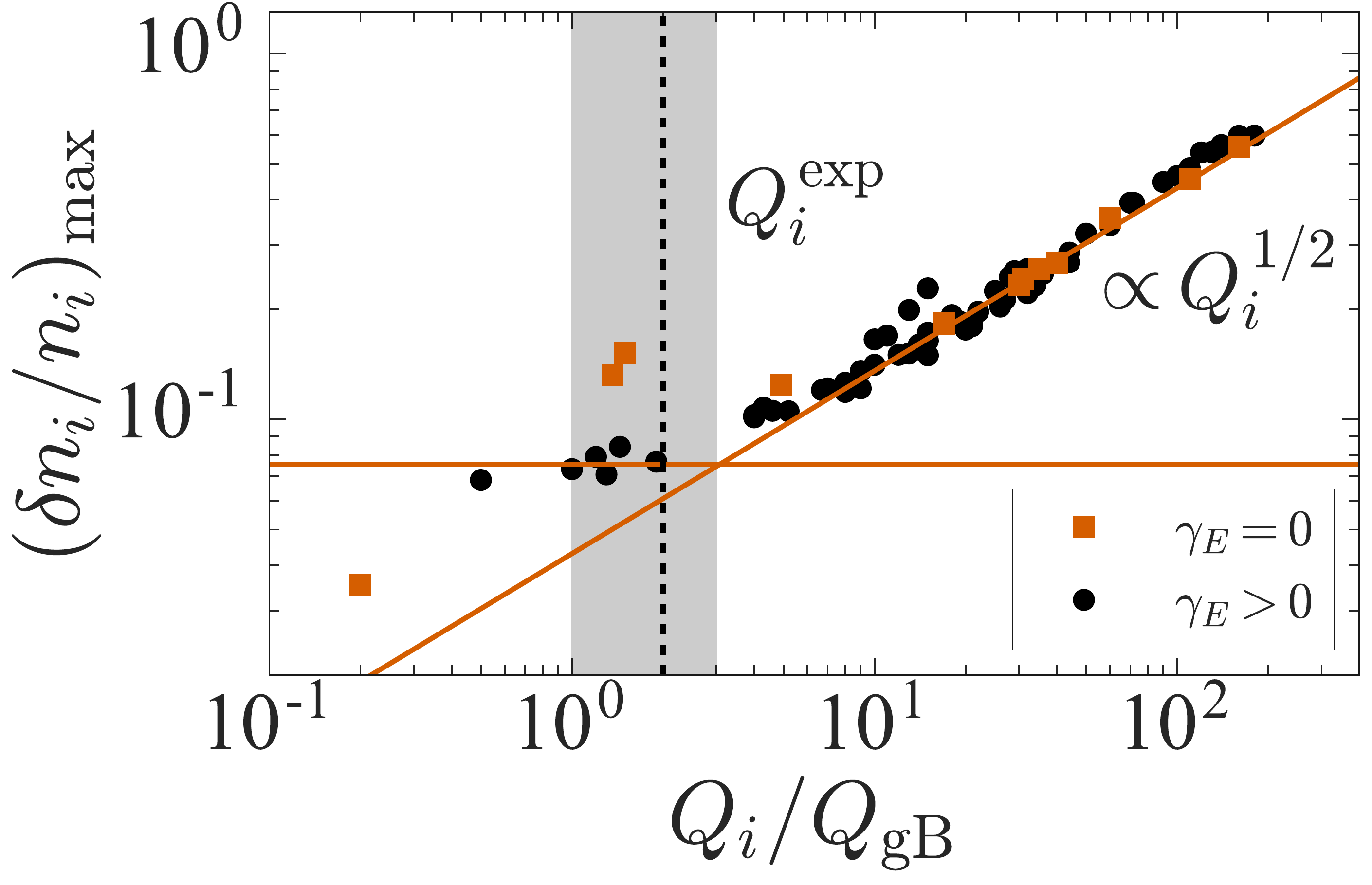}
    \caption{
      Maximum amplitude of the density fluctuations versus
      $Q_i/Q_{\mathrm{gB}}$. The naive scaling~\eqref{q_scaling}, $\delta n_i /
      n_i \propto Q_i^{1/2}$, is shown for reference and holds far from
      threshold, whereas for small values of $Q_i/Q_{\mathrm{gB}}$ (around and
      below the experimental value $Q_i^{\exp}$), the amplitude becomes
      independent of $Q_i/Q_{\mathrm{gB}}$.
    }
    \label{fig:amplitude}
  \end{figure}
  Here, we investigate how the amplitude of the density fluctuations change
  with the distance from threshold. For near-threshold
  cases, such as the one shown in \figref{marginal}, the dominant features are
  the coherent structures, which have high densities compared to the background
  fluctuations. In order to capture the amplitude of these
  structures, we measure the maximum amplitude of the density field, as opposed to an
  $(x,y)$-averaged one, which would be small because of the relatively
  small volume taken up by the coherent structures. \Figref{amplitude} shows
  the maximum amplitude ${(\delta n_i/n_i)}_{\max}$, maximised over the
  $(x,y)$-plane and averaged over time, versus $Q_i/Q_{\mathrm{gB}}$ for the
  entire set of simulations in our parameter scan.

  The striking feature of~\figref{amplitude} is that ${(\delta
  n_i/n_i)}_{\max}$ hits a finite ``floor'' as $Q_i/Q_{\mathrm{gB}}$ approaches
  and goes below its experimental value. This coincides with the appearance of
  the long-lived structures such as those shown in~\figref{marginal}. This
  floor is absent in simulations with $\gamma_E=0$, suggesting that the
  turbulence with $\gamma_E=0$ is fundamentally different close to the
  turbulence threshold (as indeed also suggested by the absence of coherent
  structures).

  Far from the turbulence threshold, we can construct from~\eqref{q_def} a
  naive estimate of the relationship between $Q_i/Q_{\mathrm{gB}}$ and $\delta
  n_i / n_i$:
  \begin{equation}
    \frac{Q_i}{Q_{\mathrm{gB}}}
    \sim
    \frac{a^2}{\rho_i^2}\frac{\delta T_i}{T_i}\frac{V_{Er}}{v_{{\mathrm{th}}i}}
    \sim
    k_y\rho_i {\left(\frac{\delta n_i}{n_i}\right)}^2,
    \label{q_scaling}
  \end{equation}
  assuming that fluctuations of $\varphi$ are related (by order
  of magnitude) to the electron (and, therefore, ion) density via the Boltzmann
  response $e\varphi/T_e \sim \delta n_e/n_e$ and that ion temperature and
  density fluctuations are approximately proportional to each other (cf.
  \figsref{density_fluctuations}{tperp_fluctuations}). The scaling
  $\delta n_i/n_i \propto Q_i^{1/2}$ that follows from \eqref{q_scaling}
  assuming that $k_y \rho_i$ is not a strong function of $Q_i$ is indicated by
  the red line in \figref{amplitude}, and describes well the scaling away from
  the threshold. We also see that $\gamma_E=0$ and $\gamma_E>0$ simulations
  are similar away from the threshold.

  The above results can be understood as follows. In the case of supercritical
  turbulence, one typically observes smaller fluctuation amplitudes all the way
  to the turbulence threshold -- there is no minimum amplitude required to
  sustain turbulence.  In contrast, \figref{amplitude} shows that, for the
  subcritical turbulence that we are investigating, the maximum fluctuation
  amplitude stays constant as we approach the threshold. This is because there is a
  critical value required in order to sustain a saturated nonlinear state --
  indeed, if the amplitude dropped below a certain value in a subcritical
  system, all perturbations would decay. However, even as the fluctuation
  amplitude stays constant, the heat flux decreases as the threshold is
  approached. The system can satisfy the requirement of finite
  amplitude while simultaneously allowing the heat flux to decrease through a
  reduction of the volume taken up by finite amplitude turbulence. As we
  demonstrate in the next section, this is achieved via a reduction in
  the number of coherent structures.

  \subsubsection{Structure counting}
  \label{sec:struc_count}
  \begin{figure}[t]
    \centering
    \begin{subfigure}{0.49\linewidth}
      \includegraphics[width=\linewidth]{density_intermediate}
      \caption{}
      \label{fig:intermediate2}
    \end{subfigure}
    \begin{subfigure}{0.49\linewidth}
      \includegraphics[width=\linewidth]{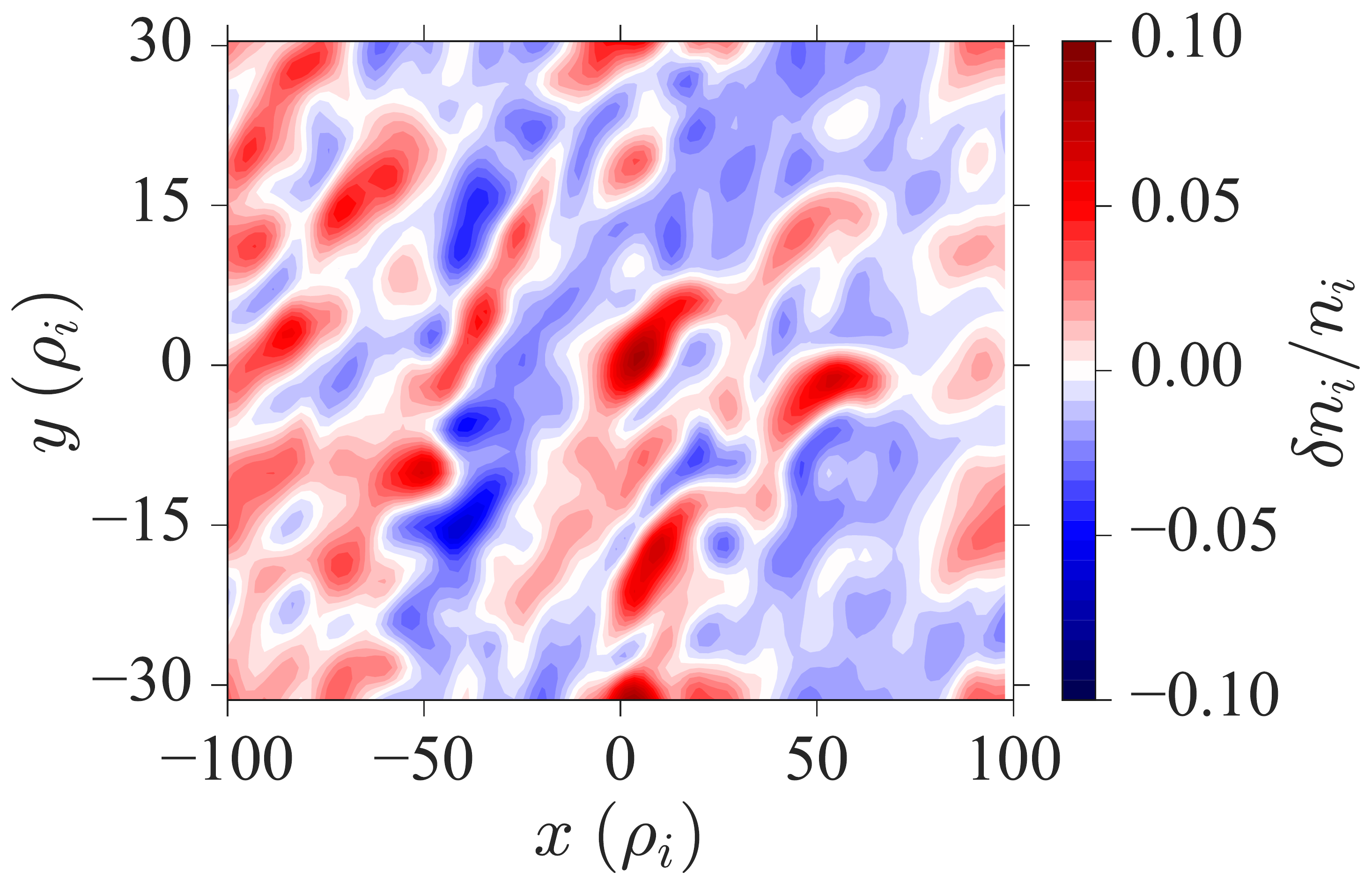}
      \caption{}
      \label{fig:post_filter}
    \end{subfigure}
    \begin{subfigure}{0.49\linewidth}
      \includegraphics[width=\linewidth]{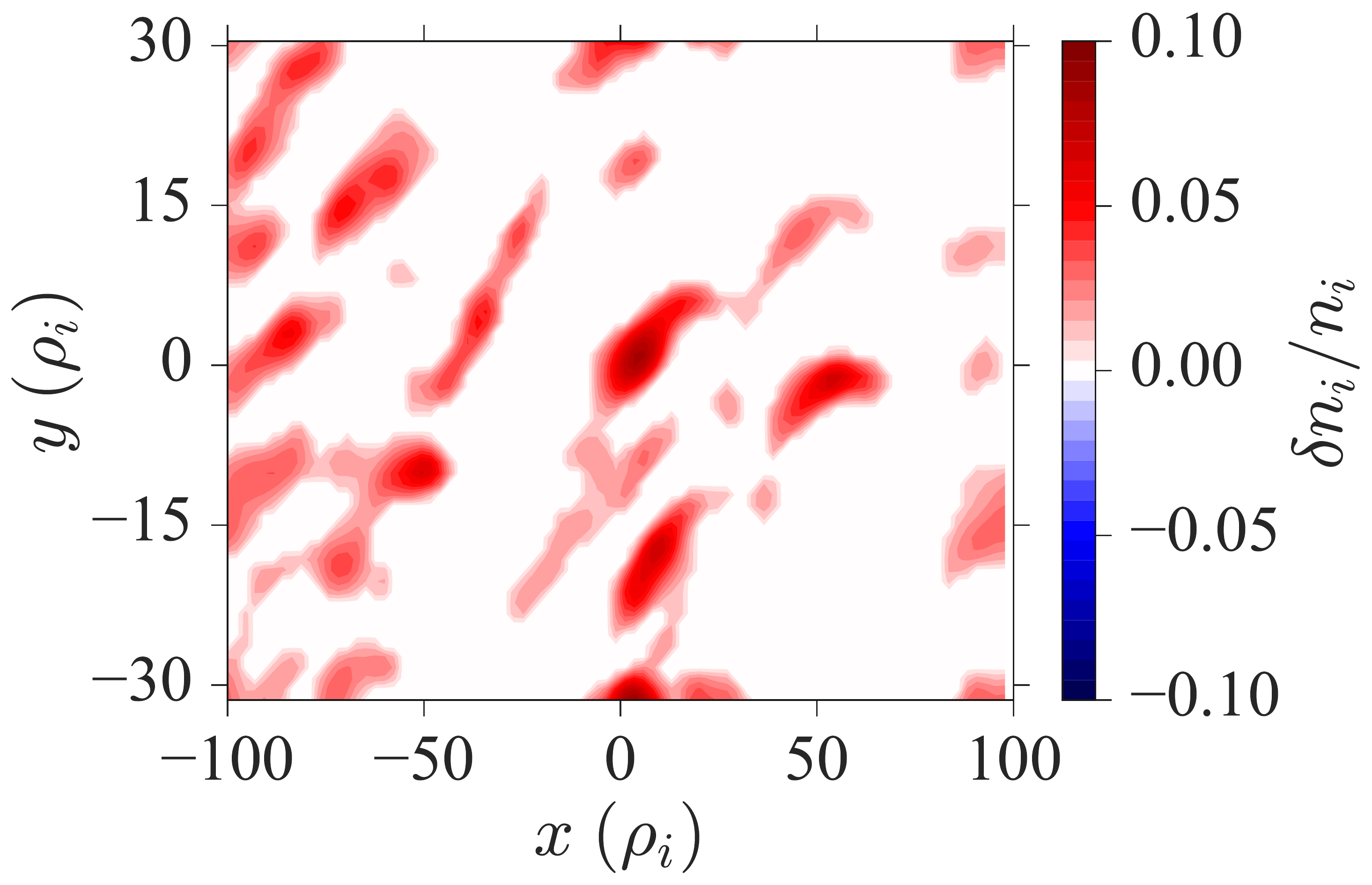}
      \caption{}
      \label{fig:post_thresh}
    \end{subfigure}
    \begin{subfigure}{0.49\linewidth}
      \includegraphics[width=\linewidth]{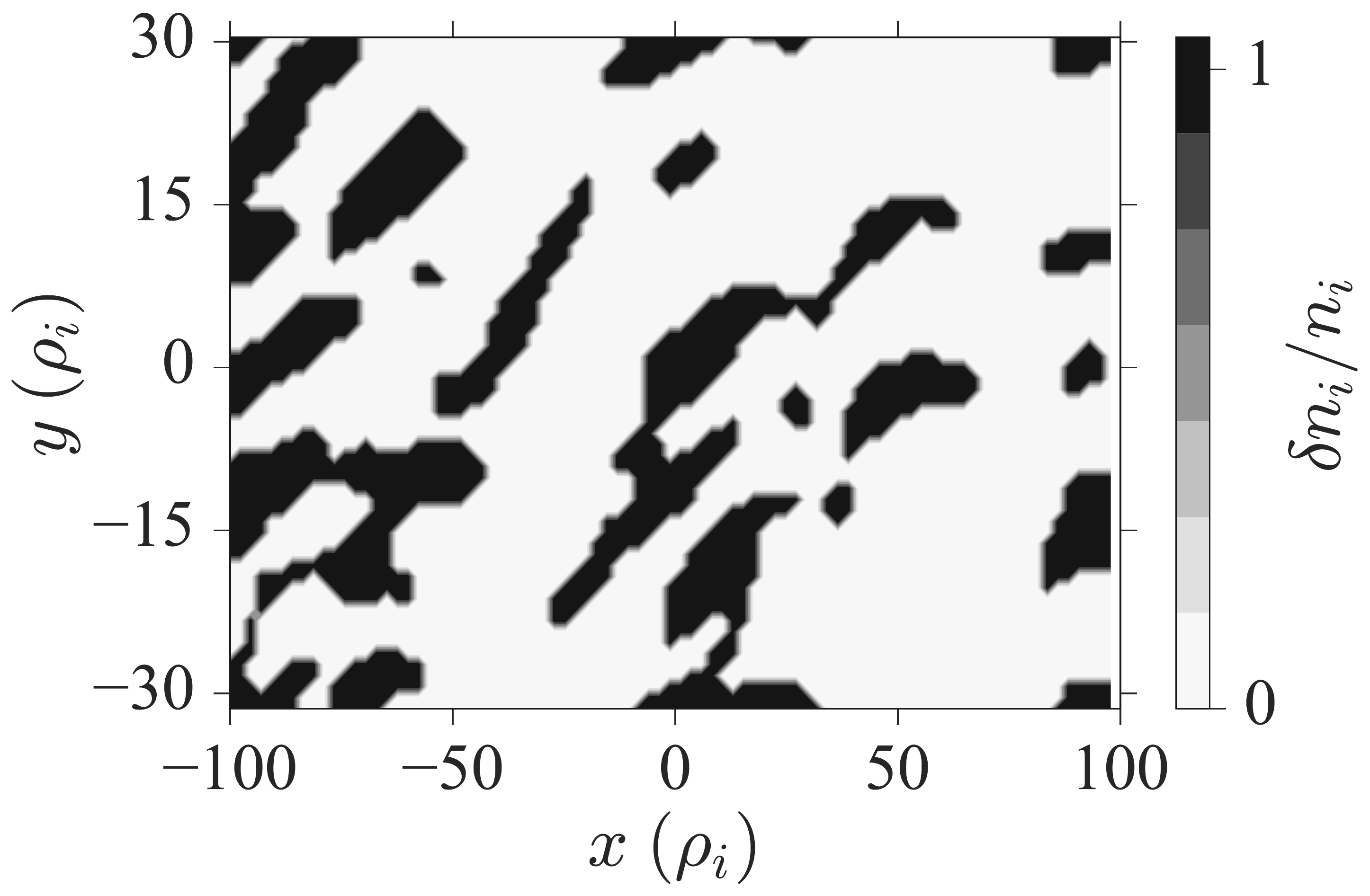}
      \caption{}
      \label{fig:struc_intermediate}
    \end{subfigure}
    \caption[Structure counting procedure]{
      Stages of the structure-counting procedure:
      \subref*{fig:intermediate2} the original density-fluctuation field
      [as in \figref{intermediate}];
      \subref*{fig:post_filter} the same field after the application of a
      Gaussian filter to smooth the structures;
      \subref*{fig:post_thresh} after the application of a 75\% threshold
      function;
      \subref*{fig:struc_intermediate} after setting $\delta n_i/n_i>0$
      values to 1 for simplicity. The image-labelling algorithm is then
      applied to \subref*{fig:struc_intermediate} and returns $19$
      structures for this case.
	}
    \label{fig:struc_count_procedure}
  \end{figure}

  We quantify the changes in volume taken up by the finite-amplitude structures
  by measuring the typical number of these structures in our simulations
  as a function of the distance from threshold. We follow the
  ``structure-counting'' methods first described in~\cite{VanWyk2016}, which
  involve the following steps, illustrated in \figref{struc_count_procedure}.

  As a pre-processing step we apply a Gaussian image filter with a standard
  deviation of the order of the grid scale. We then set all density-field
  values below a certain percentile (here 75\% of the maximum amplitude) to 0
  and above it to 1. The level of this threshold function is somewhat arbitrary
  and the exact number of structures will depend on this level, but not the
  trend as a function of our equilibrium parameters. After applying the
  threshold function, we are left with an array of 1s representing our
  structures against a background of 0s.  We then remove structures below 10\%
  of the mean structure size as a post-processing step to avoid the counting of
  spurious small isolated blobs of high density.  To count the structures, we
  employ a general-purpose image processing package
  \emph{scikit-image}~\cite{scikit-image}, which implements an efficient
  labelling algorithm~\cite{Fiorio1996}, then used by us to label connected
  regions. In \figref{struc_count_procedure}, the image-labelling algorithm
  found $19$ structures.

  \begin{figure}[t]
    \centering
    \begin{subfigure}{0.49\linewidth}
      \includegraphics[width=\linewidth]{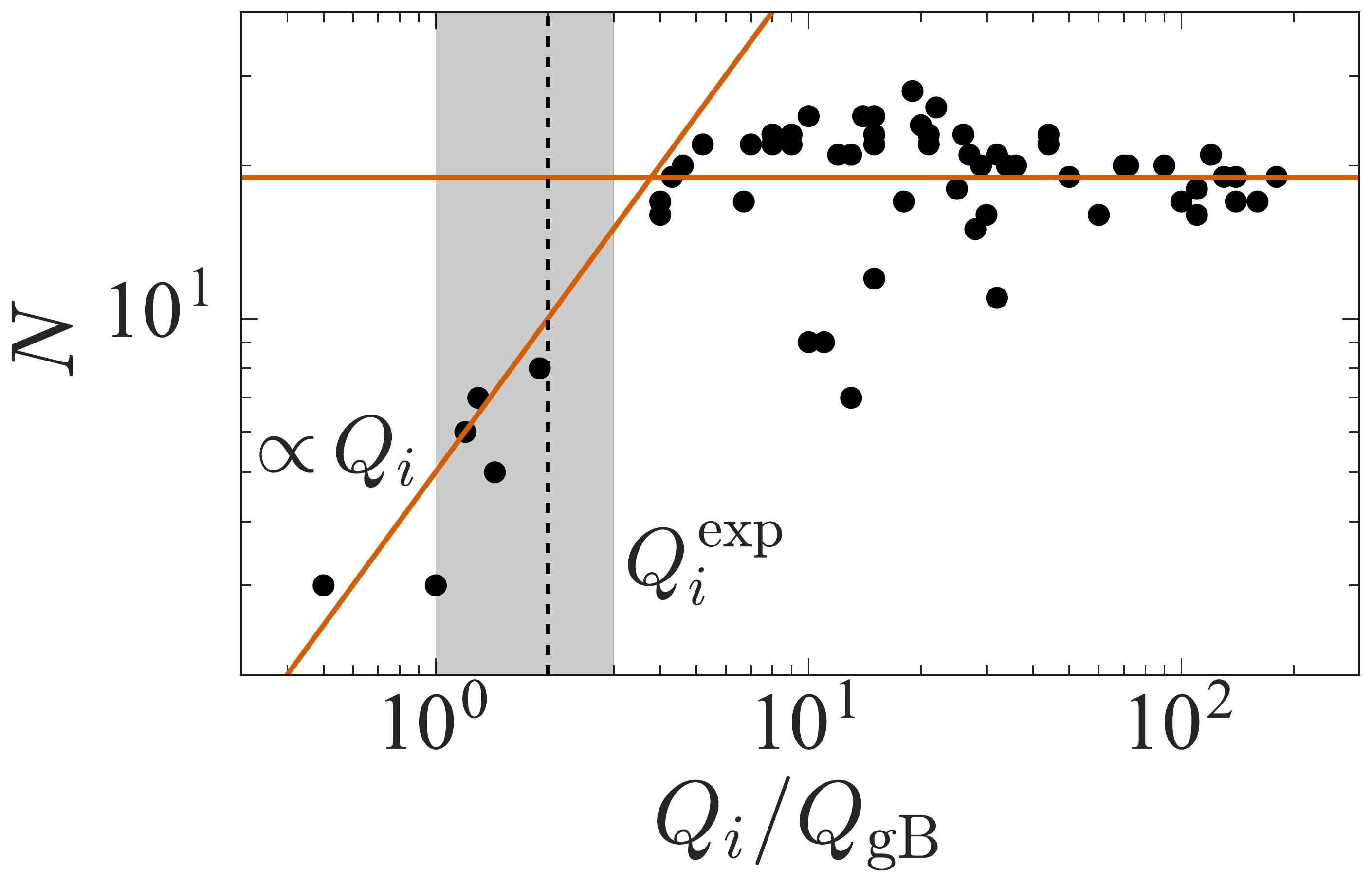}
      \caption{}
      \label{fig:nblobs}
    \end{subfigure}
    \begin{subfigure}{0.49\linewidth}
      \includegraphics[width=\linewidth]{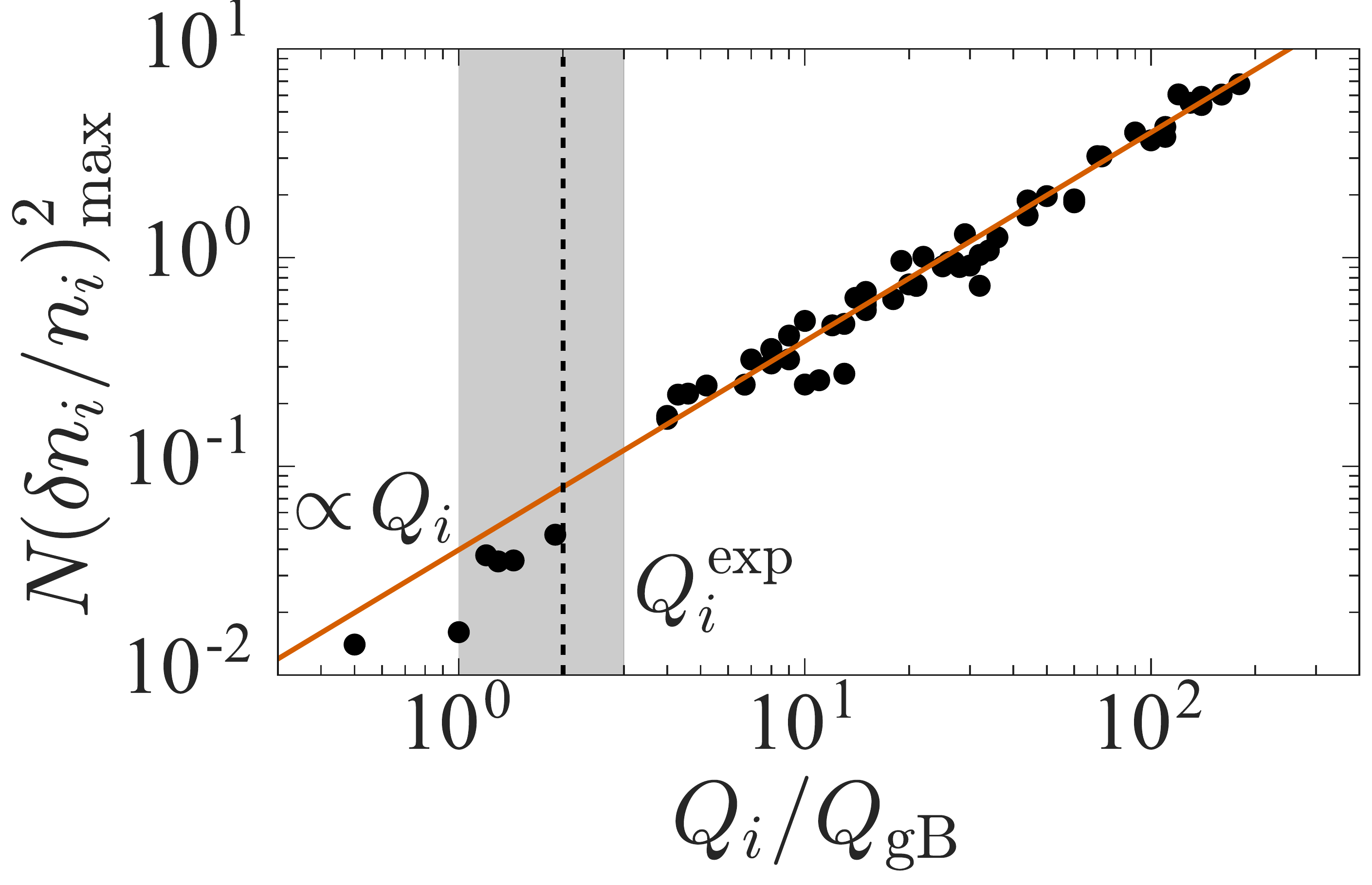}
      \caption{}
      \label{fig:n_amp_squared}
    \end{subfigure}
    \caption{
      \subref*{fig:nblobs} Number of structures (defined as instance of the
      perturbed density with an amplitude
      above 75\% of the maximum) versus $Q_i/Q_{\mathrm{gB}}$.  It grows as
      $Q_i/Q_{\mathrm{gB}}$ increases up to and slightly beyond the
      experimental value $Q_i^{\exp}$. Eventually the volume is filled with
      structures and their number tends to a constant.  The scaling $Q_i
      \propto N$ is shown for reference.
      \subref*{fig:n_amp_squared} Confirmation of the
      scaling~\eqref{n_amp_scaling}, where the red line indicates a line
      $\propto Q_i$. We note that simulations near marginality are relatively
      difficult to saturate leading to the low number of simulations around
      $Q_i^{\exp}$.  However, the trend is still clear even for those
      simulations.
    }
  \end{figure}
  \Figref{nblobs} shows the results of the above analysis applied to our entire
  set of simulations: the number of
  structures $N$ with amplitudes above the 75$^{\mathrm{th}}$ percentile versus the
  ion heat flux $Q_i/Q_{\mathrm{gB}}$.  As in \figref{amplitude}, there are two
  distinct regimes: $N$ grows with $Q_i/Q_{\mathrm{gB}}$ until the structures
  have filled the simulation domain (which happens just above the experimental
  value of the flux), whereupon $N$ tends to a constant.

  Taking \figsref{amplitude}{nblobs} in combination, we have, roughly,
  \begin{equation}
    \frac{Q_i}{Q_{\mathrm{gB}}} \sim N {\qty(\frac{\delta n_i}{n_i})}_{\max}^2,
    \label{n_amp_scaling}
  \end{equation}
  i.e., near the threshold, the turbulent heat flux increases because coherent
  structures become more numerous (but not more intense), whereas away from the
  threshold, it does so because the fluctuation amplitude increases (at a
  roughly constant number of structures). This relationship is confirmed
  by~\figref{n_amp_squared}, where the scaling \eqref{n_amp_scaling} is checked
  directly.

  \subsubsection{Shear due to zonal flows}
  \label{sec:zf_shear}

  In the conventional picture of the saturation mechanism of ITG-driven
  turbulence, zonal modes play a key role~\cite{Waltz1994, Lin1998, Dimits2000,
  Rogers2000, Diamond2005}. Zonal modes are fluctuations in the system with
  $k_y = k_\parallel = 0$ and $k_x > 0$, i.e.,\ they have finite radial extent,
  but are poloidally symmetric. They are generated by nonlinear interactions in
  the system. Previous work~\cite{Dimits2000} on the transition to turbulence
  in the case of $\gamma_E=0$ showed that near the turbulence threshold
  (approached by varying the equilibrium parameter $\kappa_T$), turbulence is
  regulated by strong zonal flows, which can cause an upshift in the critical
  $\kappa_T$ required for a saturated strongly turbulent state. However, in our
  system, the near-threshold cases the background flow shear plays an important
  role, and also has a suppressing effect on the turbulence.

  Here, we investigate the relative importance of the mean shear and the shear
  resulting from the self-generated zonal flows.  The shear due to the zonal
  flows $V'_{\mathrm{ZF}}$ is calculated from \eqref{v_er} by considering only
  the poloidally symmetric component, that is
  \begin{equation}
    V'_{\mathrm{ZF}} = \frac{c}{a B_{\mathrm{ref}}} \frac{q_0}{r_0}
    \frac{1}{|\nabla \alpha|} \pdv[2]{\varphi_{\mathrm{ZF}}}{x},
    \label{v_zf_prime}
  \end{equation}
  where $\alpha$ is the binormal coordinate, $\varphi_{\mathrm{ZF}}$ is the
  poloidally symmetric component of $\varphi$ and $V'_{\mathrm{ZF}}$ is a
  function only of $t$ and $x$.  To determine whether the zonal shear will
  dominate over the mean shear $\gamma_E$ we calculate the RMS value of the
  zonal shear, $\gamma_{\mathrm{ZF}}$:
  \begin{equation}
    \gamma_{\mathrm{ZF}} = \left< V^{\prime 2}_{\mathrm{ZF}}\right>^{1/2}_{t,x},
    \label{g_zf}
  \end{equation}
  where $\ensav{\cdots}{t,x}$ indicates an average over $t$ and $x$. We can now
  compare $\gamma_{\mathrm{ZF}}$ with $\gamma_E$ to determine their relative
  size as a function of our equilibrium parameters.

  \begin{figure}[t]
    \centering
      \begin{subfigure}{0.51\linewidth}
        \includegraphics[width=\linewidth]{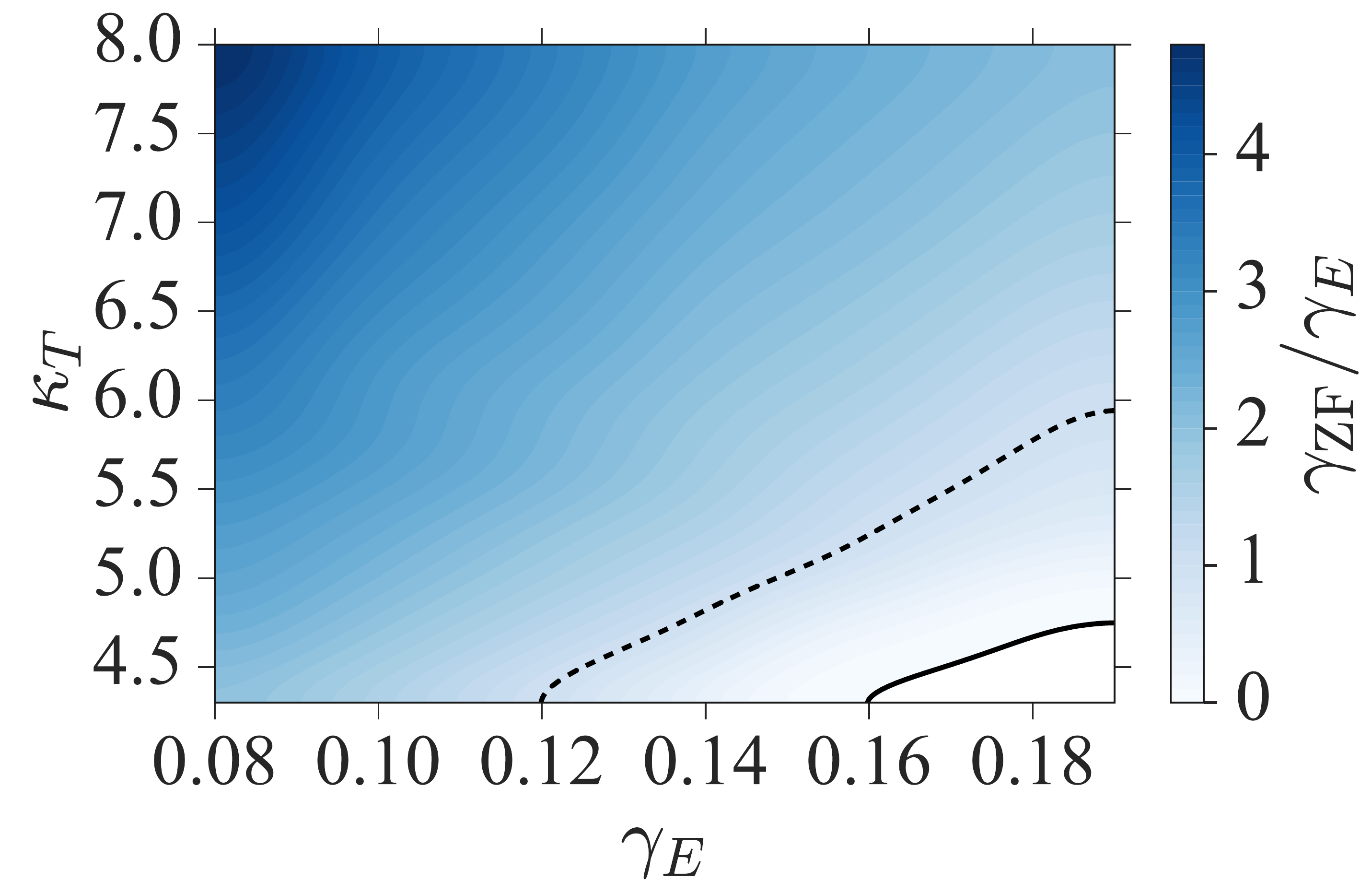}
        \caption{}
        \label{fig:zf_shear_contour}
      \end{subfigure}
      \hfill
      \begin{subfigure}{0.47\linewidth}
        \includegraphics[width=\linewidth]{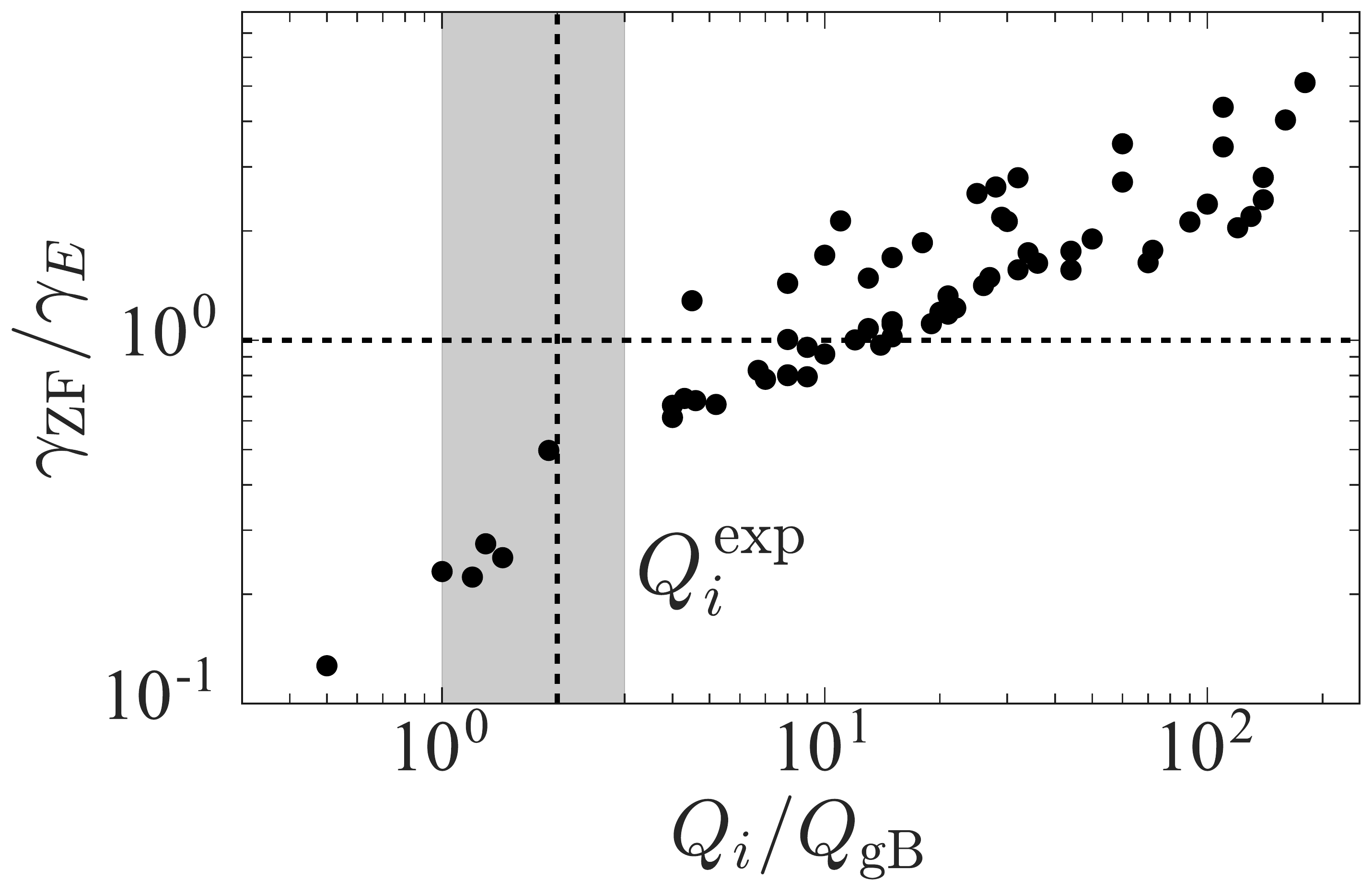}
        \caption{}
        \label{fig:zf_shear_q_scatter}
      \end{subfigure}
      \caption{
        \subref*{fig:zf_shear_contour} The ratio of zonal shear to mean
        equilibrium flow shear $\gamma_{\mathrm{ZF}}/\gamma_E$ over the same
        range of $\kappa_T$ and $\gamma_E$ as shown in
        \figref{contour_heatmap}. The zonal shear and mean flow shear are
        comparable when $\gamma_{\mathrm{ZF}}/\gamma_E \sim 1$. The white
        region in the lower right-hand corner, separated by a solid line,
        indicates the region where there is no turbulence, i.e., $Q_i=0$ [see
        \figref{contour_heatmap}], and the dashed black line indicates
        $\gamma_{\mathrm{ZF}}/\gamma_E = 1$.
        \subref*{fig:zf_shear_q_scatter} $\gamma_{\mathrm{ZF}}/\gamma_E$ as a
        function of $Q_i/Q_{\mathrm{gB}}$.  The vertical dashed line indicates
        the value of the experimental heat flux and the horizontal dashed line
        indicates $\gamma_{\mathrm{ZF}}/\gamma_E = 1$.
      }
    \label{fig:zf_shear}
  \end{figure}
  \Figref{zf_shear_contour} shows the ratio of the zonal shear to the flow
  shear, $\gamma_{\mathrm{ZF}}/\gamma_E$, as a function of $\kappa_T$ and
  $\gamma_E$ over the same parameter range as shown in
  \figref{contour_heatmap}. The magnitudes of $\gamma_{\mathrm{ZF}}$ and $\gamma_E$
  are comparable where $\gamma_{\mathrm{ZF}}/\gamma_E \sim 1$, which is
  indicated by the dashed line. We see that the regime in which
  $\gamma_{\mathrm{ZF}}$ and $\gamma_E$ become comparable occurs some distance
  away from the turbulence threshold (solid line).  Therefore, close to the
  threshold (small $\gamma_{\mathrm{ZF}}/\gamma_E$), we expect the shear due to
  the background flow to dominate, while far from the threshold (large
  $\gamma_{\mathrm{ZF}}/\gamma_E$), we expect the shear due to the zonal flows
  to dominate.

\begin{figure}[t]
     \centering
     \includegraphics[width=0.6\linewidth]{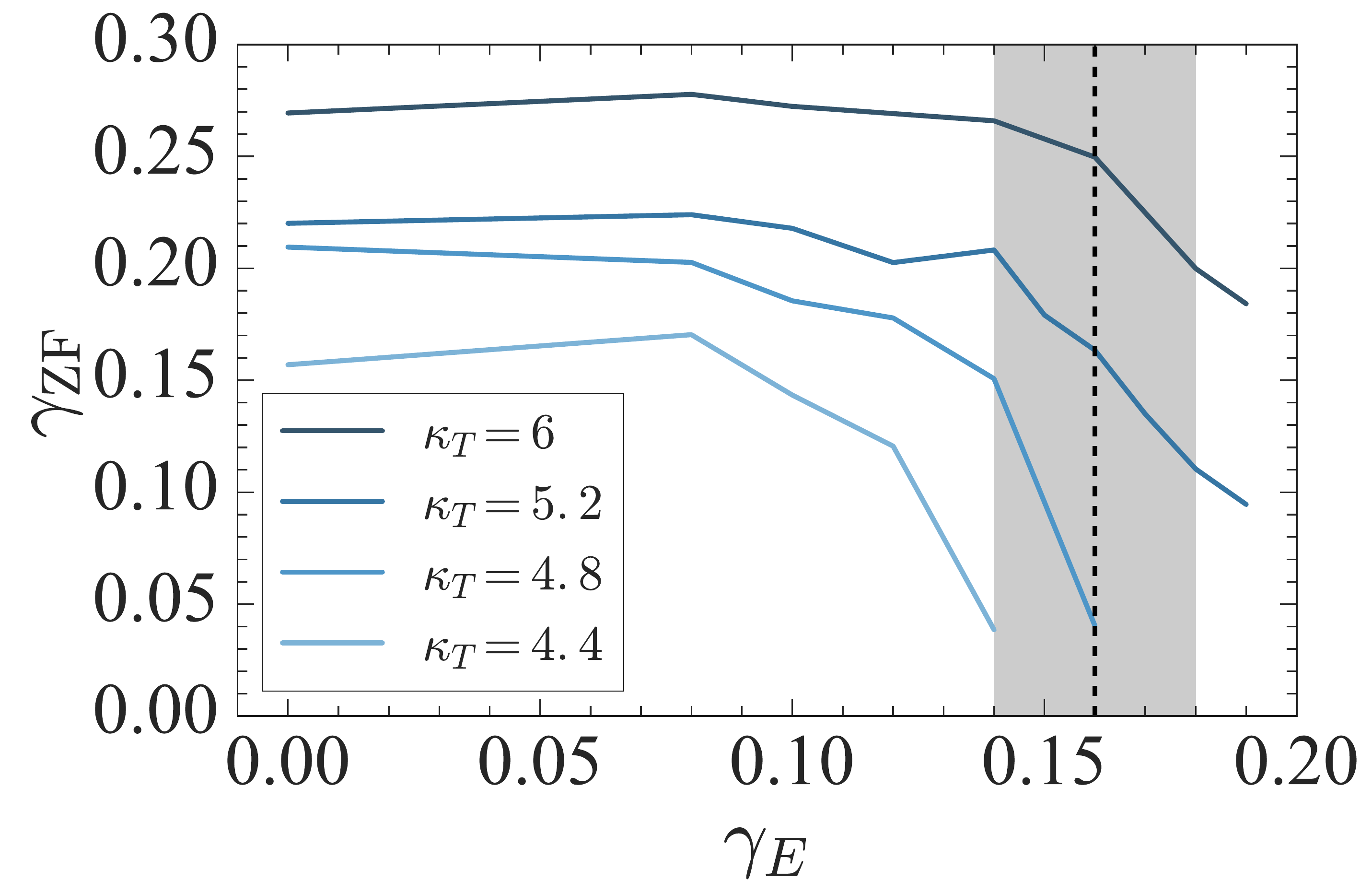}
     \caption{
       Zonal shear $\gamma_{\mathrm{ZF}}$ as a function of background flow
       shear $\gamma_E$ showing that zonal shear is comparable between low
       $\gamma_E$ (high $Q_i/Q_{\mathrm{gB}}$) cases and $\gamma_E=0$ cases.
     }
     \label{fig:zf_shear_lines}
   \end{figure}

  \Figref{zf_shear_contour} suggests that the change in
  $\gamma_{\mathrm{ZF}}/\gamma_E$ is effectively a function of the distance
  from the turbulence threshold.  \Figref{zf_shear_q_scatter} shows this
  dependence explicitly: $\gamma_{\mathrm{ZF}}/\gamma_E$ as a function of
  $Q_i/Q_{\mathrm{gB}}$.  The vertical dashed line indicates
  $Q_i^{\exp}/Q_{\mathrm{gB}}$ and we see that $\gamma_{\mathrm{ZF}}/\gamma_E$
  is quite small at this value. This suggests that zonal shear plays a weaker
  role than $\gamma_E$ in regulating experimentally relevant turbulence for
  this MAST configuration.  Therefore, near-threshold and far-from-threshold
  turbulent states are distinguished by whether it is the mean or the zonal
  shear that plays a dominant role.  Far from the threshold, the turbulence is
  likely similar to conventional ITG-driven turbulence in the absence of
  background flow shear. This is supported by \figref{zf_shear_lines}, which
  shows $\gamma_{\mathrm{ZF}}$ as a function of $\gamma_E$. We see that for low
  $\gamma_E$ and/or high $\kappa_T$ (i.e. for cases far from the threshold),
  $\gamma_{\mathrm{ZF}}$ is approximately independent of $\gamma_E$ and close
  to the value that it takes at $\gamma_E=0$.

  \subsubsection{Summary}

  In summary, we can describe the behaviour of the MAST turbulence that we
  studied as follows. For equilibrium parameters near the turbulence threshold
  (including for cases that match the experiment), the density and temperature
  fluctuations (and hence the heat flux) are concentrated in long-lived,
  intense coherent structures. As the equilibrium parameters $(\kappa_T,
  \gamma_E)$ depart slightly from their critical values into the more strongly
  driven regime, the number of the coherent structures increases rapidly while
  their amplitude stays roughly constant (in contrast to the conventional
  supercritical turbulence, where the amplitude increases with $\kappa_T$).
  Increasing $\kappa_T$ or decreasing $\gamma_E$ further leads to the
  structures filling the simulation domain and any further increase in the heat
  flux is caused by an increase in fluctuation amplitude. The latter regime is
  similar to the conventional plasma turbulence, where zonal flows are the
  dominant mechanism for regulating turbulence. In contrast, we have
  demonstrated that in the near-threshold cases, the zonal shear is small
  compared to the mean flow shear and so is unlikely to matter.

\section{Correlation analysis and comparison with BES}
\label{sec:struc_of_turb}

In the previous section, we used nonlinear simulations to demonstrate the
complicated nature of the MAST turbulence that we are studying, in particular
the details of a subcritical transition to turbulence. In this section, we seek
to establish the experimental relevance of our simulations using quantitative
comparisons between the fluctuation fields predicted numerically and those
measured by the BES diagnostic. We will review the BES diagnostic and
experimental results (section~\ref{sec:corr_exp}) from Ref.~\cite{Field2014},
and then present two types of correlation analysis of our nonlinear simulations
(the correlation-analysis techniques are described in
appendix~\ref{App:corr_overview}). The first analysis will be of GS2 density
fluctuations with a ``synthetic BES diagnostic'' applied to simulate what would
be measured by a real BES diagnostic (section~\ref{sec:corr_synth}).  We will
consider the results from nonlinear simulations with values of $(\kappa_T,
\gamma_E)$ within the experimental-uncertainty range and compare them with the
experimental results.  The second analysis will be of the raw GS2 density
fluctuations, both within the experimental-uncertainty range and, as a function
of $Q_i/Q_{\mathrm{gB}}$, for our entire parameter scan
(section~\ref{sec:corr_gs2}). In this latter case we will emphasise the extent
to which it is the distance from the turbulence threshold rather than
individual values of $\kappa_T$ or $\gamma_E$ that determines the statistical
characteristics of the density fluctuations.

\subsection{Beam emission spectroscopy diagnostic}
\label{sec:bes}
  Turbulent eddies in tokamak plasmas are anisotropic due to the strong
  background magnetic field. In the parallel
  direction, turbulent eddies have a length scale comparable to the system
  size, which in a torus is the \emph{connection length}
  $\pi qR$~\cite{Barnes2011}, i.e., $l_\parallel \sim \pi qR$ ($\approx 6 $~m
  for MAST).  In the
  direction perpendicular to the magnetic field,
  ITG-unstable turbulent structures have a typical length scale of the order of
  the ion gyroradius $l_\perp \sim \rho_i \sim 1$~cm.  Therefore, in the plane
  perpendicular to the magnetic field, we are interested in two-dimensional
  measurements of fluctuating quantities at approximately the scale of
  $\rho_i$.  Beam emission spectroscopy is a diagnostic technique that was
  developed to address this need.  Specifically, the BES diagnostic on
  MAST~\cite{Field2009, Field2012} is designed to measure ion-scale density
  fluctuations in a radial-poloidal plane.  Density fluctuations are
  inferred from D$_\alpha$ emission produced by the NBI beam as it penetrates
  the plasma. The measured fluctuating intensity of the D$_\alpha$ emission is
  proportional to the local plasma density at the corresponding viewing
  location, and the two quantities are related via point-spread functions
  (PSFs)~\cite{Ghim2012, Field2014, Fox2016}. The PSFs depend on the magnetic
  equilibrium, beam parameters, viewing location, and plasma profiles and as a
  result, have to be calculated explicitly for each
  measurement~\cite{Ghim2012}.

  Recent work~\cite{Fox2016}, based on a subset of simulations presented here,
  has shown that the PSFs play an important role in the measurement of
  turbulence and that the precise form that they take determines a lower bound
  on the BES resolution as well as affecting the measurement of the turbulent
  structures and density fluctuation levels -- effects that we will also
  consider in this work. For further details on the MAST BES system the reader
  is referred to Refs.~\cite{Field2009,Field2012,Ghim2012} and, for a detailed
  study of the effect of PSFs on the measurement of turbulent structures, to
  Ref.~\cite{Fox2016}.

\subsection{Experimental BES results}
  \label{sec:corr_exp}
  \begin{figure}[t]
    \centering
    \begin{subfigure}{0.49\linewidth}
      \includegraphics[width=\linewidth]{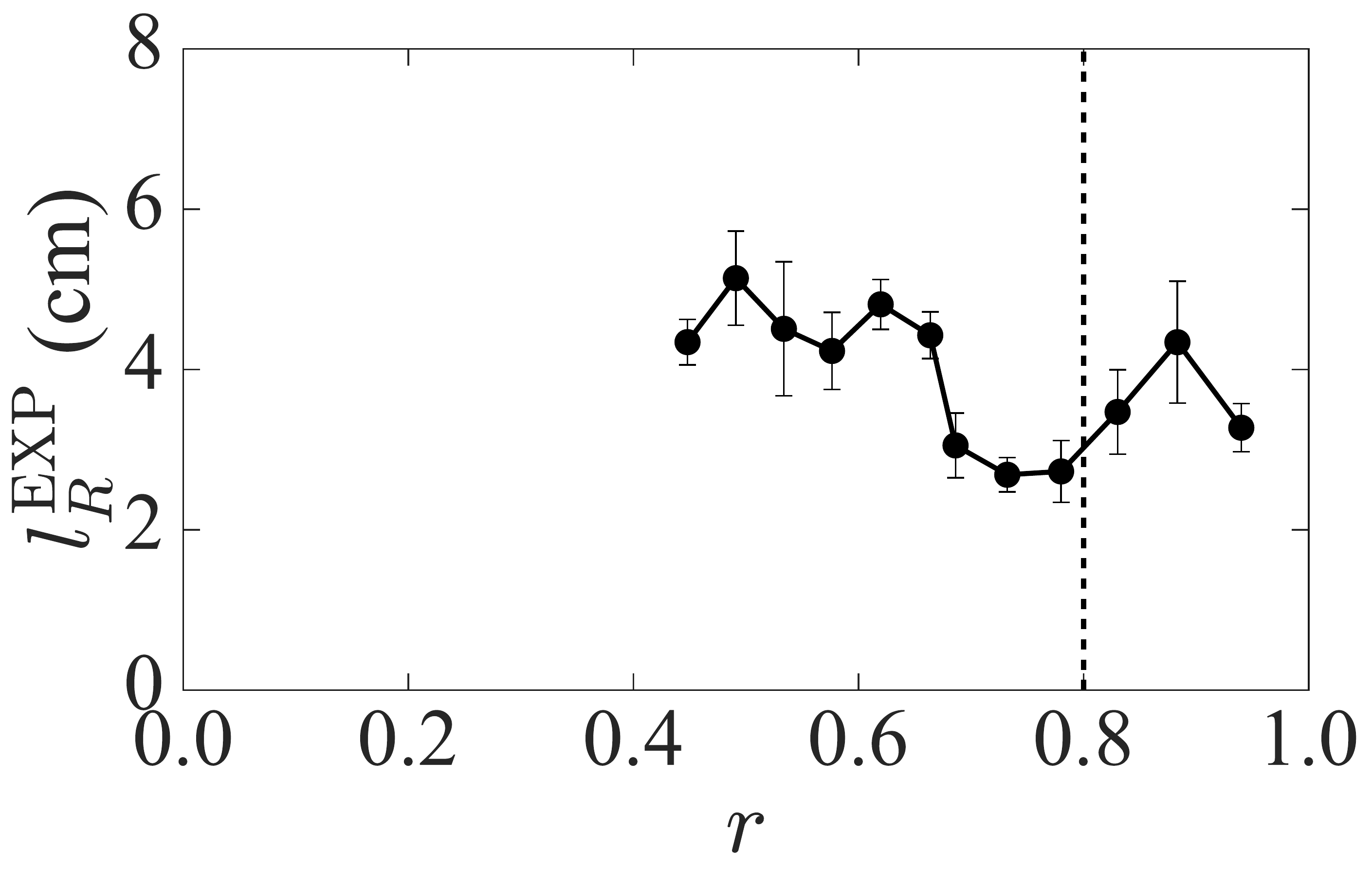}
      \caption{}
      \label{fig:lr_exp}
    \end{subfigure}
    \hfill
    \begin{subfigure}{0.49\linewidth}
      \includegraphics[width=\linewidth]{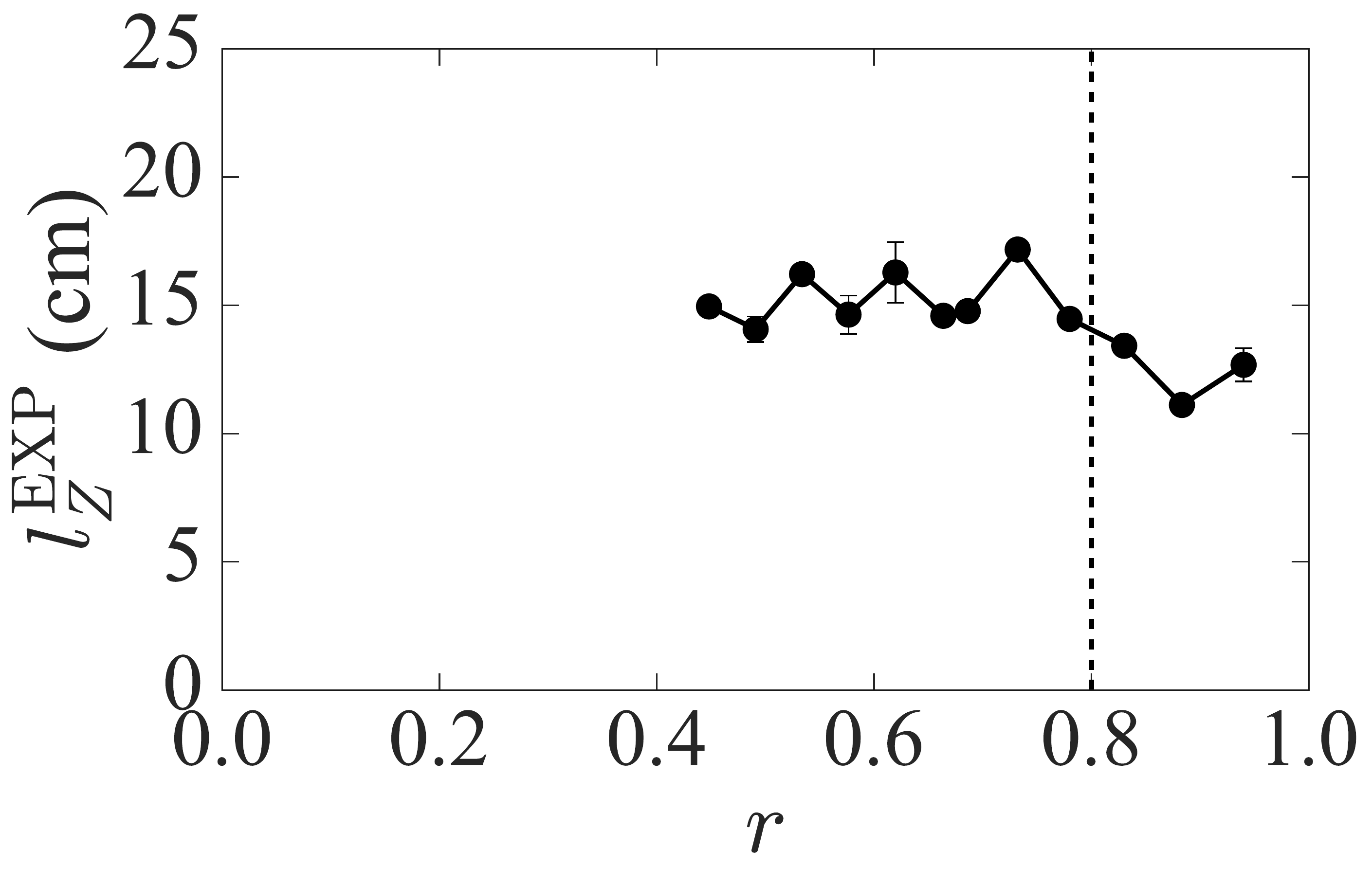}
      \caption{}
      \label{fig:lz_exp}
    \end{subfigure}
    \begin{subfigure}{0.49\linewidth}
      \includegraphics[width=\linewidth]{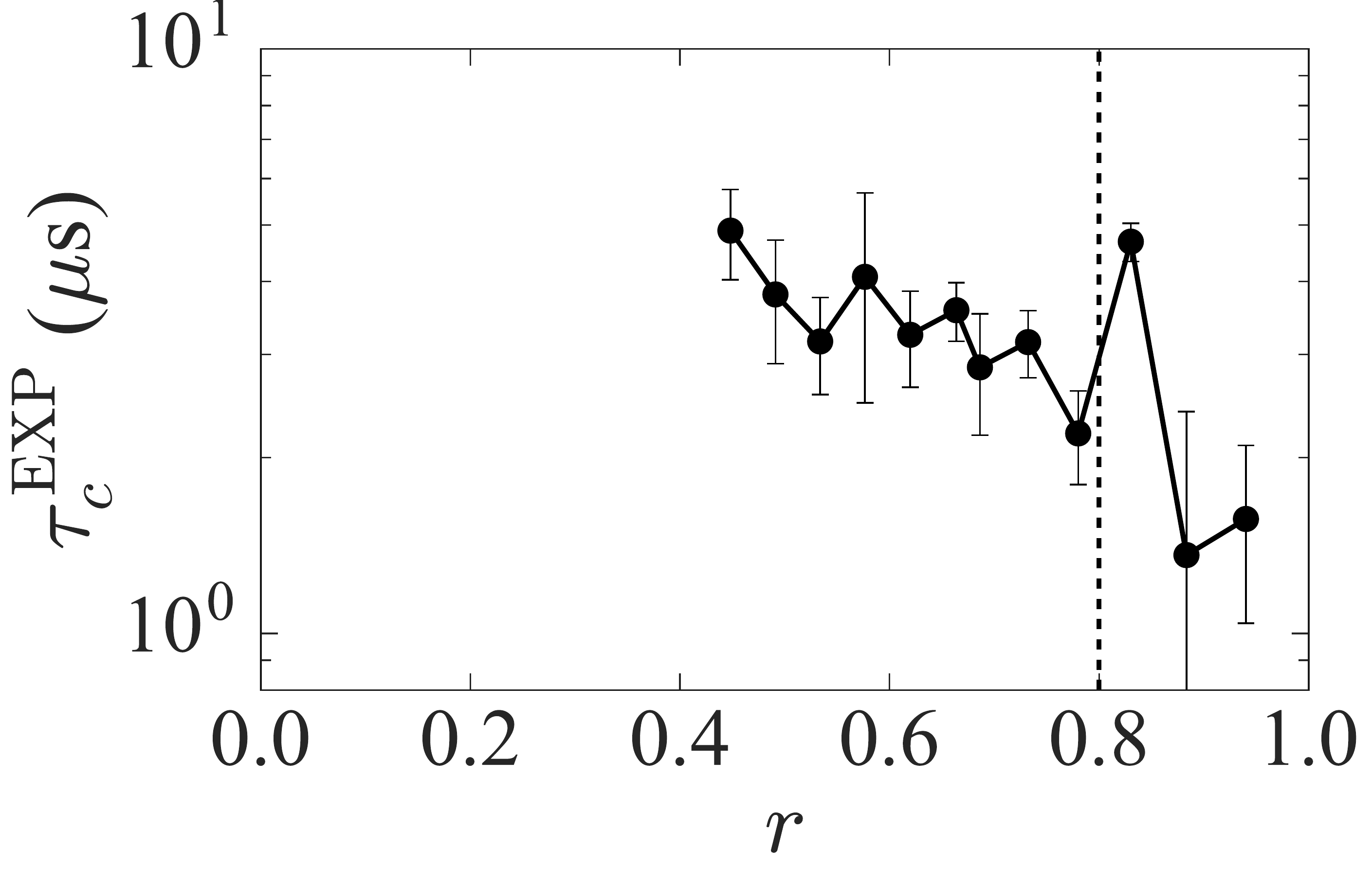}
      \caption{}
      \label{fig:tau_exp}
    \end{subfigure}
    \hfill
    \begin{subfigure}{0.49\linewidth}
      \includegraphics[width=\linewidth]{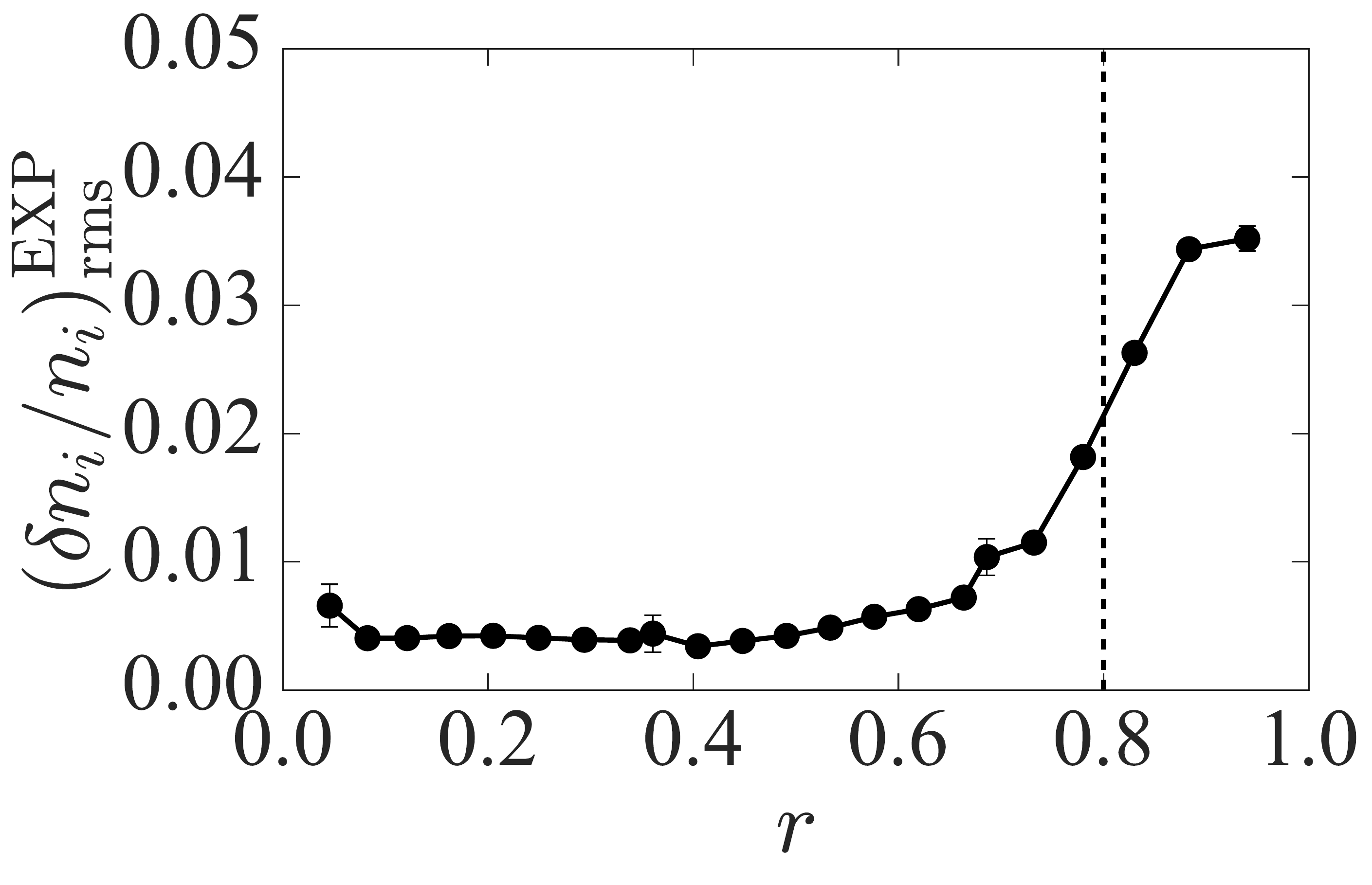}
      \caption{}
      \label{fig:n_exp}
    \end{subfigure}
    \caption[Experimental correlation results]{
      Results of the correlation analysis of BES data from MAST discharges
      \#27272, \#27268, and \#27274 combined to give correlation properties of
      the turbulence as functions of $r=D/2a$:
      \subref*{fig:lr_exp} radial correlation length $l_R^{\mathrm{EXP}}$,
      \subref*{fig:lz_exp} poloidal correlation length $l_Z^{\mathrm{EXP}}$,
      \subref*{fig:tau_exp} correlation time $\tau_c^{\mathrm{EXP}}$, and
      \subref*{fig:n_exp} RMS fluctuation amplitude $\qty( \delta n_i /
      n_i)^{\,\mathrm{EXP}}_{\mathrm{rms}}$. These quantities are defined in
      appendix~\ref{App:corr_overview}. Turbulence was suppressed for $r
      \lesssim 0.4$. The vertical dashed line indicates the radius
      corresponding to the local equilibrium configurations for which we
      performed our simulations.
    }
    \label{fig:exp_corr_results}
  \end{figure}
  Before applying the correlation analysis to our simulations, we review the
  experimental results from MAST discharge \#27274 first presented
  in Ref.~\cite{Field2014}, to which we will be comparing our own calculation.
  As discussed in section~\ref{sec:exp_profiles}, MAST discharge \#27274 forms
  part of a set of three discharges, which together allowed measurement of
  turbulence correlation properties over the whole outer radius.
  \Figref{exp_corr_results} shows the experimental results obtained for the
  radial correlation length $l_R^{\mathrm{EXP}}$, the poloidal correlation
  length $l_Z^{\mathrm{EXP}}$, the correlation time $\tau_c^{\mathrm{EXP}}$,
  and the RMS density fluctuations ${\qty(\delta n_i /
  n_i)}^{\mathrm{EXP}}_{\mathrm{rms}}$ as functions of $r = D/2a$. The vertical
  dashed line in each plot indicates the radius at which our simulations were
  done and the corresponding values of the correlation parameters.
  These target experimental values are (after interpolating between the
  experimental data points):
  \begin{align}
    \begin{split}
    l_R^{\mathrm{EXP}} &= 3 \pm 0.4~\mathrm{cm}, \\
    l_Z^{\mathrm{EXP}} &= 14.06 \pm 0.09~\mathrm{cm}, \\
    \tau_c^{\mathrm{EXP}} &= 3.2 \pm 0.4~\mu\mathrm{s}, \\
    {\qty(\frac{\delta n_i}{n_i})}^{\mathrm{EXP}}_{\mathrm{rms}} &= 0.0214 \pm 0.0006.
    \label{exp_results}
    \end{split}
  \end{align}
  We will be comparing the correlation parameters calculated from our
  simulations in the following sections to those in \eqref{exp_results}.

\subsection{Correlation analysis with synthetic diagnostic}
  \label{sec:corr_synth}
  In order to compare our simulated density field  with the BES-measured ones,
  a number of data transformations were necessary. We mapped our density
  fluctuations ``measured'' in the outboard midplane (at $\theta = 0$) from GS2
  $(x, y)$ coordinates onto a poloidal $(R,Z)$-plane and also transformed them
  from the rotating plasma frame, the frame in which our simulations were
  performed, to the laboratory frame, as explained in
  appendix~\ref{App:real_space_transform}.  We then applied a synthetic
  diagnostic to our density fluctuations, including the point-spread functions
  (described in section~\ref{sec:bes}), which models instrumentation effects and
  atomic physics, adds artificial noise similar to that found in the
  experiment, and maps the density-fluctuation field onto an $8 \times 4$
  grid similar to the arrangement of BES channels. An important feature of the
  analysis of experimental data is the application of a filter to remove
  high-energy radiation present in the experiment. We have included this filter
  for consistency with experimental measurements. The results \emph{without}
  this filter are presented in appendix~\ref{App:no_spike}.

  \Figref{synth_corr_results} shows the radial correlation length
  $l_R^{\,\mathrm{SYNTH}}$, poloidal correlation length
  $l_Z^{\,\mathrm{SYNTH}}$, correlation time $\tau_c^{\,\mathrm{SYNTH}}$, and
  RMS density fluctuation $\qty(\delta n_i /
  n_i)^{\,\mathrm{SYNTH}}_{\mathrm{rms}}$ calculated from our simulations with
  the synthetic diagnostic applied using the correlation analysis described in
  appendix~\ref{App:corr_overview}. The errors in the correlation
  parameters shown in \figref{synth_corr_results}, and elsewhere, are
  determined from the fitting procedures described in appendix~\ref{App:corr_overview}.
  We expect these values to agree with the experimentally measured correlation
  parameters in \eqref{exp_results} because the equilibrium parameters
  $\kappa_T$ and $\gamma_E$ at which the results shown in
  \figref{synth_corr_results} were obtained are strictly within the
  experimental-uncertainty range of these parameters.  The dashed lines and
  shaded areas in \figref{synth_corr_results} indicate the experimental values
  and associated errors given in \eqref{exp_results}. The circled points
  indicate the simulations that matched the experimental level of heat flux
  (listed in table~\ref{tab:exp_match_sims}).
  \begin{figure}[t!]
    \centering
    \begin{subfigure}{0.49\linewidth}
      \includegraphics[width=\linewidth]{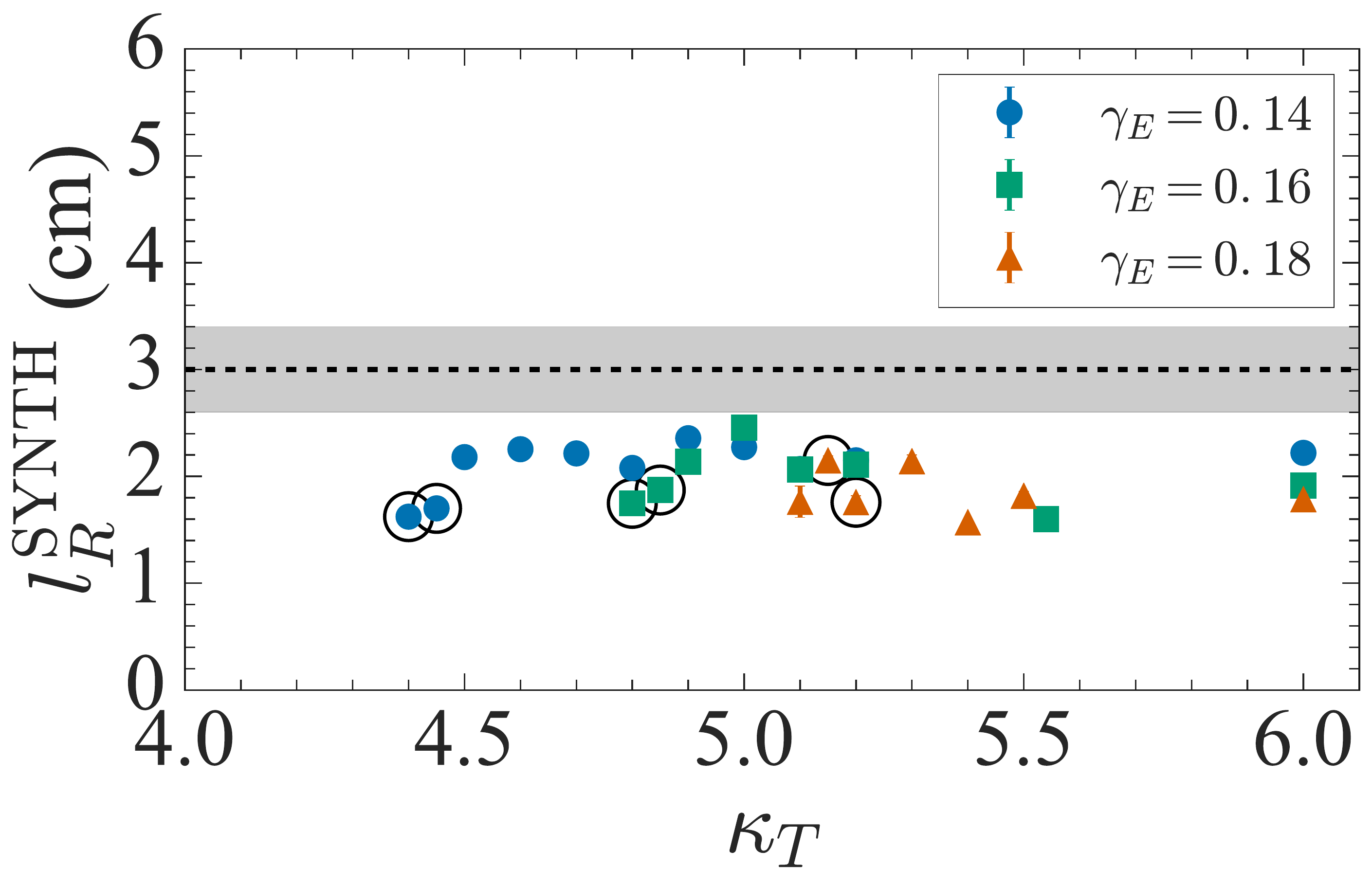}
      \caption{}
      \label{fig:lr_synth}
    \end{subfigure}
    \hfill
    \begin{subfigure}{0.49\linewidth}
      \includegraphics[width=\linewidth]{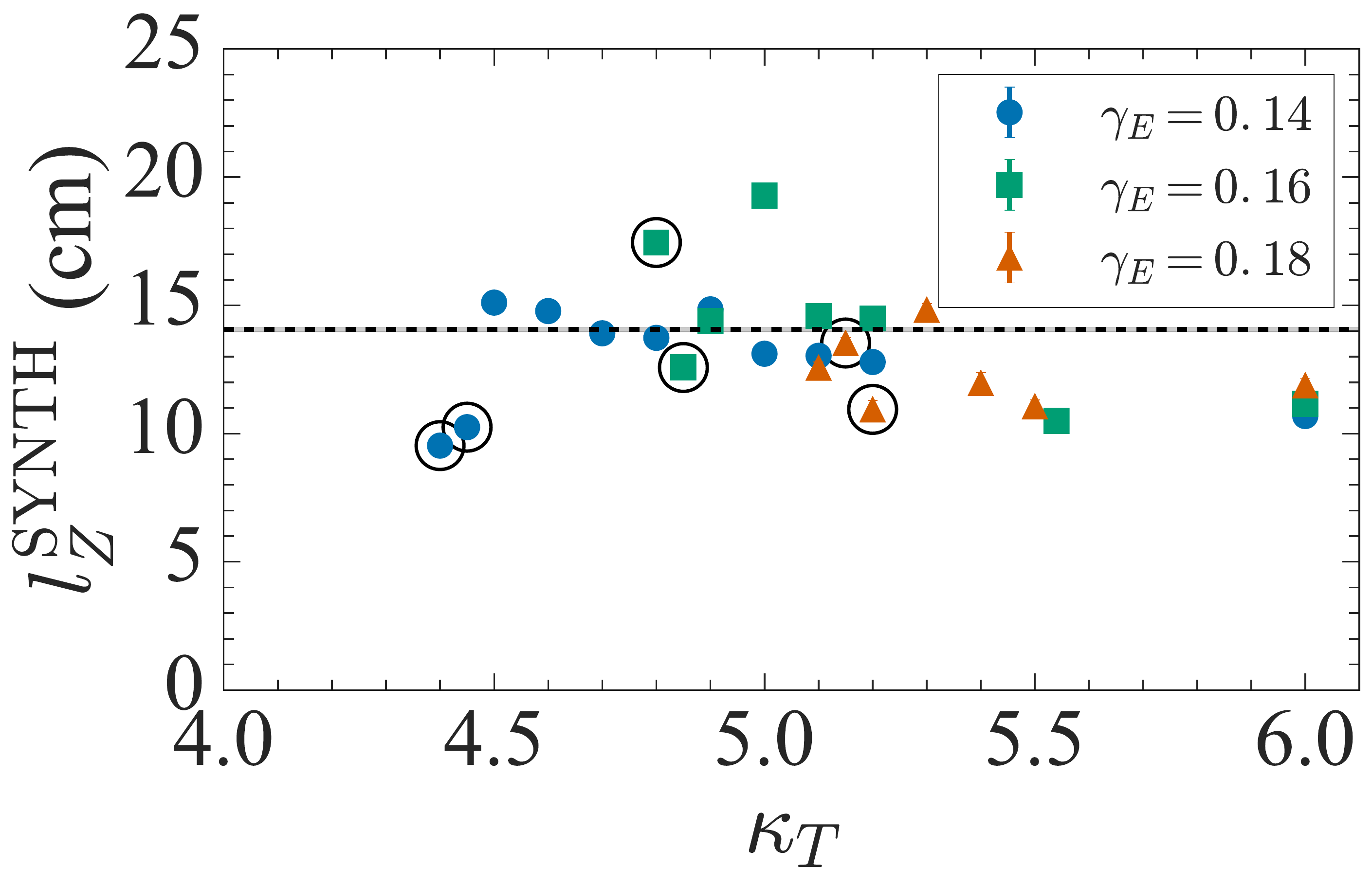}
      \caption{}
      \label{fig:lz_synth}
    \end{subfigure}
    \begin{subfigure}{0.49\linewidth}
      \includegraphics[width=\linewidth]{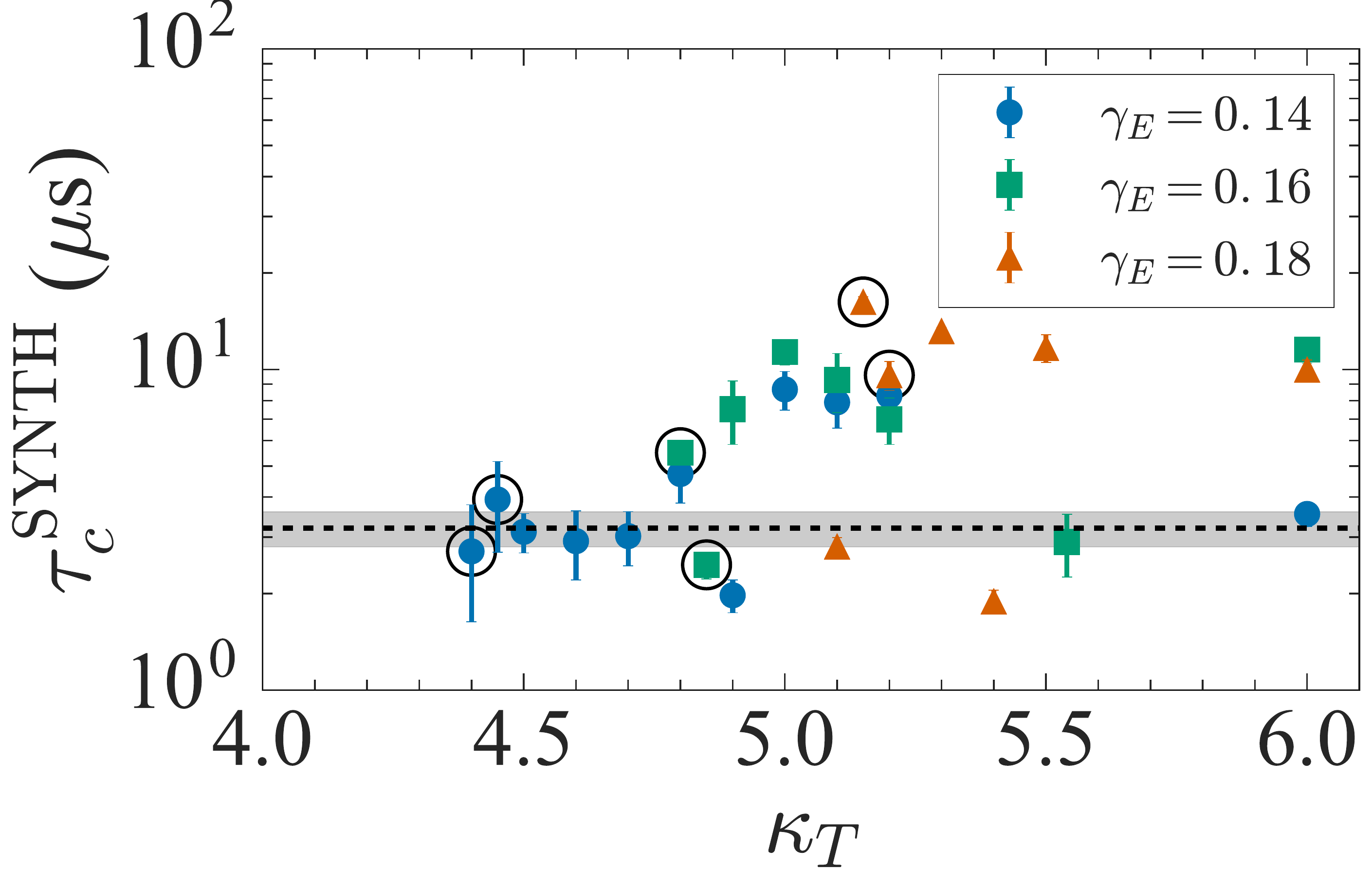}
      \caption{}
      \label{fig:tau_synth}
    \end{subfigure}
    \hfill
    \begin{subfigure}{0.49\linewidth}
      \includegraphics[width=\linewidth]{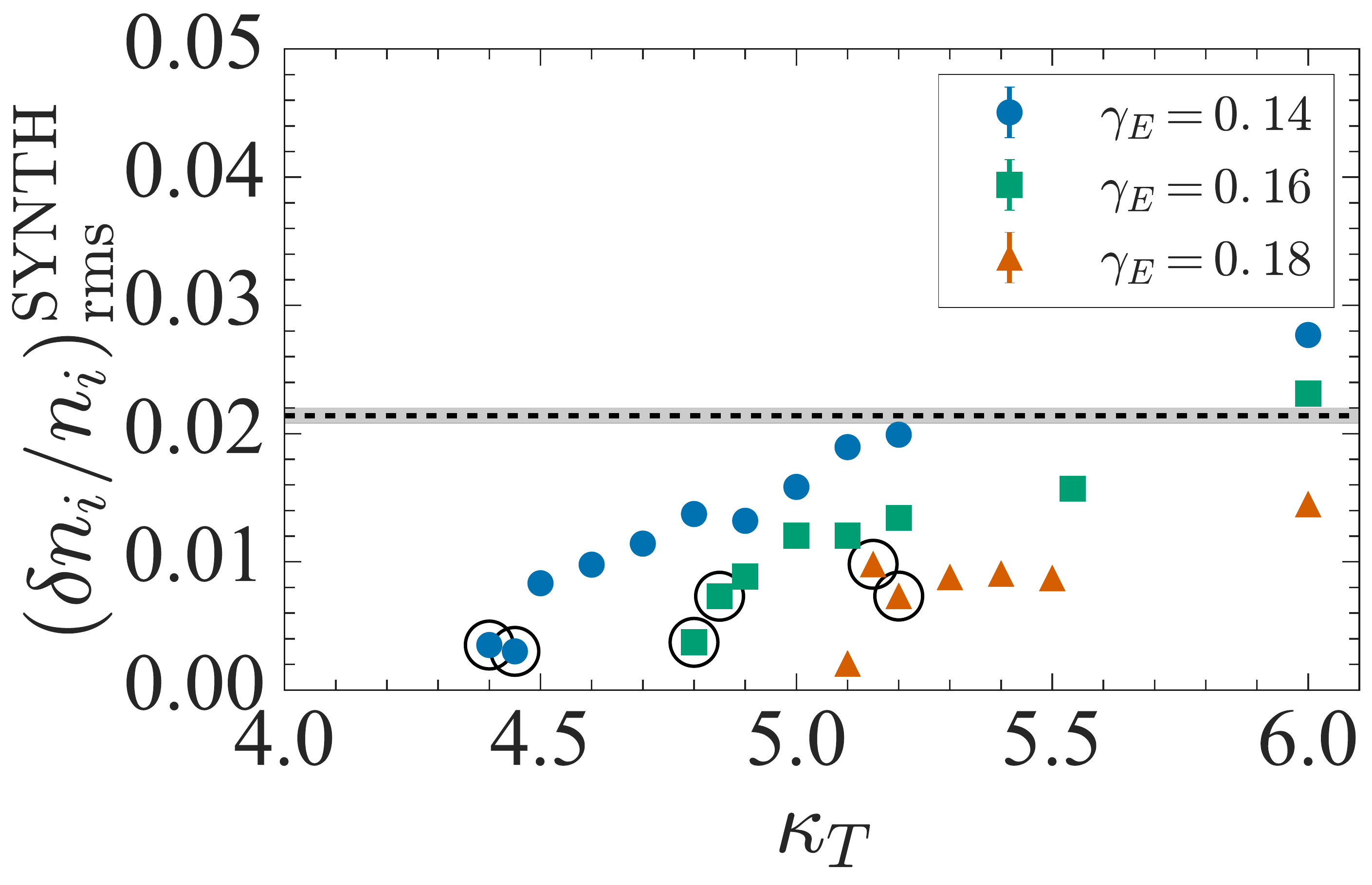}
      \caption{}
      \label{fig:n_synth}
    \end{subfigure}
    \caption[Correlation parameters of synthetic GS2 data]{
      Comparison of correlation parameters obtained via synthetic BES
      measurements of GS2-simulated density field:
      \subref*{fig:lr_synth} radial correlation length $l_R^{\mathrm{SYNTH}}$,
      \subref*{fig:lz_synth} poloidal correlation length $l_Z^{\mathrm{SYNTH}}$,
      \subref*{fig:tau_synth} correlation time $\tau_c^{\mathrm{SYNTH}}$, and
      \subref*{fig:n_synth} RMS fluctuation amplitude $\qty( \delta n_i /
      n_i)^{\,\mathrm{SYNTH}}_{\mathrm{rms}}$ as functions of $\kappa_T$ and
      for several values of $\gamma_E$ within experimental uncertainty. The
      circled points indicate the simulations match the experimental heat flux,
      given in table~\ref{tab:exp_match_sims}. The dashed lines indicate the
      experimental values and the shaded areas the associated error at $r =
      0.8$ obtained from interpolating between experimental measurements seen
      in \figref{exp_corr_results}, which correspond to the local equilibrium
      configuration studied in these simulations.
    }
    \label{fig:synth_corr_results}
  \end{figure}

  Examining \figref{lr_synth}, we see that the values of
  $l_R^{\,\mathrm{SYNTH}}$ are clustered around $2$~cm and below the
  experimental BES measurement $l_R^{\,\mathrm{EXP}} = 3\pm0.4$~cm.  The
  approximate resolution limit in the radial and poloidal directions is
  $\sim2$~cm, the physical separation between BES channels~\cite{Field2009}.
  More recent work studying the measurement effect of the PSFs, concluded that
  the radial resolution limit can be between $2$ and $4$~cm depending on the
  orientation of the PSFs for a given configuration~\cite{Fox2016}. It is,
  therefore, likely that the results shown in \figref{lr_synth} simply confirm
  the radial resolution limit of the experimental analysis and the true value
  of $l_R$ may be lower than 2~cm (as suggested in
  appendix~\ref{sec:radial_corr}). We will confirm this in
  section~\ref{sec:corr_gs2}, where we consider the correlation properties of
  the raw GS2 density fluctuations.

  Figures~\ref{fig:lz_synth}--\subref{fig:n_synth} show $l_Z^{\,\mathrm{SYNTH}}
  =$~$10$--$15$~cm, $\tau_c^{\,\mathrm{SYNTH}} =$~$2$--$15$~$\mu$s, and $\qty(
  \delta n_i / n_i)^{\,\mathrm{SYNTH}}_{\mathrm{rms}} \sim$~$0.005$--$0.03$. We
  see that these values match experimental measurements \eqref{exp_results} for
  certain combinations of $\kappa_T$ and $\gamma_E$.  The values of
  $l_Z^{\,\mathrm{SYNTH}}$ are scattered around the experimental value
  $l_Z^{\,\mathrm{EXP}} = 14.06\pm0.09$~cm, showing no clear trend. While none
  of the cases that match the experimental heat flux (circled cases) match
  $l_Z^{\,\mathrm{EXP}}$, there are several simulations within the
  experimental-uncertainty ranges of $\kappa_T$ and $\gamma_E$ that do.
  Similarly, there are several values of $\tau_c^{\,\mathrm{SYNTH}}$ that also
  match $\tau_c^{\,\mathrm{EXP}}$, including two cases that match the
  experimental level of heat flux. This is a considerable improvement over
  previous nonlinear gyrokinetic simulations of this MAST
  discharge~\cite{Field2014}, which overpredicted $\tau_c^{\,\mathrm{SYNTH}}$
  by two orders of magnitude.

  Examining \figref{n_synth}, we see that $\qty( \delta n_i /
  n_i)^{\,\mathrm{SYNTH}}_{\mathrm{rms}}$ increases with increasing
  $\kappa_T$ or decreasing $\gamma_E$ and that increasing $\gamma_E$ leads
  to a increase in the value of $\kappa_T$ required to achieve the same $\qty(
  \delta n_i / n_i)^{\,\mathrm{SYNTH}}_{\mathrm{rms}}$. The latter trend is
  consistent with \figref{q_vs_tprim}, which showed that increasing $\gamma_E$
  shifted the nonlinear turbulence threshold to higher $\kappa_T$. While
  \figref{n_synth} shows that there is agreement between $\qty( \delta n_i /
  n_i)^{\,\mathrm{SYNTH}}_{\mathrm{rms}}$ and $\qty( \delta n_i /
  n_i)^{\,\mathrm{EXP}}_{\mathrm{rms}}$ at certain combinations of $(\kappa_T,
  \gamma_E)$, we see that the circled cases, representing simulations that
  match the experimental heat flux, have values of $\qty( \delta n_i /
  n_i)^{\,\mathrm{SYNTH}}_{\mathrm{rms}}$ well below $\qty( \delta n_i /
  n_i)^{\,\mathrm{EXP}}_{\mathrm{rms}}$.

  We conclude from the above results that local gyrokinetic simulations are
  a reasonable approximation to the experimental turbulence. We showed that
  $l_Z^{\,\mathrm{SYNTH}}$ and $\tau_c^{\,\mathrm{SYNTH}}$ showed reasonable
  agreement with the experimental measurements within the
  experimental-uncertainty ranges, while there was a discrepancy in the
  predictions of $l_R^{\,\mathrm{EXP}}$ and $\qty( \delta n_i /
  n_i)^{\,\mathrm{EXP}}_{\mathrm{rms}}$. Thus, at least as far as BES
  measurements are concerned, the experimental turbulence and the synthetic
  turbulence are comparable.

  One phenomenon that was not present in our simulations but is present in the
  experiment is high-energy radiation (e.g., neutron, gamma ray, or hard X-ray)
  impinging on the BES detectors. These photons cause high-amplitude spikes in
  the time series, which are typically confined to a single detector channel
  and, therefore, uncorrelated with other channels. These radiation spikes then
  give rise to large auto-correlations at zero time delay, which are unrelated
  to the turbulent field that is being measured. A numerical ``spike filter''
  is normally used to remove radiation spikes by identifying changes above a
  certain threshold between one time point and the next, and replacing the
  high-intensity value with the value of a neighbouring point~\cite{Field2012,
  Fox2016a}. This ``spike filter'' is an important component of the
  experimental analysis of BES data and, while our simulations do not include
  spurious sources of radiation, we have included the ``spike filter'' in the
  analysis of our simulated density fluctuations for consistency with
  experimental analysis.  The results without the ``spike filter'' are given in
  appendix~\ref{App:no_spike}. These results show little difference to those
  with the ``spike filter'' except for the value of $l_Z$. We found that in
  some cases, fast-moving structures in the poloidal direction (especially the
  long-lived structures found in our simulations close to the turbulence
  threshold) were removed by the ``spike'' filter and, therefore, did not
  contribute to the poloidal correlation function, resulting in a drop in
  $l_Z$. This is an important caveat for a future programme of experimental
  detections of these structures. For a more detailed discussion, see
  appendix~\ref{App:no_spike}.

\subsection{Correlation analysis of raw GS2 data}
  \label{sec:corr_gs2}

  Having considered the structure of turbulence processed through a synthetic
  BES diagnostic, we now want to investigate the raw GS2 density fluctuations,
  which will allow us to
  \begin{inparaenum}[(i)]
    \item study the (distorting) effect of the synthetic diagnostic,
    \item study the parallel correlations using GS2 data along the field line,
      and
    \item consider our entire parameter scan to understand how the structure of
      turbulence in MAST might change with the equilibrium parameters
      $\kappa_T$ and $\gamma_E$.
  \end{inparaenum}
  This extends the previous analysis and comparison with simulations performed
  for this MAST discharge~\cite{Field2014}, which only considered the nominal
  equilibrium parameters and simulations with a synthetic diagnostic applied.
  The only operations applied here to the raw GS2 density-fluctuation field
  output are the transformation to the laboratory frame, as explained in
  appendix~\ref{App:lab_frame_transform}, and the transformation from the GS2
  parallel coordinate $\theta$ to the real-space coordinate $\lambda$, as
  explained in appendix~\ref{App:parallel_coord}. Our perpendicular correlation
  analysis is performed over a square $(R,Z)$-plane $20\times20$~cm$^2$ in
  size, located at the centre of our computational domain (see
  appendix~\ref{App:perp_domain}).  We do this to analyse a region of similar
  size to that probed by the BES diagnostic and also to avoid the real-space
  remapping effect at the edges of the radial domain inherent to the GS2
  implementation of flow shear (see appendix~\ref{App:flow_shear}).

  \subsubsection{Correlation parameters for cases within
  experimental-uncertainty range}
  \label{sec:within_uncertainty}
\begin{figure}[t]
    \centering
    \begin{subfigure}[t]{0.49\textwidth}
      \includegraphics[width=\linewidth]{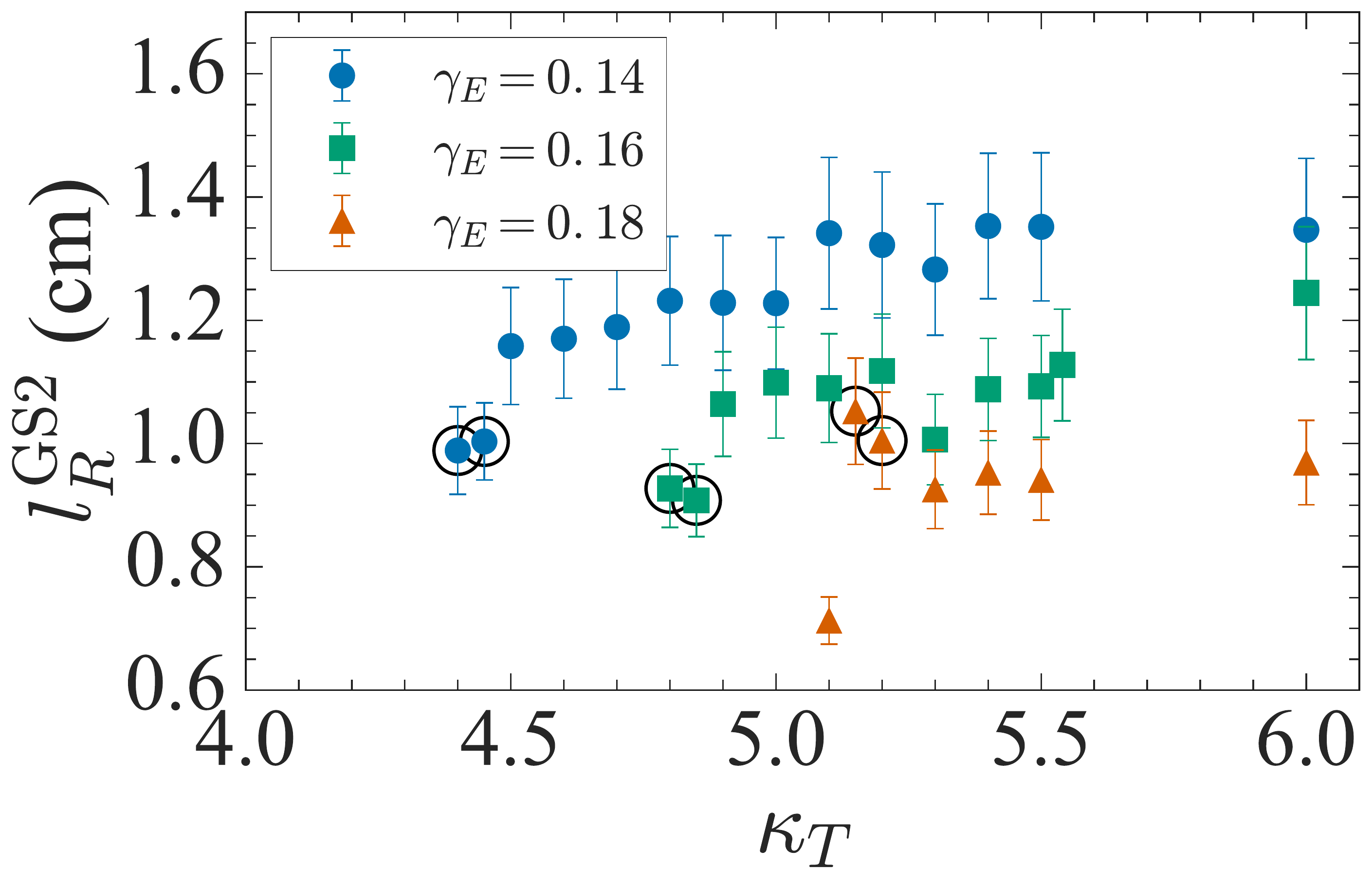}
      \caption{}
      \label{fig:lr_gs2}
    \end{subfigure}
    \begin{subfigure}[t]{0.49\textwidth}
      \includegraphics[width=\linewidth]{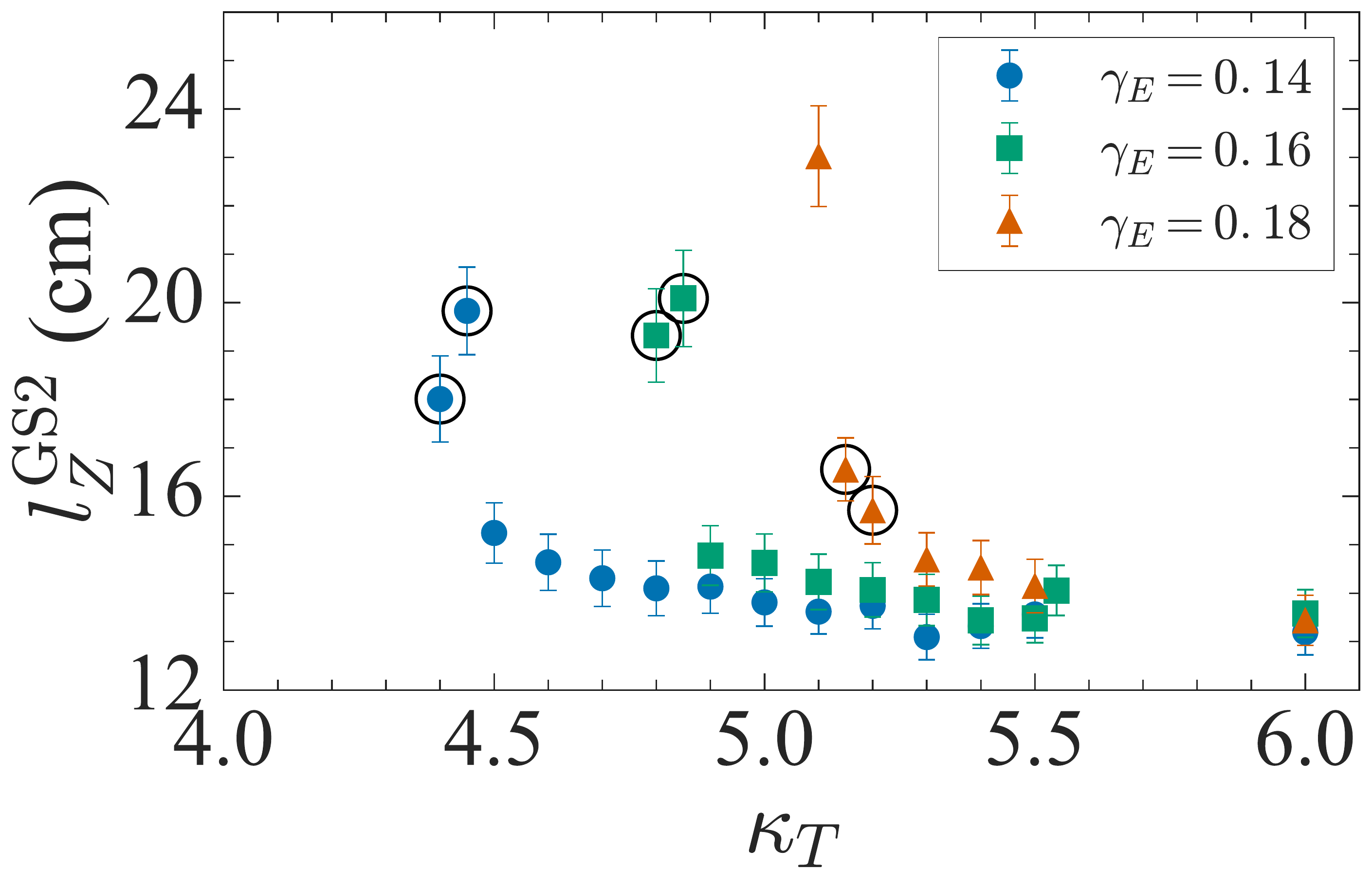}
      \caption{}
      \label{fig:lz_gs2_fixed}
    \end{subfigure}
    \\
    \begin{subfigure}[t]{0.49\textwidth}
      \includegraphics[width=\linewidth]{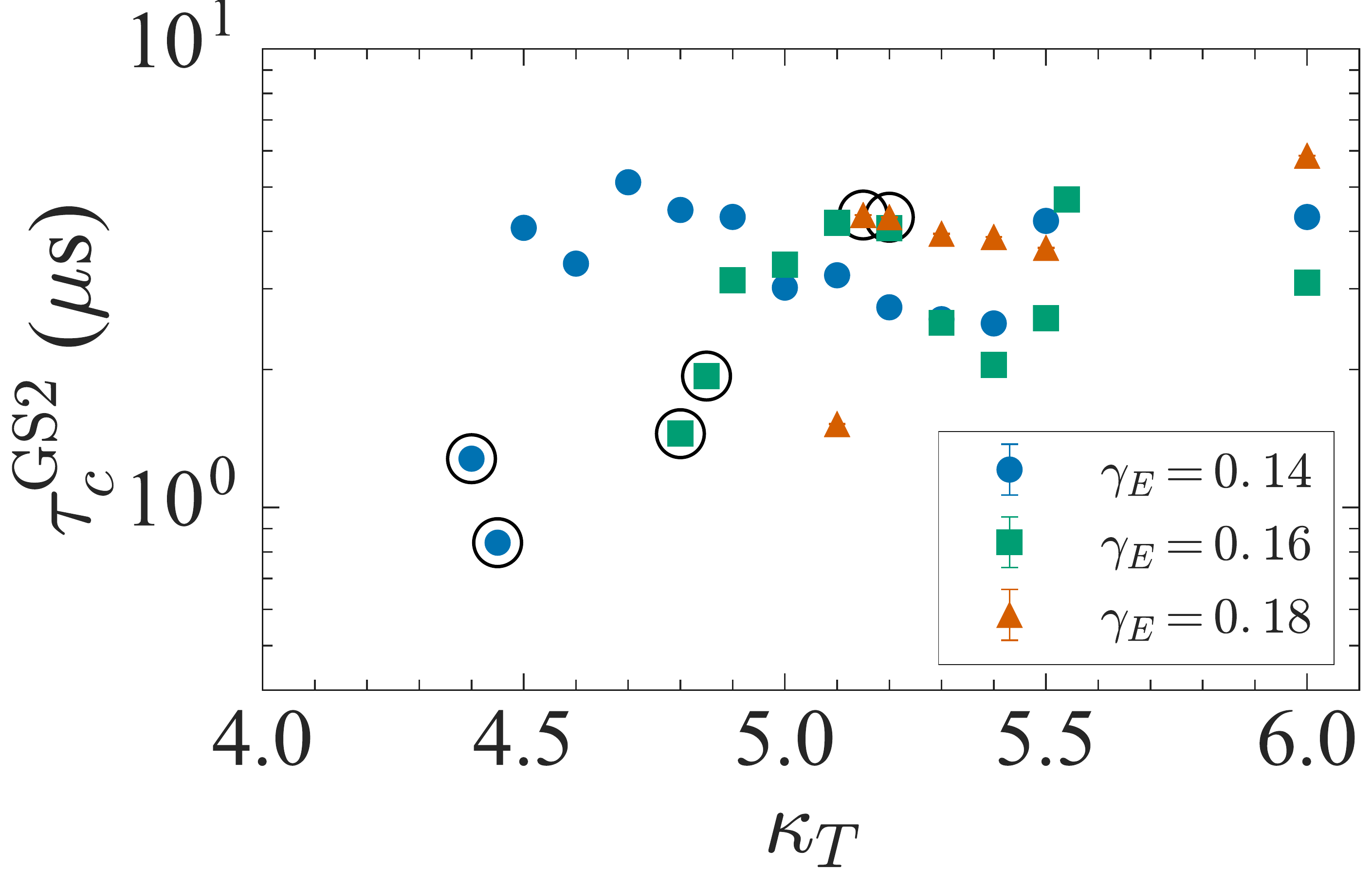}
      \caption{}
      \label{fig:tau_gs2}
    \end{subfigure}
    \begin{subfigure}[t]{0.49\textwidth}
      \includegraphics[width=\linewidth]{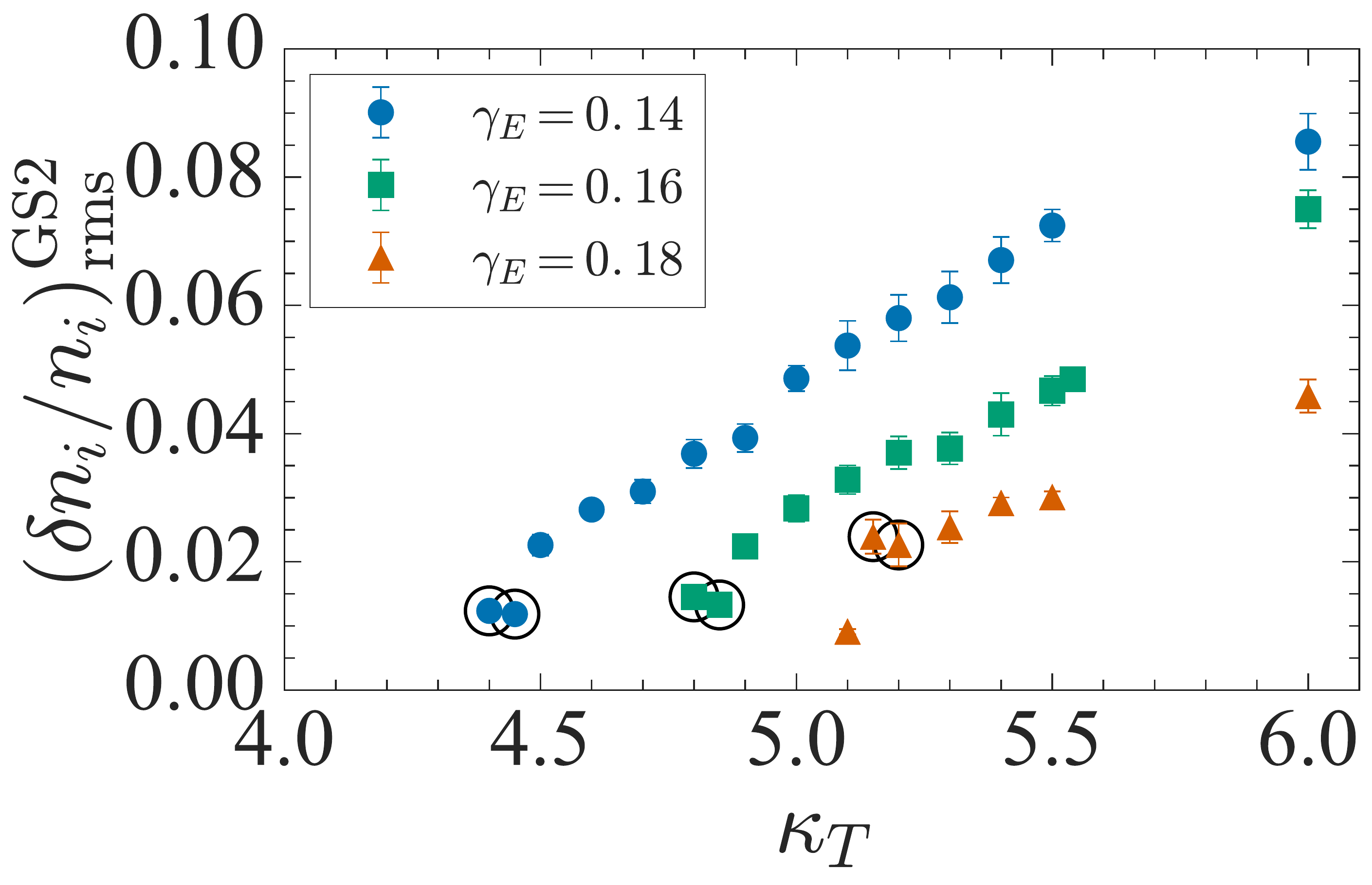}
      \caption{}
      \label{fig:n_gs2}
    \end{subfigure}
    \caption[Correlation parameters for raw GS2 density fluctuations]{
      Correlation parameters calculated for raw GS2 density fluctuations
      for $(\kappa_T, \gamma_E)$ within the range of experimental uncertainty
      indicated in \figref{contour_heatmap}:
      \subref*{fig:lr_gs2} radial correlation length $l_R^{\mathrm{GS2}}$,
      \subref*{fig:lz_gs2_fixed} poloidal correlation length
      $l_Z^{\mathrm{GS2}}$ keeping $k_y$ fixed to $k_y = 2 \pi / l_Z$,
      \subref*{fig:tau_gs2} correlation time $\tau_c^{\mathrm{GS2}}$, and
      \subref*{fig:n_gs2} RMS density fluctuations ${\qty(\delta n_i /
      n_i)}^{\mathrm{GS2}}_{\mathrm{rms}}$.
    }
    \label{fig:gs2_corr_results1}
  \end{figure}

  We start by considering the correlation analysis results for simulations with
  values of $\kappa_T$ and $\gamma_E$  within the experimental-uncertainty
  range. \Figref{gs2_corr_results1} shows the radial correlation length
  $l_R^{\mathrm{GS2}}$, the poloidal correlation length  $l_Z^{\mathrm{GS2}}$,
  the correlation time $\tau_c^{\mathrm{GS2}}$, and the RMS density fluctuation
  ${\qty(\delta n_i / n_i)}^{\mathrm{GS2}}_{\mathrm{rms}}$ calculated for our
  GS2 density-fluctuation field. The results shown in
  \figref{gs2_corr_results1} are for a range of values of $\kappa_T$ and for
  $\gamma_E = [0.14, 0.16, 0.18]$, with circled points describing the
  simulations that match the experimental value of the heat flux.

  We find that the radial correlation length is $l_R^{\mathrm{GS2}} \sim$
  $1$--$1.5$~cm, increasing with $\kappa_T$ and decreasing with $\gamma_E$.
  This suggests that $l_R^{\mathrm{GS2}}$ has a tendency to increase with
  $Q_i/Q_{\mathrm{gB}}$, as we will show explicitly later. In comparison with
  the synthetic-diagnostic results shown in \figref{lr_synth}, where
  $l_R^{\mathrm{SYNTH}} \sim 2$~cm, the true radial correlation length of the
  turbulence $l_R^{\mathrm{GS2}}$ is below $2$~cm and, therefore, below the
  resolution threshold of the BES diagnostic (discussed in
  section~\ref{sec:corr_synth}).

  \Figref{lz_gs2_fixed} shows that the poloidal correlation length is
  $l_Z^{\mathrm{GS2}} \sim$ $13$--$20$~cm (to be compared with
  $l_Z^{SYNTH} = 10$--$15$~cm), keeping the poloidal wavenumber
  $k_Z^{\mathrm{GS2}}$ fixed to $k_Z^{\mathrm{GS2}} = 2 \pi /
  l_Z^{\mathrm{GS2}}$ (giving $k_Z^{\mathrm{GS2}} \sim$~$30$--$50$~m$^{-1}$).
  We see that $l_Z^{\mathrm{GS2}}$ decreases rapidly as $\kappa_T$ is increased
  from its value at the turbulence threshold. The correlation time
  [\figref{tau_gs2}] is in the range $\tau_c^{\mathrm{GS2}}\sim$
  $1$--$6$~$\mu$s. Finally, \figref{n_gs2} shows that ${\qty(\delta n_i /
  n_i)}^{\mathrm{GS2}}_{\mathrm{rms}} \sim$~$0.01$--$0.08$ and increases with
  increasing $\kappa_T$ or decreasing $\gamma_E$, i.e., has an upward tendency
  as the heat flux increases.

  \subsubsection{Comparisons between experimental and GS2 correlation
  properties}
  We have presented the correlation parameters measured
  \begin{inparaenum}[(i)]
    \item by the BES diagnostic in section~\ref{sec:corr_exp},
    \item from GS2 density fluctuations with the synthetic diagnostic applied
      in section~\ref{sec:corr_synth}, and
    \item from the raw GS2 density fluctuations.
  \end{inparaenum}
  We show the results from all these analyses in table~\ref{tab:corr_summary}.
  We can thus summarise the comparison between simulation results and
  experimental measurements as follows. Comparing the results of the
  correlation analysis of the GS2 density fluctuations (``SYNTH'' and ``GS2''
  in table~\ref{tab:corr_summary}) with the experimental measurements
  (``EXP''), we see that all the experimental values, except for the radial
  correlation length $l_R$, fall within the ranges found for the simulation
  results. This is particularly important in the case of $\tau_c$, which was
  significantly overestimated in the previous modelling effort for this MAST
  discharge~\cite{Field2014}. It is clear that the correlation parameters vary
  with the equilibrium parameters and there is no single simulation, i.e., no
  single combination of $(\kappa_T, \gamma_E)$, that perfectly matches the BES
  measurements in all four parameters (see \figref{gs2_corr_results1}).
  \begin{table}
    \centering
    \caption{Summary of results for the correlation parameters $l_R$, $l_Z$,
      $\tau_c$, and $(\delta n_i / n_i)_{\mathrm{rms}}$ from experimental BES
      measurements (EXP), from the correlation analysis of GS2 density
      fluctuations with synthetic diagnostic applied (SYNTH; using an identical
      correlation analysis to that used on the BES data), and from the
      correlation analysis of raw GS2 density fluctuations (GS2).
    }
    \begin{tabular}{c c c c}
      \toprule
      Parameter & EXP & SYNTH & GS2 \\
      \midrule
      $l_R$ (cm) & $3 \pm 0.4$ & 1.5--2.5 & 1--1.5 \\
      $l_Z$ (cm) & $14.06 \pm 0.09$ & 10--15 & 13--20 \\
      $\tau_c$ ($\mu$s) & $3.2 \pm 0.4$ & 2--15 & 1--6 \\
      $(\delta n_i / n_i)_{\mathrm{rms}}$ & $0.0214 \pm 0.0006$ &
        0.005--0.03 & 0.01--0.08 \\
      \bottomrule
    \end{tabular}
    \label{tab:corr_summary}
  \end{table}

  Considering the difference between the GS2 density fluctuations with (SYNTH)
  and without (GS2) the synthetic diagnostic gives us an indication of the
  effect of the PSFs on the measurement of turbulence correlation properties.
  Given that the value of $l_R$ measured from the raw GS2 density fluctuations
  is below the approximate resolution threshold, it is unclear what effect the
  PSFs have on the radial correlation length $l_R$. We see from
  table~\ref{tab:corr_summary} that the ranges of values of the poloidal
  correlation length $l_Z$ are comparable in the SYNTH and GS2 cases. However,
  \figref{lz_synth} shows that, with the synthetic diagnostic applied, there
  are not the clear trends with $\kappa_T$ that we see in
  \figref{lz_gs2_fixed}. This may be because the limited poloidal resolution,
  while sufficient to resolve the measured correlation lengths, is not
  sensitive enough to recover the trend of decreasing $l_Z$ with $\kappa_T$
  seen in \figref{lz_gs2_fixed}. The measurement of the correlation time
  $\tau_c$ is, again, less certain in the case of the correlation analysis of
  density fluctuations with a synthetic diagnostic applied, but there is
  reasonable agreement with the correlation time measured from the raw GS2
  density fluctuations.  Finally, the application of the synthetic diagnostic
  leads to a reduction of roughly $50$\% of the RMS fluctuation amplitude,
  i.e., from ${(\delta n_i / n_i)}^{\mathrm{GS2}}_{\mathrm{rms}}
  \sim$~$0.01$--$0.08$ for the raw density fluctuations to ${(\delta
  n_i/n_i)}^{\mathrm{SYNTH}}_{\mathrm{rms}}~\sim$~$0.005$--$0.03$.  This
  observation is consistent with a recent detailed analysis~\cite{Fox2016} of
  the effect of PSFs on the measurement of MAST turbulence using a subset of
  GS2 simulations reported here.

  \subsubsection{Poloidal and parallel correlation parameters}
  \label{sec:pol_par_corr}
  We now consider two further diagnostics, which were not available to us
  experimentally: poloidal and parallel correlation functions parametrised by
  the poloidal and parallel correlation lengths and wavenumbers (defined in
  appendix~\ref{App:corr_overview}) independently fitted as free parameters
  (although the poloidal correlation length was calculated previously in
  section~\ref{sec:within_uncertainty}, the resolution of the BES diagnostic
  necessitated fixing the poloidal wavenumber $k_Z$).

\begin{figure}[t]
    \centering
    \begin{subfigure}[t]{0.49\textwidth}
      \includegraphics[width=\linewidth]{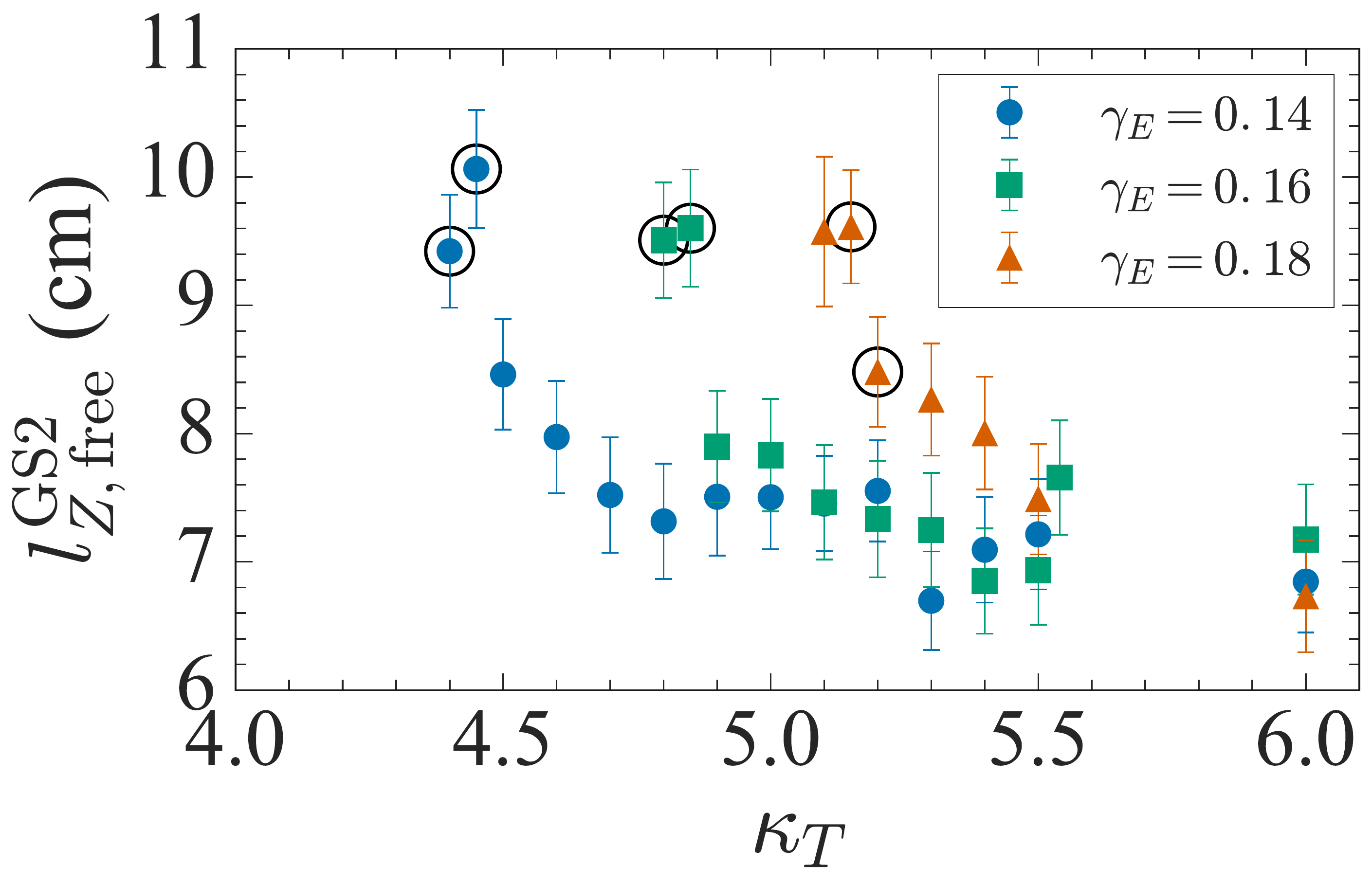}
      \caption{}
      \label{fig:lz_gs2_free}
    \end{subfigure}
    \begin{subfigure}[t]{0.49\textwidth}
      \includegraphics[width=\linewidth]{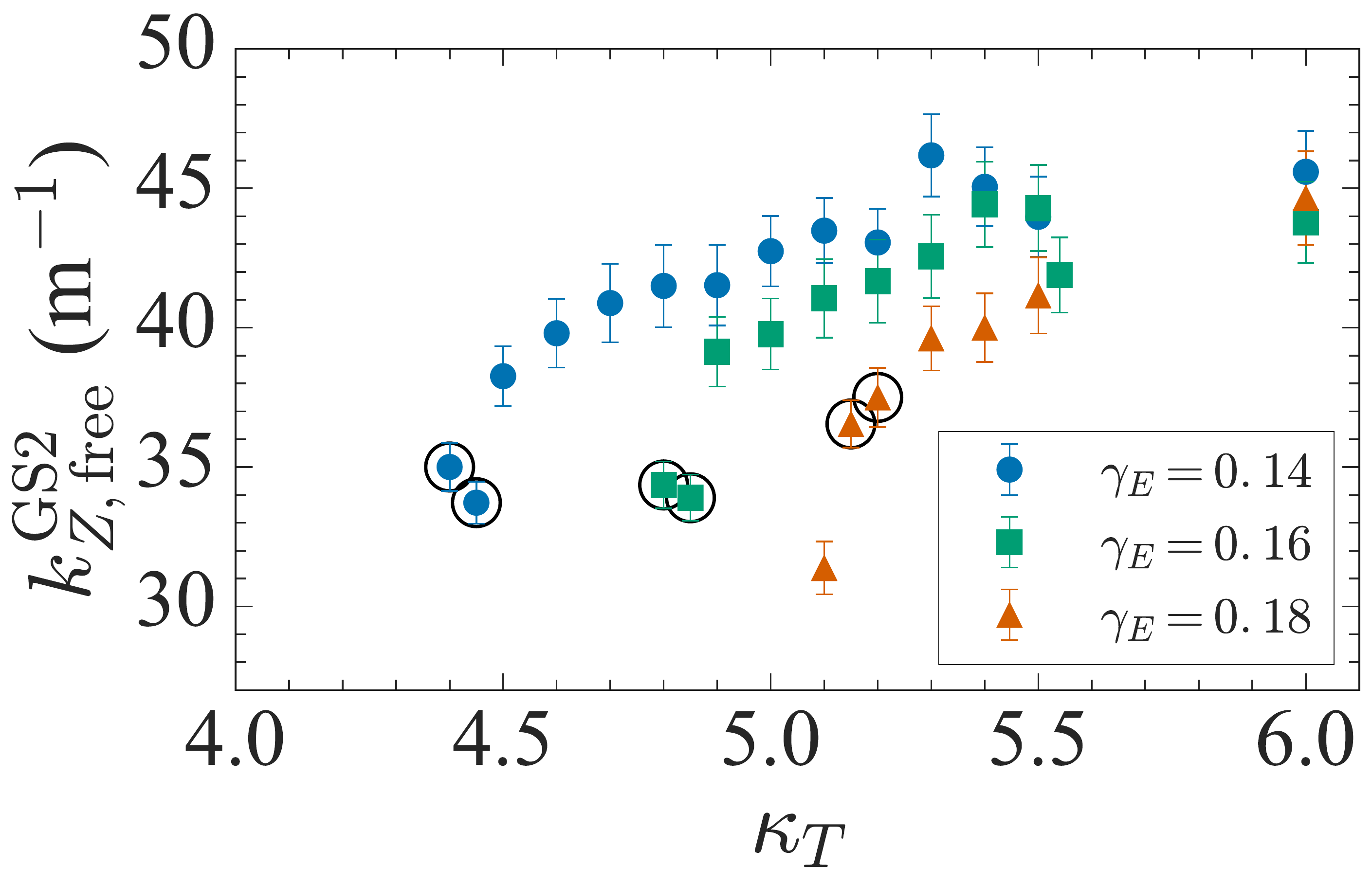}
      \caption{}
      \label{fig:kz_gs2}
    \end{subfigure}
    \\
    \begin{subfigure}[t]{0.49\textwidth}
      \includegraphics[width=\linewidth]{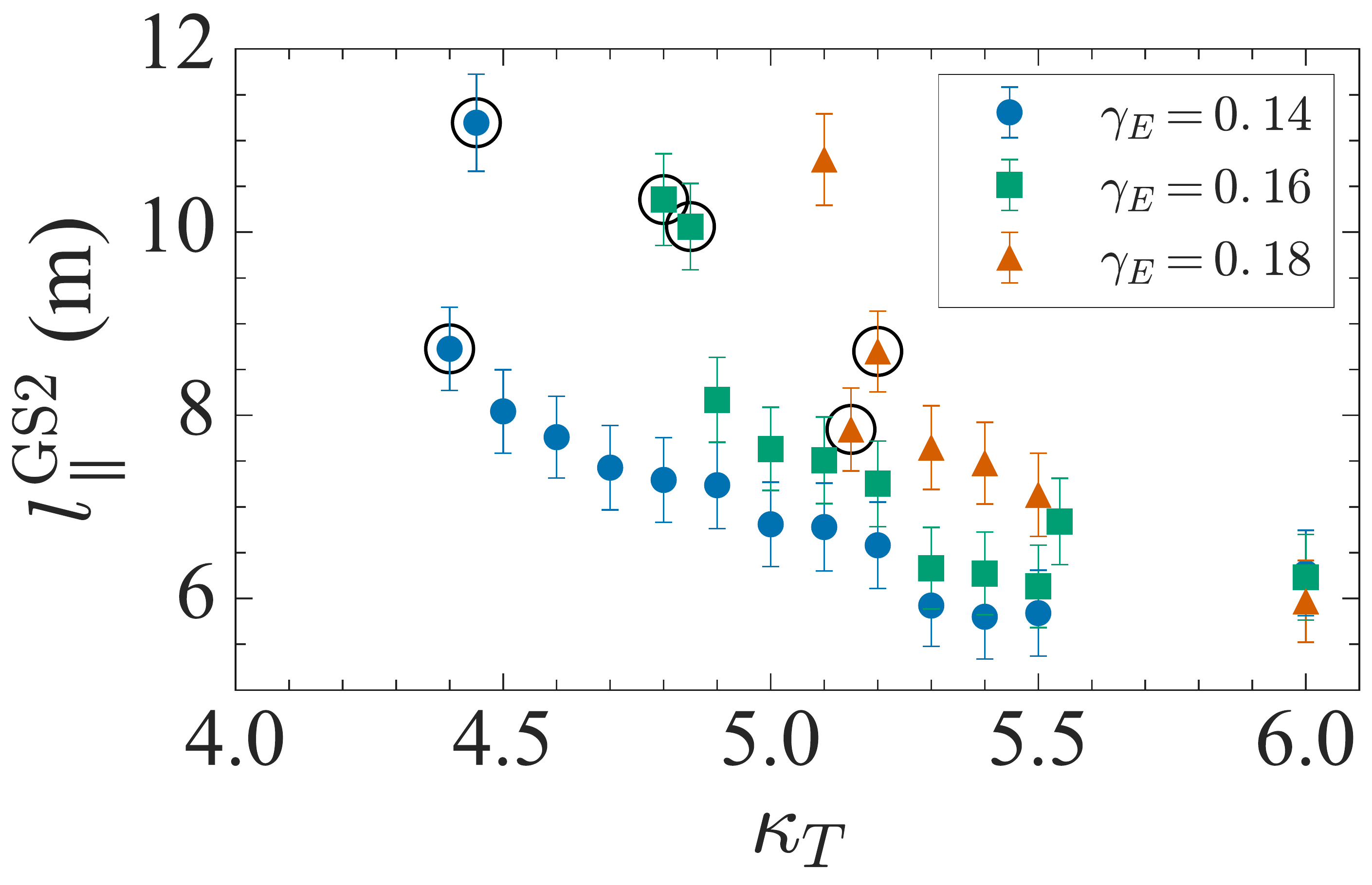}
      \caption{}
      \label{fig:lpar_gs2}
    \end{subfigure}
    \begin{subfigure}[t]{0.49\textwidth}
      \includegraphics[width=\linewidth]{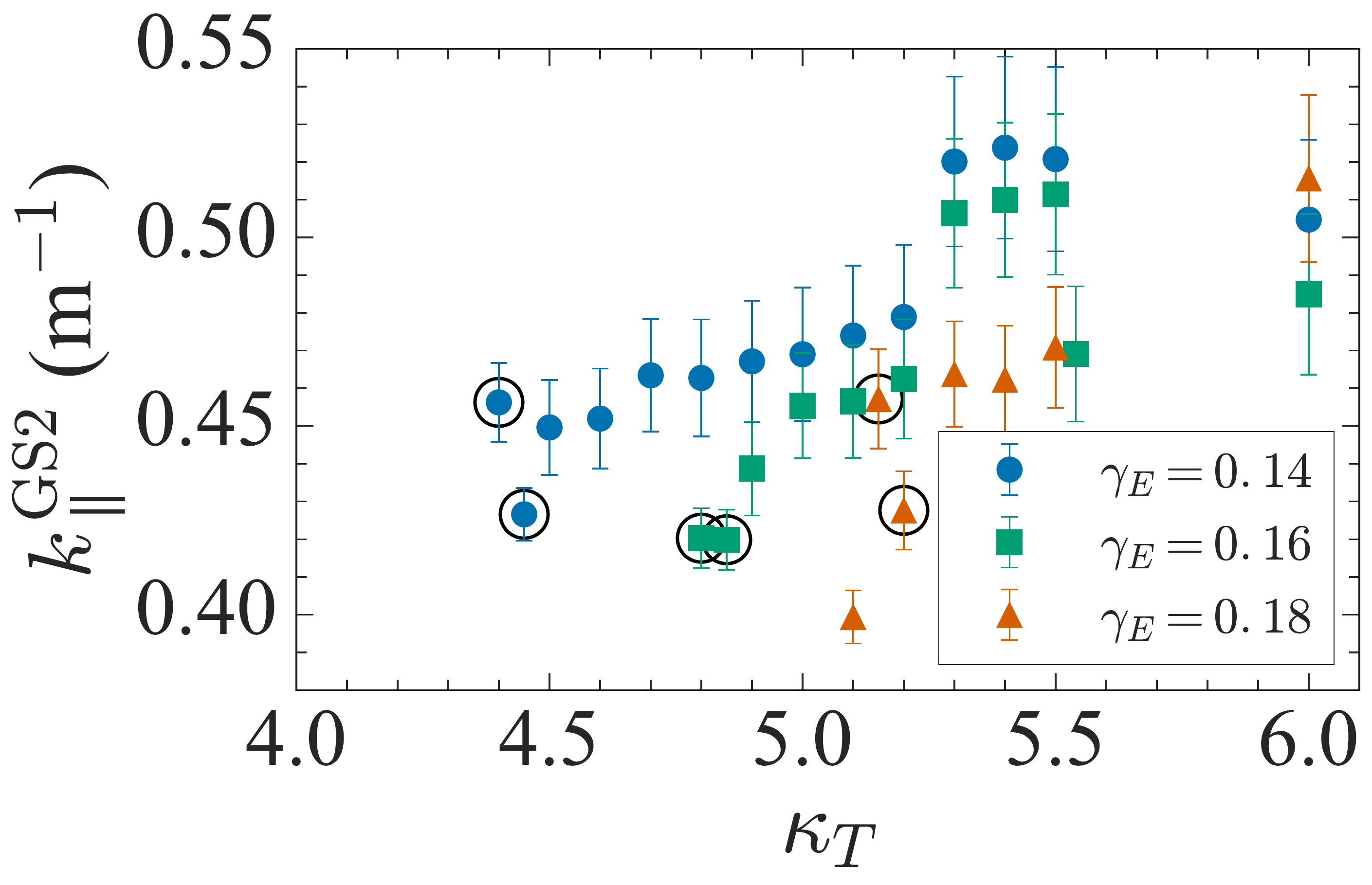}
      \caption{}
      \label{fig:kpar_gs2}
    \end{subfigure}
    \caption{
      Correlation parameters calculated for raw GS2 density fluctuations
      for $(\kappa_T, \gamma_E)$ within the range of experimental uncertainty
      indicated in \figref{contour_heatmap}:
      \subref*{fig:lz_gs2_free} poloidal correlation length
      $l_{Z,\mathrm{free}}^{\mathrm{GS2}}$ with $k_y$ as a free fitting
      parameter,
      \subref*{fig:kz_gs2} poloidal wavenumber
      $k_{Z,\mathrm{free}}^{\mathrm{GS2}}$ (section~\ref{sec:poloidal_corr}),
      \subref*{fig:lpar_gs2} parallel correlation length
      $l_{\parallel}^{\mathrm{GS2}}$, and
      \subref*{fig:kpar_gs2} parallel wavenumber $k_{\parallel}^{\mathrm{GS2}}$
      (section~\ref{sec:par_corr}).
    }
    \label{fig:gs2_corr_results2}
  \end{figure}

  Figures~\ref{fig:lz_gs2_free} and \subref{fig:kz_gs2} show the result of such
  fitting for the poloidal correlations: $l_{Z,\mathrm{free}}^{\mathrm{GS2}}$
  and $k_Z^{\mathrm{GS2}}$ versus $\kappa_T$. As supported by
  \figref{poloidal_fit} in appendix~\ref{sec:poloidal_corr}, we see a roughly
  $50$\% decrease in $l_{Z,\mathrm{free}}^{\mathrm{GS2}}$ compared to
  $l_Z^{\mathrm{GS2}}$ [\figref{lz_gs2_fixed}], from $13$--$20$~cm to
  $7$--$10$~cm, again decreasing as $\kappa_T$ increases or $\gamma_E$
  decreases. Surprisingly, the value of $k_{Z,\mathrm{free}}^{\mathrm{GS2}}$ is
  still in the $35$--$45$~m$^{-1}$ range -- comparable to one obtained via the
  fitting procedure where $k_Z = 2\pi/l_Z$.  Regardless of the fitting method,
  \figref{lz_gs2_fixed} and \figref{lz_gs2_free} show a similar dependence of
  $l_Z$ on $\kappa_T$ and $\gamma_E$.

  The results of the parallel correlation analysis, given in
  \figref{lpar_gs2} and \subref{fig:kpar_gs2}, are the values
  $l_\parallel^{\mathrm{GS2}}$ and $k_\parallel^{\mathrm{GS2}}$ versus
  $\kappa_T$ and $\gamma_E$. We see that $l_\parallel^{\mathrm{GS2}}
  \sim$~$6$--$12$~m and decreases with increasing $\kappa_T$ and decreasing
  $\gamma_E$. Based on this measurement of the parallel correlation length, it
  is clear that the turbulence is highly anisotropic, i.e., $l_\parallel \gg
  l_\perp$, as it is expected to be~\cite{Abel2013}.

  Using the measurement of $l_\parallel^{\mathrm{GS2}}$, we can return to, and
  confirm, the assumption upon which the calculation of $\tau_c$ depends.
  Namely, in appendix~\ref{sec:time_corr}, we assume that reliably estimating
  the correlation time requires that the temporal decorrelation be dominant
  over effects due to the finite parallel correlation length
  [condition~\eqref{time_assumption}]. Using the value of $l_\parallel$
  reported above, we estimate $l_\parallel \cos \vartheta / u_\phi \sim$
  $80$--$160$~$\mu$s, where we have used the experimental parameters $R = 1.32$
  m, $\omega = 4.71 \times 10^4$ $\mathrm{s}^{-1}$, and $\vartheta \approx
  0.6$. This confirms that the values of $\tau_c$ summarised in
  table~\ref{tab:corr_summary} are smaller than $l_\parallel \cos \vartheta /
  u_\phi$ by more than an order of magnitude and that our time correlation
  analysis is valid in this MAST configuration.

  Currently the BES diagnostic on MAST is not capable of determining both $l_Z$
  and $k_Z$, but the above estimates may be used for future comparisons between
  experimental measurements and numerical results if higher-resolution BES
  measurements become available. Similarly, there is currently no diagnostic on
  MAST capable of measuring the parallel correlation length, but our
  estimates may guide future attempts at designing diagnostics to measure it.

  \subsubsection{Comparison between linear and nonlinear time scales}
  \label{sec:time_scales}
  With the knowledge of the correlation parameters, we can return to the
  comparison of the transient-growth time $t_0$ and nonlinear time
  $\tau_{\mathrm{NL}}$ discussed in section~\ref{sec:subcritical}.  In
  particular, we want to determine one of the two conditions for the onset of
  subcritical turbulence [equation~\eqref{schek_t0}] proposed
  in~\cite{Schekochihin2012}. We also want to compare $\tau_{\mathrm{NL}}$ with
  the correlation time of the turbulence $\tau_c$; we then discuss the
  corresponding experimental results in~\cite{Field2014}.

\begin{figure}[t]
    \centering
    \begin{subfigure}{0.49\linewidth}
      \includegraphics[width=\linewidth]{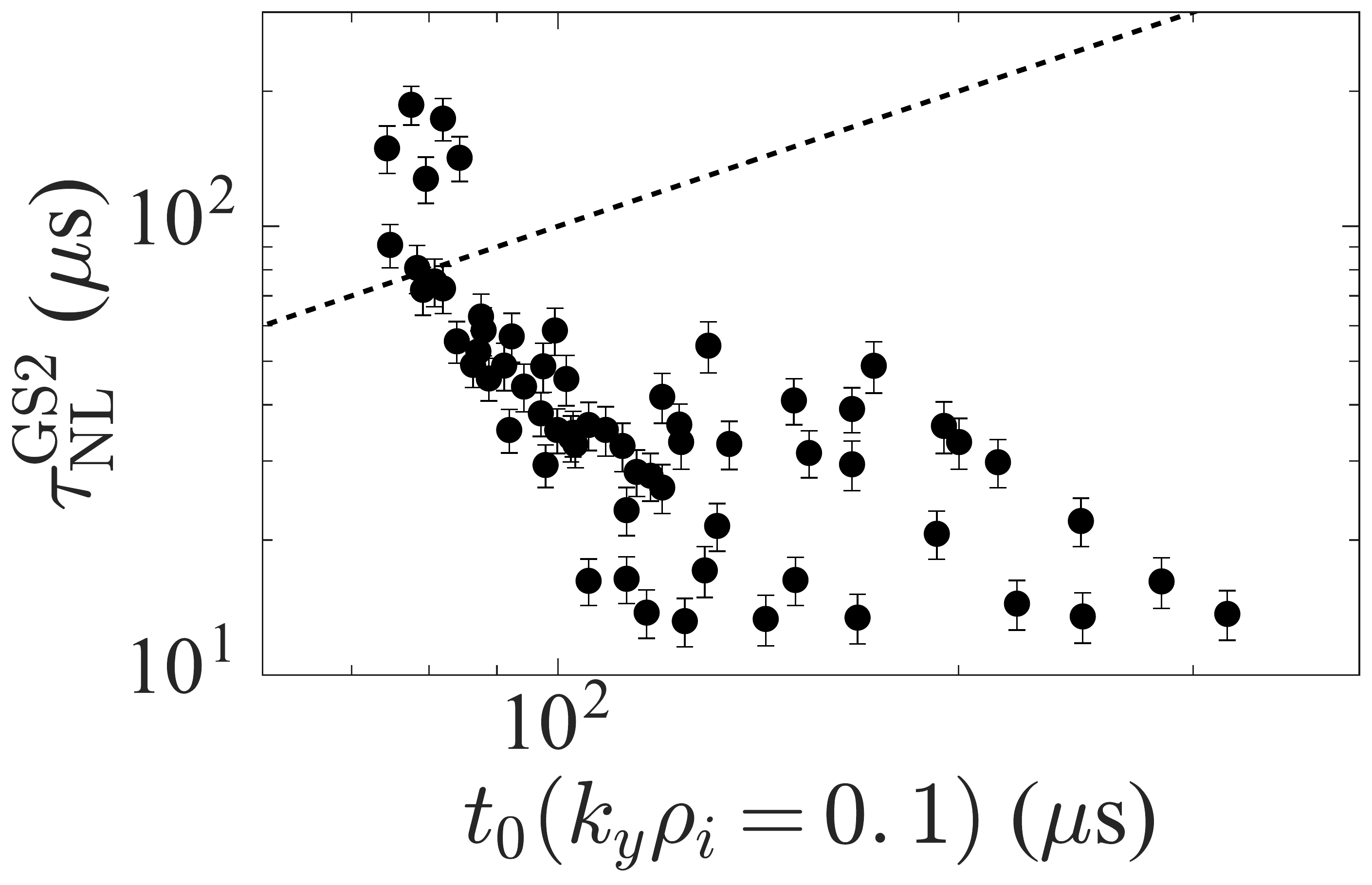}
      \caption{}
      \label{fig:tnl_t0}
    \end{subfigure}
    \hfill
    \begin{subfigure}{0.49\linewidth}
      \includegraphics[width=\linewidth]{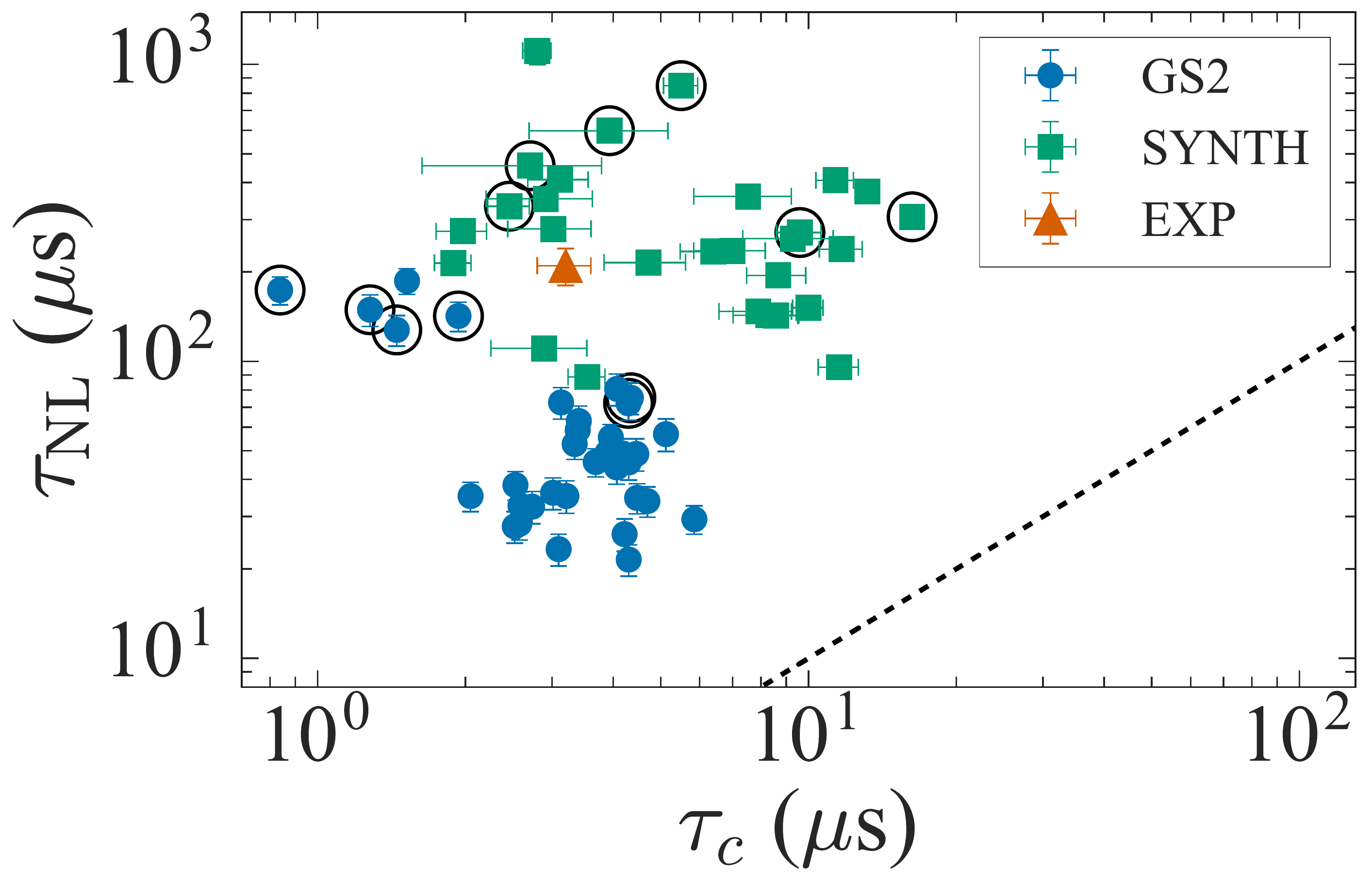}
      \caption{}
      \label{fig:tnl_tauc}
    \end{subfigure}
    \caption{
      \subref*{fig:tnl_t0} Nonlinear interaction time of the raw GS2 density
      fluctuations $\tau_{\mathrm{NL}}^{\mathrm{GS2}}$, calculated
      using~\eqref{tau_nl}, versus the transient-growth time $t_0$, defined by
      \eqref{amp_exponent}.
      We have taken $t_0$ at $k_y \rho_i = 0.1$, where $t_0$ is largest (see
      \figref{N_and_t0_16}). We show here all simulations in our parameter scan
      with $\gamma_E>0$.
      \subref*{fig:tnl_tauc} $\tau_{\mathrm{NL}}$ versus $\tau_c$ for
      the correlation parameters calculated from the raw GS2 density
      fluctuations (GS2), from density fluctuations with the synthetic
      diagnostic applied (SYNTH), and from experimental measurements (EXP).
      The cases shown are for values of $(\kappa_T, \gamma_E)$ within
      the experimental-uncertainty range and the circled simulations indicate
      the simulations that match the experimental heat flux. The dashed lines
      in each plot indicate where the time scales are equal.
    }
    \label{fig:time_scales}
  \end{figure}

  The non-zonal nonlinear interaction time is estimated to be~\cite{Ghim2013}:
  \begin{equation}
    \tau_{\mathrm{NL}}^{-1} =
    \frac{v_{\mathrm{th}i} \rho_i}{l_R l_Z} \frac{T_e}{T_i}
      \qty(\frac{\delta n_i}{n_i})_{\mathrm{rms}},
    \label{tau_nl}
  \end{equation}
  where we have assumed $l_Z \approx l_y$ (where $l_y$ is the correlation
  length in the binormal direction as defined in~\cite{Ghim2013}) because $l_Z =
  l_y \cos \vartheta$, where $\vartheta$ is the magnetic field pitch-angle, and
  $\cos \vartheta \sim 1$ for this magnetic equilibrium.  The transient-growth
  time $t_0$ was calculated from linear simulations and plotted in
  \figref{N_and_t0_16}, showing that, at ion scales, the longest transient
  growth occurred at $k_y \rho_i \sim 0.1$. \Figref{tnl_t0} shows
  $\tau_{\mathrm{NL}}^{\mathrm{GS2}}$ versus $t_0$ (at $k_y \rho_i = 0.1$) for
  all simulations with $\gamma_E > 0$, where the dashed line indicates
  $\tau_{\mathrm{NL}}^{\mathrm{GS2}} = t_0$. We see that the majority of
  simulations are below the line defined by $\tau_{\mathrm{NL}}^{\mathrm{GS2}}
  = t_0$, showing that at the very least the condition \eqref{schek_t0} is
  approximately satisfied in most turbulent states, i.e., that $t_0 \gtrsim
  \tau_{\mathrm{NL}}$. The inverse relationship between $\tau_{\mathrm{NL}}$
  and $t_0$ is explained simply: as the strength of the drive increases and we
  move away from the turbulence threshold, the growth time of the transient
  amplification increases, the nonlinear interaction becomes more
  vigorous, and so the nonlinear time goes down.

  \Figref{tnl_tauc} shows $\tau_{\mathrm{NL}}$ versus $\tau_c$ for
  nonlinear simulations with values of $(\kappa_T, \gamma_E)$ within
  experimental uncertainty. The values of $\tau_{\mathrm{NL}}$ were calculated
  from correlation parameters of raw GS2 density fluctuations (GS2), from
  correlation parameters calculated from GS2 density fluctuations with a
  synthetic diagnostic applied (SYNTH), and from the experimental BES
  measurements at $r=0.8$ (EXP). The dashed line corresponds to
  $\tau_{\mathrm{NL}} = \tau_c$. First, we see that $\tau_{\mathrm{NL}} >
  \tau_c$ for both the GS2 and SYNTH cases, and for the experimental
  case (EXP). Secondly, $\tau_{\mathrm{NL}}$ for the raw
  GS2 density fluctuations tends to be below the experimental value, whereas
  the SYNTH cases are comparable. The results shown in \figref{tnl_tauc} are
  consistent with the experimental results in~\cite{Ghim2013, Field2014} that
  showed $\tau_{\mathrm{NL}} \gg \tau_c$ for this and other experimental cases.

  It might appear strange that eddies would be able
  to interact nonlinearly with each other over a much longer time than it takes
  them to decorrelate. These results suggest either that \eqref{tau_nl} is a
  gross overestimate of the nonlinear interactions and/or that the value of
  $\tau_c$ that we measure from the turbulent density field captures some
  decorrelation process that is otherwise invisible to our synthetic diagnostic
  (\cite{Ghim2013} speculated that this might involve zonal flows). However,
  what is reassuring about these results is that GS2 simulations and BES
  measurements of turbulence in MAST appear to be consistent with each other
  in exhibiting this thus far unexplained feature.

  \subsubsection{Correlation parameters versus $Q_i/Q_{\mathrm{gB}}$}

  The correlation analysis results shown in figures~\ref{fig:gs2_corr_results1}
  and~\ref{fig:gs2_corr_results2}, in particular $l_Z^{\mathrm{GS2}}$, $(
  \delta n_i / n_i)^{\mathrm{GS2}}_{\mathrm{rms}}$, and
  $l_\parallel^{\mathrm{GS2}}$, exhibit similar trends versus $\kappa_T$ for
  different values of $\gamma_E$. As we showed in \figref{contour_heatmap},
  increasing $\kappa_T$ or decreasing $\gamma_E$ effectively amounts to
  controlling the distance from the turbulence threshold. Furthermore, our
  investigations of the transition to turbulence (see~\cite{VanWyk2016} and
  section~\ref{sec:subcritical}) and the effect of flow shear on its
  structure~\cite{Fox2016a} suggest that the key determining factor is the
  distance from the threshold. This can be quantified by the ion heat flux
  $Q_i/Q_{\mathrm{gB}}$, which increases monotonically with this distance and
  can be interpreted as an order parameter for our system. Here we describe the
  results of our correlation analysis of raw GS2 density fluctuations recast as
  a function of this parameter.
  \begin{figure}[t]
    \centering
    \begin{subfigure}[t]{0.49\textwidth}
      \includegraphics[width=\linewidth]{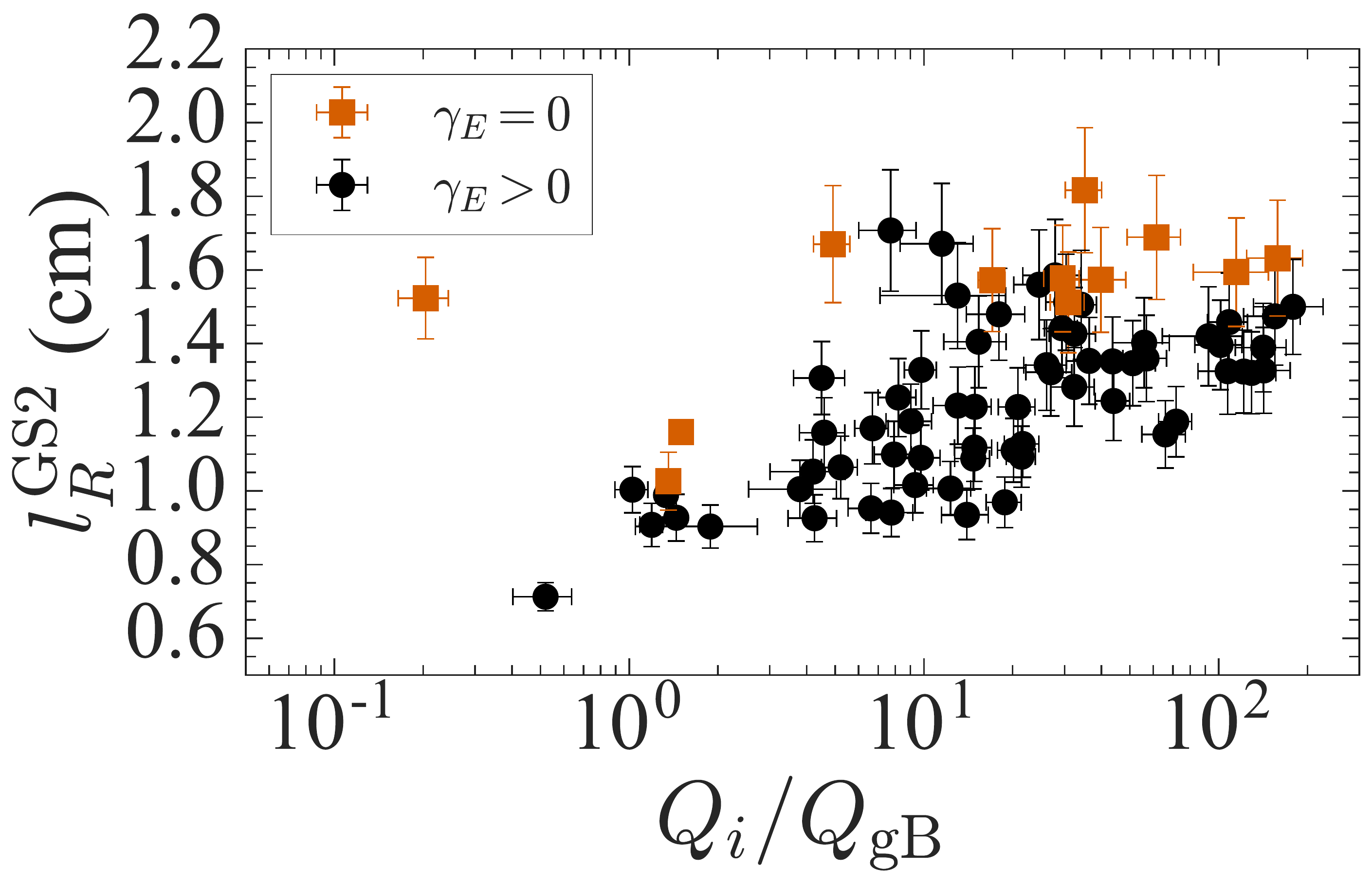}
      \caption{}
      \label{fig:lr_q}
    \end{subfigure}
    \begin{subfigure}[t]{0.49\textwidth}
      \includegraphics[width=\linewidth]{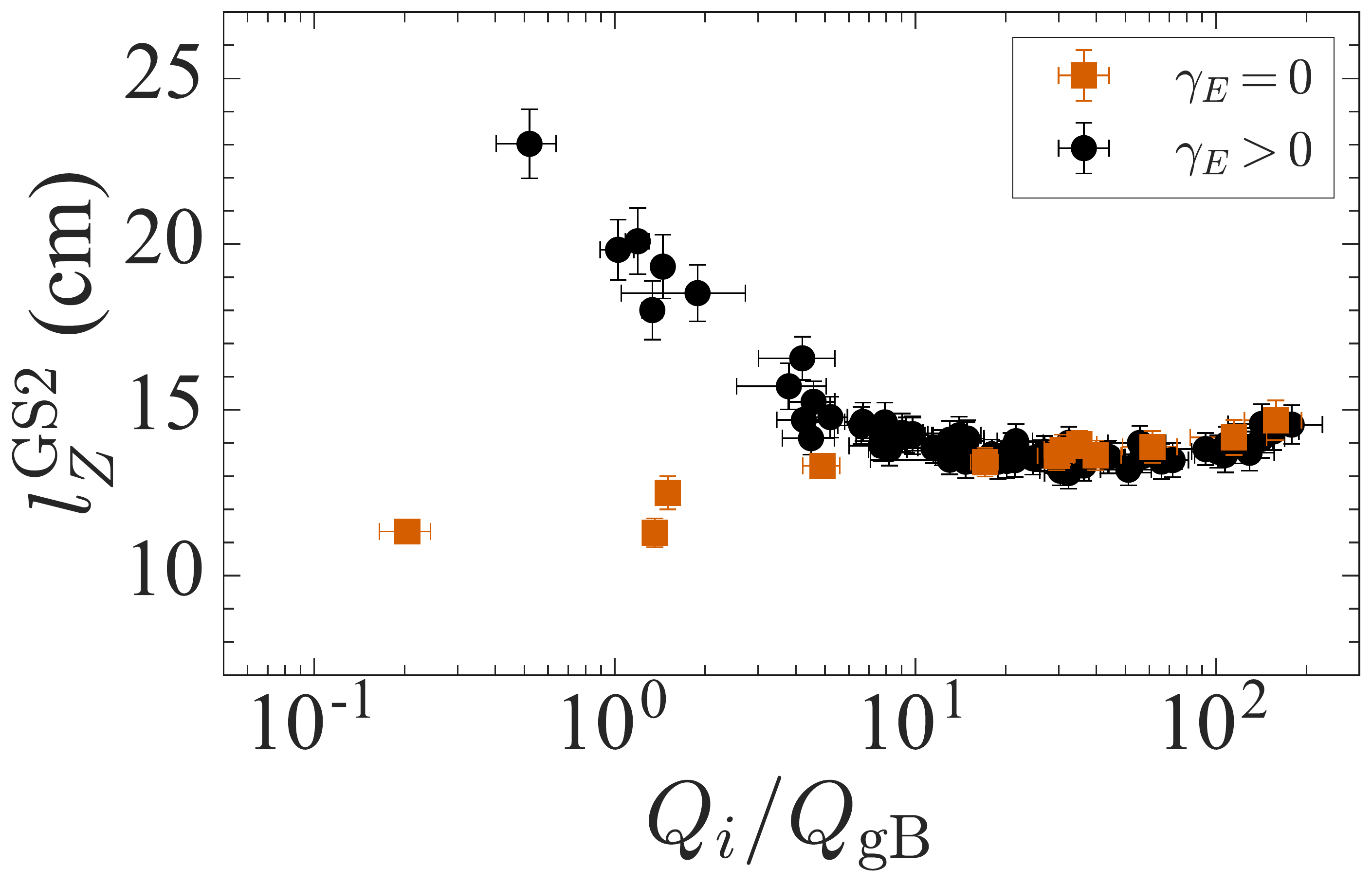}
      \caption{}
      \label{fig:lz_q}
    \end{subfigure}
    \\
    \begin{subfigure}[t]{0.49\textwidth}
      \includegraphics[width=\linewidth]{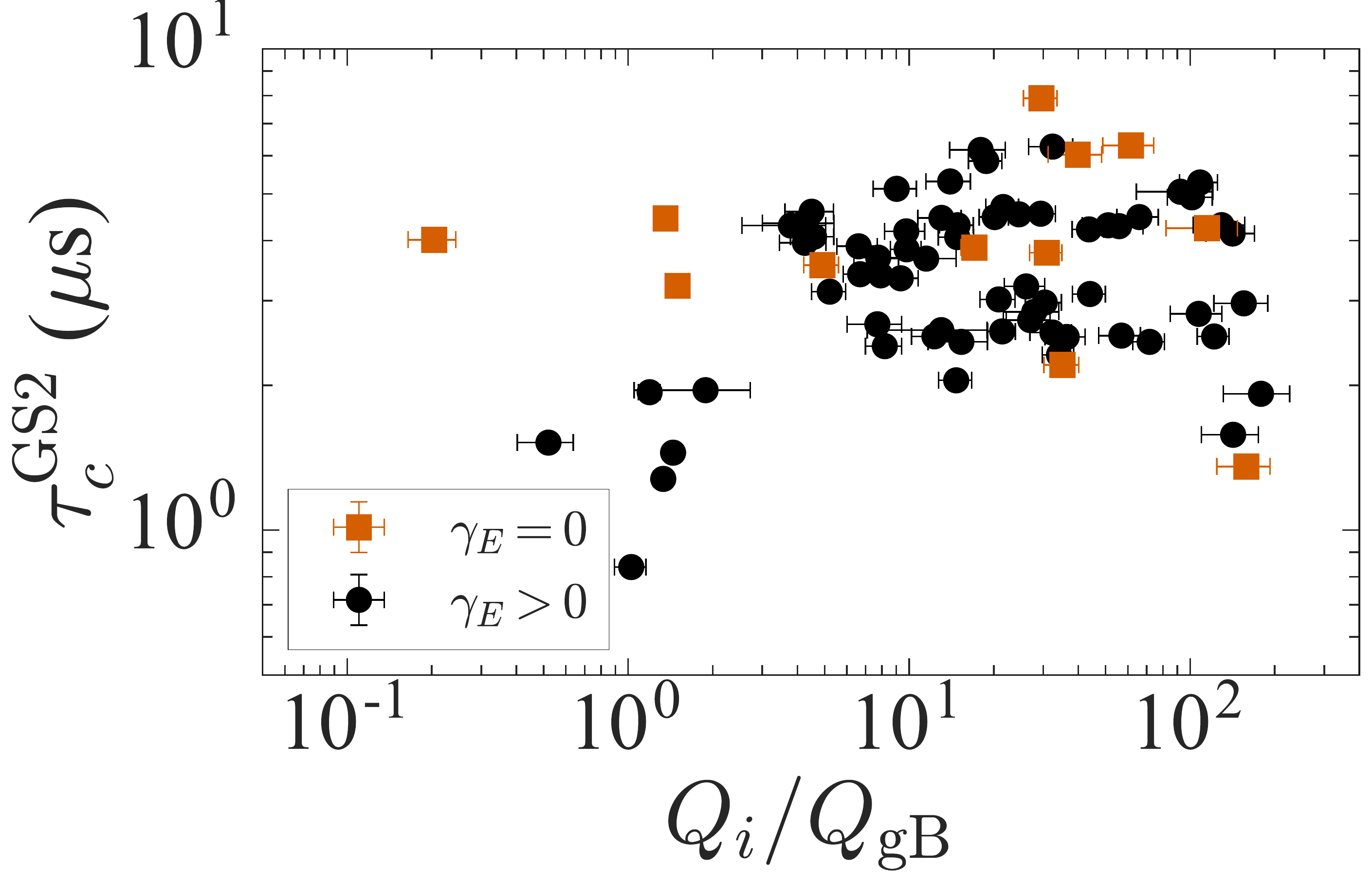}
      \caption{}
      \label{fig:tau_q}
    \end{subfigure}
    \begin{subfigure}[t]{0.49\textwidth}
      \includegraphics[width=\linewidth]{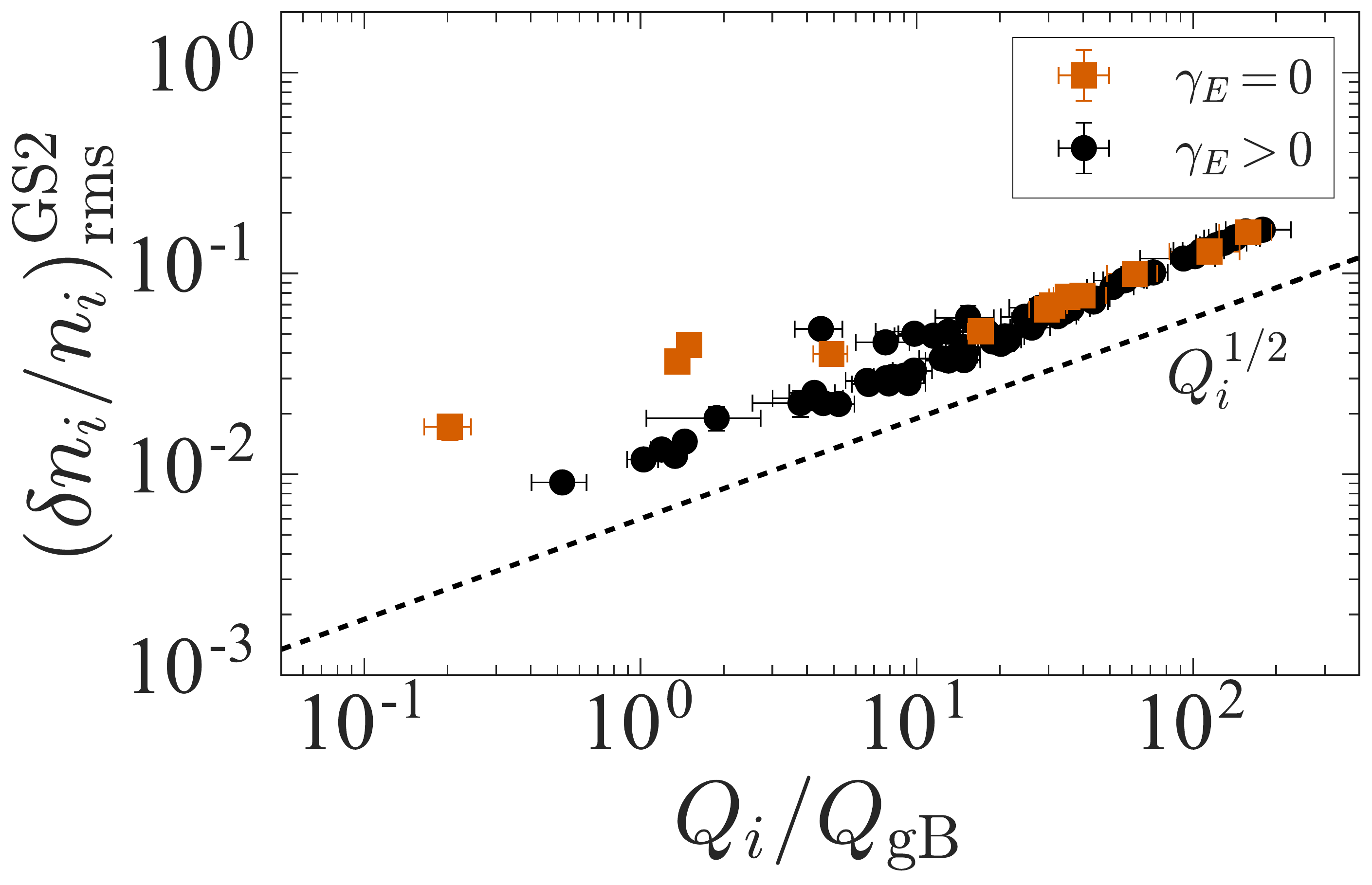}
      \caption{}
      \label{fig:n_q}
    \end{subfigure}
    \caption{
      Correlation parameters as functions of $Q_i/Q_{\mathrm{gB}}$, calculated
      for raw GS2 density fluctuations
      for the entire parameter scan:
      \subref*{fig:lr_gs2} radial correlation length $l_R^{\mathrm{GS2}}$;
      \subref*{fig:lz_gs2_fixed} poloidal correlation length;
      $l_Z^{\mathrm{GS2}}$ keeping $k_y$ fixed to $k_y = 2 \pi / l_Z$;
      \subref*{fig:tau_gs2} correlation time $\tau_c^{\mathrm{GS2}}$; and
      \subref*{fig:n_gs2} RMS density fluctuations $(\delta n_i /
      n_i)^{\mathrm{GS2}}_{\mathrm{rms}}$,
      where the dashed line indicates the scaling \eqref{q_scaling}.
    }
    \label{fig:gs2_q_scatter1}
  \end{figure}
  \begin{figure}[t]
    \centering
    \begin{subfigure}[t]{0.49\textwidth}
      \includegraphics[width=\linewidth]{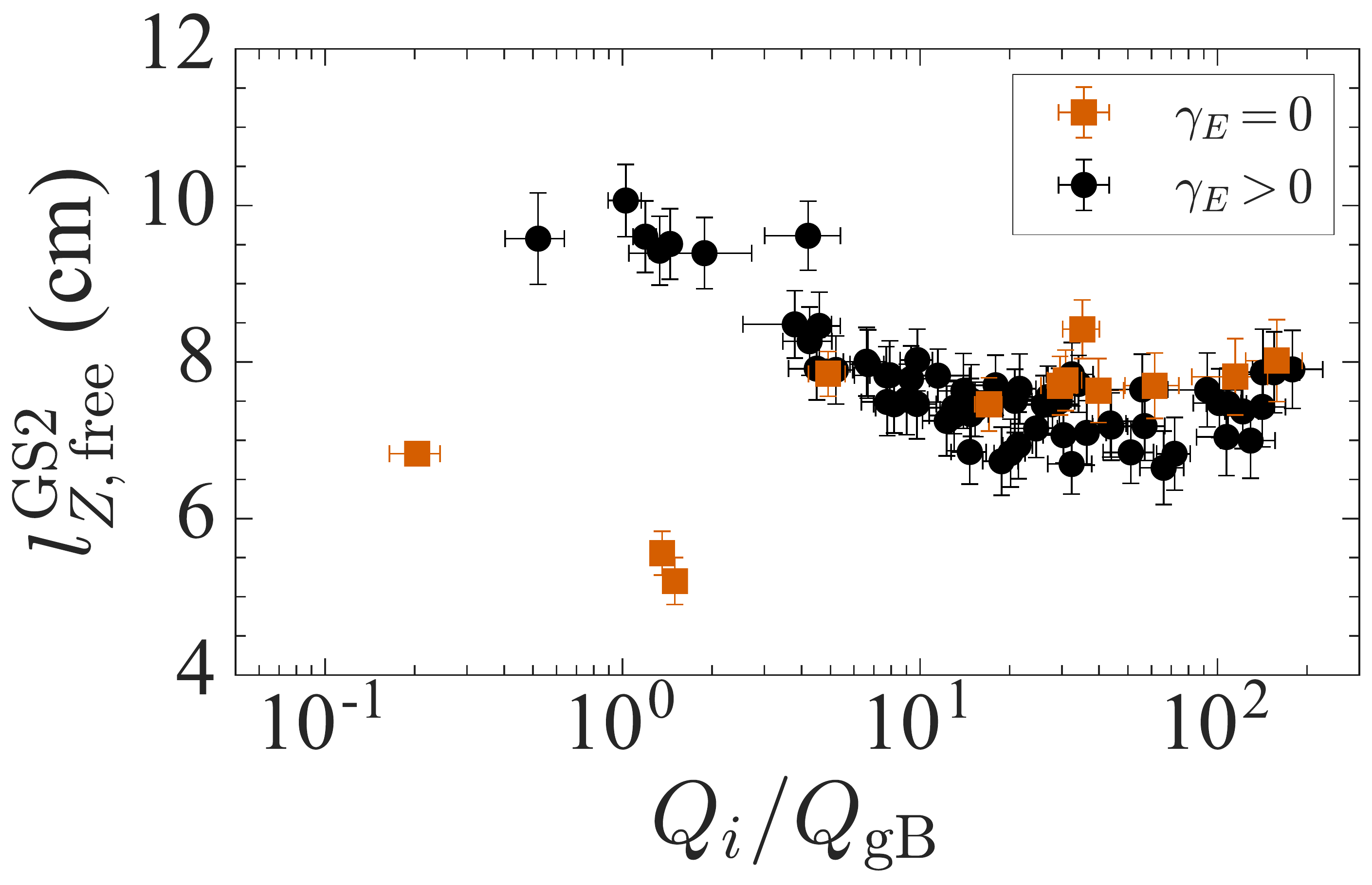}
      \caption{}
      \label{fig:lz_free_q}
    \end{subfigure}
    \begin{subfigure}[t]{0.49\textwidth}
      \includegraphics[width=\linewidth]{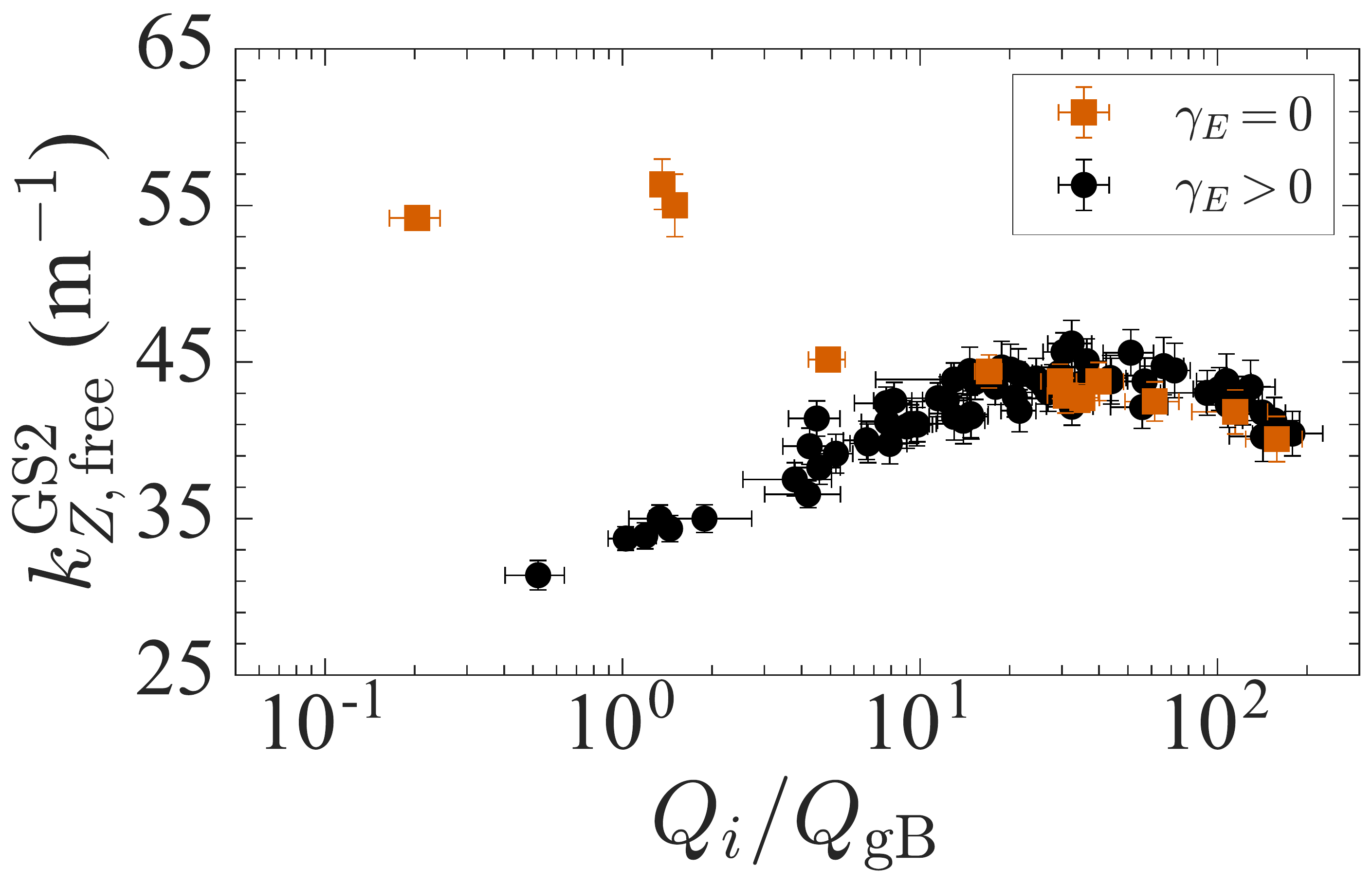}
      \caption{}
      \label{fig:kz_free_q}
    \end{subfigure}
    \\
    \begin{subfigure}[t]{0.49\textwidth}
      \includegraphics[width=\linewidth]{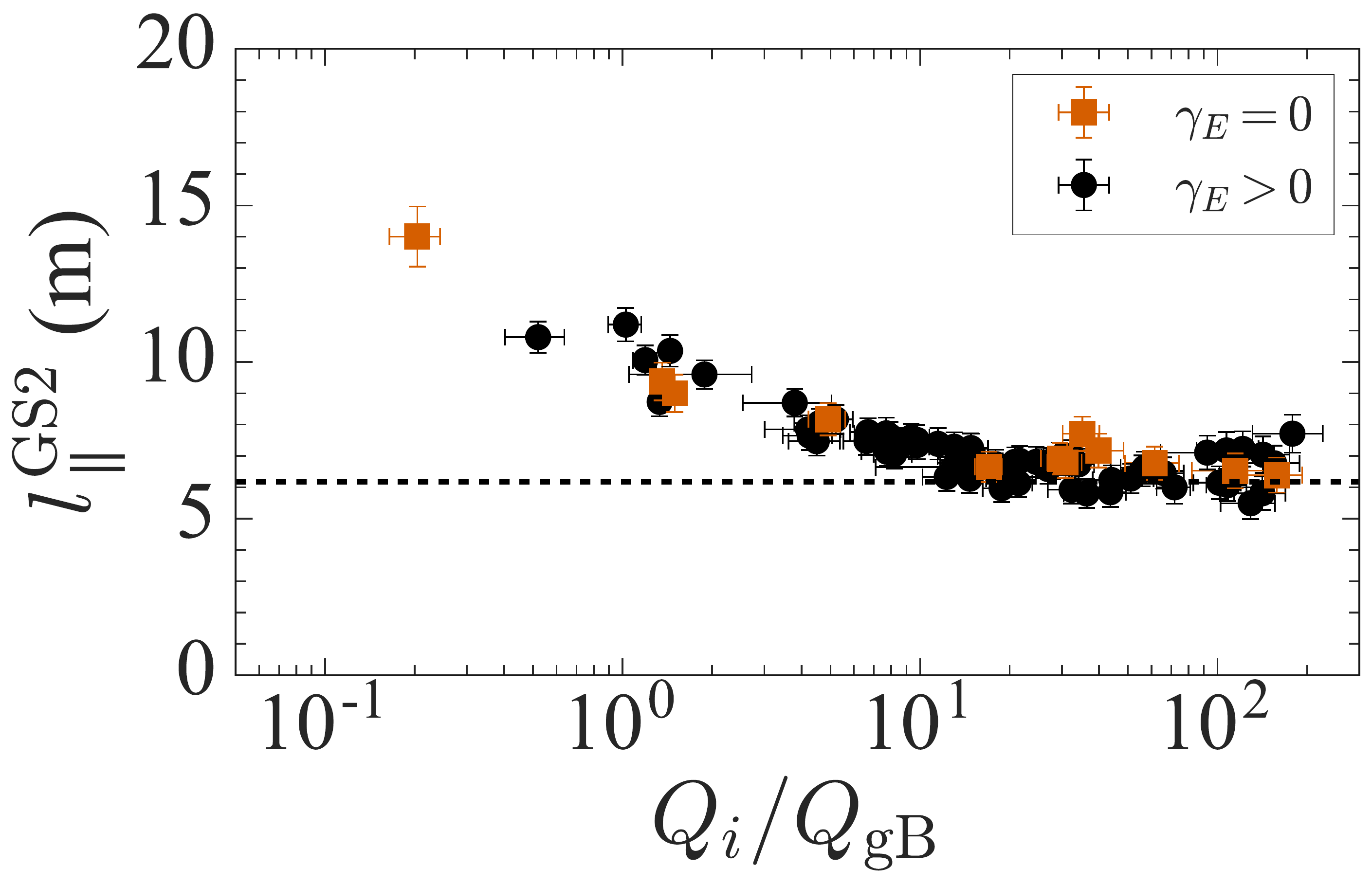}
      \caption{}
      \label{fig:lpar_q}
    \end{subfigure}
    \begin{subfigure}[t]{0.49\textwidth}
      \includegraphics[width=\linewidth]{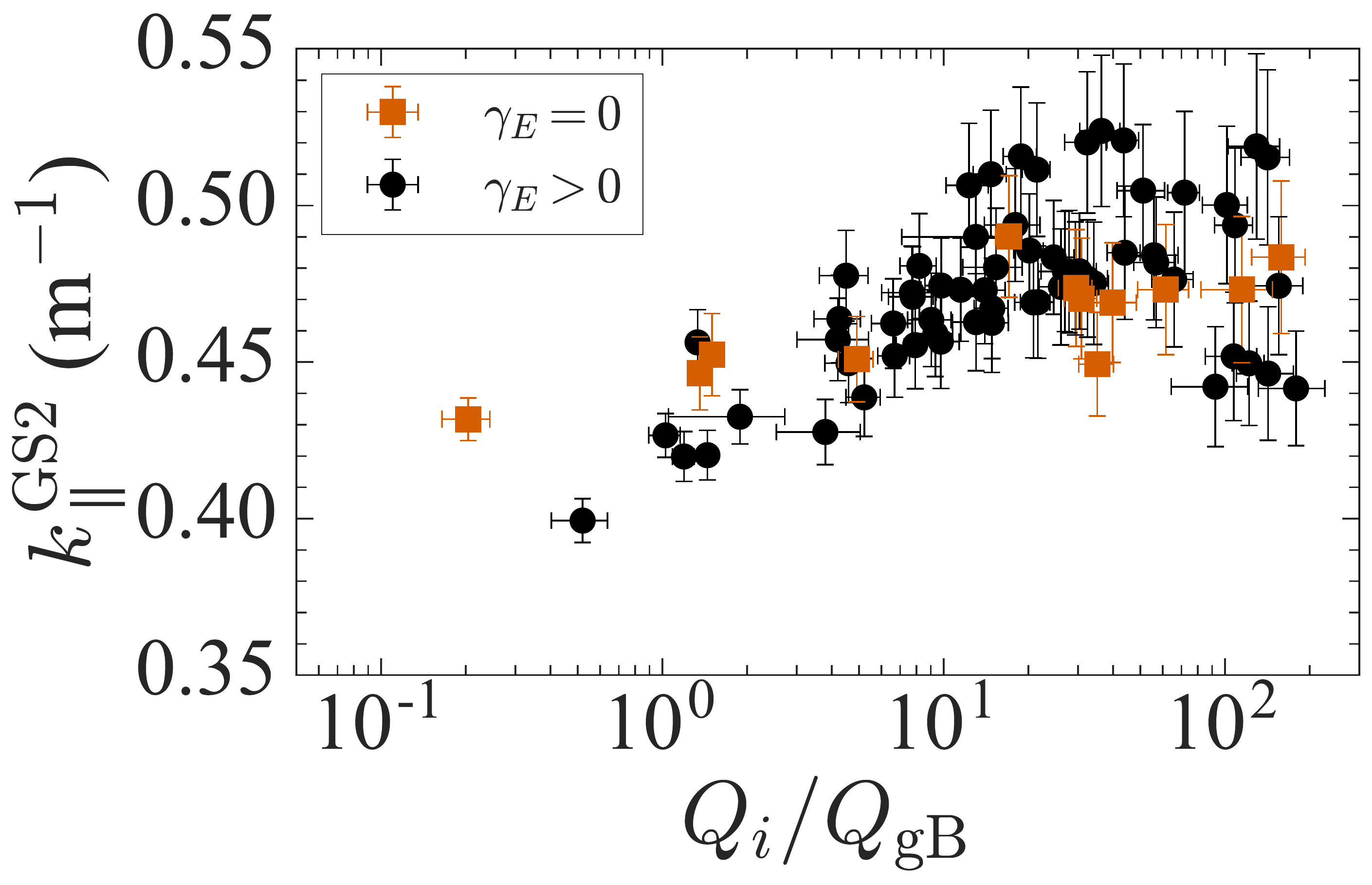}
      \caption{}
      \label{fig:kpar_q}
    \end{subfigure}
    \caption{
      Correlation parameters as functions of $Q_i/Q_{\mathrm{gB}}$, calculated
      for raw GS2 density fluctuations for the entire parameter scan:
      \subref*{fig:lz_gs2_free} poloidal correlation length
      $l_{Z,\mathrm{free}}^{\mathrm{GS2}}$ with $k_y$ as a free fitting
      parameter and
      \subref*{fig:kz_gs2} poloidal wavenumber
      $k_{Z,\mathrm{free}}^{\mathrm{GS2}}$;
      \subref*{fig:lpar_gs2} parallel correlation length
      $l_{\parallel}^{\mathrm{GS2}}$ and
      \subref*{fig:kpar_gs2} parallel wavenumber
      $k_{\parallel}^{\mathrm{GS2}}$. The dashed
      line in \subref*{fig:lpar_gs2} indicates a line of $l_\parallel \sim \pi
      qR$ (see main text and~\cite{Barnes2011}).
    }
    \label{fig:gs2_q_scatter2}
  \end{figure}

  Figures~\ref{fig:gs2_q_scatter1} and~\ref{fig:gs2_q_scatter2} show the
  correlation parameters from figures~\ref{fig:gs2_corr_results1}
  and~\ref{fig:gs2_corr_results2} as functions of $Q_i/Q_{\mathrm{gB}}$ for our
  entire parameter scan, including the cases with $\gamma_E = 0$.  These
  figures support the notion that it is distance from threshold that determines
  the structure of turbulence and characterise this structure for a realistic
  MAST configuration and for a large range of $Q_i/Q_{\mathrm{gB}}$. We start
  by discussing the $\gamma_E>0$ cases.

  In \figref{lr_q}, we see a roughly monotonic increase in the radial
  correlation length $l_R^{\mathrm{GS2}}$ with increasing
  $Q_i/Q_{\mathrm{gB}}$, this makes sense because the formation of larger
  radial structures is one way the turbulence can transport heat more
  effectively.

  \Figref{lz_q} [along with figures~\ref{fig:lz_free_q}
  and~\subref{fig:kz_free_q}] shows the poloidal correlation length
  $l_Z^{\mathrm{GS2}}$ decreasing with increasing $Q_i/Q_{\mathrm{gB}}$.
  This is consistent with~\eqref{q_scaling} (note that, according to
  \figref{kz_free_q}, $k_Z$, and therefore $k_y$, increases as $l_Z$
  decreases). Though \figref{lz_q}
  shows that $l_Z^{\mathrm{GS2}}$ decreases to roughly $14$~cm for
  $Q_i/Q_{\mathrm{gB}} \sim O(10)$, it starts \emph{increasing} again for
  $Q_i/Q_{\mathrm{gB}} \sim O(100)$. This is perhaps in line with the
  theoretical and numerical estimates of the scaling of $l_Z$ far from the
  turbulence threshold predicting that $l_Z \sim q \kappa_T$~\cite{Barnes2011}.
  It stands to reason that at such high values of $Q_i/Q_{\mathrm{gB}}$, the
  system is entering a strongly driven regime, but further simulations at
  higher $\kappa_T$ are necessary to confirm whether our simulations adhere to
  the scaling predicted by~\cite{Barnes2011}.

  The RMS density fluctuations $(\delta n_i /
  n_i)^{\mathrm{GS2}}_{\mathrm{rms}}$ shown in \figref{n_q} increase as
  $(Q_i/Q_{\mathrm{gB}})^{1/2}$ far from threshold, as expected from the
  scaling \eqref{q_scaling}. However, in contrast to the results in
  \figref{amplitude}, we do not see a flattening of $( \delta n_i /
  n_i)^{\mathrm{GS2}}_{\mathrm{rms}}$ at low $Q_i/Q_{\mathrm{gB}}$ for
  $\gamma_E>0$ cases. This is due to the relatively little volume taken up by
  the coherent structures and, hence, their small contribution to the RMS
  value. We verified this by calculating the RMS density fluctuations while
  excluding varying numbers of the turbulence structures (near the threshold)
  and found that the RMS value did not change very much, confirming that for
  the cases near the threshold the RMS value is dominated by the low-amplitude
  density fluctuations.

  Finally, in \figref{lpar_q}, we see that the parallel correlation length
  $l_\parallel^{\mathrm{GS2}}$ decreases towards a constant value as the system
  is taken away from the turbulence threshold. Theoretical and numerical
  estimates of $l_\parallel$ for strongly driven ITG
  turbulence~\cite{Barnes2011} indeed predicted that $l_\parallel$ should be
  constant and proportional to the connection length, viz. $l_\parallel \sim
  \pi qR$.  This estimate is indicated by the dashed line in \figref{lpar_q}
  and shows reasonably good agreement with the data.

  We have included both the cases for which $\gamma_E=0$ (red) and those with
  $\gamma_E>0$ (black) in \figsref{gs2_q_scatter1}{gs2_q_scatter2} to highlight
  two important features of unsheared versus sheared turbulence previously
  discussed in section~\ref{sec:struc_analysis}. First, close to the turbulence
  threshold, the cases with $\gamma_E=0$ represent a different regime of
  turbulence to those with $\gamma_E>0$. In particular,
  $l_Z^{\mathrm{GS2}}$ shown in \figref{lz_q} [as well as in
  figures~\ref{fig:lz_free_q} and~\subref{fig:kz_free_q}], shows an
  \emph{increasing} trend for cases with $\gamma_E=0$: from $\sim 10$~cm near
  the turbulence threshold (which is significantly lower than the sheared case
  at the experimentally relevant $Q_i/Q_{\mathrm{gB}}$) to $\sim 15$~cm far
  away from the threshold. In contrast,
  $l_Z^{\mathrm{GS2}}$ in cases with $\gamma_E>0$
  \emph{decreases} from $\sim 23$~cm near marginality to $\sim 15$~cm far away
  from it, before starting to increase in a similar trend to the $\gamma_E=0$
  cases. Furthermore, \figref{tau_q} shows that $\tau_c^{\mathrm{GS2}}$
  predicted by $\gamma_E=0$ simulations stays roughly constant over a large
  range of $Q_i/Q_{\mathrm{gB}}$ whereas for $\gamma_E>0$ simulations,
  $\tau_c^{\mathrm{GS2}}$ diminishes rapidly for small $Q_i/Q_{\mathrm{gB}}$.
  Secondly, we see that far from the threshold, the $\gamma_E=0$ and
  $\gamma_E>0$ cases for \emph{all} correlation parameters show the same
  dependence on $Q_i/Q_{\mathrm{gB}}$. This suggests that far from the
  threshold there is little difference between sheared and unsheared (by a
  background flow) turbulence.

  The above two observations highlight an important finding of this work: close
  to the turbulence threshold, the background flow shear has a significant
  effect on the turbulence leading to reduced heat transport, whereas far from
  the threshold, the turbulence appears to be similar to the conventional
  ITG-driven turbulence in the absence of flow shear.  This result is
  consistent with the results in section~\ref{sec:struc_analysis} and with the
  conclusions of the related work in Ref.~\cite{Fox2016a}, which argued a
  similar case in terms of the reflection (up-down) symmetry of the turbulence
  being broken by the flow shear close to the threshold but effectively
  restored far from it.

\section{Conclusions}
\label{sec:conclusion}

We have simulated the plasma microturbulence in an equilibrium configuration
corresponding to MAST discharge \#27274 using local gyrokinetic simulations and
performed a systematic parameter scan in the ion temperature gradient length
scale $\kappa_T$ and the flow shear $\gamma_E$.  We demonstrated in
section~\ref{sec:heat_flux} that, within experimental uncertainty, our
simulations reproduce the experimental ion heat flux, and that the
experimentally measured equilibrium gradients lie close to the turbulence
threshold inferred from the simulations.  This is one of the first numerical
demonstrations that a MAST plasma is close to the turbulence threshold.  The
parameter scan performed in this work has clearly shown that $\kappa_T$ and
$\gamma_E$ are useful control parameters (and in particular, that experimental
values of $\gamma_E$ do matter for the turbulent state), supporting several
previous experimental and numerical studies~\cite{Dimits1996, Mantica2009,
Ritz1990, Burrell1997}.

We showed in section~\ref{sec:subcritical} that the system is subcritical
for all values of $\gamma_E>0$, i.e., in the linear approximation,
perturbations grow only transiently before decaying, and in the nonlinear
system, finite initial perturbations, which we assume are available in the
experiment, are required in order to achieve a non-zero saturated state.
Subcriticality is a defining feature of this system: for $\gamma_E>0$, even
cases with the largest values of $\kappa_T$ that we considered required large
initial perturbations to ignite turbulence. Using linear and nonlinear
simulations, we have estimated the conditions necessary for the onset of
subcritical turbulence: we require that maximum transient-amplification factor
be $N_{\gamma,\max} \gtrsim 0.4$, and that the transient-growth time $t_0$ be
approximately greater than the nonlinear interaction time, i.e., $t_0 \gtrsim
\tau_{\mathrm{NL}}$ (section~\ref{sec:time_scales}).  These conditions were
comparable to those in previous work for simpler geometric
configurations~\cite{Schekochihin2012, Highcock2012}. Furthermore, we have
showed that the linear dynamics do not change in any quantitatively obvious way
as the turbulence threshold is passed, and so nonlinear simulations are
essential in predicting the onset of subcritical turbulence and mapping out
the turbulence threshold in the parameter space.

Our simulations have shown that, near the turbulence threshold, a previously
unreported turbulent state exists in which fluctuation energy is concentrated
into a few coherent, long-lived structures, which have a finite minimum
amplitude (section~\ref{sec:coherent_strucs}). We have argued that this
phenomenon is due to the subcriticality of the system, which cannot support
arbitrarily small-amplitude perturbations (in contrast to supercritical
turbulence). We have investigated the changes in the nature of these nonlinear
structures by tracking the maximum fluctuation amplitude
(section~\ref{sec:max_amp}) and the number of structures
(section~\ref{sec:struc_count}) as we changed our equilibrium parameters, and
have arrived at the following picture of the transition to turbulence. Near the
turbulence threshold, the system is comprised of just a few finite-amplitude
structures, which are not volume-filling. As the system is taken away from the
turbulence threshold, the number of these structures increases (while their
amplitude stays the same). Upon increasing in number sufficiently to fill the
spatial simulation domain, they begin to increase in amplitude (at a roughly
constant number of structures). They also become much less long-lived in time,
presumably breaking up against each other as they overlap and interact.
Interestingly, this scenario of evolution of our system as it is taken away
from the turbulence threshold is reminiscent of the transition to subcritical
turbulence via localised structures in pipe flows~\cite{Barkley2015}.

We have further shown that, in contrast to conventional ITG-driven turbulence
regulated by zonal flows~\cite{Dimits2000} (and their associated shear), in our
system, close to the turbulence threshold, the shear due to the mean toroidal
flow dominates over the shear due to the zonal flows
(section~\ref{sec:zf_shear}).  We have shown that the
nominal experimental gradients lie close to the threshold, meaning that it is
essential to include the background flow shear in simulations of MAST plasmas.
Only reasonably far from the turbulence threshold do the zonal
shear and the flow shear due to the background flow become comparable, and
further still the turbulence becomes similar to ITG-driven turbulence in the
absence of background flow shear (cf.~\cite{Fox2016a}).

We have made quantitative comparisons between density fluctuations in our
simulations and those measured by the MAST BES diagnostic~\cite{Field2009,
Field2012} (section~\ref{sec:struc_of_turb}). A correlation
analysis~\cite{Ghim2012} was previously performed on the measurements of
density fluctuations from the BES diagnostic~\cite{Field2014}
(section~\ref{sec:corr_exp}), focusing on the following properties of the
turbulence: the radial correlation length $l_R$, the poloidal correlation
length $l_Z$, and the correlation time $\tau_c$. We have performed two types of
correlation analysis on our simulated density fluctuations: one after applying
a synthetic BES diagnostic (section~\ref{sec:corr_synth}), and one directly on
the raw GS2-generated density fluctuations (section~\ref{sec:corr_gs2}). We
have compared these results to experimental measurements and found reasonable
agreement of the correlation lengths and correlation time, except
for the radial correlation length, which was predicted by us to be lower than
the resolution limit of the BES diagnostic. Notably, the simulated and
experimentally measured correlation times were in good agreement, unlike in
previous global, gyrokinetic simulations of the same MAST
discharge~\cite{Field2014}. However, we showed that, for simulations that
match the experimental heat flux, GS2 underpredicts the fluctuation
amplitude [\figref{n_synth}], similar to previous studies comparing gyrokinetic
turbulence to BES measurements~\cite{Holland2009, Holland2011, Shafer2012,
Gorler2014}. While a 10--20\% increase in $\kappa_T$ (i.e., within experimental
uncertainty) leads to turbulence that matches the fluctuation amplitude, it also
leads to significantly higher transport than experimental levels. It is not
clear as yet why GS2 (or other) codes systematically predict lower fluctuation
amplitudes than observed experimentally.

Finally, we have argued that the nature of the turbulence is effectively a
function of the distance from the turbulence threshold. We have quantified
this distance from threshold by the magnitude of the ion heat flux
$Q_i/Q_{\mathrm{gB}}$, and have shown that it is this quantity, rather than the
specific values of the equilibrium parameters $\kappa_T$ and $\gamma_E$, that
determines the properties of the turbulence. Throughout this work, we have
presented our data as functions of the distance from threshold to highlight the
two distinct nonlinear regimes that we have identified: close to the threshold,
where coherent structures dominate the dynamics, and far from the threshold,
where the turbulence appears to be similar to conventional strongly driven ITG
turbulence in the absence of flow shear (e.g.~\cite{Dimits2000, Barnes2011a}).
The experiment appears to be located at the boundary of these two regimes in
parameter space. This suggests that this boundary is the most experimentally
relevant one, as opposed to the boundary separating the laminar and turbulent
states---the so-called ``zero-turbulence manifold''~\cite{Highcock2012}.

Using the local gyrokinetic code GS2, we have been able to reproduce both the
experimental heat flux and the quantitative measurements of turbulence
obtained using the BES diagnostic. This should perhaps give some credence to
the conclusion from the simulations that do not (yet) have direct experimental
backing. More broadly, this should serve to increase one's confidence in the
future use of local gyrokinetic simulations in predicting turbulence and
transport in high-aspect-ratio spherical tokamaks such as MAST.

A key open question that results from this study is of the
experimental existence of the long-lived, coherent structures near the
turbulence threshold. Recent work on this topic provided some tentative but
encouraging indications that such a regime might manifest itself in terms of
experimentally observed skewed probability distributions of density
fluctuations \cite{Fox2016a}. More extensive analysis of MAST BES measurements,
and indeed more BES measurements are needed in future to identify these
structures, if they do indeed exist experimentally. While we have focused on
the ion heat flux in this study, future work may strive to include other
transport channels, such as electron heat flux, momentum, and particle
transport. This will require computationally expensive multiscale gyrokinetic
simulations and will also describe the fine-scale ETG turbulence that is likely
responsible for the majority of the heat transport in this plasma. Resolving
electron scales, as well as scales intermediate to the ion and electron scales,
would also allow comparison with other fluctuation diagnostics, such as the
recently installed Doppler backscattering diagnostic on
MAST~\cite{Hillesheim2015}.

\section*{Acknowledgements}
  We would like to thank M.~Barnes, J.~Ball, G.~Colyer, and M.~Fox for many
  useful discussions. Our simulations were carried out using the HELIOS
  supercomputer system at International Fusion Energy Research Centre, Aomori,
  Japan, under the Broader Approach collaboration between Euratom and Japan,
  implemented by Fusion for Energy and JAEA.\@ Further computational resources
  were in part provided by the Plasma HEC Consortium (EP/L000237/1), the
  Collaborative Computational Project in Plasma Physics funded by UK EPSRC
  (EP/M022463/1), and the RCUK Energy Programme (EP/1501045).  This work has
  been carried out within the framework of the EUROfusion Consortium and has
  received funding from the Euratom research and training programme 2014-2018
  under grant agreement No 633053 and from the RCUK Energy Programme
  (EP/P012450/1). The views and opinions expressed herein do not
  necessarily reflect those of the European Commission.  AAS's work was funded
  in part by grants from UK EPSRC and STFC.

\newpage
\begin{appendices}
\section{Full parameter scan}
  \label{App:parameter_scan}

\begin{figure}[t]
    \centering
    \includegraphics[width=0.6\linewidth]{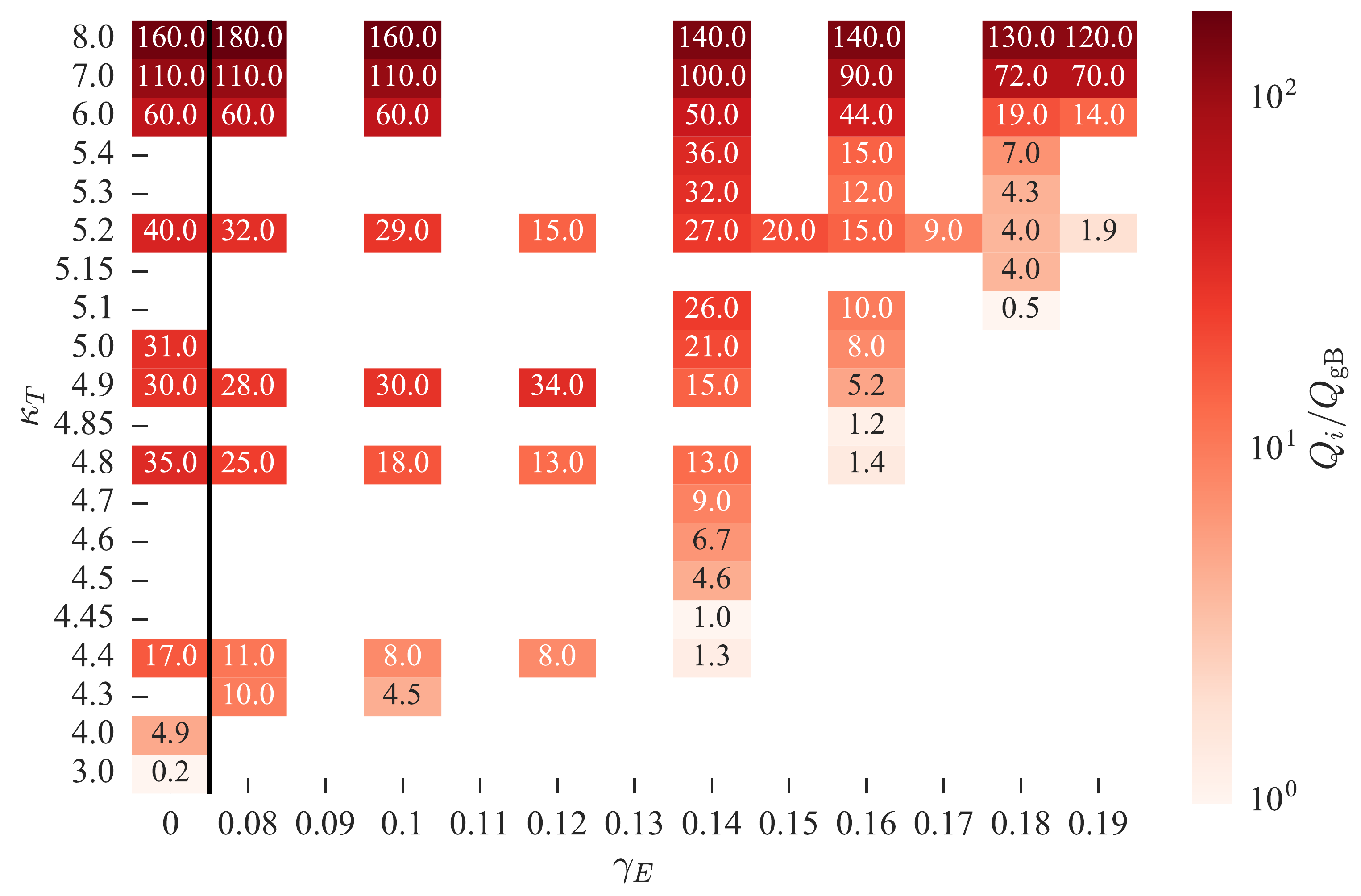}
    \caption{Ion heat flux $Q_i/Q_{\mathrm{gB}}$ as a function $\kappa_T$ and
      $\gamma_E$ explicitly showing the values of $Q_i/Q_{\mathrm{gB}}$.
      \Figref{contour_heatmap} is a smoothed version of this figure.
    }
    \label{fig:parameter_scan}
  \end{figure}

  Our parameter scan in the equilibrium parameters $\kappa_T$ and $\gamma_E$
  covered approximately $\kappa_T \in [3.0, 8.0]$ and $\gamma_E \in [0, 0.19]$,
  and consisted of approximately 76 simulations. \Figref{parameter_scan} shows
  the parameter values of $\kappa_T$ and $\gamma_E$ that we simulated, and the
  associated ion heat flux $Q_i/Q_{\mathrm{gB}}$. To produce
  \figref{contour_heatmap}, we interpolated between the parameter values in
  \figref{parameter_scan}.

\section{Correlation analysis}
  \label{App:corr_overview}
  In this section, we give an overview of the correlation-analysis techniques
  used in Refs.~\cite{Ghim2013,Field2014}, which motivated the experimental
  comparisons presented in this paper. We will also present an alternative
  measurement of the poloidal correlation length $l_Z$, taking advantage of the
  increased resolution available in the poloidal direction from our
  simulations. While there is no experimental estimate of the parallel
  correlation length $l_\parallel$ available from the BES data, we are able to
  use the three-dimensional data available from GS2 to extend the correlation
  analysis to the parallel direction (see section~\ref{sec:pol_par_corr}).

  The two-point spatio-temporal correlation function is defined to be
  \begin{multline}
    C(\Delta R, \Delta Z, \Delta \lambda, \Delta t) = \\
    \frac{\left< \delta n_i/n_i(R, Z, \lambda, t) \delta n_i/n_i(R+\Delta R, Z+\Delta  Z, \lambda+\Delta \lambda, t+\Delta t)\right>}
    {{\qty[\left<{(\delta n_i/n_i)}^2(R, Z, \lambda, t) \right>
    \left<{(\delta n_i/n_i)}^2(R+\Delta R, Z+\Delta  Z, \lambda+\Delta \lambda, t+\Delta t)\right>]}^{1/2}},
    \label{corr_fn}
  \end{multline}
  where $\delta n_i/n_i$ is the density-fluctuation field, which has a mean of
  zero, calculated by GS2 and $\Delta R$, $\Delta Z$, $\Delta \lambda$ are the
  radial, poloidal, and parallel point separations, $\Delta t$ is the time lag,
  and $\left<\ldots\right>$ is an ensemble average, viz. an average over all
  possible pairs of points with the same separation and time lag.  Note that
  the ensemble averages in the plane perpendicular to the magnetic field are
  calculated at $\theta=0$, i.e., they are not averaged over $\theta$. This is
  because in the laboratory coordinate system, correlation properties vary
  strongly with $\theta$ owing to the twisting of the magnetic field lines.
  Here we wish to capture what occurs at the outboard midplane, which is the
  location of the BES diagnostic. Note also that we divide our data in the time
  domain into windows of $\sim 100$--$400$~$\mu$s, and calculate separate
  ensemble averages in each time window. We define the error bars on our
  correlation parameters by calculating variances between those values
  calculated over these time windows.

  Instead of calculating the full correlation function
  \eqref{corr_fn}, we will estimate individual correlation lengths and times
  (which are defined below) by performing one-dimensional correlation
  analyses separately in each direction, i.e., we calculate 1D versions of the
  correlation function \eqref{corr_fn}, with respect to only one of its
  arguments.  All of the representative correlation functions that are plotted
  in the sections that follow will be for the equilibrium parameters
  $(\kappa_T, \gamma_E) = (5.1, 0.16)$ over a real-space domain of
  $20\times20$~cm$^2$ (see \figref{marginal_rz}).

  \subsection{Radial correlation length}
  \label{sec:radial_corr}
\begin{figure}[t]
    \centering
    \includegraphics[width=0.6\linewidth]{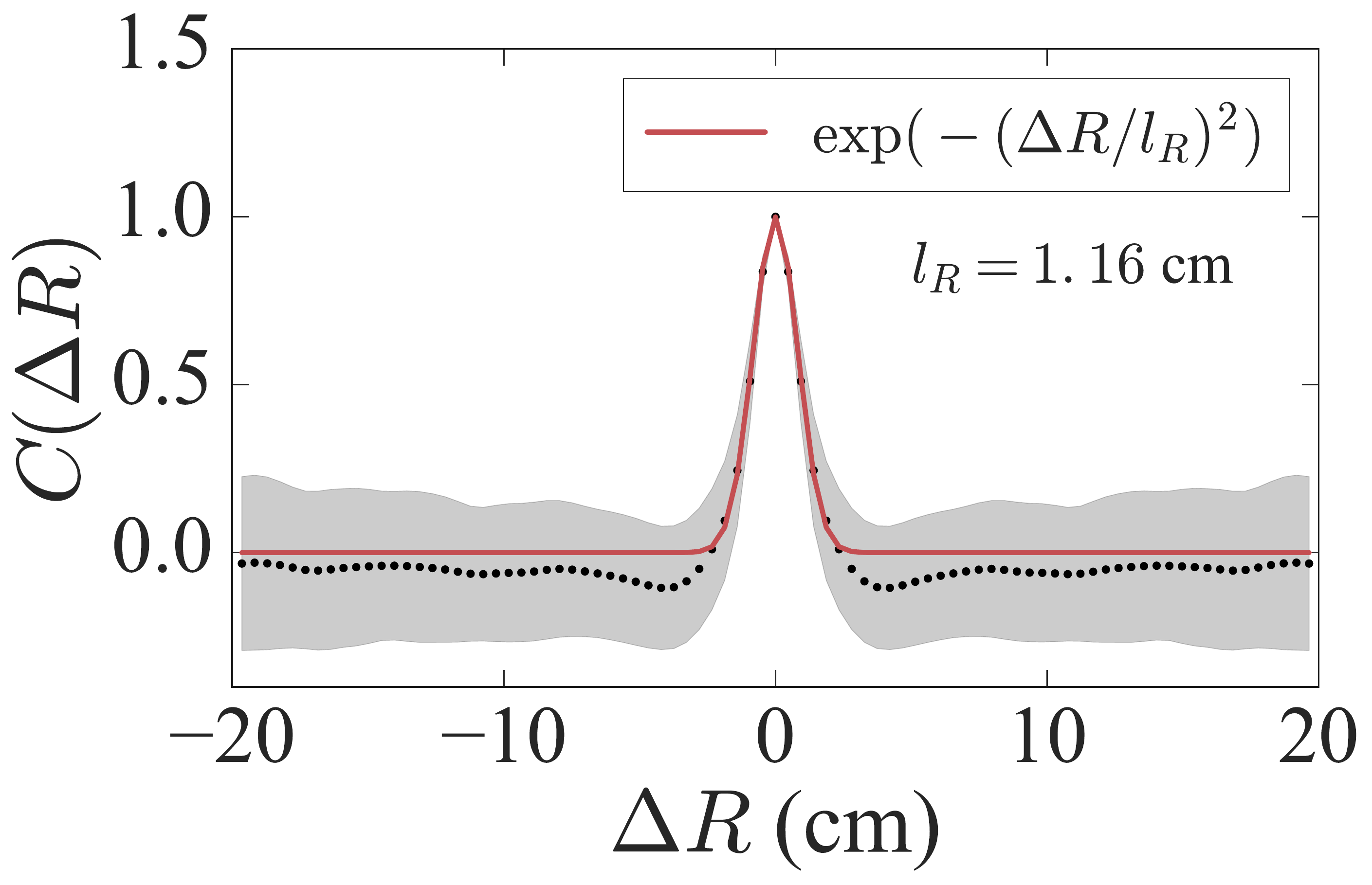}
    \caption{
      A representative radial correlation function fitted with the
      function~\eqref{radial_fit} (red line). The points show the measured
      correlation function $C(\Delta R)$ averaged over $t$ and $Z$ and the
      shaded region shows the associated standard deviation.
    }
    \label{fig:radial_fit}
  \end{figure}

  The radial correlation length $l_R$ is estimated by fitting the correlation
  function $C(\Delta R, \Delta Z = 0, \lambda(\theta=0), \Delta t = 0)$ with a
  Gaussian function:
  \begin{equation}
    f_R(\Delta R) = \exp \qty[- {\qty(\frac{\Delta R}{l_R})}^2].
    \label{radial_fit}
  \end{equation}
  Following~\cite{Ghim2013,Field2014}, this fitting function is adopted on the
  assumption that fluctuations have no wave-like structure in the radial
  direction. Unlike in the treatment of experimental
  data~\cite{Ghim2013,Field2014}, no fitting parameters are necessary here to
  account for global offsets in density fluctuations, which are usually due to
  large-scale, global MHD modes: in our simulations, the mean density
  fluctuation over the whole domain is zero. A representative example of
  \eqref{radial_fit} for the radial correlation function is shown
  in~\figref{radial_fit}.
  The points show the measured correlation function and the red line the fit
  \eqref{radial_fit}. The ensemble average is over $t$ and $Z$ and we assume
  that radial correlations do not change with $t$ and $Z$, i.e., that the
  system is statistically homogeneous in time and in the poloidal direction.
  The shaded region in \figref{radial_fit} indicates the standard deviation
  calculated over the sum of $t$ and $Z$. We expect that $C(\Delta R) \to 0$ as
  $\Delta R \to \infty$ (and similarly for subsequent correlation functions
  in the other directions).

  \subsection{Poloidal correlation length}
  \label{sec:poloidal_corr}
\begin{figure}[t]
    \centering
    \begin{subfigure}{0.49\linewidth}
      \includegraphics[width=\linewidth]{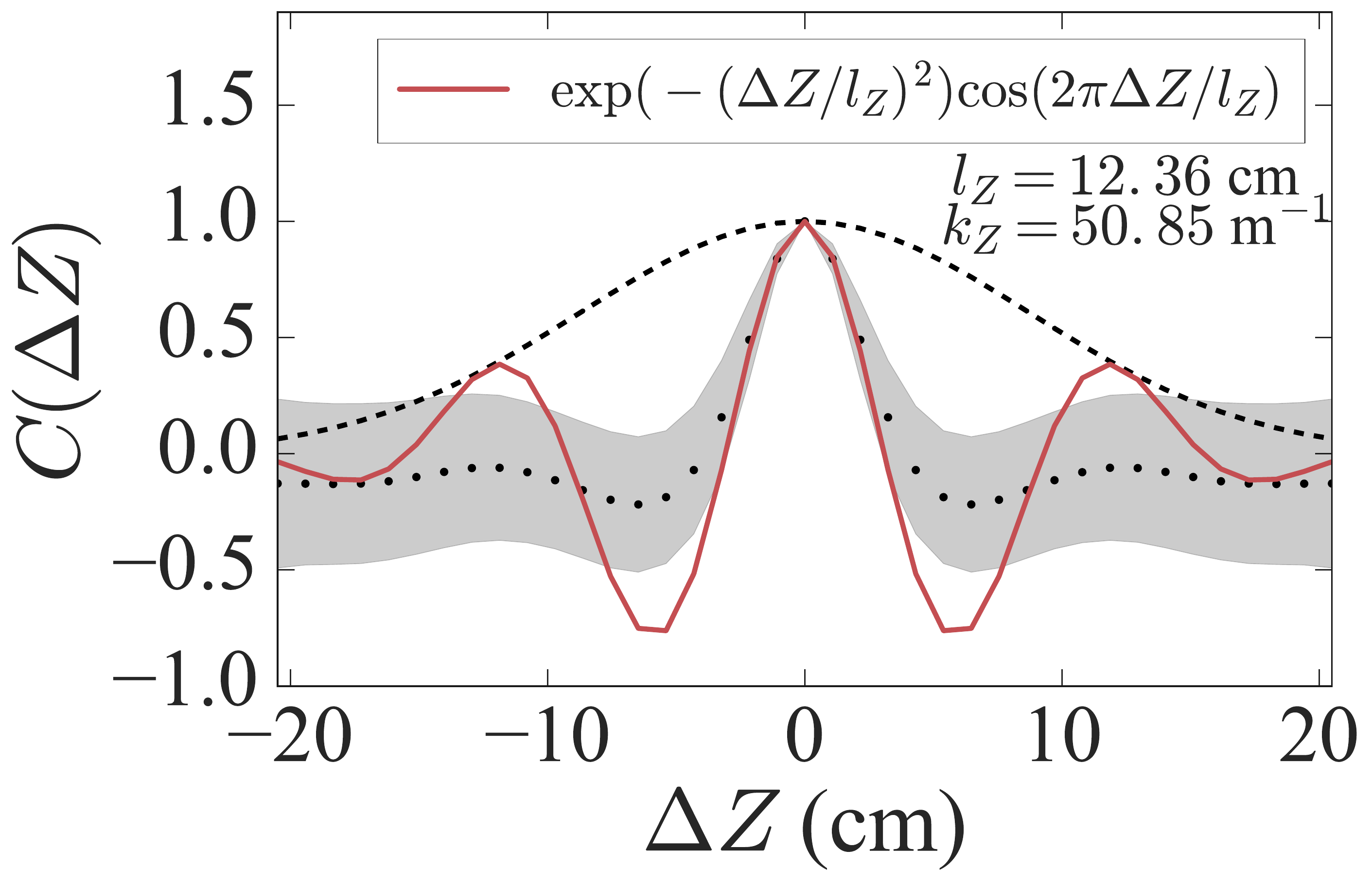}
      \caption{}
      \label{fig:poloidal_fixed_fit}
    \end{subfigure}
    \hfill
    \begin{subfigure}{0.49\linewidth}
      \includegraphics[width=\linewidth]{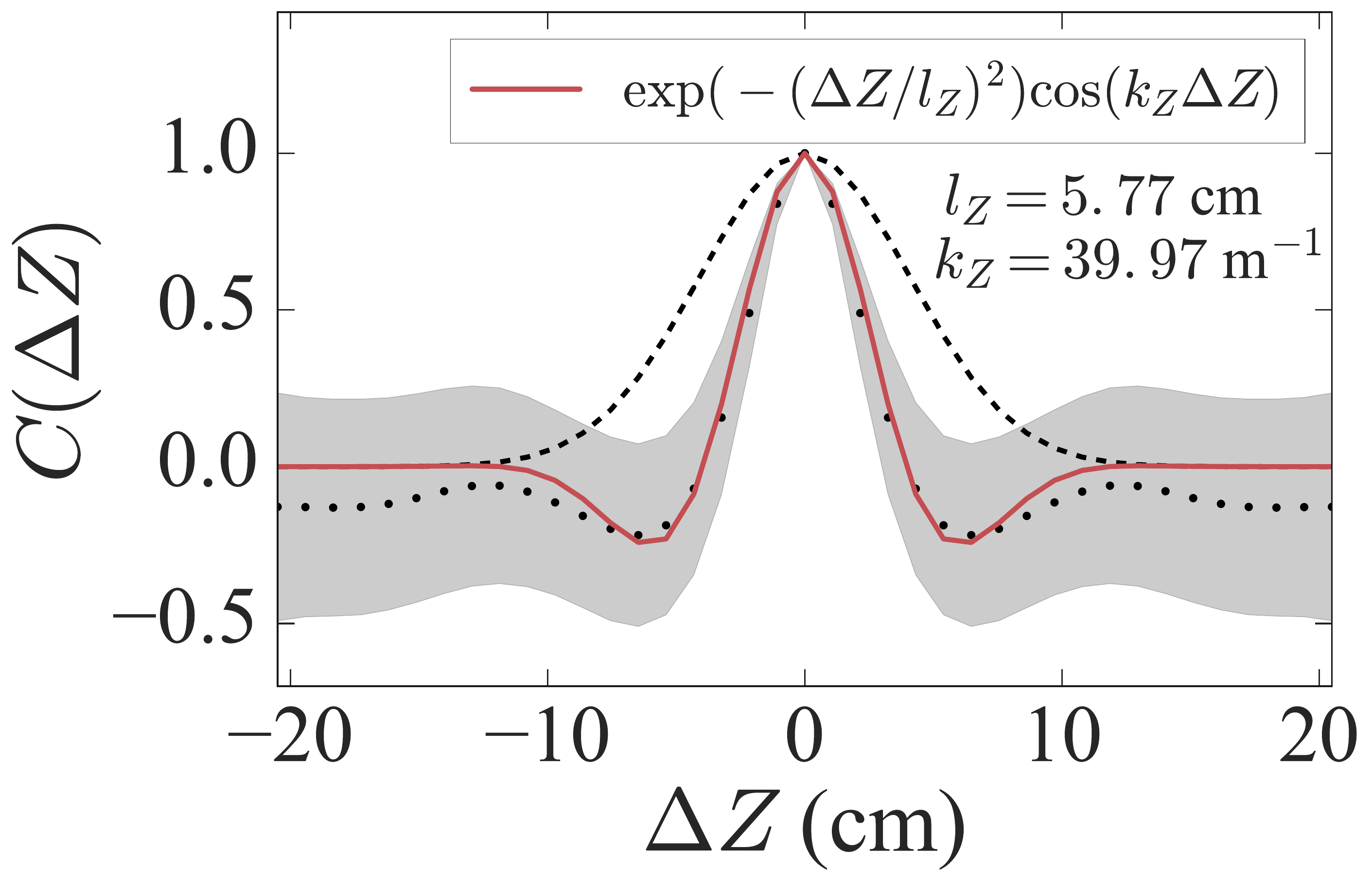}
      \caption{}
      \label{fig:poloidal_free_fit}
    \end{subfigure}
    \caption[Poloidal correlation function]{
      Representative poloidal correlation function fitted with the
      function~\eqref{poloidal_fit} (red line) keeping the poloidal wavenumber
      $k_Z$ \subref*{fig:poloidal_fixed_fit} fixed to $k_Z = 2 \pi / l_Z$,
      \subref*{fig:poloidal_free_fit} as an independent fitting parameter.  The
      points in each plot show the correlation function $C(\Delta Z)$ averaged
      over $t$ and $R$ and the shaded regions show the associated standard
      deviation. The dashed lines indicate the Gaussian envelope
      $\exp[-{(\Delta Z/l_Z)}^2]$.
    }
    \label{fig:poloidal_fit}
  \end{figure}

  The poloidal correlation length is calculated by assuming wave-like
  fluctuations in the poloidal direction and fitting
  $C(\Delta R = 0, \Delta Z, \lambda(\theta=0), \Delta t = 0)$ with an
  oscillating Gaussian function of the form
  \begin{equation}
    f_Z(\Delta Z) = \cos \qty(2 \pi k_Z \Delta Z)
                    \exp \qty[-{\qty(\frac{\Delta Z}{l_Z})}^2],
    \label{poloidal_fit}
  \end{equation}
  where $k_Z$ is the poloidal wavenumber. Refs.~\cite{Ghim2013,Field2014}
  found that with only four poloidal channels, there was insufficient data from
  the BES diagnostic to fit $l_Z$ and $k_Z$ independently. As
  a result, when fitting experimental data (and the data with the synthetic
  diagnostic applied), the wavenumber is fixed to the value $k_Z = 2 \pi /
  l_Z$.  In the direct output of our GS2 simulations, on the other hand, there
  is a sufficient number of data points in the poloidal direction, to allow us
  to compare fits both with $k_Z$ as a free fitting parameter and fixed in the
  way described above.  \Figref{poloidal_fit} shows a representative poloidal
  correlation function from our simulations along with a fitted
  function~\eqref{poloidal_fit}, both with fixed $k_Z = 2 \pi / l_Z$
  [\figref{poloidal_fixed_fit}] and free $k_Z$ [\figref{poloidal_free_fit}].
  The red lines in each plot indicate the fit \eqref{poloidal_fit} and the
  dashed lines indicate the Gaussian envelope $\exp[-{(\Delta Z/l_Z)}^2]$. The
  ensemble average is over $t$ and $R$. We see that the fit with $k_Z$ as a
  free parameter approximates the correlation function better and predicts a
  shorter $l_Z$.  For consistency with previous work, we show the correlation
  results for both fitting schemes in section~\ref{sec:corr_gs2}.

  \subsection{Correlation time}
  \label{sec:time_corr}
\begin{figure}[t]
    \centering
    \includegraphics[width=0.6\linewidth]{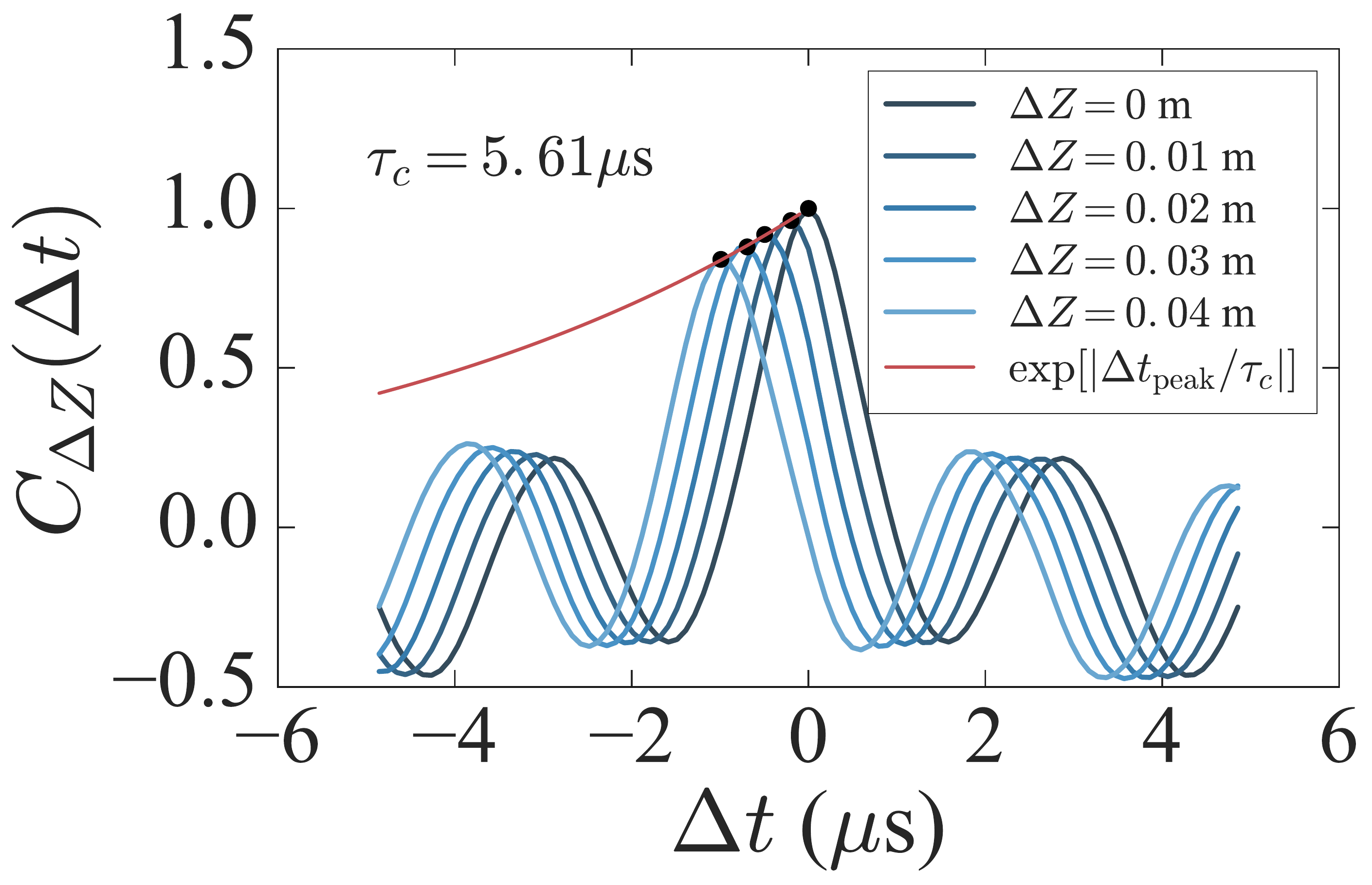}
    \caption[Time correlation function]{
      Time correlation functions $C_{\Delta Z} (\Delta t)$ for several poloidal
      separations $\Delta Z$. The points indicate the maximum value of
      $C(\Delta t)$ for a given $\Delta Z$, and the red line indicates the
      function~\eqref{time_fit} fitted to those points.
    }
    \label{fig:time_fit}
  \end{figure}

  In the presence of toroidal rotation, turbulent structures are advected in
  the poloidal direction with an apparent velocity $v_Z$ given
  by~\cite{Ghim2012}
  \begin{equation}
    v_{Z} = R \omega_0 \tan \vartheta,
    \label{v_z}
  \end{equation}
  where $\vartheta$ is the magnetic-field pitch-angle (the angle that the field
  line makes with the midplane on the outboard side of the flux surface).
  Following~\cite{Ghim2012}, we can use this to calculate the correlation time
  $\tau_c$ by tracking turbulent structures as they move poloidally and
  measuring their temporal decorrelation.
  This ``cross-correlation time delay'' technique~\cite{Durst1992, Ghim2012,
  Fox2016} is as follows. We calculate the correlation function $C_{\Delta
  Z}(\Delta t) = C(\Delta R = 0, \Delta Z, \lambda(\theta=0), \Delta t)$ for
  several poloidal separations $\Delta Z$, as shown in \figref{time_fit}. As
  the structures are advected poloidally, they decorrelate and the peak of the
  correlation function at a given $\Delta Z$, i.e., the value of $C_{\Delta
  Z}(\Delta t)$, decreases for increasing $\Delta Z$.  The correlation time
  $\tau_c$ is then defined as the characteristic exponential decay time of the
  peaks of the correlation functions. Namely, we fit $C_{\Delta Z}(\Delta t =
  \Delta t_{\mathrm{peak}})$ with the function
  \begin{equation}
    f_\tau(\Delta Z) =
      \exp \qty[- \qty|\frac{\Delta t_{\mathrm{peak}}(\Delta Z)}{\tau_c}|],
    \label{time_fit}
  \end{equation}
  as shown for a representative correlation function in \figref{time_fit},
  where correlation functions $C_{\Delta Z}(\Delta t)$ for different poloidal
  separations are shown and the red line shows the fit \eqref{time_fit}.

  This method assumes that the temporal decorrelation dominates over any
  effects due to the finite parallel correlation length, as we will now
  explain. While turbulent structures are extended along the field lines, they
  rotate rapidly in the toroidal direction. After accounting for the apparent
  poloidal motion, a measurement of the correlation time using data from a
  single poloidal plane will conflate two separate effect:
  \begin{inparaenum}[(i)]
    \item true decorrelation of turbulent structures in time, and
    \item structures of finite parallel length moving past the measurement
      point.
  \end{inparaenum}
  With only data from a single poloidal plane, these two effects are
  indistinguishable. In order for the true decorrelation of structures in time
  (which is what we are interested in) to dominate over the movement of
  structures past the detector, it must be the case that~\cite{Ghim2013}
  \begin{equation}
    \tau_c \ll l_\parallel \cos \vartheta / R \omega_0.
    \label{time_assumption}
  \end{equation}
  Since from GS2, unlike from the BES measurements, we can obtain the full 3D
  turbulence data, we are able to confirm in section~\ref{sec:pol_par_corr}
  that this condition is indeed satisfied.

  \subsection{Parallel correlation length}
  \label{sec:par_corr}
\begin{figure}[t]
    \centering
    \includegraphics[width=0.6\linewidth]{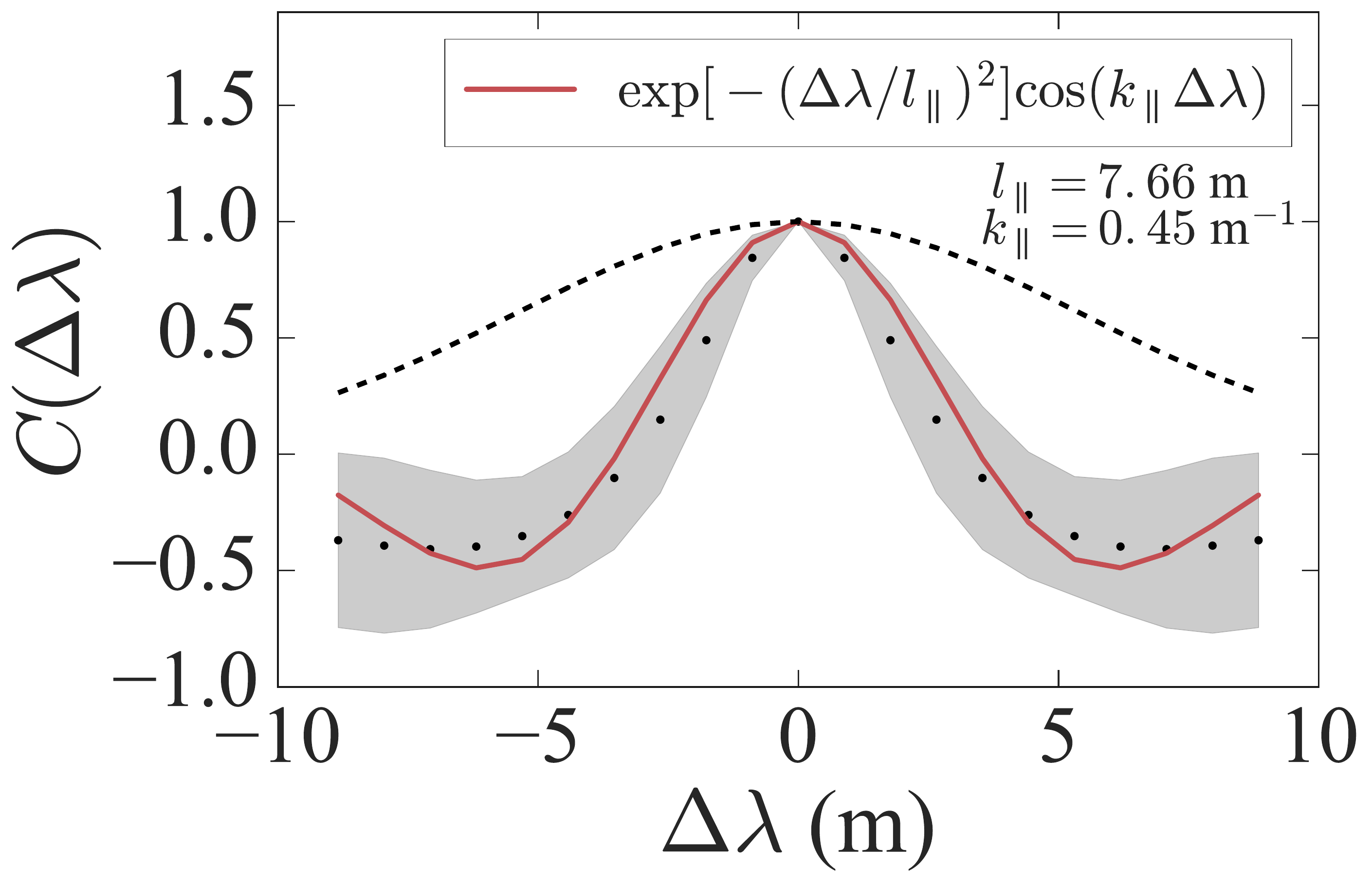}
    \caption[Parallel correlation function]{
      Representative parallel correlation function fitted with the oscillating
      Gaussian function~\eqref{parallel_fit} (red line). The points show the
      correlation function $C(\Delta \lambda)$ averaged over $(t,R,Z)$ and the
      shaded region shows the associated standard deviation. The dashed line
      shows the Gaussian envelope $\exp[-{(\Delta \lambda/k_\parallel)}^2]$.
    }
    \label{fig:parallel_fit}
  \end{figure}

  Since GS2 simulations supply the full 3D density-fluctuation field, we are
  able to study the parallel structure of the
  turbulence. To do this, we convert the fluctuation field from the GS2
  parallel coordinate $\theta$ to a real-space coordinate $\lambda(\theta)$
  along the field line, as discussed in Appendix~\ref{App:parallel_coord}.
  We then calculate the correlation function $C(\Delta R=0, \Delta Z=0, \Delta
  \lambda, \Delta t = 0)$ and take an average over $(R, Z, t)$.
  We fit the correlation function with an oscillating Gaussian function of the
  form
  \begin{equation}
    f_\parallel(\Delta \lambda) = \cos \qty(2 \pi k_\parallel \Delta \lambda)
                  \exp \qty[- {\qty(\frac{\Delta \lambda}{l_\parallel})}^2],
    \label{parallel_fit}
  \end{equation}
  where $k_\parallel$ is the parallel wavenumber. A representative example of
  the fitting procedure for the parallel correlation function is shown
  in~\figref{parallel_fit}, where the red line indicates the fit
  \eqref{parallel_fit} and the dashed line shows the Gaussian envelope
  $\exp[-{(\Delta \lambda/k_\parallel)}2]$.

  \subsection{Density-fluctuation amplitude}
  \label{sec:rms_density}
  The final simulation prediction we can compare with the experimental results
  in~\cite{Field2014}, is the RMS density fluctuation at the outboard midplane
  ($\theta = 0$) averaged over $(t,R,Z)$:
  \begin{equation}
    \qty(\frac{\delta n_i}{n_i})_{\mathrm{rms}} =
    \left< \frac{\delta n_i^{2}(t,R,Z)}{n_i^2} \right>_{t,R,Z}^{1/2}.
    \label{dn_rms}
  \end{equation}
  Formally in gyrokinetics, the quantity $\delta n_i/n_i$ is infinitesimal, and
  throughout this work, we have written $\delta n_i/n_i$ to mean
  $(\rho_i/a) \delta n_i / n_i$, i.e., the physical density-fluctuation field
  predicted by GS2.

\section{Transforming to real space and laboratory frame}
  \label{App:real_space_transform}

  GS2 solves the gyrokinetic equation~\eqref{gk} in curvilinear
  coordinates~\cite{Beer1995a} in a domain known as a ``flux tube''
  (see~\figref{flux_tube}), which rotates with the plasma. In order to analyse
  the real-space structure of turbulence and compare with BES measurements, we
  need to transform our data from the rotating plasma frame to the laboratory
  frame and from flux-tube geometry to real-space geometry, i.e., from the GS2
  coordinates $(x,y,\theta)$ to $(R, Z, \lambda)$ where $R$ is the major
  radius, $Z$ is poloidal height above the midplane of the machine, and
  $\lambda$ is the distance along the field line (not the toroidal direction
  because correlations are meaningfully long-scale in the direction parallel to
  the field; see, e.g.,~\cite{Abel2013}).

  \subsection{Laboratory frame}
  \label{App:lab_frame_transform}

  GS2 simulations are carried out in a frame rotating with the plasma, with
  toroidal rotation frequency $\omega_0$, whereas the BES diagnostic measures
  turbulence in the laboratory frame. In order to make realistic comparisons
  with BES measurements, we applied the following transformation to the
  GS2-calculated distribution function, to transform from the rotating to the
  laboratory frame~\cite{Holland2009}:
  \begin{equation}
    \left(\frac{\delta n_i}{n_i}\right)_{\mathrm{LAB}}(t, k_x, k_y, \theta) =
    \left(\frac{\delta n_i}{n_i}\right)_{\mathrm{GS2}}(t, k_x, k_y, \theta)
                                        e^{- i n \omega_0 t},
    \label{lab_transform}
  \end{equation}
  where $(\delta n_i/n_i)_{\mathrm{LAB}}$ is the the fluctuating density field
  calculated by GS2 in the laboratory frame, $(\delta n_i/n_i)_{\mathrm{GS2}}$
  is the density field in the rotating plasma frame, and
  \begin{equation}
    n = k_y \rho_i \dv{\psi_N}{r} \frac{a}{\rho_i}
    \label{tor_mode_no}
  \end{equation}
  is the toroidal mode number of a given $k_y$ mode, $\psi_N$ is the normalised
  poloidal magnetic flux, $r=D/2a$ is the Miller~\cite{Miller1998} radial
  coordinate, $D$ is the diameter of the flux surface, $a$ is half the
  diameter of the last closed flux surface (LCFS), and $\rho_i$ is the ion
  gyroradius.

  \subsection{Radial-poloidal domain size}
  \label{App:perp_domain}

\begin{figure}[t]
    \centering
    \includegraphics[width=0.6\linewidth]{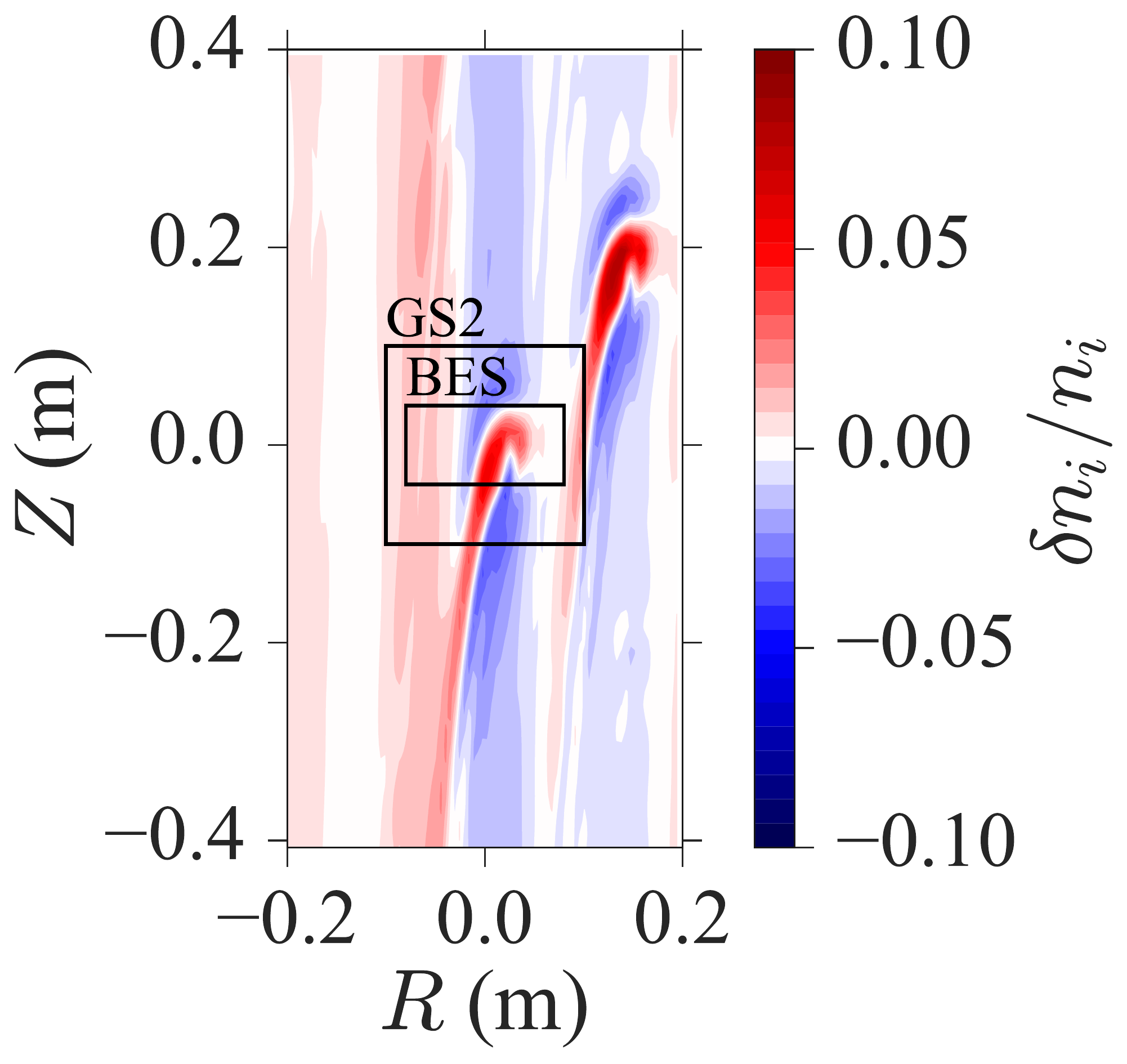}
    \caption{
      Density-fluctuation field $\delta n_i/n_i$ as a function $R$ and $Z$ for
      the same near-marginal case as shown in \figref{marginal} for the
      equilibrium parameters $(\kappa_T,\gamma_E) = (4.8, 0.16)$. The indicated
      domains are those used for the correlation analysis of raw GS2 density
      fluctuations (GS2) and the approximate size of the BES viewing window
      (BES).
	}
    \label{fig:marginal_rz}
  \end{figure}

  The GS2 flux tube is approximately rectangular at the outboard midplane.
  Converting from the local GS2 coordinates $(x,y)$ to the radial-poloidal
  coordinates $(R,Z)$ (the plane of the BES measurement window) is a non-trivial
  procedure and the reader is referred to~\cite{VanWykThesis} for a detailed
  explanation. Here, we only note that the radial size of the domain is $L_R
  \approx 0.4$~m and the poloidal size of the domain is $L_Z \approx 0.81$~m.
  \Figref{marginal_rz} shows the same plot as in \figref{marginal} at $\theta =
  0$ in terms of the real-space poloidal coordinates $R$ and $Z$.  Also
  indicated in \figref{marginal_rz} are the domains used for the correlation
  analysis of BES data and raw GS2 data in sections~\ref{sec:corr_exp}
  and~\ref{sec:corr_gs2}, respectively.

  \subsection{Parallel coordinate}
  \label{App:parallel_coord}

  Finally, we calculate the parallel distance along the magnetic field line at
  the centre of our flux tube. This procedure is non-trivial for a
  general geometry because a uniform grid in $\theta$ does not map to a uniform
  spatial grid along the field line (as it would have done for circular flux
  surfaces). For our D-shaped geometry, we want to find $\lambda(\theta)$, the
  distance along the field line parametrised by the poloidal angle $\theta$.
  The differential arc length of a line element along the field line in terms
  of $(R,Z,\phi)$ is
  \begin{equation}
    d \lambda^2 = dR^2 + dZ^2 + {(R d\phi)}^2,
    \label{line_element}
  \end{equation}
  where $R = R(\theta)$ and $Z = Z(\theta)$ are the coordinates of the magnetic
  field line at the centre of the flux tube, and $\phi$ is the toroidal angle.
  Therefore,
  \begin{equation}
    \lambda(\theta) = \int_0^\theta d \theta' \sqrt{{\qty(\dv{R}{\theta'})}^2 +
                      {\qty(\dv{Z}{\theta'})}^2 +
                      {\qty(R \dv{\phi}{\theta'})}^2}.
    \label{l_theta}
  \end{equation}
  The quantities $R(\theta)$, $Z(\theta)$, $\dv*{\phi}{\theta}$ are obtained
  from the specification of the equilibrium and we then calculate their
  numerical derivatives with respect to $\theta$ and then the
  integral~\eqref{l_theta} to determine $\lambda(\theta)$. With the knowledge
  of the real-space parallel grid, we can calculate correlation lengths in the
  parallel direction.

\section{Real-space effect of flow shear}
  \label{App:flow_shear}

\begin{figure}[t]
    \centering
    \includegraphics[width=0.7\linewidth]{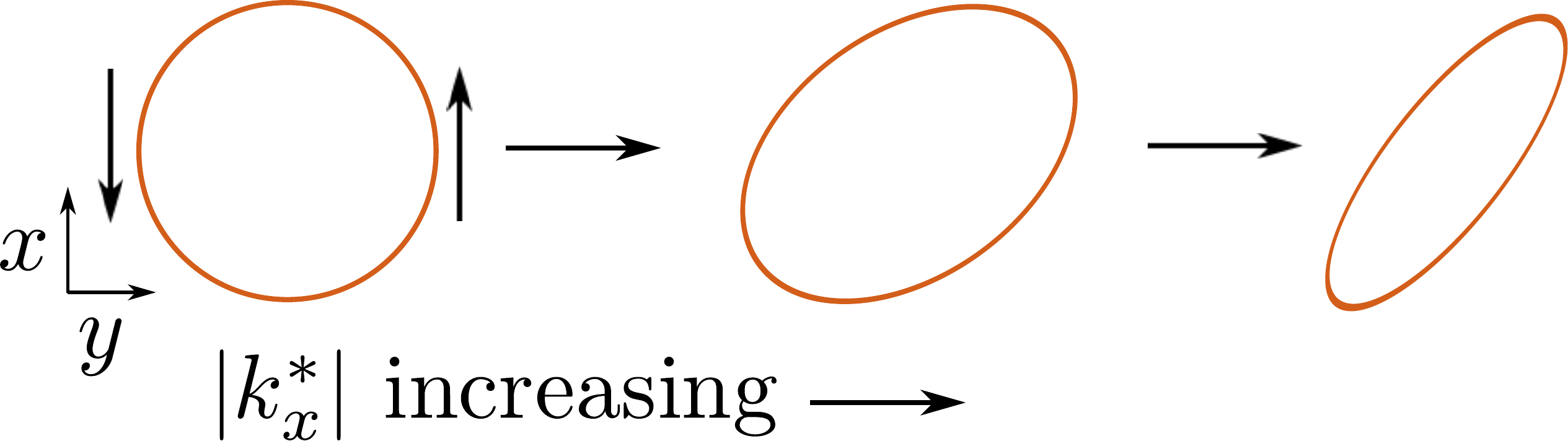}
    \caption[Physical effect of $\gamma_E$]{
      Illustration of the effect of flow shear of turbulent structures.  As
      $k_x^*$ increases in time there is increased radial structure and
      displacement in the $y$ direction.
    }
    \label{fig:flow_shear_effect}
  \end{figure}

  Flow shear is implemented in GS2 by allowing the radial wavenumber of each
  Fourier mode to vary with time~\cite{Hammett2006}:
  \begin{equation}
    k_x^*(t) = k_x - \gamma_E k_y t,
    \label{kx_time}
  \end{equation}
  where $k_x$ would be constant radial Fourier mode in the absence of flow
  shear.  In simplified terms, GS2 shifts the fluctuation fields along the
  $k_x$ dimension as a function of time (see~\cite{HighcockThesis} for a
  complete review of the GS2 flow shear algorithm). This leads to finer radial
  structure and a displacement of fluctuations in the $y$ direction, as
  illustrated in \figref{flow_shear_effect}. However, complications arise in
  this implementation as a result of the fixed $k_x$ grid in GS2, which causes
  jumps in the displacement of fluctuations in the $y$ direction at the radial
  extremes of the box as we will now explain.

  When $k_x^*$ changes by $\delta k_x = \gamma_E k_y \Delta t$, where $\Delta
  t$ is a GS2 time step, the value of the GS2 fluctuation fields at $k_x$ would
  ideally be shifted to $k_x \pm \delta k_x$. However, the $k_x$ grid
  is fixed in GS2 (with a grid separation of $\Delta k_x$) and so the
  fluctuation fields cannot be shifted by less than $\Delta k_x$. This issue is
  resolved in GS2 by keeping track of the difference between the exact shift in
  $k_x$ and the grid spacing $\Delta k_x$: when the exact shift is less than
  $\Delta k_x/2$, no shifting takes place but the value is recorded and added
  to the size of the shift at the next time step. This process is repeated
  until the shift is greater than or equal to $\Delta k_x/2$, at which point
  all fluctuation fields are shifted by $\Delta k_x$. We will now calculate
  the effect of those shifts.

  The distribution function calculated by GS2 is of the form
  \begin{equation}
    h \propto \exp[i (k_x^* x + k_y y)] =
    \exp[i (k_x x + k_y y - \gamma_E k_y x t)],
    \label{gs2_h}
  \end{equation}
  where we have substituted for $k_x^*$ using~\eqref{kx_time}. We can identify
  the wave frequency $\omega_h = \gamma_E k_y x$ to calculate the group
  velocity
  \begin{equation}
    \vb*{v}_g = \pdv{\omega_h}{\vb*{k}} = -\gamma_E x \vu*{y}.
    \label{v_group}
  \end{equation}
  Writing $\vb*{v}_g = \Delta y / \Delta t$, we find the local $x$-dependent
  displacement of fluctuations in the $y$ direction, for an ideal $k_x$ shift
  of $\delta k_x = \gamma_E k_y \Delta t$,
  \begin{equation}
   \Delta y =  - \frac{ \delta k_x x}{k_y}.
   \label{y_disp}
  \end{equation}
  However, $\delta k_x$ is forced to match the fixed $k_x$
  grid with the spacing $\Delta k_x = 2 \pi / L_x$, where $L_x$ is the size of
  the box in the $x$ direction. Using $k_y = 2 \pi / \lambda_y$,
  where $\lambda_y$ is the wavelength of a given $k_y$ mode, we can finally
  write the displacement due to the flow shear as,
  \begin{equation}
   \Delta y = \lambda_y \frac{x}{L_x}.
   \label{y_disp_final}
  \end{equation}
  This means that at the edges of the radial domain, where $x = \pm L_x/2$, the
  displacement in the $y$ direction for every shift in $k_x$ due to the flow
  shear is $\Delta y = \pm \lambda_y/2$. The rate of shifting depends on
  $k_y$ according to~\eqref{kx_time} and so the largest modes (smallest
  $k_y$'s) will be acted on more infrequently than smaller modes
  (larger $k_y$'s). However, the largest modes are then shifted by half the
  size of their wavelength according to~\eqref{y_disp_final}. This causes
  visual separation (or multiplication) of structures at the edges of the GS2
  domain in real space in a way that may affect the correlation analyses
  performed in section~\ref{sec:struc_of_turb}. For this reason, we have only
  analysed an area at the centre of the computational domain, shown in
  \figref{marginal_rz}.

  We emphasise that the separation of turbulent structures that we have
  described above is only present in the real-space representation of the GS2
  distribution function. Given that GS2 performs calculations (apart from the
  calculation of nonlinear interactions) in Fourier space, this does not
  present a problem to the overall calculation. We note that the implementation
  of flow shear in GS2 is correct in the limit of infinitely small $\Delta k_x$
  and so it is sufficient to check convergence with $\Delta k_x$ to be
  confident of our results. We performed convergence checks by varying
  $\Delta k_x$ for a fixed maximum $k_x$ and confirmed that our results were
  approximately the same. Ideally, some form of interpolation could be used to
  smooth out these shifts in $k_x$ and a future program of work is planned to
  implement this in GS2.

\section{Correlation analysis without the spike filter}
  \label{App:no_spike}

\begin{figure}[t]
    \centering
    \begin{subfigure}[t]{0.49\textwidth}
      \includegraphics[width=\linewidth]{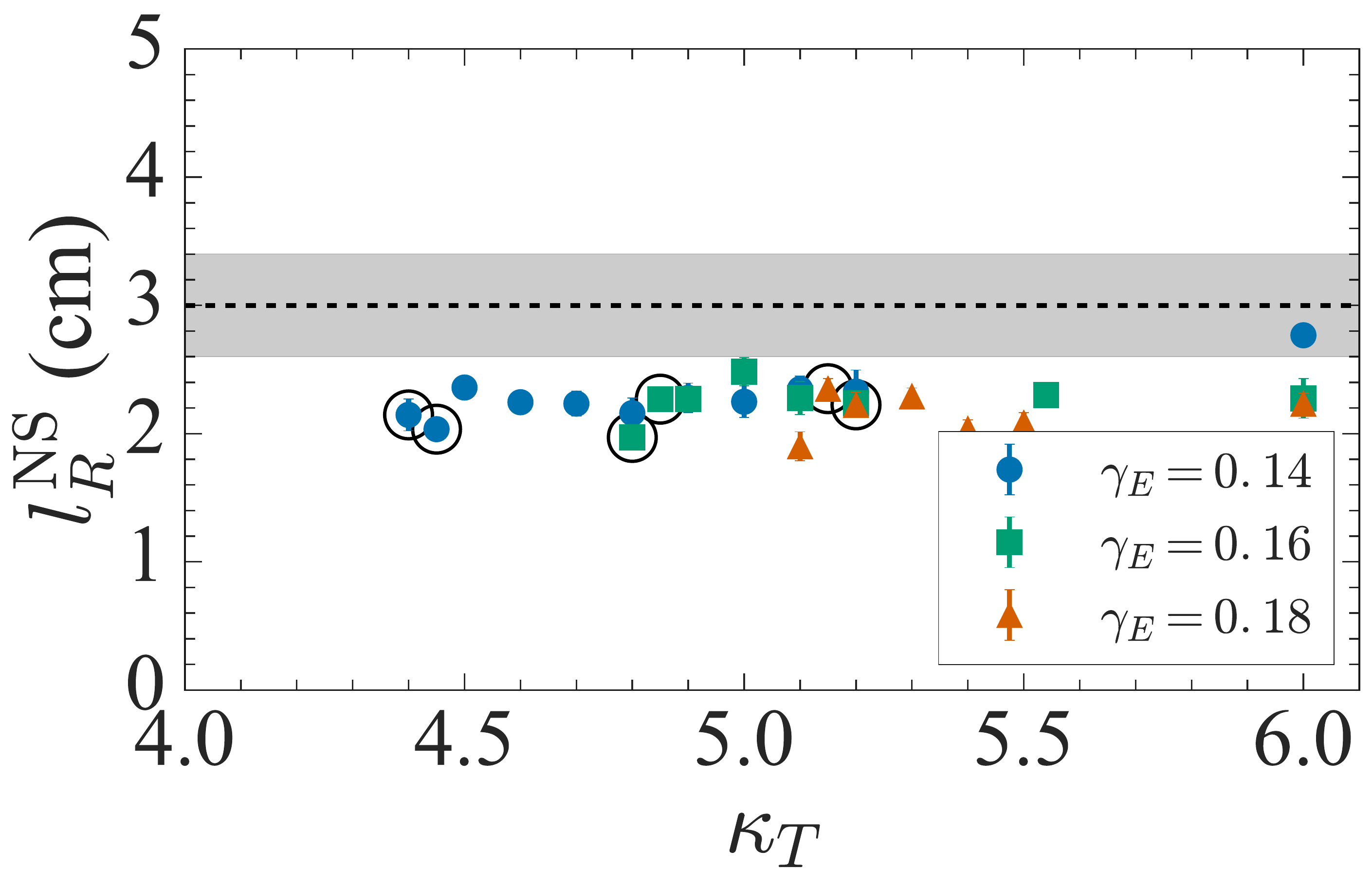}
      \caption{}
      \label{fig:lr_bes_ns}
    \end{subfigure}
    \hfill
    \begin{subfigure}[t]{0.49\textwidth}
      \includegraphics[width=\linewidth]{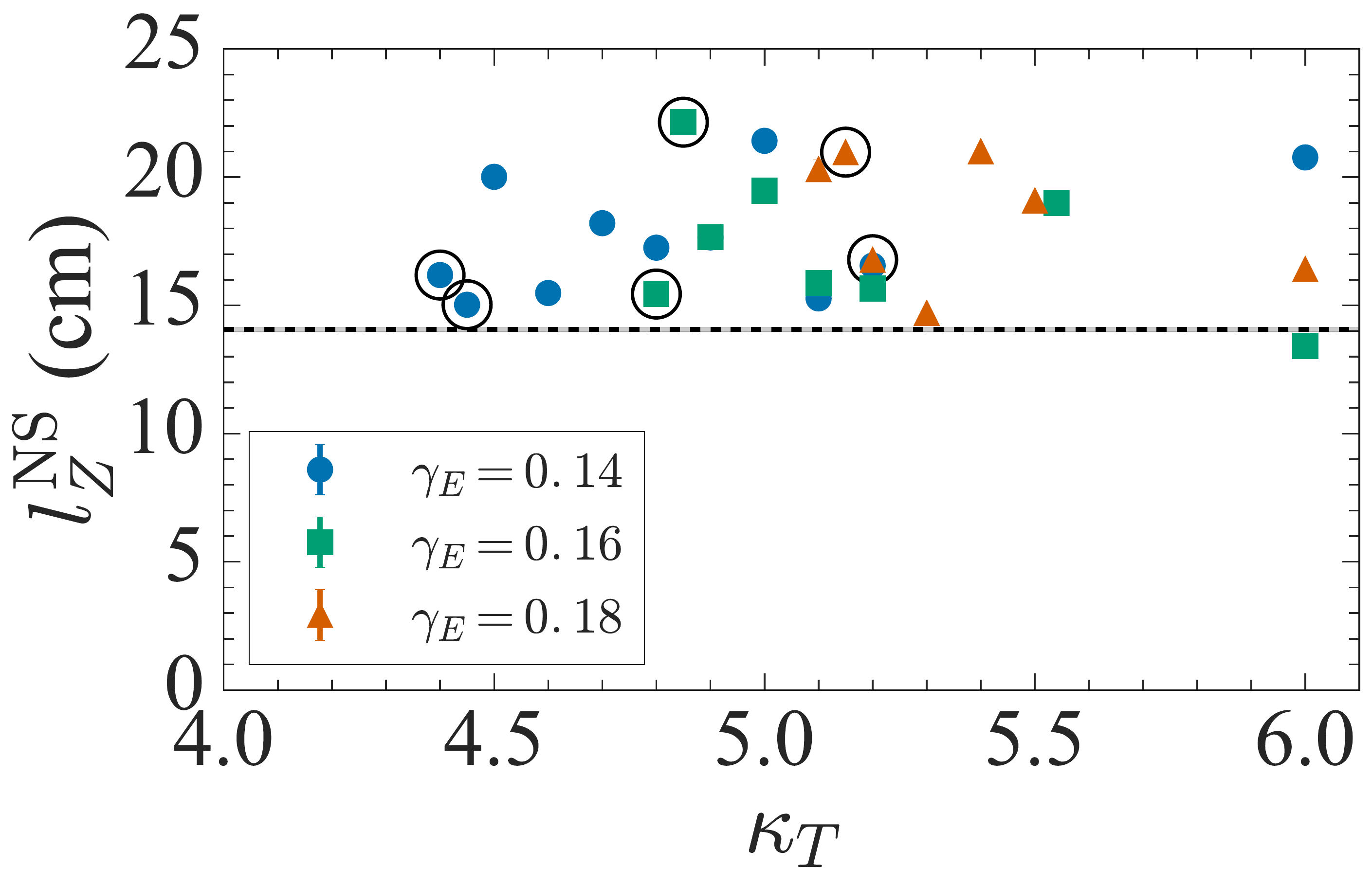}
      \caption{}
      \label{fig:lz_bes_ns}
    \end{subfigure}
    \\
    \begin{subfigure}[t]{0.49\textwidth}
      \includegraphics[width=\linewidth]{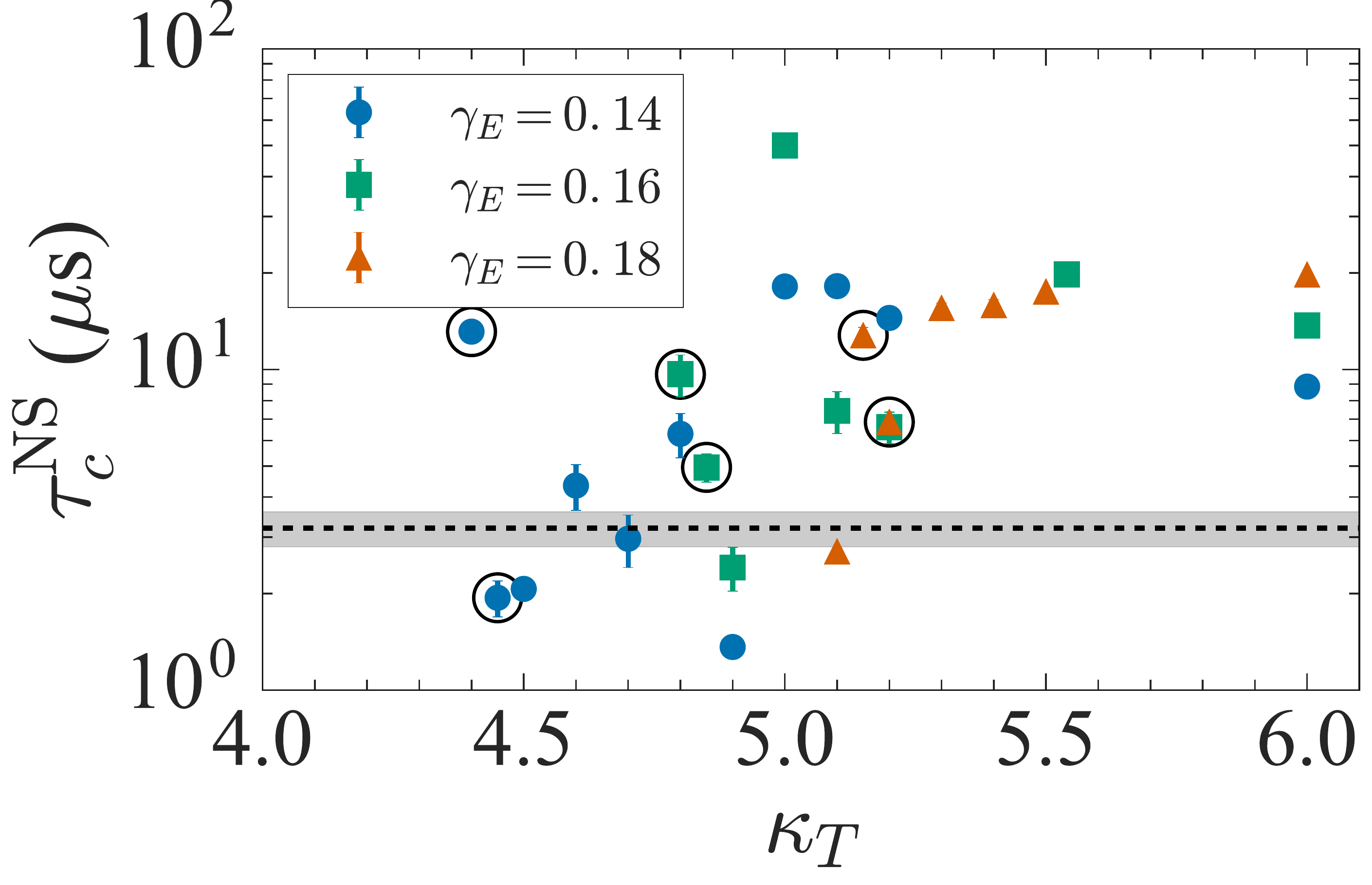}
      \caption{}
      \label{fig:tau_bes_ns}
    \end{subfigure}
    \hfill
    \begin{subfigure}[t]{0.49\textwidth}
      \includegraphics[width=\linewidth]{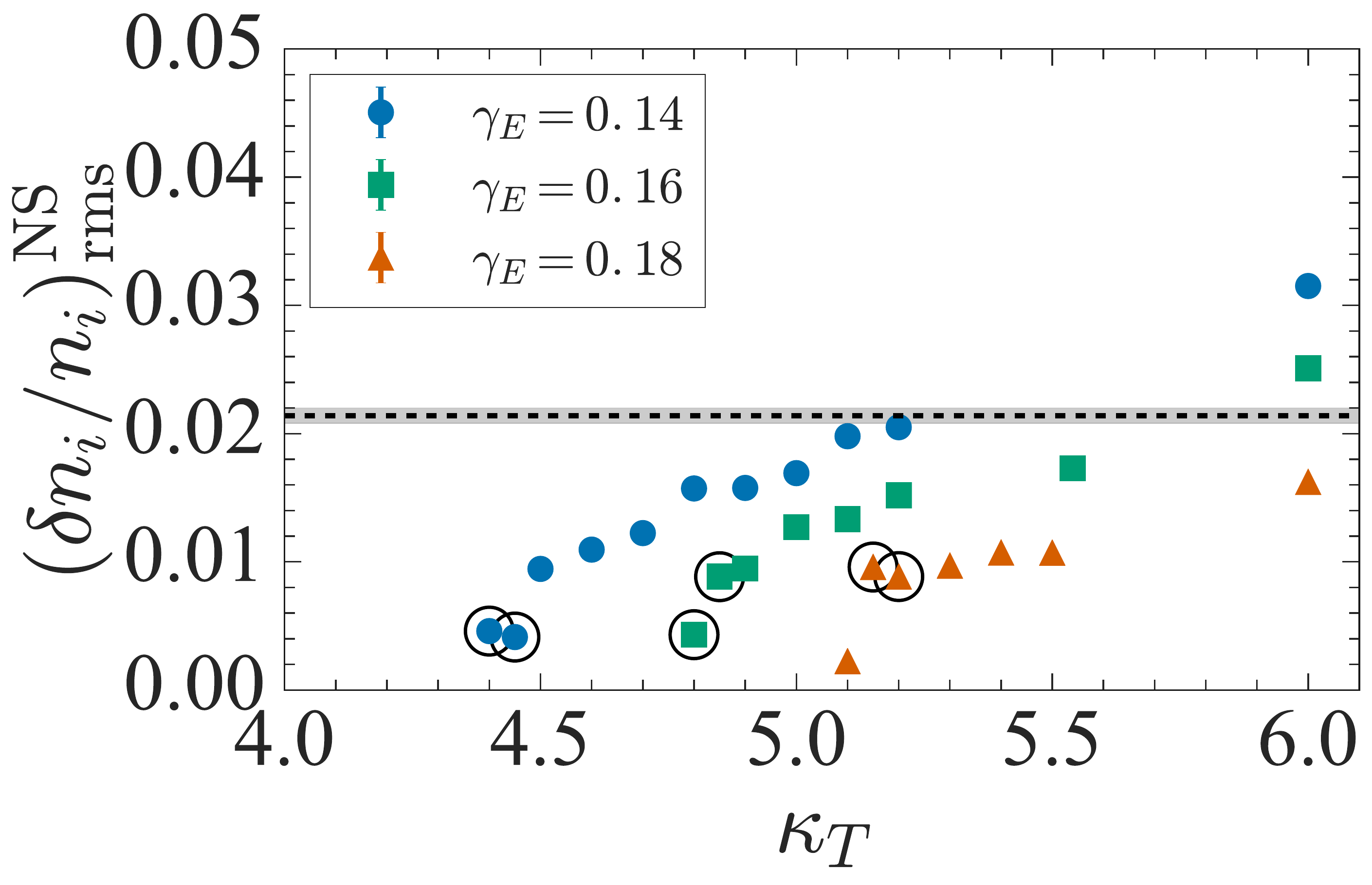}
      \caption{}
      \label{fig:n_bes_ns}
    \end{subfigure}
    \caption{
      Correlation-analysis results calculated from the analysis of GS2
      fluctuation data (within the region of experimental uncertainty) after
      applying the synthetic diagnostic, but without the spike filter normally
      applied to experimental data (see \figref{synth_corr_results}):
      \subref*{fig:lr_bes_ns} radial correlation length $l_R^{\mathrm{NS}}$
      (section~\ref{sec:radial_corr}),
      \subref*{fig:lz_bes_ns} poloidal correlation length $l_Z^{\mathrm{NS}}$
      (section~\ref{sec:poloidal_corr}),
      \subref*{fig:tau_bes_ns} correlation time $\tau_c^{\mathrm{NS}}$
      (section~\ref{sec:time_corr}), and
      \subref*{fig:n_bes_ns} RMS fluctuation amplitude $\qty( \delta n_i /
      n_i)^{\,\mathrm{NS}}_{\mathrm{rms}}$ (section~\ref{sec:rms_density}).
      The simulations that matched the experimental heat flux are circled. The
      quantities plotted here are discussed in appendix~\ref{App:corr_overview}.
    }
    \label{fig:bes_ns}
  \end{figure}

  An important step in the analysis of experimental data involves the removal of
  high-energy radiation (e.g., neutron, gamma ray, or hard X-ray) impinging on
  the BES detector. This radiation manifests itself as
  delta-function-like spikes in time, typically only on a single BES channel.
  These are removed via a numerical ``spike filter''~\cite{Field2012,Fox2016a},
  which was included in the main analysis (where the synthetic diagnostic was
  applied; see section~\ref{sec:corr_synth}) for consistency with experimental
  analysis.  Here, we show the results of a correlation analysis of GS2 density
  fluctuations with the synthetic diagnostic applied, but without this spike
  filter.  \Figref{bes_ns} shows the correlation results for values of
  $\kappa_T$ and $\gamma_E$ within the experimental-uncertainty range: the
  radial correlation length $l^{\mathrm{NS}}_R$ [\figref{lr_bes_ns}], the
  poloidal correlation length $l^{\mathrm{NS}}_Z$ [\figref{lz_bes_ns}], the
  correlation time $\tau^{\mathrm{NS}}_c$ [\figref{tau_bes_ns}], the RMS
  density fluctuation ${(\delta n_i/n_i)}^{\mathrm{NS}}_{\mathrm{rms}}$
  [\figref{n_bes_ns}].

  Comparing these results to the results in section~\ref{sec:corr_synth} with
  the spike filter, we see that it is mainly the poloidal correlation
  length that is affected: $l^{\mathrm{NS}}_Z$ is several centimetres lower
  with the spike filter compared to cases without it. We found that in some
  cases, fast-moving structures in the poloidal direction (especially the
  long-lived structures found in our near-marginal simulations; see
  section~\ref{sec:coherent_strucs}) were removed by
  the spike filter and, therefore, did not affect the poloidal
  correlation function thus copmputed, resulting in a drop in
  $l^{\mathrm{NS}}_Z$. In particular, \figref{lz_bes_ns} shows that
  $l^{\mathrm{NS}}_Z$ increased significantly in near-marginal simulations
  compared to its values obtained with the spike filter. This
  observation of the vulnerability of coherent structures predicted by our
  simulations to the spike filter should inform future attempts to observe
  these structures experimentally.

\end{appendices}

\clearpage
\bibliographystyle{unsrt}
\bibliography{paper}

\end{document}